\documentclass[final,3p,times]{elsarticle}

\usepackage{amssymb}
\usepackage{amsmath}
\usepackage{graphicx}
\usepackage{algorithm2e}
\RestyleAlgo{ruled}
\usepackage{caption}
\usepackage{subcaption}
\usepackage{pdflscape}
\usepackage{booktabs}
\DeclareMathOperator*{\argmin}{\arg\!\min}
\biboptions{sort&compress}

\usepackage[dvipsnames]{xcolor}
\usepackage{soul}
\usepackage{ulem}
\usepackage{dutchcal}
\usepackage{float}
\usepackage{multirow}
\usepackage{fancyhdr}
\usepackage{lineno}
\usepackage[table]{xcolor}
\usepackage{accents}
\usepackage[pscoord]{eso-pic}
\usepackage{enumitem}

\newcolumntype{C}[1]{>{\centering\arraybackslash}m{#1}}

\usepackage[english]{babel}
\addto\captionsenglish{}

\usepackage{xurl}

\begin{document}
\rfoot{\thepage}

\begin{frontmatter}

\title{Efficient model-order reduction via graph autoencoders and operator learning for systems with sharp gradients using a novel point cloud error metric for performance assessment}

\author[1]{Liam K. Magargal}

\author[1]{Parisa Khodabakhshi}
\ead{pak322@lehigh.edu}

\affiliation[1]{organization={Department of Mechanical Engineering and Mechanics, Lehigh University},
            city={Bethlehem},
            state={PA},
            country={United States}}

\begin{abstract}
This study investigates the use of graph autoencoders, a class of graph neural networks (GNNs), to perform model-order reduction (MOR) on high-dimensional systems characterized by sharp gradients for which conventional linear approximations, such as proper orthogonal decomposition (POD) bases, are less effective. Specifically, this study introduces graph neural network latent space dynamics identification (GNN-LaSDI), a MOR framework that uses an operator learning strategy to directly evaluate the temporal evolution of the graph autoencoder's latent representation. In addition to using standard error metrics, this study presents a novel point cloud error metric tailored to directly evaluate the locations of sharp gradients, such as shocks in compressible fluid simulations and solid-liquid interfaces in phase-field simulations. The performance of GNN-LaSDI is compared to that of geometric deep least-squares Petrov-Galerkin (GD-LSPG), which uses a nonlinear manifold least-squares Petrov-Galerkin projection to evaluate the temporal evolution of the graph autoencoder's latent representation, and POD latent space dynamics identification (POD-LaSDI), which combines POD-based dimensionality reduction with operator learning. The effectiveness of GNN-LaSDI and the point cloud error metric are demonstrated using two numerical experiments: Kobayashi's solidification model for two-dimensional (2D) nucleus growth and a Riemann solver for the 2D Euler equations modeling a bow shock generated by flow past a cylinder. For the Euler equations experiment, a constraint-enforcement strategy is incorporated into the graph autoencoder to ensure compliance with physical laws. For the studied problems, GNN-LaSDI substantially reduces the computational cost relative to GD-LSPG by eliminating repeated evaluations of the Jacobian of the decoder, while achieving significantly greater predictive accuracy than POD-LaSDI, thereby providing a balance between predictive accuracy and computational efficiency. Additionally, the proposed point cloud error provides a more informative measure of reduced-order model accuracy in capturing the locations of sharp gradients than conventional error metrics.
\end{abstract}


\end{frontmatter}

\section{Introduction}

Outer-loop applications in computational mechanics, such as optimization, uncertainty quantification, and control, require repeated evaluations of a numerical model. When the underlying numerical model is large-scale, such as those obtained from the semidiscretization of partial differential equations (PDEs) using finite difference, finite volume, or finite element methods, the associated computational cost can become prohibitively expensive. Model-order reduction (MOR) addresses this challenge by constructing efficient low-dimensional approximations of the original high-dimensional system. MOR techniques generally involve two stages. In the offline stage, high-dimensional solution data generated by the full-order model (FOM) are collected and used to construct a low-dimensional representation of the system dynamics. This reduced representation is then employed to derive a reduced-order model (ROM) that approximates the behavior of the original system. In the subsequent online stage, the ROM is deployed to efficiently evaluate new queries in outer-loop applications. 

Traditionally, low-dimensional representations of the high-dimensional state solutions are constructed using linear manifolds, most notably through proper orthogonal decomposition (POD) \cite{sirovich1987turbulence}. While linear manifolds have demonstrated success across a wide range of engineering applications \cite{lieu2006aircraft,wentland2023scalablePROM,mcquarrie2021data}, it is well known that MOR approaches based on linear manifolds often struggle to achieve practical dimensionality reduction for systems characterized by a slow-decaying Kolmogorov $n$-width \cite{welper2017interpolation,peherstorfer2020transport,peherstorfer2020kolmogorov}, including solutions containing sharp gradients, shocks, and moving interfaces. To address this limitation of linear manifolds, several studies have sought to employ nonlinear manifolds, such as quadratic manifolds \cite{barnett2022quadratic,geelen2023operator}, closure models \cite{Barnett2023NNAugmentedPROM,ardesdeparga2026pmor}, nonlinear probabilistic manifold decompositions \cite{guo2026probabilistic}, and autoencoder-based methods \cite{lee2020deeplspg,lee2021deepconservation,magargal2025projection,wiewel2019latent,fresca2021comprehensive,fu2023aerom,fu2025paerom}. Such studies demonstrate that nonlinear manifold approximations of high-dimensional state solutions with a slow-decaying Kolmogorov $n$-width can more significantly reduce the dimension of the system than linear manifold approximations.

In the authors' previous study \cite{magargal2025projection}, a graph-based nonlinear MOR framework was developed to enable dimensionality reduction of systems represented on unstructured meshes. The framework leveraged the natural graph representation of unstructured meshes to construct a graph autoencoder capable of learning low-dimensional nonlinear representations of systems discretized on unstructured meshes, providing greater flexibility than conventional convolutional neural network (CNN) based autoencoders, which are mainly restricted to structured meshes. The graph autoencoder was integrated within a nonlinear manifold least-squares Petrov-Galerkin (LSPG) projection \cite{carlberg2011lspg,carlberg2017galerkinvslspg}, resulting in the geometric deep least-squares Petrov-Galerkin (GD-LSPG) framework for evolving the latent representation in time \cite{magargal2025projection}. Although GD-LSPG achieved substantial dimensionality reduction for systems exhibiting sharp gradients, it did not yield computational savings because the required evaluation of the Jacobian of the decoder at each Newton iteration incurred a computational cost that scaled with the FOM dimension. Hyper-reduction schemes \cite{barrault2004eim,chaturantabut2010deim,carlberg2013gnat,farhat2015ecsw,drmac2016deimerror} may be used to alleviate such computational bottlenecks by restricting the evaluation of the nonlinear terms and the Jacobian of the decoder at selected FOM degrees of freedom; however, their efficient application to autoencoder-based nonlinear manifolds has primarily been demonstrated for shallow decoder architectures \cite{kim2022masked}. In contrast, GD-LSPG employs a deep graph decoder, making conventional hyper-reduction strategies ill-suited for efficiently evaluating the Jacobian of the decoder.

As an alternative to hyper-reduction, operator learning strategies, such as dynamic mode decomposition \cite{schmid2010dmd} and operator inference (OpInf) \cite{peherstorfer2016data}, have been employed to efficiently model the temporal evolution of the low-dimensional representation. OpInf, which is particularly suited for systems with polynomial nonlinearities, leverages the fact that the polynomial structure is preserved under linear projection \cite{mcquarrie2023popinf} to infer low-dimensional operators directly from FOM solutions. For systems that do not naturally admit a polynomial nonlinearity, lifting transformations can be introduced to reformulate the dynamics in a polynomial form amenable to OpInf \cite{gu2011qlmor,kramer2019lifting,qian2020lift,swischuk2020learning,mcquarrie2021data,khodabakhshi2022opinf}. More recently, operator learning strategies have been extended to settings using both linear and nonlinear manifolds. For example, latent space dynamics identification (LaSDI) constructs a ROM using a general library of functions by learning latent state dynamics from the training data recovered from the FOM projected onto the latent space \cite{fries2022lasdi,he2023glasdi}. While the use of linear manifolds with OpInf has been shown to preserve the polynomial structure of the dynamical system \cite{qian2022reduced,mcquarrie2023popinf}, LaSDI extends operator learning to nonlinear manifolds, where the latent dynamics generally do not retain the original form of the FOM dynamics and are approximated using a general library of functions. This study introduces a regularized LaSDI scheme, referred to as graph neural network latent space dynamics identification (GNN-LaSDI), to model the temporal evolution of the graph autoencoder's latent representation without the repeated evaluation of the Jacobian of the decoder (as in the case of GD-LSPG), thereby achieving a computationally efficient autoencoder-based MOR framework that enables significant cost savings.

Conventional error metrics used to assess ROM predictions typically quantify discrepancies in the predicted state solutions \cite{lee2020deeplspg,khodabakhshi2022opinf,geelen2023operator}. However, such metrics may not adequately characterize ROM performance for systems exhibiting sharp gradients \cite{magargal2025projection},  as they do not directly assess errors in the locations of localized features, such as interfaces and shocks. Relatively few studies have developed error metrics specifically tailored to such features \cite{khodabakhshi2022opinf,geelen2022localized,chmiel2025unified}. To address this limitation, this study introduces a point cloud error metric that directly quantifies errors in the predicted locations of sharp gradients, providing a more informative assessment of ROM accuracy in capturing these features than conventional state-based error metrics.

Through two numerical experiments with sharp gradient features, the predictive accuracy and computational efficiency of GNN-LaSDI are evaluated against GD-LSPG and POD latent space dynamics identification (POD-LaSDI). In the first experiment, the studied MOR frameworks are applied to Kobayashi's solidification model for two-dimensional (2D) homogeneous phase evolution to assess their ability to capture the evolution of the solid-liquid interface. In the second numerical experiment, the MOR frameworks are applied to a Riemann solver for the 2D Euler equations modeling the bow shock generated by flow past a cylinder, with emphasis placed on accurately predicting the shock location. For the latter problem, a strategy for ensuring the physical admissibility of the state solutions in the graph autoencoder-based ROM predictions is presented. The primary contributions of this work are summarized as follows:
\begin{enumerate}[leftmargin=*]
    \item An operator learning framework based on LaSDI is developed to approximate the latent space dynamics of graph autoencoders, enabling a computationally efficient autoencoder-based MOR method.
    \item A novel point cloud error metric is presented to directly quantify the accuracy of the predicted sharp gradient locations, such as interfaces and shocks, in ROM solutions.
    \item In the second numerical example, a physics-informed strategy is proposed to enforce physical admissibility (e.g., preservation of positive pressure) in graph autoencoder predictions. 
\end{enumerate}

The remainder of the paper is organized as follows. Section \ref{sec:problemFormulation} introduces the form of the FOM considered in this study and provides background on the graph autoencoder. Section \ref{sec:lasdi} formulates the regularized operator learning procedure used to construct the LaSDI ROMs. Section \ref{sec:evaluation} discusses the error metrics employed in this study, including the novel point cloud error specifically suited to better capture discrepancies near sharp gradients. Section \ref{sec:experiments} presents two numerical experiments used to evaluate both the accuracy of the presented ROM framework and the effectiveness of the point cloud error for sharp gradients. Finally, Section \ref{sec:conclusions} summarizes the main conclusions of this work and outlines directions for future work.

\section{Background and Preliminaries} \label{sec:problemFormulation}
This section presents the foundation for the MOR methods studied in this manuscript. Section \ref{ssec:FOM} presents the semi-discrete form of the considered FOMs. Section \ref{ssec:ae} provides background on the graph autoencoder employed in this study for nonlinear dimensionality reduction, along with a discussion on the authors' prior work to incorporate graph autoencoders into MOR frameworks.

\subsection{Full-order model} \label{ssec:FOM}
Consider a system of $n_q \in \mathbb{N}$ PDEs. Upon spatial discretization of the physical domain using a mesh with $N_c \in \mathbb{N}$ spatial points, the resulting semi-discrete FOM is represented by a system of time-continuous ordinary differential equations of the form:
\begin{equation} \label{eq:ode1}
    \frac{\mathrm{d} \mathbf{x}}{\mathrm{dt}} = \mathbf{f}\left(\mathbf{x}, t; \boldsymbol{\mu}\right), \quad \quad \mathbf{x}\left(0; \boldsymbol{\mu}\right) = \mathbf{x}^0 \left(\boldsymbol{\mu} \right),
\end{equation}
where $\mathbf{x} \in \mathbb{R}^{N}$ denotes the semi-discrete state vector, with $N = n_q N_c$ representing the dimension of the FOM. The state vector $\mathbf{x}$ is constructed as $\mathbf{x} = [ \mathbf{x}_1^\top, \mathbf{x}_2^\top, \ldots, \mathbf{x}_{n_q}^\top ]^\top$ where subvector $\mathbf{x}_i$ corresponds to the semi-discretized solution of the $i^{\mathrm{th}}$ PDE variable, $\boldsymbol{\mu} \in \mathcal{D}$ denotes the parameters, $\mathbf{f}: \mathbb{R}^{N} \times (0, T_f] \times \mathcal{D} \rightarrow \mathbb{R}^{N}$ represents the semi-discrete velocity function, and $\mathbf{x}^0 : \mathcal{D} \rightarrow \mathbb{R}^N$ denotes the initial conditions. Temporal discretization of \eqref{eq:ode1} yields a sequence of discrete-time solutions evaluated over $N_t \in \mathbb{N}$ time steps using an appropriate time-integration scheme. For simplicity, this study assumes a constant time step size, $\Delta t \in \mathbb{R}_+$, such that $T_f = N_t \Delta t$.

\subsection{Nonlinear dimensionality reduction via graph autoencoders} \label{ssec:ae}
A wide range of nonlinear mappings has been proposed in recent years to construct low-dimensional latent spaces for MOR on nonlinear systems. Existing approaches include quadratic manifolds \cite{barnett2022quadratic,geelen2023operator}, closure models \cite{Barnett2023NNAugmentedPROM,ardesdeparga2026pmor}, and autoencoder-based techniques \cite{lee2020deeplspg,wiewel2019latent,fresca2021comprehensive,fu2023aerom}. In this study, graph autoencoders are employed to perform nonlinear dimensionality reduction and extract low-dimensional latent representations from high-dimensional state solution data. A detailed description of the graph autoencoder architecture used in this study is provided in \cite{magargal2025projection}; here, only key concepts relevant to the present framework are briefly reviewed. 

Graph autoencoders are a class of deep learning architecture designed to learn low-dimensional representations of graph-structured data. A graph is defined as the tuple $\mathcal{G} = \{\mathcal{V}, \mathcal{E}\}$, where $\mathcal{V}$ denotes the set of nodes, and $\mathcal{E}$ denotes the set of edges that encode relationships between nodes. A node feature matrix, $\mathbf{X}\in\mathbb{R}^{\vert \mathcal{V} \vert \times N_F}$, can be used to represent information about each node in the graph, where the $i^{\mathrm{th}}$ row of $\mathbf{X}$ corresponds to the feature vector associated with the $i^{\mathrm{th}}$ node in $\mathcal{V}$, $\vert \mathcal{V} \vert$ represents the number of nodes in node set $\mathcal{V}$, and $N_F \in \mathbb{N}$ denotes the number of features assigned to each node. 

\begin{figure}[!htb]
    \centering
    \centerline{\includegraphics[scale=.7]{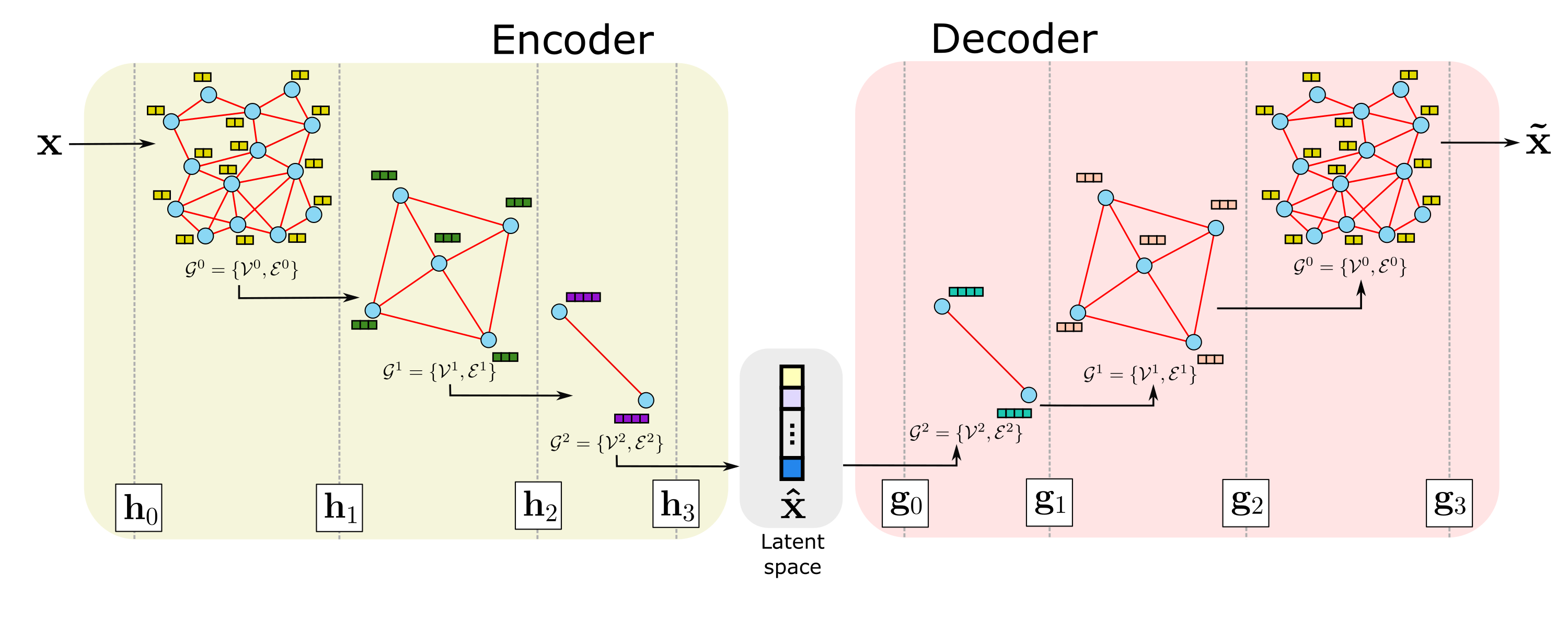}}
    \captionsetup{justification=centering}
    \caption{The graph autoencoder employed in this study, adapted from \cite{magargal2025projection}. The figure corresponds to the case $n_{\ell}=3$, yielding the graph hierarchy $\mathcal{G}^0,\mathcal{G}^1,\mathcal{G}^2$, where $\mathcal{G}^0$ denotes the original graph with a node set that represents the collocation points of the physical domain. The encoder maps the high-dimensional state vector, $\mathbf{x}\in\mathbb{R}^N$, to a low-dimensional latent representation, $\hat{\mathbf{x}} \in \mathbb{R}^n$, where $n \ll N$, through a sequence of nonlinear functions applied across the graph hierarchy. The decoder then reconstructs an approximation of the original state, $\tilde{\mathbf{x}} \in \mathbb{R}^N$, by traversing the hierarchy in reverse order. The vertical dotted lines with overlaid boxes denote each of the layers used in the encoder and decoder. Through the end-to-end training of the network's trainable parameters, the autoencoder learns a latent representation capable of accurately reconstructing the high-dimensional state solution. (Online version in color.)} 
    \label{fig:ae_arch}
\end{figure}

In the context of large-scale dynamical systems arising in engineering applications, graphs provide a natural representation of both structured and unstructured meshes by modeling collocation points as nodes in the graph and prescribing edges to connect neighboring collocation points according to the mesh topology. Representing the mesh as a graph enables the direct application of graph autoencoders for nonlinear dimensionality reduction of high-dimensional state solution data, yielding low-dimensional latent representations. A graph autoencoder consists of two main components: an encoder and a decoder (see Figure \ref{fig:ae_arch}). The encoder, $\mathrm{\mathbf{Enc}} : \mathbf{x} \mapsto \hat{\mathbf{x}}$ with $\mathrm{\mathbf{Enc}} : \mathbb{R}^N \rightarrow \mathbb{R}^n$, where $n \ll N$, compresses the high-dimensional input state $\mathbf{x}\in \mathbb{R}^N$ into a low-dimensional latent representation $\hat{\mathbf{x}}\in \mathbb{R}^n$. On the other hand, the decoder, $\mathrm{\mathbf{Dec}} : \hat{\mathbf{x}} \mapsto \tilde{\mathbf{x}}$ with $\mathrm{\mathbf{Dec}} : \mathbb{R}^n \rightarrow \mathbb{R}^N$, reconstructs an approximation $\tilde{\mathbf{x}}\in \mathbb{R}^N$ of the original high-dimensional state from the latent representation $\hat{\mathbf{x}}\in \mathbb{R}^n$. More specifically, the encoder and decoder each consist of a sequence of layers. In this study, the encoder and decoder architectures are assumed to comprise the same number of layers, denoted by $n_{\ell}\in\mathbb{N}$, 
\begin{align}
    &\mathbf{Enc}: (\mathbf{x}; \theta) \mapsto \mathbf{h}_{n_{\ell}}(\hspace{1mm} \boldsymbol{\cdot} \hspace{1mm} ; \boldsymbol{\Theta}_{n_{\ell}}) \circ \mathbf{h}_{n_{\ell}-1}(\hspace{1mm} \boldsymbol{\cdot} \hspace{1mm} ; \boldsymbol{\Theta}_{n_{\ell}-1}) \circ \ldots \circ \mathbf{h}_{0}(\mathbf{x} ; \boldsymbol{\Theta}_{0}), \\
    &\mathbf{Dec}: (\mathbf{\hat{x}}; \omega) \mapsto \mathbf{g}_{n_{\ell}}(\hspace{1mm} \boldsymbol{\cdot} \hspace{1mm}; \boldsymbol{{\Omega}}_{n_{\ell}}) \circ \mathbf{g}_{n_{\ell}-1}(\hspace{1mm} \boldsymbol{\cdot} \hspace{1mm}; \boldsymbol{{\Omega}}_{n_{\ell}-1}) \circ \ldots \circ \mathbf{g}_{0}(\mathbf{\hat{x}}; \boldsymbol{{\Omega}}_{0}),
\end{align}
where $\mathbf{h}_i(\hspace{1mm} \boldsymbol{\cdot} \hspace{1mm} ; \boldsymbol{\Theta}_i)$ and $\mathbf{g}_i(\hspace{1mm} \boldsymbol{\cdot} \hspace{1mm} ; \boldsymbol{\Omega}_i), i=0,\ldots,n_{\ell}$, denote the function(s) that act on the input of the $i^{\mathrm{th}}$ layer of the encoder and decoder, respectively. The trainable weights and biases for each layer in the encoder and decoder are denoted by $\boldsymbol{\Theta}_i$ and $\boldsymbol{\Omega}_i$ for $i=0,\ldots,n_{\ell}$, respectively. In particular, $\mathbf{h}_0$ and $\mathbf{g}_{n_{\ell}}$ correspond to preprocessing and postprocessing functions applied to the encoder input and decoder output, respectively, and contain no trainable parameters.

The graph autoencoder achieves compression using a hierarchy of $n_{\ell}$ graphs, $\mathcal{G}^0, \ldots, \mathcal{G}^{n_{\ell}-1}$, where $\mathcal{G}^0$ denotes the original graph with a node set that represents the collocation points of the physical domain. Each successive graph in the hierarchy contains fewer nodes than its predecessors, i.e., $\vert \mathcal{V}^{i+1} \vert < \vert \mathcal{V}^{i} \vert$, where $\vert \mathcal{V}^{i} \vert$ denotes the number of nodes in $\mathcal{G}^i$. This hierarchical reduction in graph size enables progressive compression of the state representation. The functions in the encoder use this hierarchy to reduce the dimension of the state solution to its low-dimensional latent representation, while the functions in the decoder use the hierarchy of graphs in reverse order to reconstruct the high-dimensional state solution from its low-dimensional representation. In other words, the $i^{\mathrm{th}}$ graph in the encoder serves as the $(n_\ell-i-1)^{\mathrm{th}}$ graph in the decoder. Figure \ref{fig:ae_arch} demonstrates the use of the hierarchy of graphs within the graph autoencoder architecture. Several methods can be used to construct the hierarchy of graphs \cite{karypis1998kway,barwey2023multiscale,tran2026scalable,magargal2026topology}. This study uses a hierarchical spectral clustering algorithm, also used in \cite{magargal2025projection}. At each level of the hierarchy, the graph autoencoder performs a message passing (MP) operation to perform filtering (i.e., extracting useful information), where the MP operation contains trainable weights and biases. Numerous choices for MP operations exist in the literature, but in this study, the mean-aggregation MP layer found in \cite{hamilton2018graphsage} is deployed. The final layer of the encoder and the first layer of the decoder consist of a fully-connected (FC) layer to encode the node feature matrix associated with the graph $\mathcal{G}^{n_{\ell}-1}$ to the latent state vector, $\hat{\mathbf{x}}$, and vice versa. Further details regarding each layer in the graph autoencoder are provided in \ref{sec:gdlspg}. The trainable parameters of the graph autoencoder (from the MP operations and FC layers) are optimized by minimizing the reconstruction error over the training dataset. Accordingly, the loss function is defined as
\begin{equation} \label{eq:loss}
    \mathcal{L} =  \sum_{i=1}^{N_{\mathrm{train}}} \Big\vert \Big\vert \mathbf{x}^i - \mathbf{Dec}\left( \cdot \right) \circ \mathbf{Enc}\left( \mathbf{x}^i \right) \Big\vert \Big\vert ^2 _2,
\end{equation}
where $N_{\mathrm{train}}$ is the number of training snapshots, and $\vert \vert \cdot \vert \vert_2$ represents the $L_2$-norm. 

To evaluate the temporal evolution of the graph autoencoder's latent representation, GD-LSPG \cite{magargal2025projection} was previously developed using a nonlinear manifold LSPG projection to project the high-dimensional system onto the low-dimensional nonlinear manifold defined by the graph autoencoder. A brief overview of GD-LSPG is provided in \ref{sec:gdlspg}. Here, we focus on the key computational limitation of GD-LSPG that motivates the development of the present work. Although GD-LSPG was shown to achieve significant dimensionality reduction, the repeated evaluation of the Jacobian of the decoder impeded cost savings. In particular, the operation count associated with evaluating the Jacobian of the decoder scaled with the dimension of the FOM. In conventional projection-based ROMs employing linear trial bases, this is typically addressed through hyper-reduction techniques, which sparsely sample terms in the high-dimensional system. Hyper-reduction has also been previously applied to autoencoder-based ROMs with shallow decoder architectures \cite{kim2022masked}. By masking output layers in a shallow decoder, terms in the high-dimensional system can be sparsely sampled to enable hyper-reduction. Although this strategy is effective for shallow decoder architectures, the graph autoencoder presented in this study uses a deep decoder architecture, rendering efficient hyper-reduction more challenging. As discussed in \cite{farhat2015ecsw}, in the absence of hyper-reduction, ROMs employing LSPG schemes achieve computational cost savings primarily in two scenarios. In the first scenario, savings may be achieved if the cost of the FOM is dominated by the cost of inverting the Jacobian of the time-discrete high-dimensional residual, while the cost to set up and solve the low-dimensional system at each Newton iteration is considerably lower. In the second scenario, savings may be obtained when the LSPG ROM remains stable and accurate using larger time step sizes than the FOM, thereby reducing the total number of time steps required for simulation. Therefore, if neither condition is satisfied, GD-LSPG may not achieve cost savings. Motivated by this limitation, the remainder of this manuscript focuses on alternative strategies for approximating the temporal evolution of the low-dimensional latent representation of the graph autoencoder.

\section{Latent space dynamics identification for graph autoencoders} \label{sec:lasdi}
In this section, the operator learning framework used to evaluate the temporal evolution of the graph autoencoder's latent representation is presented. Specifically, this work builds upon LaSDI, originally introduced in \cite{fries2022lasdi,he2023glasdi}. Similar to OpInf \cite{peherstorfer2016data,qian2020lift} and sparse identification of nonlinear dynamics (SINDy) \cite{brunton2016discovering}, LaSDI seeks the governing dynamics directly from data. In particular, LaSDI seeks the optimal coefficient matrices associated with a specified library of functions to approximate the temporal evolution of the low-dimensional latent variables. To this end, LaSDI assumes the latent space dynamics to exhibit the form
\begin{equation}\label{eq:lowDimDynamics}
    \frac{\mathrm{d}\mathbf{\hat{x}} }{\mathrm{dt}}  = \mathbf{O} \boldsymbol{\Lambda}(\mathbf{\hat{x}}) ,
\end{equation}
where $\boldsymbol{\Lambda} : \mathbb{R}^n \rightarrow \mathbb{R}^{n_f}$ denotes a library of functions, in which $n_f \in \mathbb{N}$ is the dimension of the library of functions $\boldsymbol{\Lambda}$, and $\mathbf{O} \in \mathbb{R}^{n \times n_f}$ is the learned coefficient matrix. Importantly, the computational complexity associated with \eqref{eq:lowDimDynamics} is independent of the FOM dimension, $N$, enabling efficient online evaluations. To identify the operators appearing in \eqref{eq:lowDimDynamics}, the encoded training snapshots and their corresponding time derivatives are first assembled
\begin{equation}
    \begin{split}
    \hat{\mathbf{X}} = \begin{bmatrix}
        \hat{\mathbf{X}}(\boldsymbol{\mu}_1) & \hat{\mathbf{X}}(\boldsymbol{\mu}_2) & \cdots & \hat{\mathbf{X}}(\boldsymbol{\mu}_{n_p})
    \end{bmatrix} \in \mathbb{R}^{n \times (N_t+1) n_p}, \\ \dot{\hat{\mathbf{X}}} = \begin{bmatrix}
        \dot{\hat{\mathbf{X}}}(\boldsymbol{\mu}_1) & \dot{\hat{\mathbf{X}}}(\boldsymbol{\mu}_2) & \cdots & \dot{\hat{\mathbf{X}}}(\boldsymbol{\mu}_{n_p})
    \end{bmatrix} \in \mathbb{R}^{n \times (N_t+1) n_p},
    \end{split}
\end{equation}
where $n_p \in \mathbb{N}$ denotes the number of training parameter instances used to generate FOM trajectories, and
\begin{equation}
    \begin{split}
    \hat{\mathbf{X}}(\boldsymbol{\mu}_i) = \begin{bmatrix}
        \vert  & \vert &  & \vert \\ 
        \hat{\mathbf{x}}^0(\boldsymbol{\mu}_i) & \hat{\mathbf{x}}^1(\boldsymbol{\mu}_i) & \ldots & \hat{\mathbf{x}}^{N_t}(\boldsymbol{\mu}_i) \\
        \vert  & \vert &  & \vert
    \end{bmatrix} \in \mathbb{R}^{n\times (N_t+1)}, \\ \dot{\hat{\mathbf{X}}}(\boldsymbol{\mu}_i) = \begin{bmatrix}
        \vert  & \vert &  & \vert \\ 
        \dot{\hat{\mathbf{x}}}^0(\boldsymbol{\mu}_i) & \dot{\hat{\mathbf{x}}}^1(\boldsymbol{\mu}_i) & \ldots & \dot{\hat{\mathbf{x}}}^{N_t}(\boldsymbol{\mu}_i) \\
        \vert  & \vert &  & \vert \\ 
    \end{bmatrix} \in \mathbb{R}^{n\times (N_t+1)},
    \end{split}
\end{equation}
denote the low-dimensional trajectory associated with the $i^{\mathrm{th}}$ training parameter instance, obtained by applying the encoder to the corresponding high-dimensional FOM solution trajectory, i.e., $\hat{\mathbf{x}}^j(\boldsymbol{\mu}_i) = \mathrm{\mathbf{Enc}}(\mathbf{x}^j(\boldsymbol{\mu}_i))$ for $j=0,\ldots,N_t$, along with the corresponding latent space velocity trajectory, approximated using finite difference approximation of the temporal derivatives of the encoded state trajectory. Note that the solution trajectory includes the initial condition and therefore consists of $N_t+1$ time instances. Next, a data matrix $\mathbf{D} \in \mathbb{R}^{(N_t+1)n_p  \times n_f}$ is constructed using the library of functions,
\begin{equation}
    \mathbf{D} = \begin{bmatrix}
        (\mathbf{D}(\boldsymbol{\mu}_1))^\top \\
        (\mathbf{D}(\boldsymbol{\mu}_2))^\top \\
        \vdots \\
        (\mathbf{D}(\boldsymbol{\mu}_{n_p}))^\top
    \end{bmatrix} \in \mathbb{R}^{(N_t+1)n_p \times n_f},
\end{equation}
with
\begin{equation}
    \mathbf{D}(\boldsymbol{\mu}_i) = \begin{bmatrix}
        \vert  & \vert &  & \vert \\ 
        \boldsymbol{\Lambda}(\hat{\mathbf{x}}^0(\boldsymbol{\mu}_i)) & \boldsymbol{\Lambda}(\hat{\mathbf{x}}^1(\boldsymbol{\mu}_i)) & \cdots & \boldsymbol{\Lambda}(\hat{\mathbf{x}}^{N_t}(\boldsymbol{\mu}_i)) \\
        \vert  & \vert &  & \vert \\ 
    \end{bmatrix} \in \mathbb{R}^{n_f \times (N_t+1)}.
\end{equation}
The coefficient matrix can then be obtained by solving the least-squares problem
\begin{equation}\label{eq:minLaSDI}
    \underset{\mathbf{O} \in \mathbb{R}^{n \times n_f}}{\text{minimize}} \, \big \vert \big \vert \mathbf{\dot{\hat{X}}}^\top - \mathbf{D} \mathbf{O}^\top \big \vert \big \vert_F^2,
\end{equation}
where $\vert \vert \cdot \vert \vert_F$ denotes the Frobenius norm. Note that \eqref{eq:minLaSDI} may equivalently be interpreted as $n$ independent least-squares problems \cite{peherstorfer2016data} 
\begin{equation*}
    \underset{\boldsymbol{o}_i \in \mathbb{R}^{n_f}}{\text{minimize}}\, \vert \vert \dot{\hat{\boldsymbol{\chi}}}_i - \mathbf{D}\boldsymbol{o}_i\vert \vert_2^2, \quad i=1,\ldots,n,
\end{equation*}
where $\dot{\hat{\boldsymbol{\chi}}}_i$ and $\boldsymbol{o}_i$ denote the $i^{\mathrm{th}}$ column of $\dot{\hat{\mathbf{X}}}^\top$ and $\mathbf{O}^\top$, respectively. Therefore, the latent-space dynamics are learned independently for each latent variable, resulting in a set of decoupled regression problems that can be solved efficiently. Previous studies in the context of OpInf \cite{peherstorfer2016data,mcquarrie2021data,swischuk2020learning} have shown that directly solving \eqref{eq:minLaSDI} through the normal equations can lead to ill-conditioned systems and overfitting. To mitigate these issues, this work employs $L_2$-regularization (i.e., Tikhonov regularization) by introducing a bias term, yielding the modified optimization problem
\begin{equation}\label{eq:minLaSDI_reg}
    \underset{\mathbf{O} \in \mathbb{R}^{n \times n_f}}{\text{minimize}}\, \big \vert \big \vert \mathbf{\dot{\hat{X}}}^\top - \mathbf{D} \mathbf{O}^\top \big \vert \big \vert_F^2 + \big \vert \big \vert \boldsymbol{\Gamma} \mathbf{O}^\top \big \vert \big \vert_F^2,
\end{equation}
where $\boldsymbol{\Gamma} \in \mathbb{R}^{n_{f} \times n_{f}}$ is a diagonal regularization matrix. The resulting regularized normal equations are given by
\begin{equation} \label{eq:regLasdi}
    \left( \mathbf{D}^\top \mathbf{D} + \boldsymbol{\Gamma}^\top \boldsymbol{\Gamma} \right) \mathbf{O}^\top = \mathbf{D}^\top \mathbf{\dot{\hat{X}}}^\top.
\end{equation}

\section{Evaluation of model prediction accuracy and model efficiency} \label{sec:evaluation}

Evaluating the accuracy of ROM predictions for systems exhibiting sharp gradients, such as interfaces and shocks, has recently emerged as an active area of research \cite{chmiel2025unified}. Conventional error metrics used in the MOR literature \cite{peherstorfer2016data,lee2020deeplspg,geelen2022localized} are often not well-representative of a ROM's predictive accuracy in such settings, since even small discrepancies in the location of a sharp gradient can produce large global errors despite the solution remaining qualitatively accurate \cite{magargal2025projection,chmiel2025unified}. Furthermore, these metrics often fail to adequately capture the severity of errors associated with diffused or poorly resolved sharp gradients, making it difficult to distinguish between solutions that accurately preserve sharp features and those that do not. To address these limitations, this study introduces a point cloud error metric particularly suited for quantifying the accuracy of the predicted sharp gradient location. Section \ref{ssec:traditional_error} summarizes the conventional error metrics employed in this study, while Section \ref{ssec:pc_error} presents the proposed point cloud error metric. In addition to predictive accuracy, the computational efficiency of each ROM is assessed using the speedup factor
\begin{equation}\label{eq:speedUpFactor}
    \text{speedup factor} = \frac{\text{wall-clock time for FOM}}{\text{wall-clock time for ROM}},
\end{equation}
which compares the online wall-clock time of the ROM against that of the FOM. Larger speedup factors indicate higher computational efficiency.

\subsection{Conventional error metrics} \label{ssec:traditional_error}
The first error metric considered is the reconstruction error. The autoencoder reconstruction error is defined as
\begin{equation}\label{eq:ae_error}
    \text{graph autoencoder reconstruction error} = \frac{ \Big\vert\Big\vert \mathbf{X} \left( \boldsymbol{\mu} \right) -  \left(\underaccent{\tilde}{\mathbf{X}} (\boldsymbol{\mu})\right)_{\mathrm{recon,AE}} \Big\vert\Big\vert_F^2} { \Big\vert\Big\vert \mathbf{X} \left( \boldsymbol{\mu} \right)   \Big\vert\Big\vert_F^2},
\end{equation}
where $\mathbf{X}(\boldsymbol{\mu}) = \left[ \mathbf{x}^0 \left( \boldsymbol{\mu}\right) \; \mathbf{x}^1 \left( \boldsymbol{\mu}\right) \; \ldots \; \mathbf{x}^{N_t} \left( \boldsymbol{\mu}\right) \right] \in \mathbb{R}^{N\times (N_t+1)}$ represents the FOM solution trajectory (including initial conditions), and $\left(\underaccent{\tilde}{\mathbf{X}} (\boldsymbol{\mu})\right)_{\mathrm{recon,AE}} \in \mathbb{R}^{N\times (N_t+1)}$ denotes the reconstructed solution obtained by encoding and subsequently decoding the FOM snapshots, i.e.,  $\left(\underaccent{\tilde}{\mathbf{X}} (\boldsymbol{\mu})\right)_{\mathrm{recon,AE}} = \left[ \left(\mathbf{Dec} \circ \mathbf{Enc} \left(\mathbf{x}^0 \left( \boldsymbol{\mu} \right) \right) \right) \; \left(\mathbf{Dec} \circ \mathbf{Enc} \left(\mathbf{x}^1 \left( \boldsymbol{\mu} \right) \right) \right) \; \ldots \; \left(\mathbf{Dec} \circ \mathbf{Enc} \left(\mathbf{x}^{N_t} \left( \boldsymbol{\mu} \right) \right) \right)\right]$. Likewise, the POD reconstruction error is defined as
\begin{equation}\label{eq:pod_error}
    \text{POD reconstruction error} = \frac{ \Big\vert\Big\vert \left( \mathbf{I}_N - \boldsymbol{\Phi} \boldsymbol{\Phi}^\top \right) \mathbf{X} \left( \boldsymbol{\mu} \right) \Big\vert\Big\vert_F^2}{ \Big\vert\Big\vert \mathbf{X} \left( \boldsymbol{\mu} \right)  \Big\vert\Big\vert_F^2},
\end{equation}
where $\mathbf{\Phi}\in \mathbb{R}^{N\times n}$ is the matrix of reduced basis vectors from a linear POD approximation constructed based on the method of snapshots \cite{sirovich1987turbulence}, and $\mathbf{I}_N \in \mathbb{R}^{N\times N}$ is the identity matrix of dimension $N$. 

The second error metric is the state prediction error, which is expressed as
\begin{equation}\label{eq:state_err}
    \text{state prediction error} = \frac{\Big\vert\Big\vert \mathbf{X} \left( \boldsymbol{\mu} \right) - \mathbf{\tilde{X}} \left( \boldsymbol{\mu}\right) \Big\vert\Big\vert_F^2}{\Big\vert\Big\vert \mathbf{X} \left( \boldsymbol{\mu} \right)  \Big\vert\Big\vert_F^2},
\end{equation}
where $\tilde{\mathbf{X}}(\boldsymbol{\mu}) \in \mathbb{R}^{N \times (N_t+1)}$ is the approximate solution trajectory generated by the ROM (including initial conditions). As will be demonstrated in the numerical experiments, these conventional error metrics often fail to adequately characterize ROM accuracy for problems dominated by sharp gradients, since they primarily measure global state discrepancies rather than the accuracy of localized structures such as interfaces or shocks. 

\subsection{Point cloud error for sharp gradients} \label{ssec:pc_error}

Inspired by \cite{cucchiara2024convexDisplacement}, the proposed point cloud error metric seeks to assess the accuracy of the predicted sharp gradient location by constructing a point cloud representation of the sharp gradient regions in both the ROM and FOM solutions. The procedure used to generate the point cloud representation is application-dependent. Additional details for the specific problems considered in this work are provided in Section \ref{sec:experiments}. Representative examples include Canny edge detection \cite{canny1986edge} for detecting interfaces in phase-field simulations, and Ducros sensors \cite{ducros1999sensor} for identifying shocks in compressible flow simulations. Here, the procedure for computing the point cloud error from a general point cloud representation of the sharp gradient is described.

Let $\mathcal{X}^m_{\mathrm{ROM}} \subset \mathbb{R}^{n_d}$ and $\mathcal{X}^m_{\mathrm{FOM}} \subset \mathbb{R}^{n_d}$ denote the set of locations of collocation points corresponding to the sharp gradient regions identified in the ROM and FOM solutions, respectively, where $n_d \in \mathbb{N}$ denotes the spatial dimension of the problem (e.g., $n_d=2$ for 2D problems). The point cloud error metric is then defined as
\begin{equation}\label{eq:pc_error}
    \epsilon_{\mathrm{pc}} (t;\boldsymbol{\mu}) = \frac{d_\infty(\mathcal{X}^m_{\mathrm{ROM}}, \,  \mathcal{X}^m_{\mathrm{FOM}})}{L}
\end{equation}
where $L\in\mathbb{R}_+$ is a characteristic length scale, and $d_\infty$ denotes the Hausdorff distance \cite{rockafellar1998variational,taha2015hausdorff}, defined as
\begin{equation}
    d_\infty (\mathcal{X}, \,  \mathcal{Y}) = \max\left \{ \breve{d}(\mathcal{X}, \, \mathcal{Y}), \, \breve{d}(\mathcal{Y}, \, \mathcal{X}) \right \},
\end{equation}
with 
\begin{equation}
    \breve{d}(\mathcal{X}, \, \mathcal{Y}) = \underset{\mathbf{x} \in \mathcal{X}}{\max} \left \{ \underset{\mathbf{y} \in \mathcal{Y}}{\min} \left \{ \big\vert\big\vert \mathbf{x} - \mathbf{y} \big\vert\big\vert_2 \right \} \right \}.
\end{equation}
Qualitatively, the point cloud error measures the largest minimum distance between the point cloud representations of the sharp gradient extracted from the FOM and ROM solutions. Consequently, unlike traditional error metrics, the proposed error metric directly quantifies discrepancies in the predicted location of interfaces and shocks.

\section{Numerical experiments}\label{sec:experiments}

The numerical experiments presented in this section assess the accuracy and efficiency of GNN-LaSDI compared against GD-LSPG and POD-LaSDI. A summary of the three ROM frameworks is provided in Table \ref{table:methodsSummary}. GNN-LaSDI and GD-LSPG employ the graph autoencoder (described in Section \ref{ssec:ae}) to obtain a low-dimensional representation of the high-dimensional state, whereas POD-LaSDI uses a POD basis constructed via the method of snapshots \cite{sirovich1987turbulence}. To evolve the reduced state in time, GNN-LaSDI and POD-LaSDI utilize the regularized operator learning framework presented in Section \ref{sec:lasdi}, while GD-LSPG advances the solution using a nonlinear manifold LSPG projection \cite{lee2020deeplspg,magargal2025projection}. 

\begin{table}[h]
\centering
\caption{Summary of MOR methods considered in this study. The first column identifies each method, with GNN-LaSDI denoting the primary contribution of this work. The second column specifies the dimensionality reduction technique employed by each method, while the third column indicates the approach used to model the temporal evolution of the reduced representation.}
\begingroup
\small
\begin{tabular}{C{2cm} C{4.5cm} C{6cm}}
    \hline
    method & dimensionality reduction & temporal evolution\\
    \hline
    \textbf{GNN-LaSDI} & graph autoencoder (Section \ref{ssec:ae}) & LaSDI (Section \ref{sec:lasdi}) \\
    GD-LSPG & graph autoencoder (Section \ref{ssec:ae}) & nonlinear manifold LSPG (\ref{sec:gdlspg}) \\
    POD-LaSDI & POD basis & LaSDI (Section \ref{sec:lasdi}) \\
    \hline
\end{tabular}
\endgroup
\label{table:methodsSummary}
\end{table}

Two numerical experiments are considered in this study. The first example examines MOR for Kobayashi's solidification model for 2D isotropic nucleus growth \cite{kobayashi1993dendritic}, where the feature of interest is the moving interface between the solid and liquid phases. A Canny edge filter \cite{canny1986edge} is used to generate a point cloud representation of the solid-liquid interface. The second experiment considers a bow shock generated by flow past a cylinder governed by the 2D Euler equations. In this case, a point cloud representation of the bow shock is extracted using a Ducros sensor \cite{ducros1999sensor}. In both experiments, the ability of GNN-LaSDI, GD-LSPG, and POD-LaSDI to accurately predict the shock location is then evaluated using the point cloud error metric \eqref{eq:pc_error}. 

In both numerical experiments, the GNN-LaSDI and POD-LaSDI models employ a quadratic library of functions, $\boldsymbol{\Lambda}$, to model the latent space dynamics. Specifically, the reduced dynamics are assumed to take the form 
\begin{equation}\label{eq:quadLowDimDynamics}
    \frac{\mathrm{d}}{\mathrm{dt}} \hat{\mathbf{x}} = \hat{\mathbf{C}} + \hat{\mathbf{A}}\hat{\mathbf{x}} + \hat{\mathbf{F}} (\hat{\mathbf{x}} \otimes \hat{\mathbf{x}}),
\end{equation}
where $\hat{\mathbf{C}} \in \mathbb{R}^n$, $\hat{\mathbf{A}}\in\mathbb{R}^{n \times n}$, $\hat{\mathbf{F}} \in \mathbb{R}^{n\times \frac{(n+1)n}{2}}$, and $\otimes$ denotes the non-repeating Kronecker product (i.e., $\hat{\mathbf{x}} \otimes \hat{\mathbf{x}} \in \mathbb{R}^{\frac{(n+1)n}{2}}$) \cite{peherstorfer2016data}. The corresponding operator learning problem can be expressed in the form of \eqref{eq:lowDimDynamics} by taking
\begin{equation}
    \mathbf{O} = \begin{bmatrix}
        \hat{\mathbf{C}} & \hat{\mathbf{A}} & \hat{\mathbf{F}}
    \end{bmatrix},
    \quad
    \boldsymbol{\Lambda}: (\hat{\mathbf{x}}) \mapsto \begin{bmatrix}
        \mathbf{1} \\
        \hat{\mathbf{x}} \\
        \hat{\mathbf{x}} \otimes \hat{\mathbf{x}}
    \end{bmatrix}.
\end{equation}
With this choice, the library of functions, $\boldsymbol{\Lambda}$, contains $n_f = 1+n+\frac{(n+1)n}{2}$ terms. The optimal operator coefficients are obtained through the $L_2$-regularized least-squares formulation described in Section \ref{sec:lasdi}. Following \cite{swischuk2020learning,mcquarrie2021data,geelen2022localized,mcquarrie2023popinf}, separate regularization parameters are employed for the constant-linear and quadratic components of the dynamics. Specifically, $\gamma_1 \in \mathbb{R}_+$ regularizes the constant and linear terms, $\hat{\mathbf{C}}$ and $\hat{\mathbf{A}}$, while $\gamma_2 \in \mathbb{R}_+$ regularizes the quadratic term, $\hat{\mathbf{F}}$. The resulting regularization matrix is
\begin{equation}
    \boldsymbol{\Gamma} = \begin{bmatrix}
        \gamma_1 \mathbf{I}_{n+1} & \mathbf{0} \\
        \mathbf{0} & \gamma_2 \mathbf{I}_{\frac{n(n+1)}{2}}
    \end{bmatrix},
\end{equation}
where $\mathbf{I}_p$ denotes the identity matrix of dimension $p$. In this study, $\gamma_1$ and $\gamma_2$, for each model, are obtained from a grid search.

To evaluate the computational efficiency of each of the studied methods, the speedup factors are evaluated using the implementation details and machine specifications provided in \ref{sec:aeTraining}. To mitigate the influence of small fluctuations in time measurements, all reported wall-clock times are averaged over 10 independent runs of each solver.

\subsection{Kobayashi's solidification model for 2D isotropic nucleus growth}
\begin{figure}[ht!]
    \centering
    \begin{tabular}{cc}
        \includegraphics[height=0.3\textwidth]{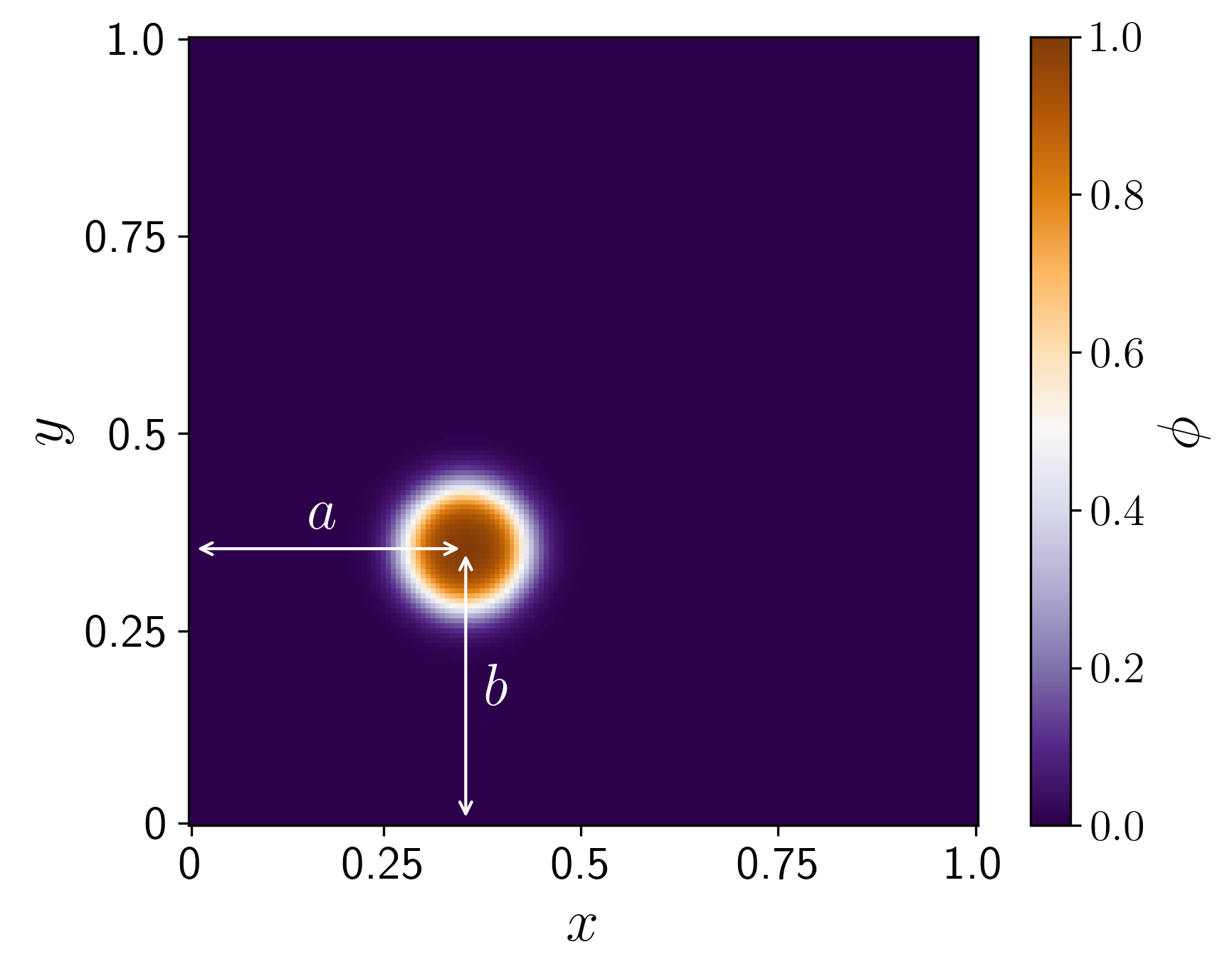}
        & \includegraphics[height=0.3\textwidth]{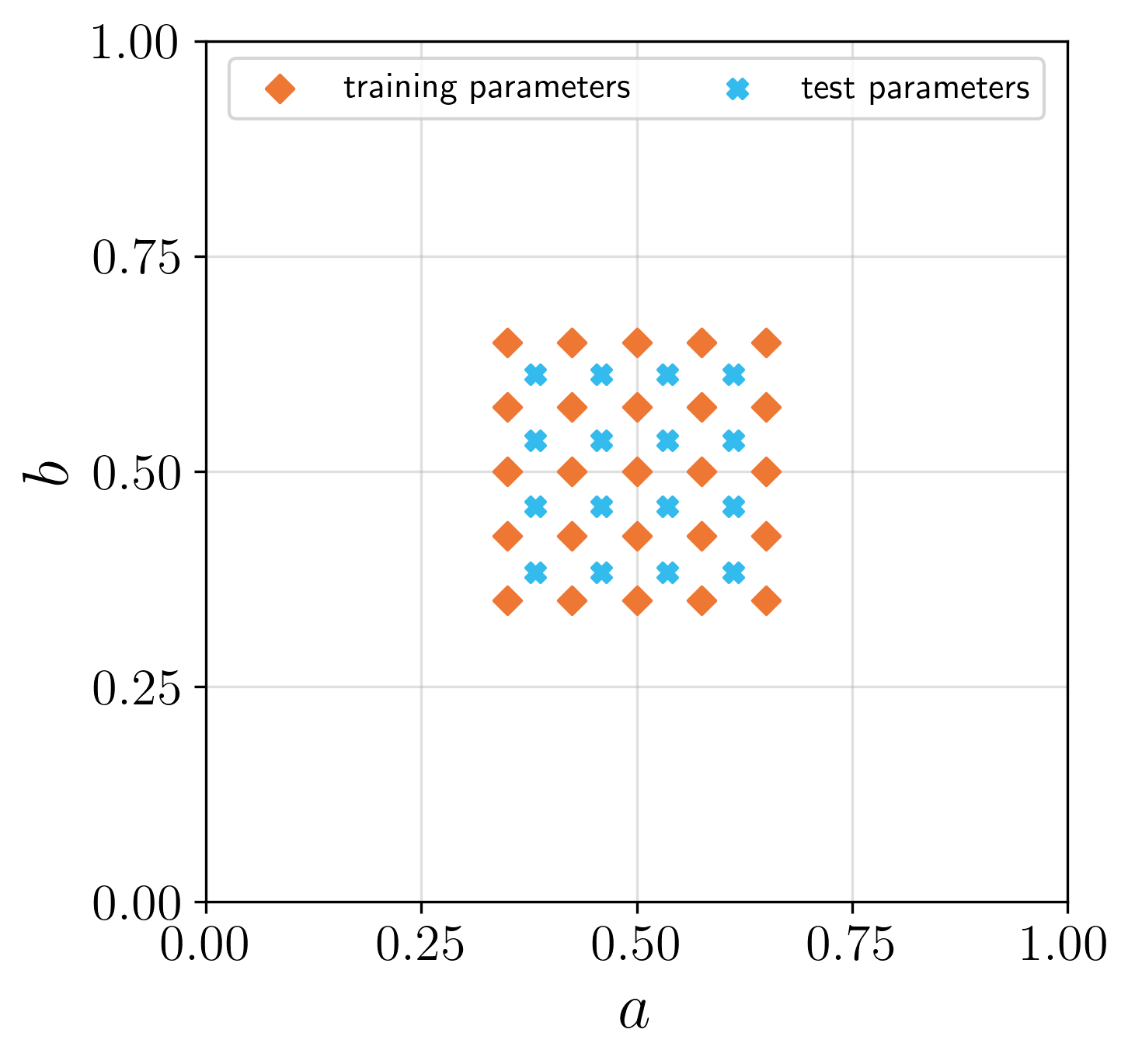}
    \end{tabular}
    \captionsetup{justification=centering}
    \caption{Initial condition and parameter-space sampling for Kobayashi's solidification model. The left figure presents the initial conditions for the order parameter, $\phi$, for parameter instance $\boldsymbol{\mu}=(a,b)=(0.35,0.35)$. The figure on the right shows the distribution of the training and test parameter sets used in Kobayashi's solidification model experiment. (Online version in color.)}
    \label{fig:ac_initialConditions}
\end{figure}

The governing equations for Kobayashi's phase field model of 2D isotropic nucleus growth \cite{kobayashi1993dendritic} are given by
\begin{equation} \label{eq:kobayashi}
    \begin{aligned}
    0.0003 \frac{\partial \phi}{\partial t} &= 0.0001 \Delta \phi + \phi \left(1 - \phi \right) \left( \phi - \frac{1}{2} + q \right), \\\
    \frac{\partial T}{\partial t} &= \Delta T + \frac{\partial \phi}{\partial t},
\end{aligned}
\end{equation}
where the system is solved over the time interval $t=(0,0.01]$ on a square physical domain with $x,y \in (0,1)$, $\phi(x,y,t;\boldsymbol{\mu}) \in \mathbb{R}$ is the phase-field order parameter, with $\phi=0$ and $1$ corresponding to the liquid and solid phases, respectively, $T(x,y,t;\boldsymbol{\mu}) \in \mathbb{R}$ denotes the temperature field, and $q=\frac{0.9}{\pi}\mathrm{tan}^{-1}\left(10(1-T))\right)$ is the thermodynamic driving force. Together with the double-well potential, $q$ drives the phase transition from liquid to solid and governs the solidification process. The spatial domain is discretized using a $160 \times 160$ finite difference mesh, resulting in $N_c=25600$ collocation points. Applying a second-order central difference discretization in space results in a semi-discrete system of the form \eqref{eq:ode1}, where the state vector is defined as $\mathbf{x}(t;\boldsymbol{\mu}) = [(\boldsymbol{\phi}(t;\boldsymbol{\mu}))^{\top}, (\mathbf{T}(t;\boldsymbol{\mu}))^{\top}]^{\top}$ with $\boldsymbol{\phi}$ and $\mathbf{T}$ denoting the semi-discrete representations of the order parameter and temperature, respectively. Consequently, the system contains $n_q=2$ state variables and has a total dimension of $N=n_qN_c = 51200$. Temporal discretization is performed using a backward Euler time integration scheme with $\Delta t=2\times 10^{-5}$, resulting in $N_t=500$ time steps over the simulation horizon. In this study, the system is parameterized by the initial location of the nucleus. The initial conditions are given by
\begin{equation}\label{eq:kobayashiInit}
    \begin{split}
    \phi^0 (x,y;\boldsymbol{\mu})= T^0(x,y;\boldsymbol{\mu}) &= \frac{1}{2}\left[ 1 - \mathrm{tanh}\left(\frac{r(x,y;\boldsymbol{\mu}) - 0.075}{0.02\sqrt{2}} \right) \right], \\
    \text{with}\;r(x,y;\boldsymbol{\mu}) &= \sqrt{\left(x-a\right)^2 + \left(y-b\right)^2},
    \end{split}
\end{equation}
where $\boldsymbol{\mu} = [a,b]^\top$, with $a,b\in\mathbb{R}_+$, specifying the spatial location of the center of the initial solid nucleus, which has a radius of $0.075$. These initial conditions define a circular solid nucleus embedded within the liquid phase with a diffuse interface along the solid and liquid regions. Figure \ref{fig:ac_initialConditions} presents the initial conditions for the order parameter, $\phi$, corresponding to the parameter instance $\boldsymbol{\mu}=(0.35,0.35)$. Dirichlet boundary conditions are applied at $x=0$, $x=1$, $y=0$, and $y=1$. To generate the training data used for graph autoencoder training in GNN-LaSDI and GD-LSPG, POD basis construction for POD-LaSDI, and operator learning in both GNN-LaSDI and POD-LaSDI, FOM solutions are computed for 25 training parameter sets (i.e., $n_p=25$) with $\boldsymbol{\mu} = (0.35+0.075i, 0.35+0.075j)$ for $i,j=0,\ldots,4$. In the online stage, the performance of GNN-LaSDI, GD-LSPG, and POD-LaSDI is evaluated for 16 test parameter sets given by $\boldsymbol{\mu} = (0.3875+0.075i, 0.3875+0.075j)$ with $i,j=0,\ldots,3$. The distributions of the training and test parameter sets within the parameter space are schematically illustrated in Figure \ref{fig:ac_initialConditions}. Additional details regarding the training of the graph autoencoder are provided in \ref{sec:aeTraining}. After training the graph autoencoder and constructing the POD basis, the low-dimensional operators for the GNN-LaSDI and POD-LaSDI frameworks are obtained using the procedure outlined in Section \ref{sec:lasdi}. For consistency with the FOM and GD-LSPG, the backward Euler scheme is used for time integration during the online stage for the GNN-LaSDI and POD-LaSDI models. A fourth-order central difference scheme is used to determine the latent state derivative matrix, $\dot{\hat{\mathbf{X}}}$, with special treatment applied at the first two and last two time steps of each trajectory\footnote{A second-order forward difference scheme is used for the first time instance of each trajectory, a second-order backward difference scheme is used for the last time instance of each trajectory, and a second-order central difference scheme is used for the second and second to last time instances in each trajectory.}. The resulting training dataset contains a total of $25(N_t+1)=12525$ latent state snapshots and corresponding latent space velocity samples.

\begin{figure}[ht!]
    \centering
    \begin{tabular}{cc}
        \includegraphics[height=0.3\textwidth]{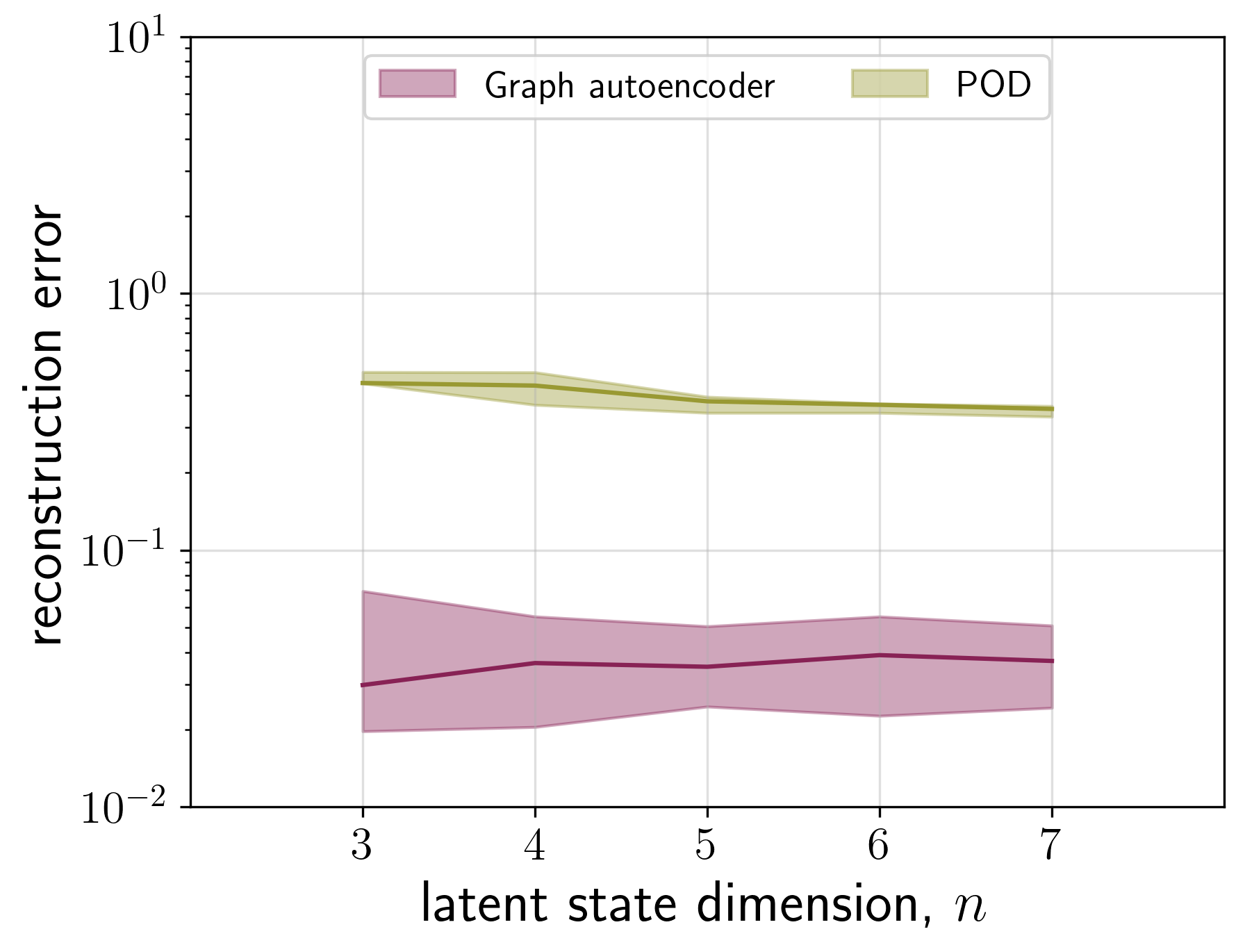}
        &  \includegraphics[height=0.3\textwidth]{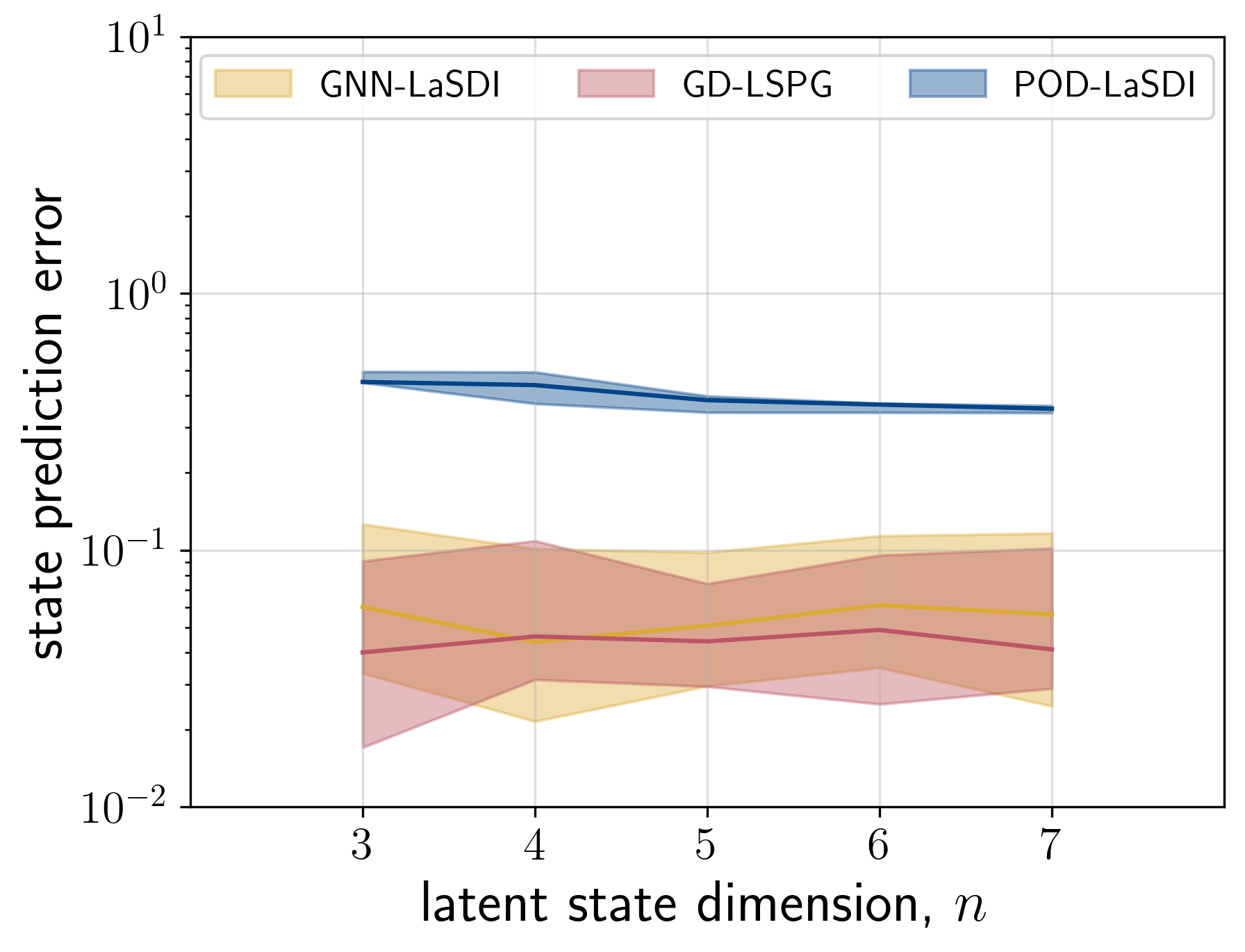}
    \end{tabular}
    \captionsetup{justification=centering}
    \caption{Comparison of reconstruction and state prediction errors as a function of latent state dimension. The left figure shows the range of graph autoencoder and POD reconstruction errors across all 16 test parameter sets, plotted against the latent state dimension. The figure on the right shows the range of state prediction errors for GNN-LaSDI, GD-LSPG, and POD-LaSDI on the same test set. In each figure, solid lines indicate the median error for each model, while the shaded regions indicate the range of errors observed across the test parameter instances. Across the considered latent state dimensions, the graph autoencoder consistently achieves reconstruction errors approximately one order of magnitude lower than those obtained with POD. Similarly, GNN-LaSDI and GD-LSPG yield substantially lower state prediction errors than POD-LaSDI. (Online version in color.)}
    \label{fig:ac_errorRange}
\end{figure}

Figure \ref{fig:ac_errorRange} presents the graph autoencoder reconstruction errors, POD reconstruction errors, and state prediction errors obtained for GNN-LaSDI, GD-LSPG, and POD-LaSDI across all test parameter sets. The results indicate that GD-LSPG achieves state prediction errors nearly identical to the graph autoencoder reconstruction errors. GNN-LaSDI produces state prediction errors comparable to those of GD-LSPG and considerably lower than those of POD-LaSDI, demonstrating the effectiveness of learning the latent space dynamics associated with the graph autoencoder representation. Notably, the reconstruction and state prediction errors remain relatively insensitive to the latent state dimension $n$ over the range of dimensions considered.

\begin{figure}[ht!]
    \centering
     \begin{tabular}{cccc}
        \vspace{-.1cm}
        & {\footnotesize{\quad GNN-LaSDI}} & {\footnotesize{\;\; GD-LSPG}} & {\footnotesize{\;\; POD-LaSDI}} \\
        \raisebox{2.3em}{\rotatebox[origin=lb]{90}{\parbox{2cm}{\centering \footnotesize{$n=3$}}}}&
        \includegraphics[trim =0.3cm 0 0.8cm 1cm, clip,height=0.2\textwidth]{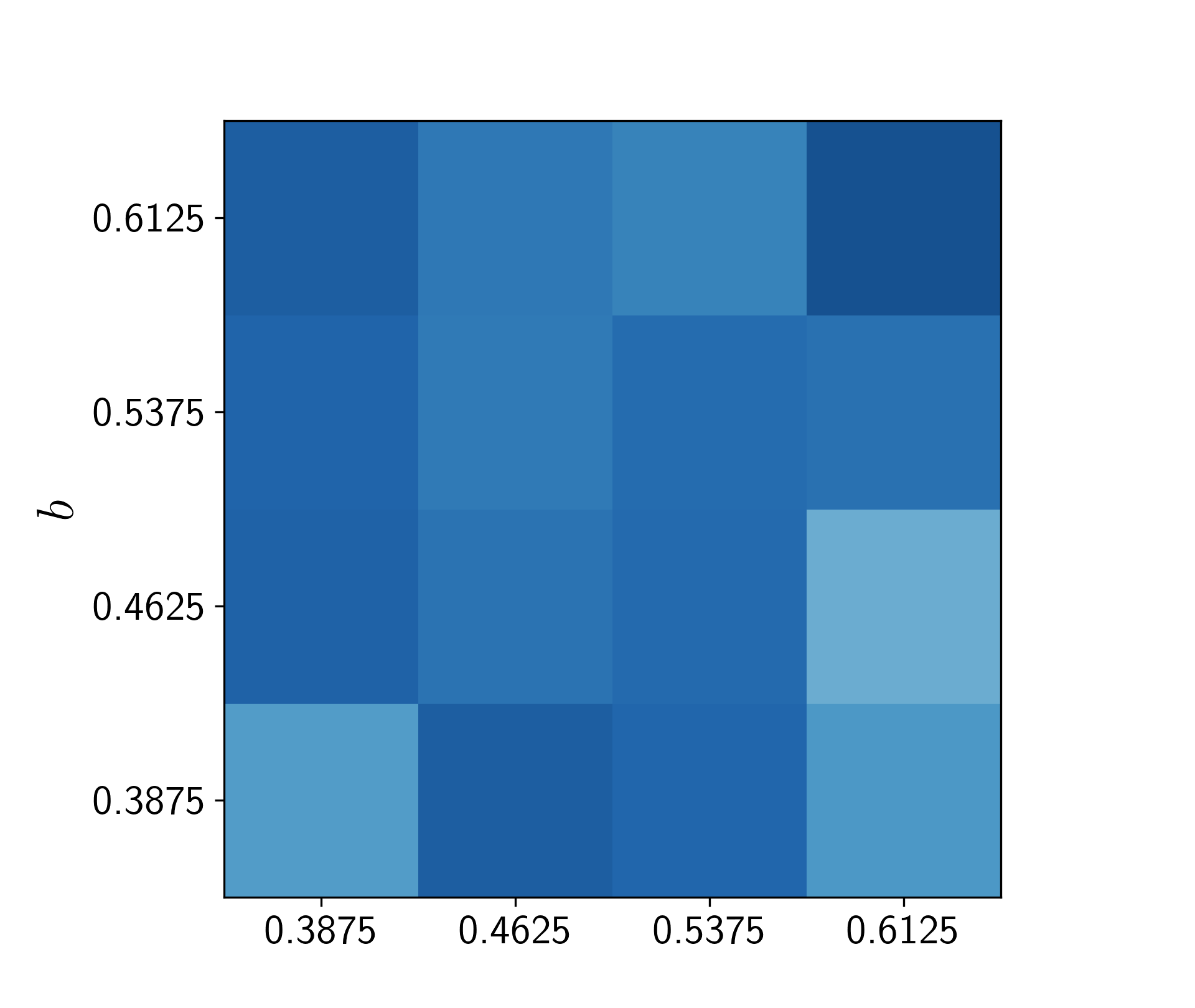}&
        \includegraphics[trim=0.8cm 0 0.8cm 1cm, clip,height=0.2\textwidth]{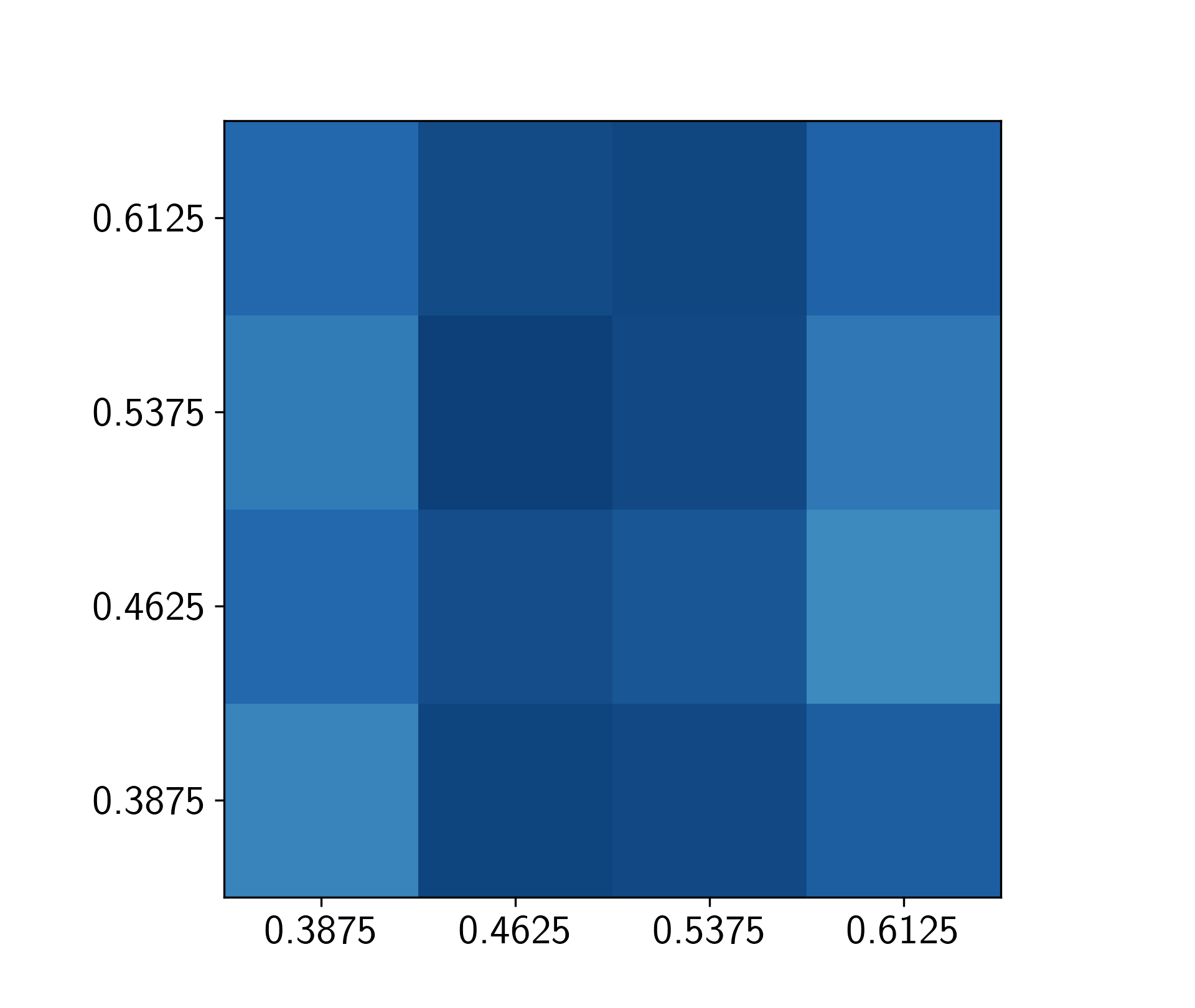}&
        \includegraphics[trim=0.8cm 0 0.8cm 1cm, clip,height=0.2\textwidth]{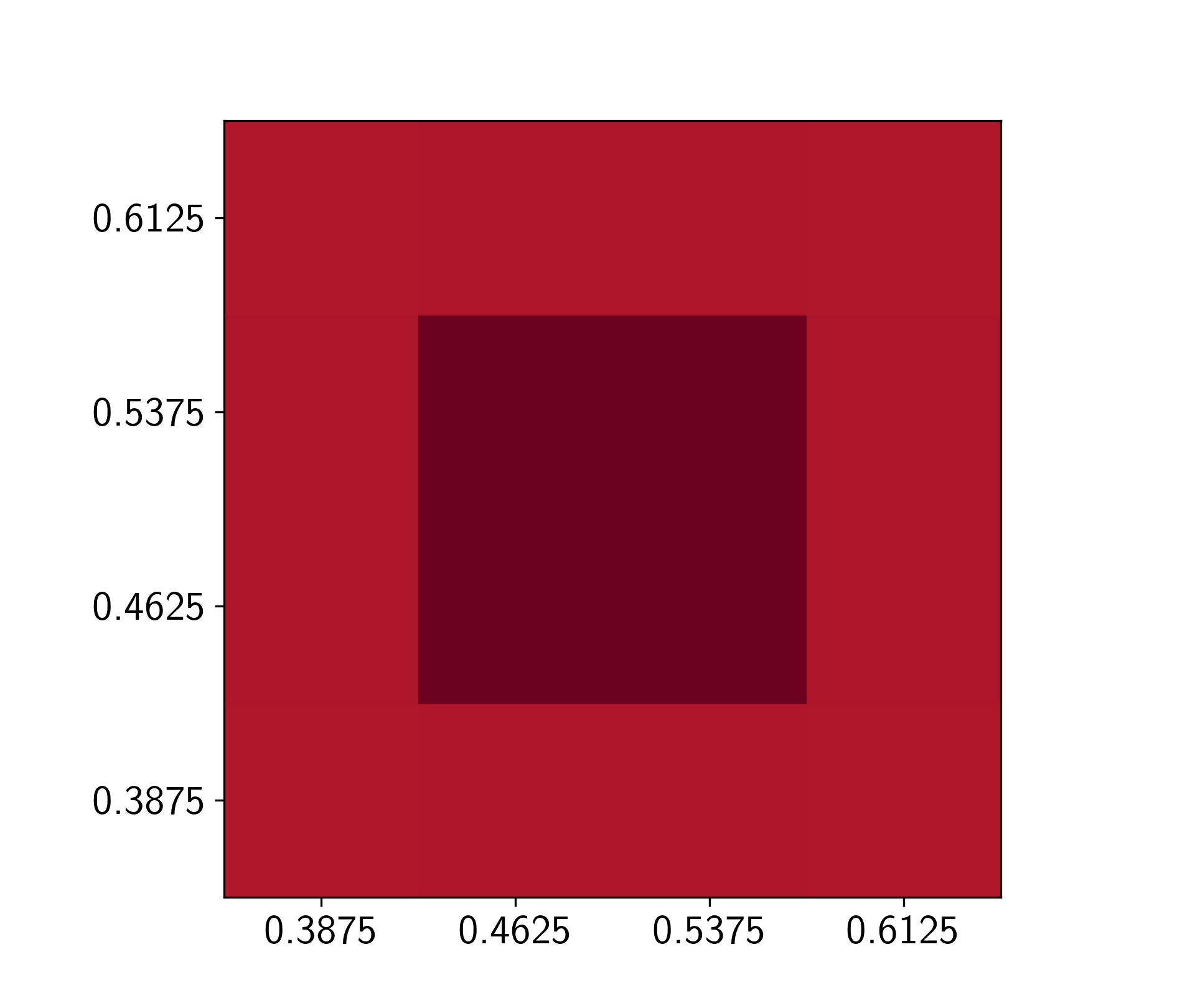}\\ 
        \raisebox{2.3em}{\rotatebox[origin=lb]{90}{\parbox{2cm}{\centering \footnotesize{$n=5$}}}}&
        \includegraphics[trim=0.3cm 0 0.8cm 1cm, clip,height=0.2\textwidth]{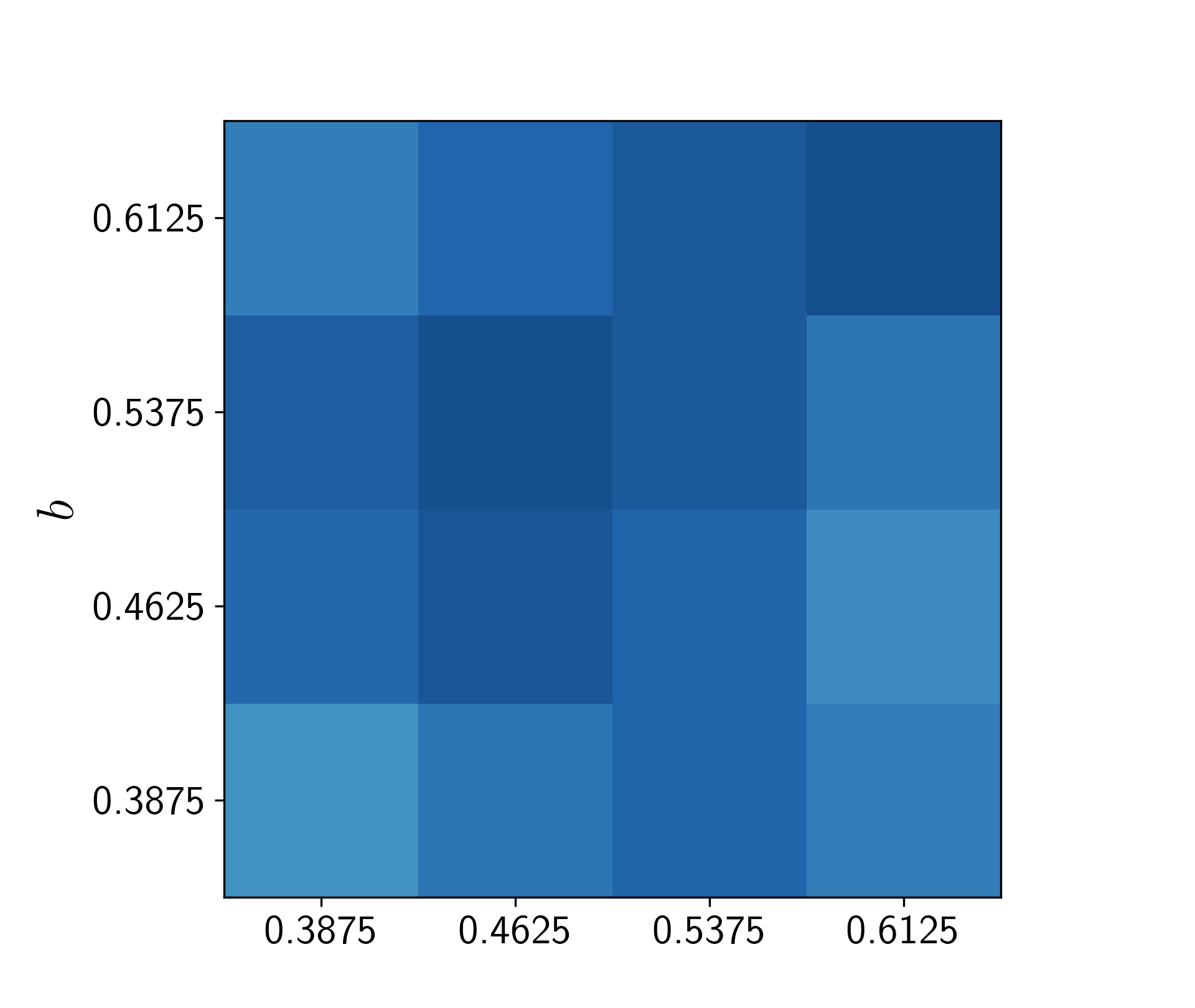}&
        \includegraphics[trim=0.8cm 0 0.8cm 1cm, clip,height=0.2\textwidth]{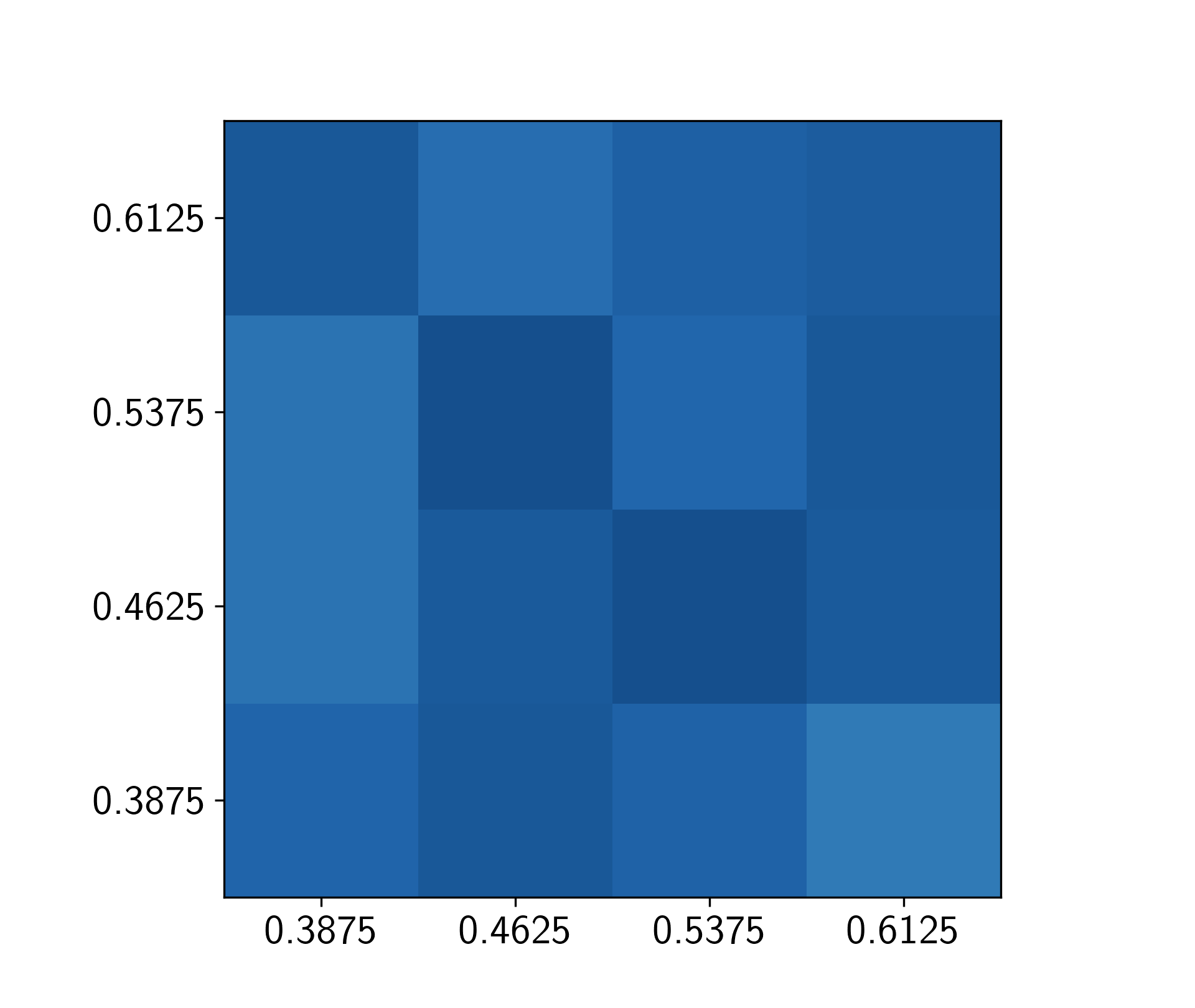}&
        \includegraphics[trim=0.8cm 0 0.8cm 1cm, clip,height=0.2\textwidth]{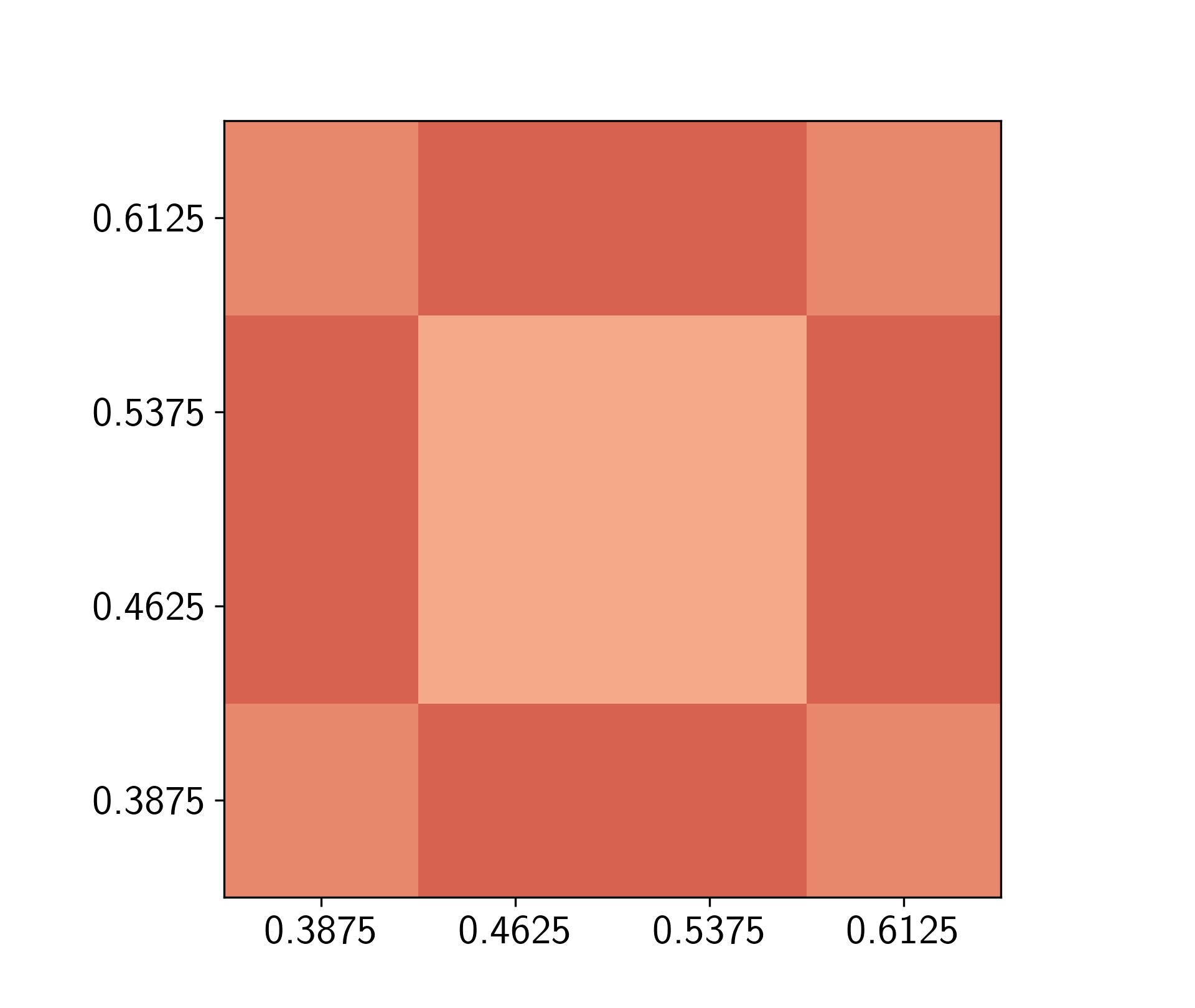}\\
        \raisebox{2.3em}{\rotatebox[origin=lb]{90}{\parbox{2cm}{\centering \footnotesize{$n=7$}}}}&
        \includegraphics[trim=0.3cm 0 0.8cm 1cm, clip,height=0.2\textwidth]{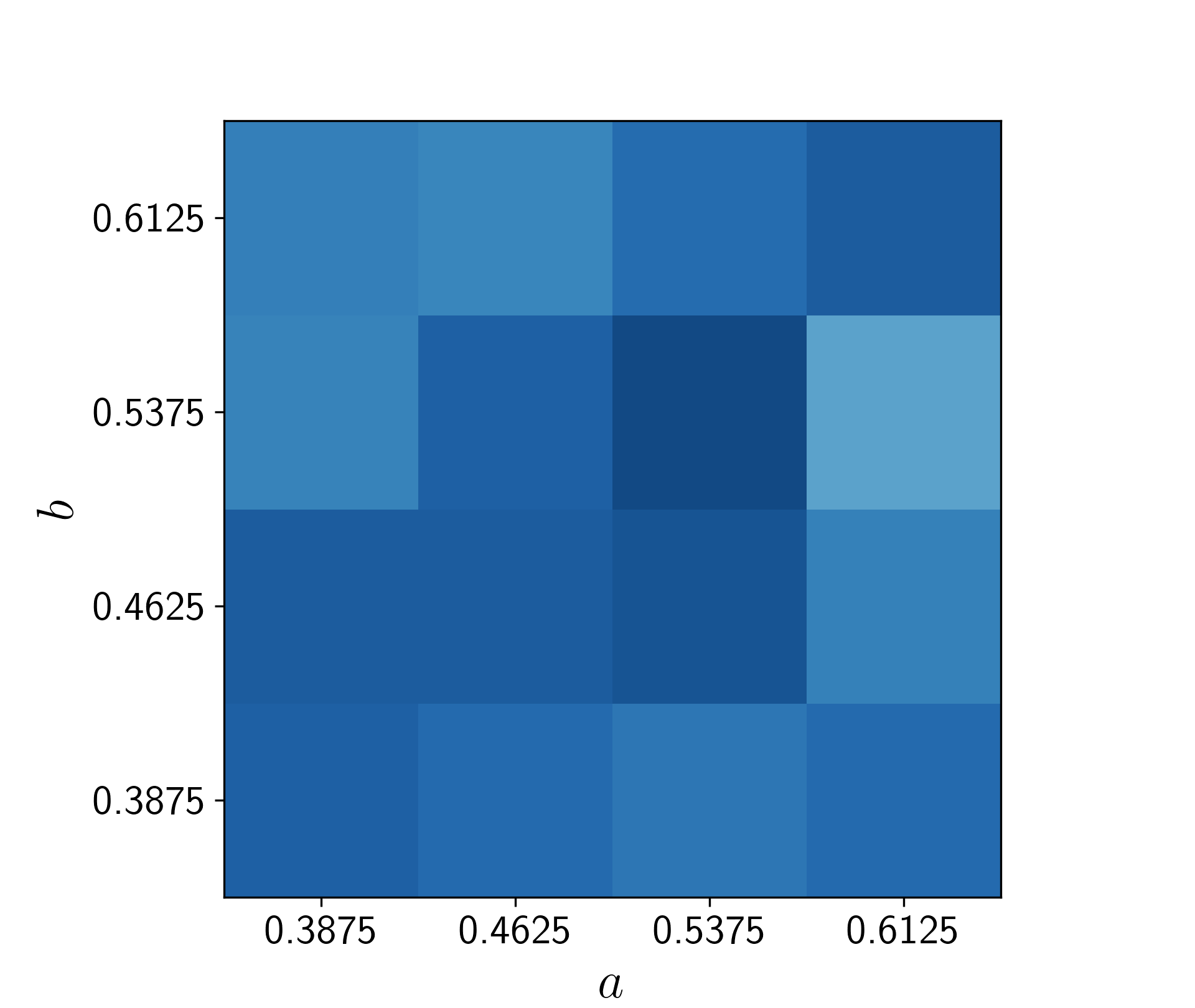}&
        \includegraphics[trim=0.8cm 0 0.8cm 1cm, clip,height=0.2\textwidth]{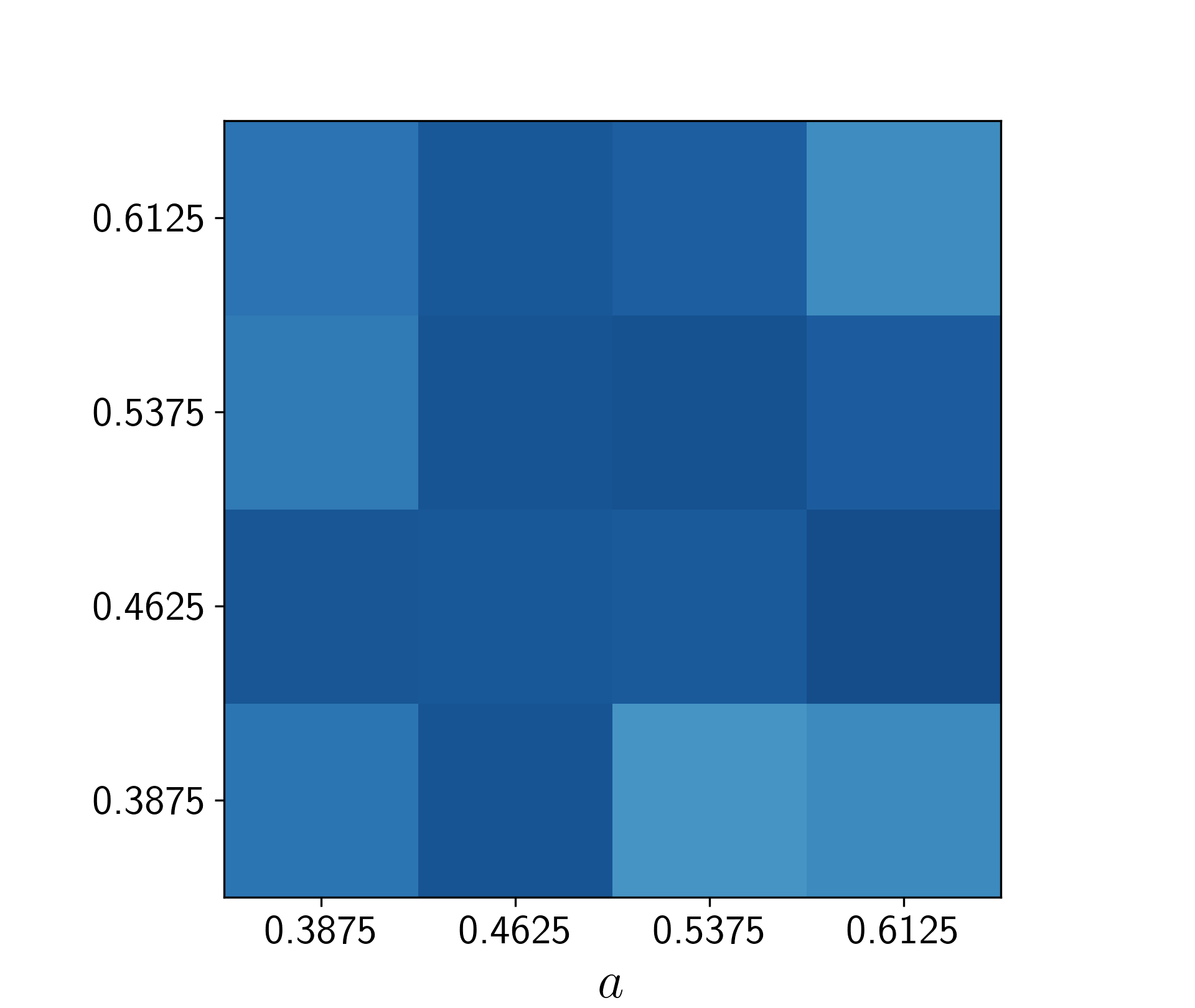}&
        \includegraphics[trim=0.8cm 0 0.8cm 1cm, clip,height=0.2\textwidth]{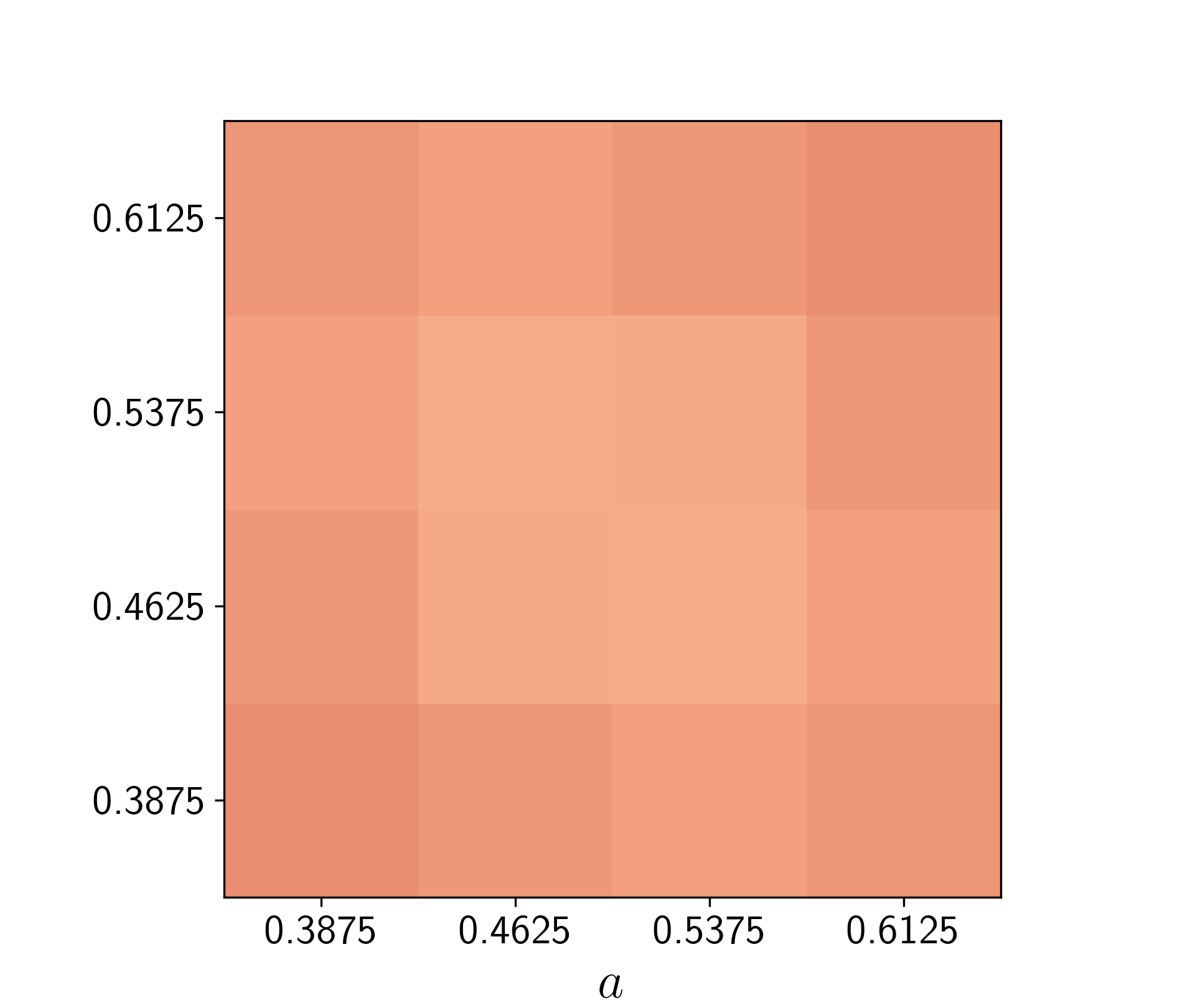}\\
        & \multicolumn{3}{c}{\includegraphics[scale=.4]{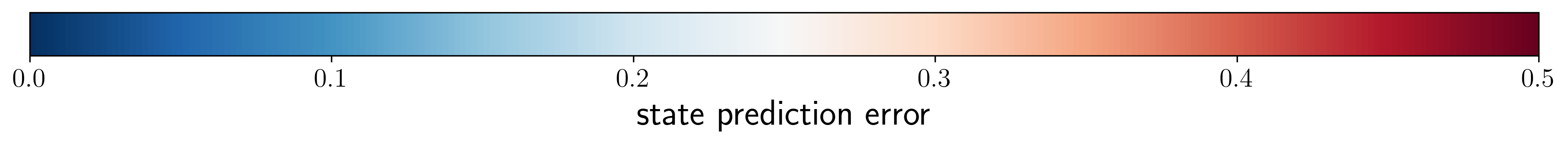}}
    \end{tabular}
    \captionsetup{justification=centering}
    \caption{State prediction errors for GNN-LaSDI (left column), GD-LSPG (middle column), and POD-LaSDI (right column) across the test parameter space. Results are shown for latent state dimensions $n=3$ (top row), $n=5$ (middle row), and $n=7$ (bottom row). The horizontal and vertical axes correspond to the parameters $a$ and $b$, which define the location of the initial solid nucleus. The results indicate that both graph-based approaches generalize well to unseen parameter instances and significantly outperform the linear POD-based approach. (Online version in color.) }
    \label{fig:ac_ab_sp}
\end{figure}

To assess how the state prediction errors vary throughout the parameter space, Figure \ref{fig:ac_ab_sp} plots the state prediction errors of GNN-LaSDI, GD-LSPG, and POD-LaSDI for each test parameter set at latent state dimensions $n=3$, $5$, and $7$. The results show that GD-LSPG achieves consistently low state prediction errors across the parameter domain for all considered latent space dimensions. Similarly, GNN-LaSDI maintains uniformly low errors across the parameter space, indicating strong generalization and consistent accuracy across all test parameter instances. In contrast, POD-LaSDI exhibits substantially larger state prediction errors, highlighting the limitations of the linear manifold for this problem.

\begin{figure}[ht!]
    \centering
     \begin{tabular}{ccc|cc}
        & \multicolumn{2}{c|}{\footnotesize{$t=0.005$}} & \multicolumn{2}{c}{\footnotesize{$t=0.01$}} \\
        \hline & & & &\\
        & \hspace{-0.5cm} \footnotesize{state solution} & \hspace{-1.25cm} \footnotesize{point cloud} & \footnotesize{\;\, state solution} & \hspace{-1.25cm} \footnotesize{\; point cloud} \\ \vspace{-.25cm}
        \raisebox{2.1em}{\rotatebox[origin=lb]{90}{\parbox{2cm}{\centering \footnotesize{FOM}}}}& \hspace{-.5cm}
        \includegraphics[trim=0.2cm 0 0.2cm 1cm, clip,height=0.18\textwidth]{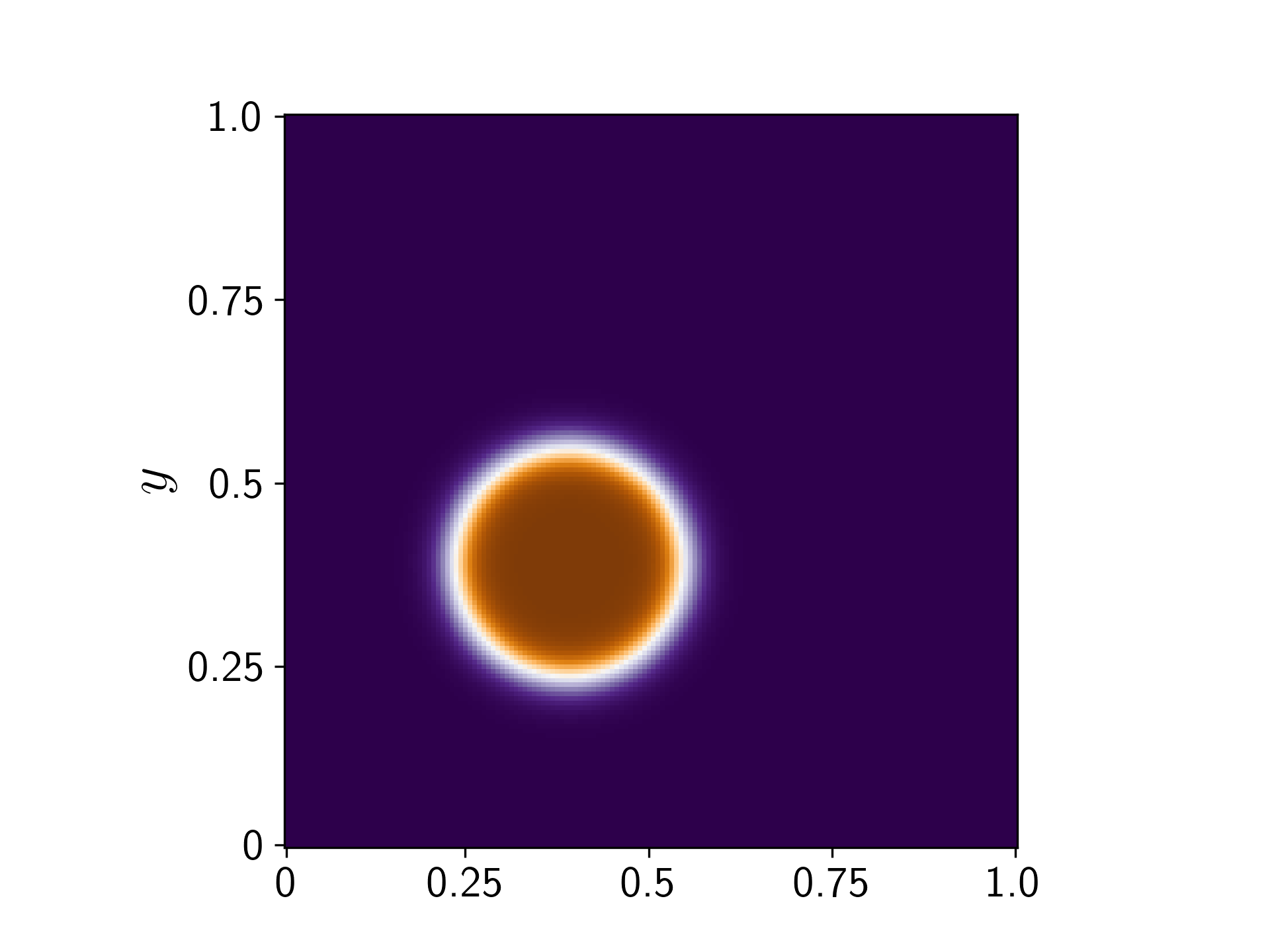}&  \hspace{-1.25cm}
        \includegraphics[trim=0.2cm 0 0.2cm 1cm, clip,height=0.18\textwidth]{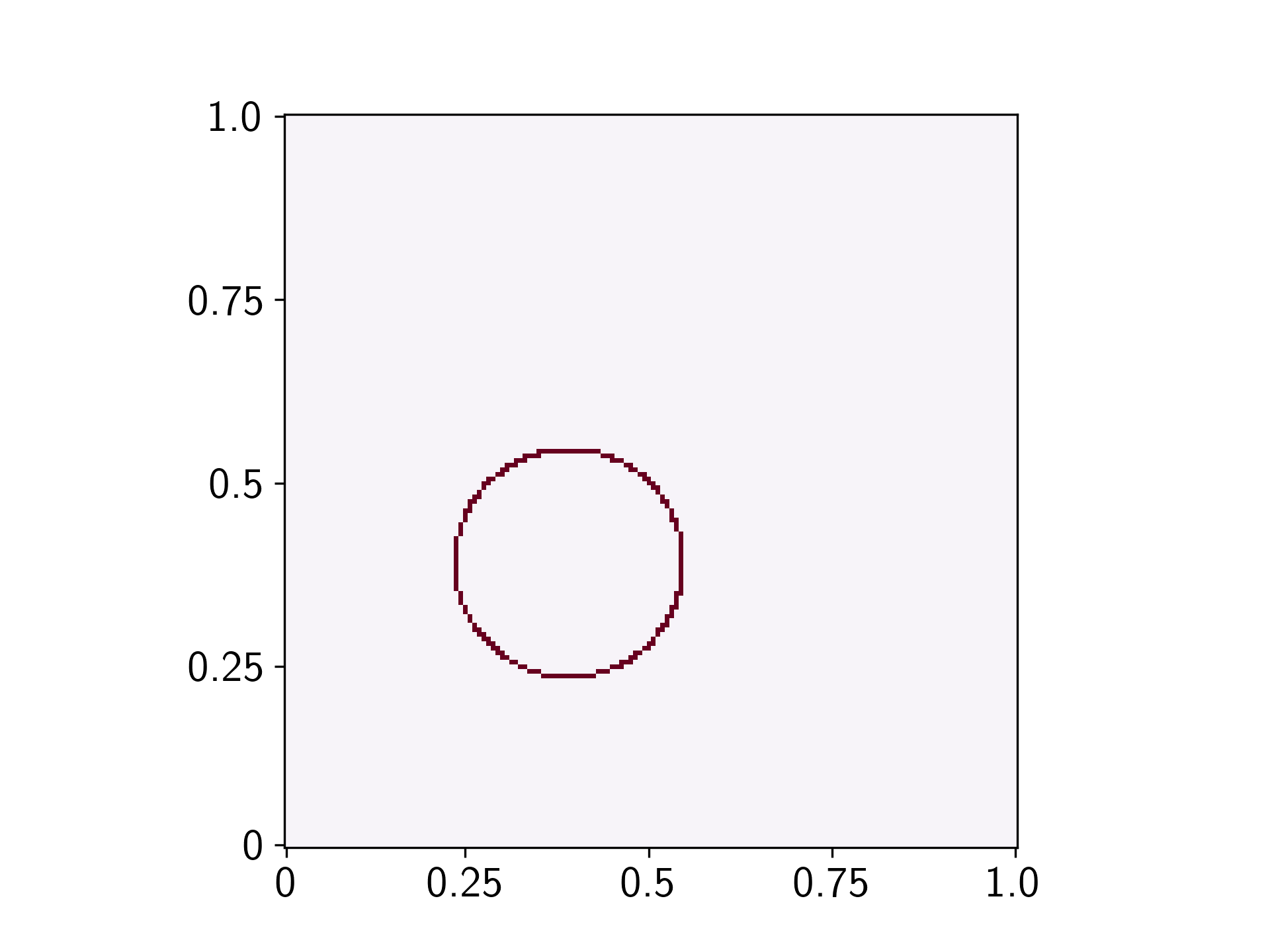}&
        \includegraphics[trim=0.2cm 0 0.2cm 1cm, clip,height=0.18\textwidth]{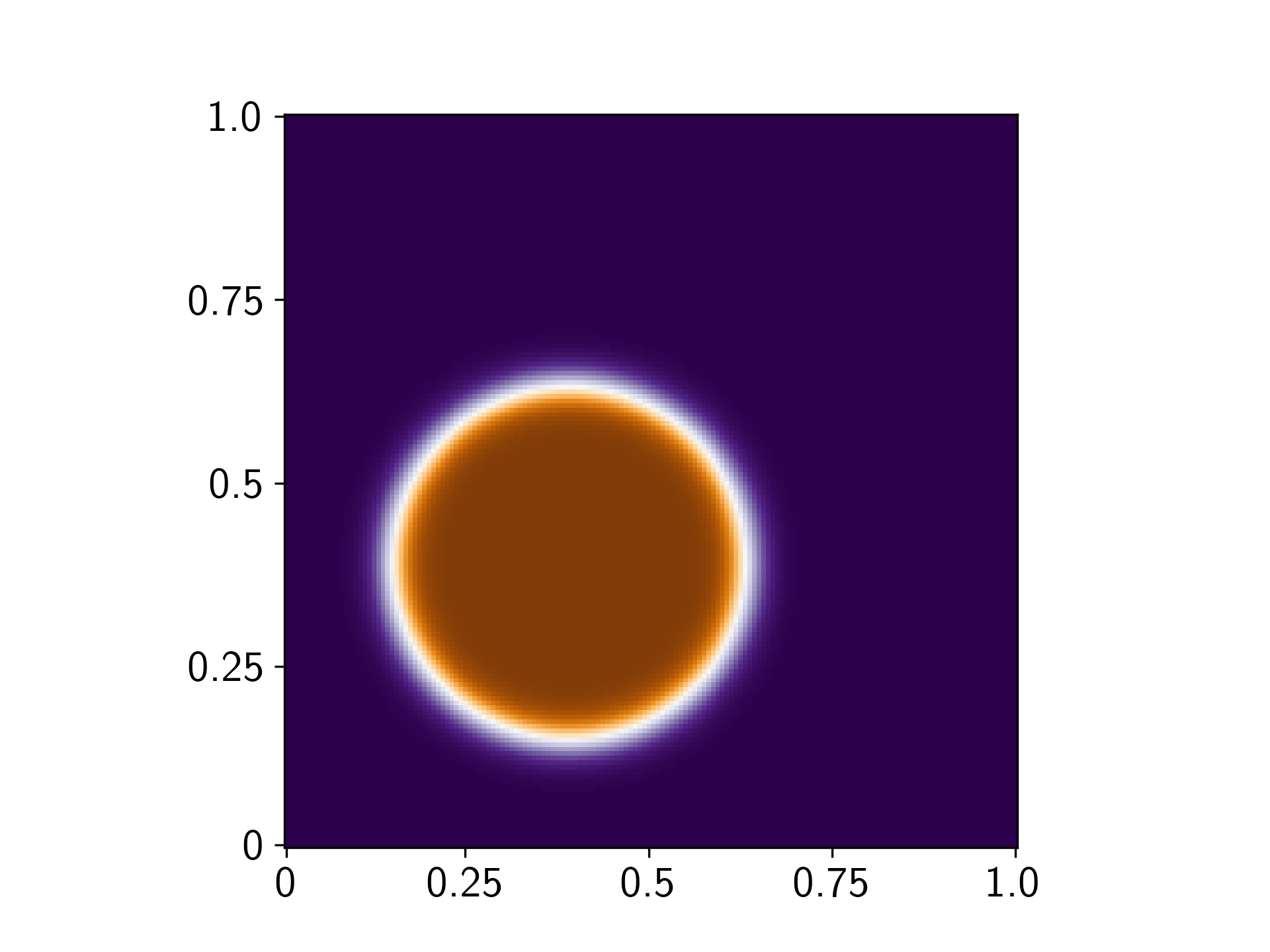}&  \hspace{-1.25cm}
        \includegraphics[trim=0.2cm 0 0.2cm 1cm, clip,height=0.18\textwidth]{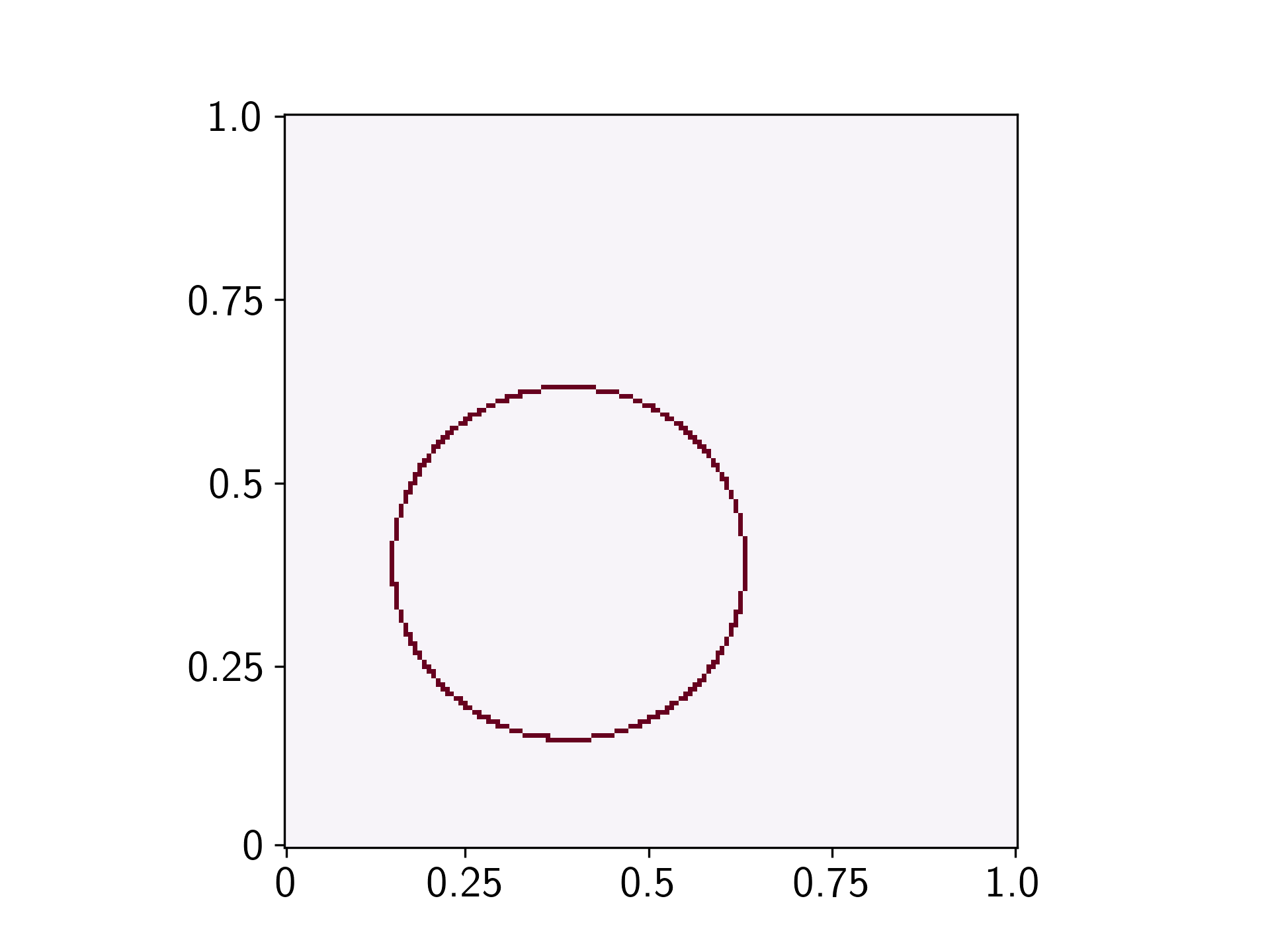}\\ \vspace{-.25cm}
        \raisebox{2.0em}{\rotatebox[origin=lb]{90}{\parbox{2cm}{\centering \footnotesize{GNN-LaSDI}}}}& \hspace{-.5cm}
        \includegraphics[trim=0.2cm 0 0.2cm 1cm, clip,height=0.18\textwidth]{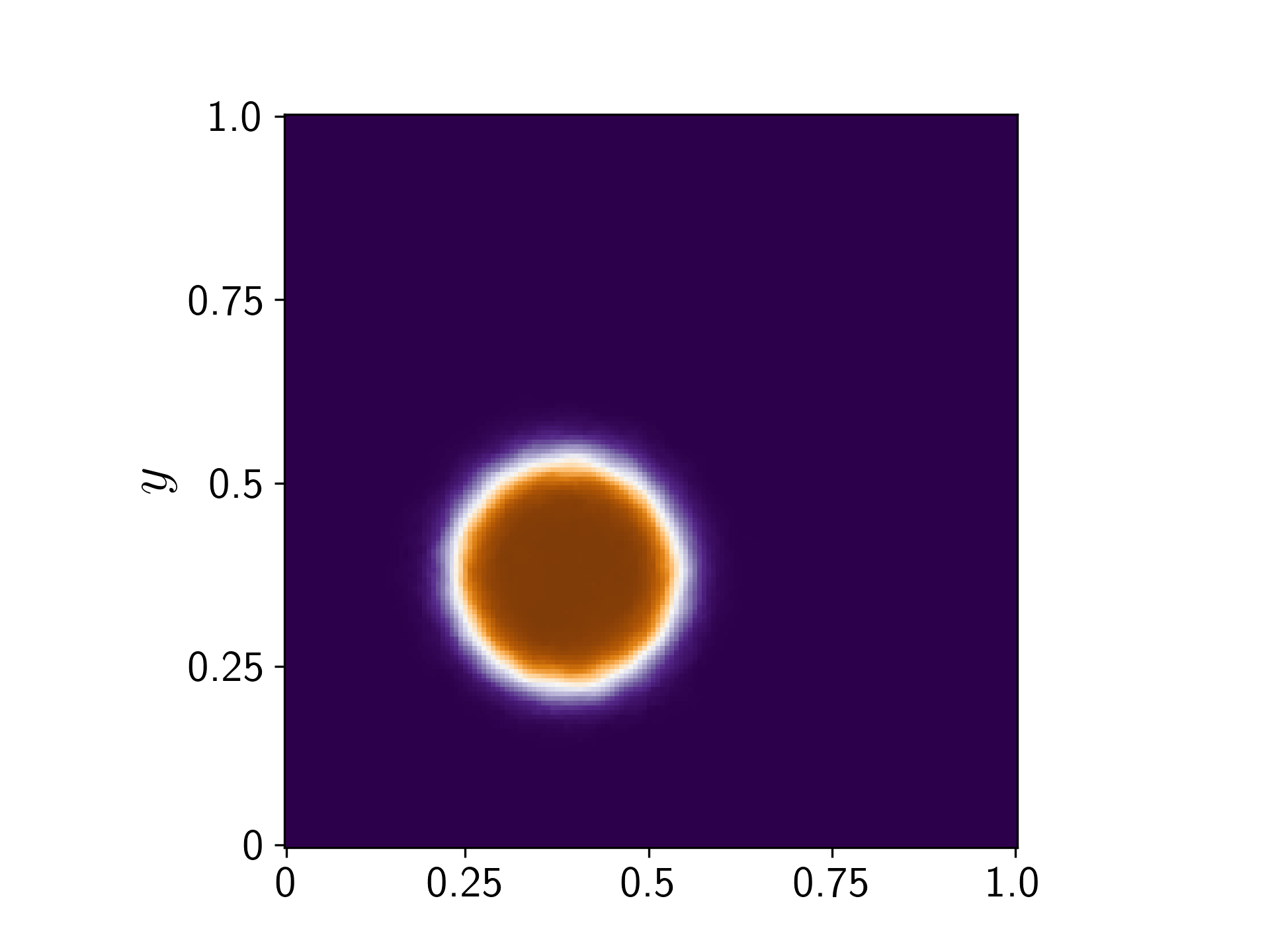}&  \hspace{-1.25cm}
        \includegraphics[trim=0.2cm 0 0.2cm 1cm, clip,height=0.18\textwidth]{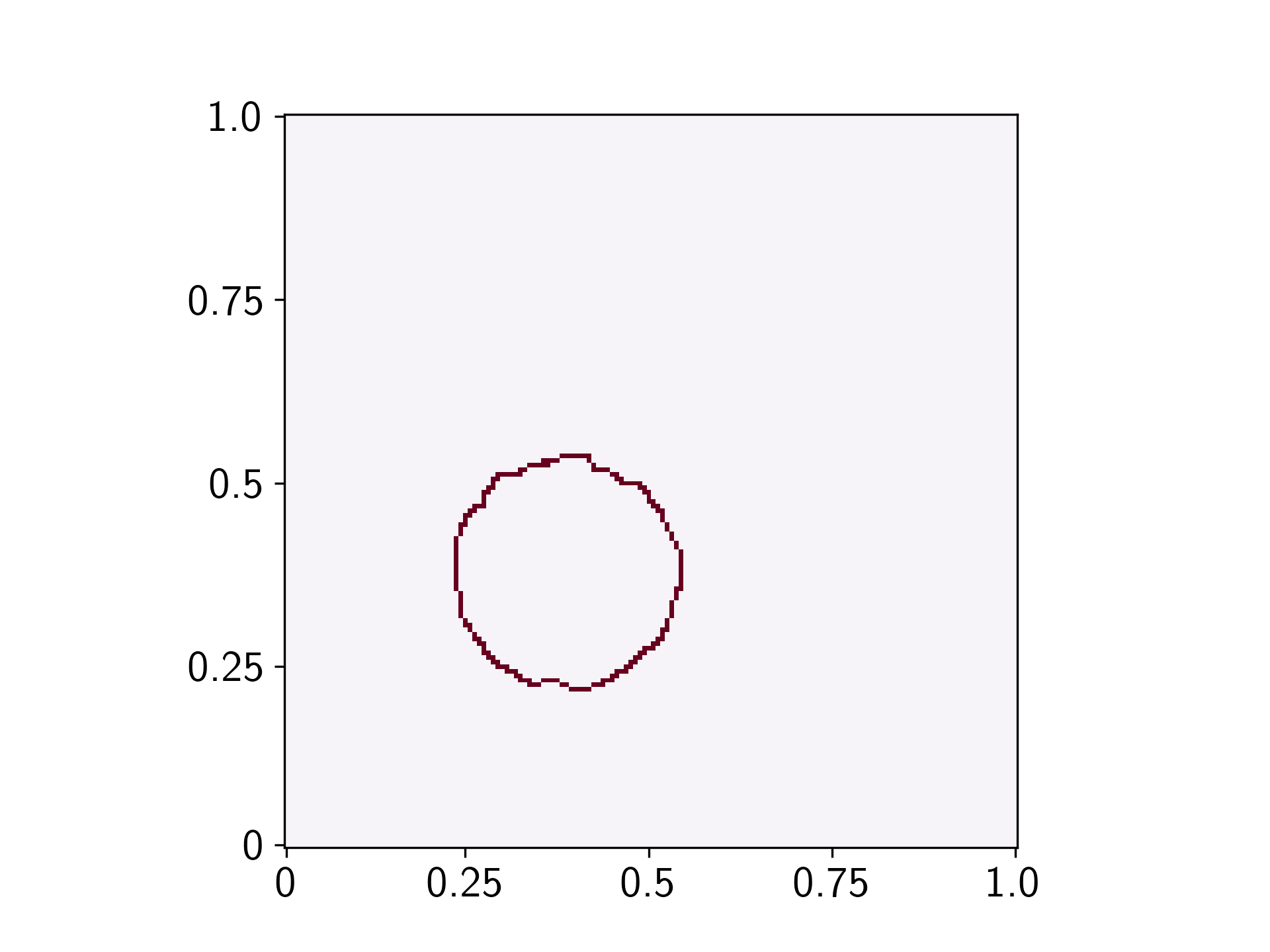}& 
        \includegraphics[trim=0.2cm 0 0.2cm 1cm, clip,height=0.18\textwidth]{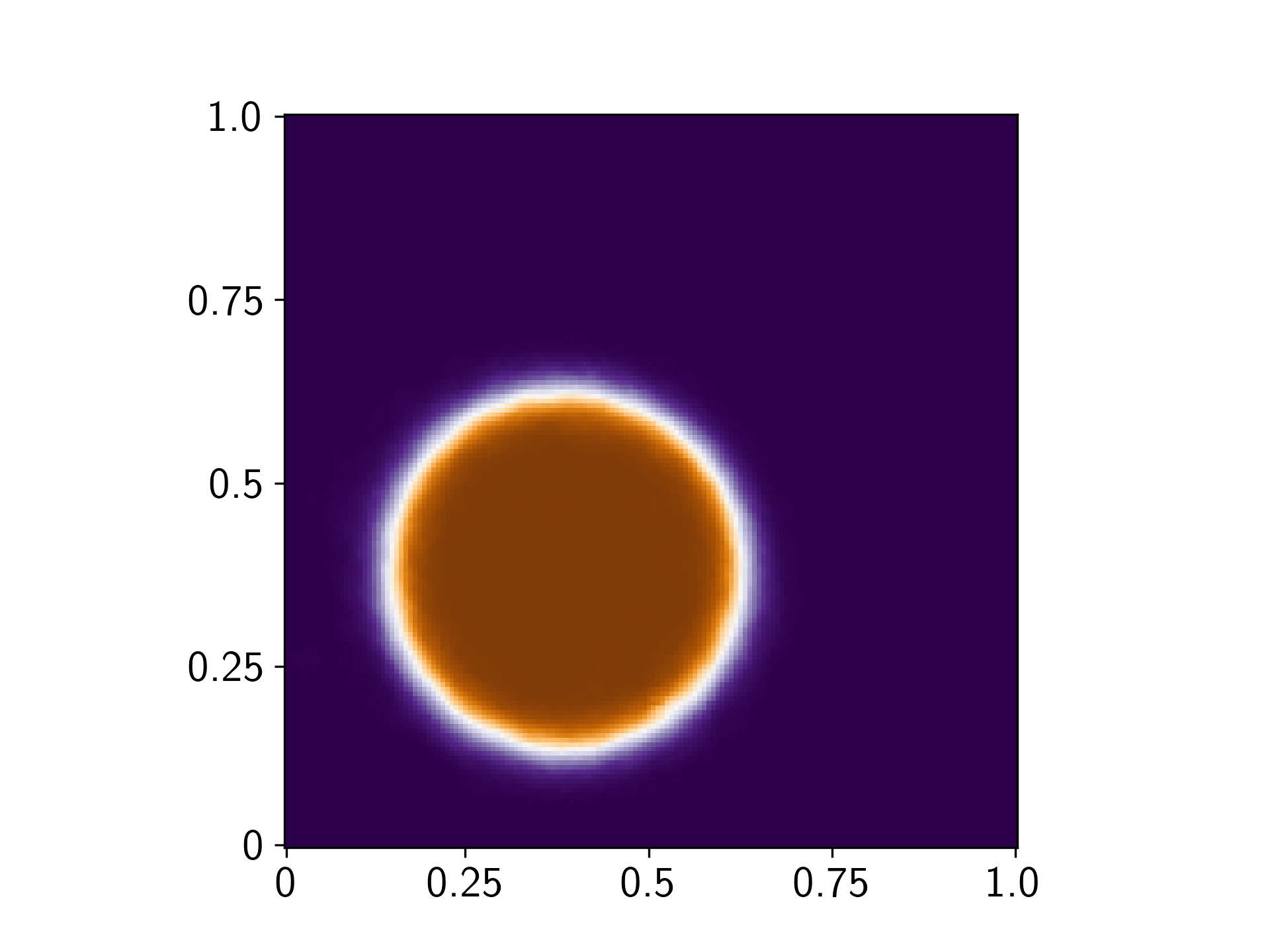}& \hspace{-1.25cm}
        \includegraphics[trim=0.2cm 0 0.2cm 1cm, clip,height=0.18\textwidth]{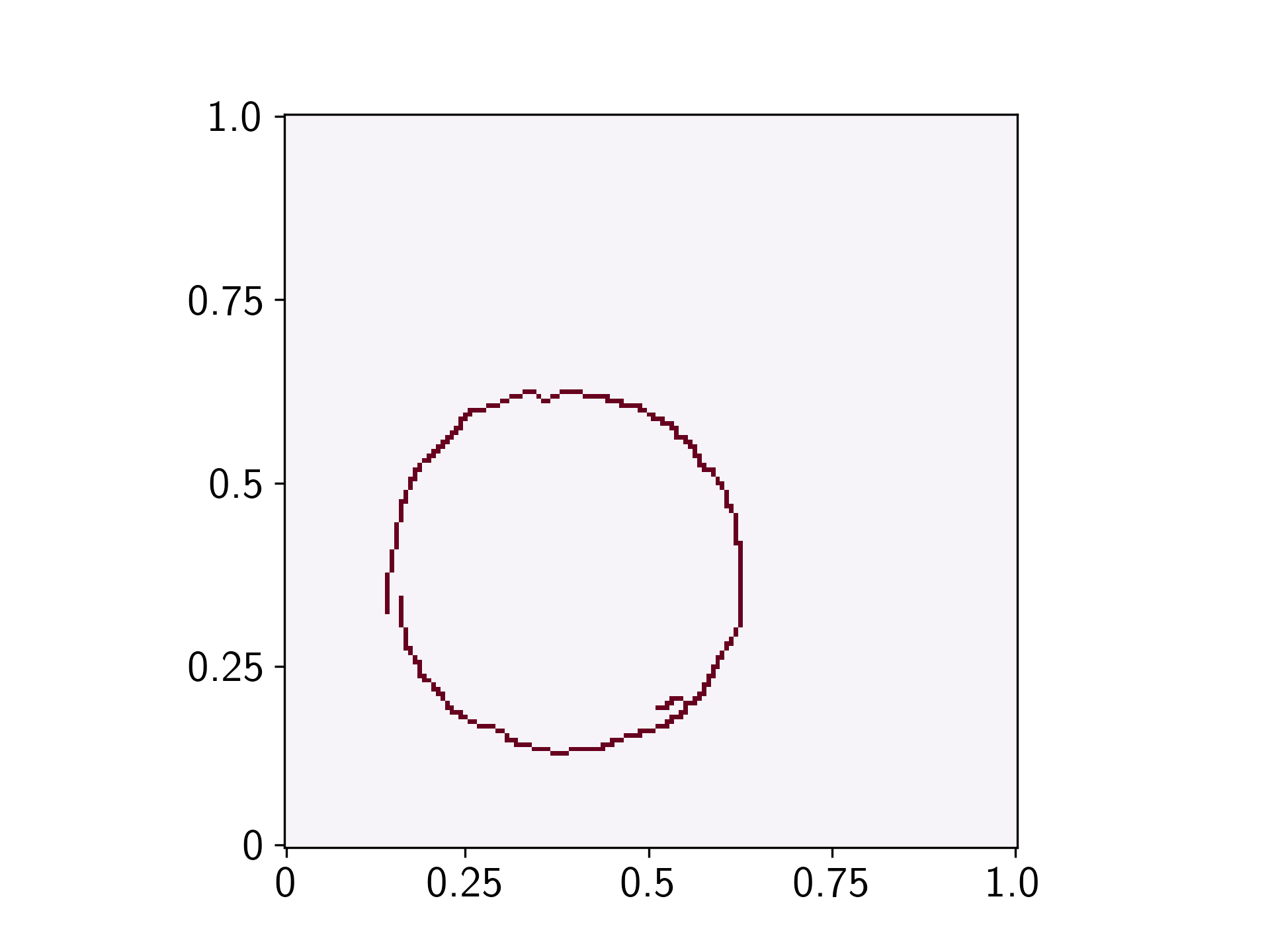}\\ \vspace{-.25cm}
        \raisebox{2.0em}{\rotatebox[origin=lb]{90}{\parbox{2cm}{\centering \footnotesize{GD-LSPG}}}}&  \hspace{-.5cm}
        \includegraphics[trim=0.2cm 0 0.2cm 1cm, clip,height=0.18\textwidth]{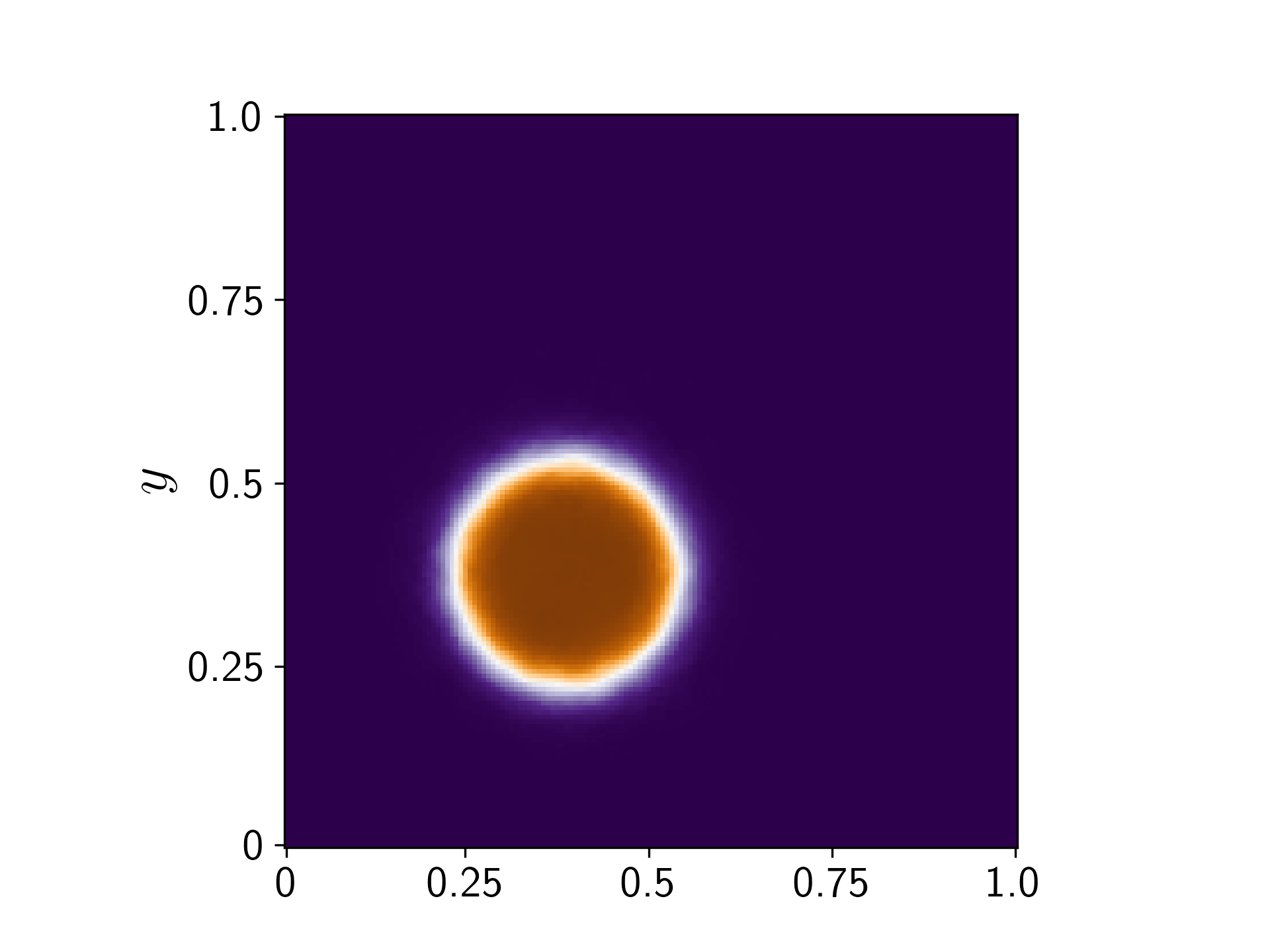}& \hspace{-1.25cm}
        \includegraphics[trim=0.2cm 0 0.2cm 1cm, clip,height=0.18\textwidth]{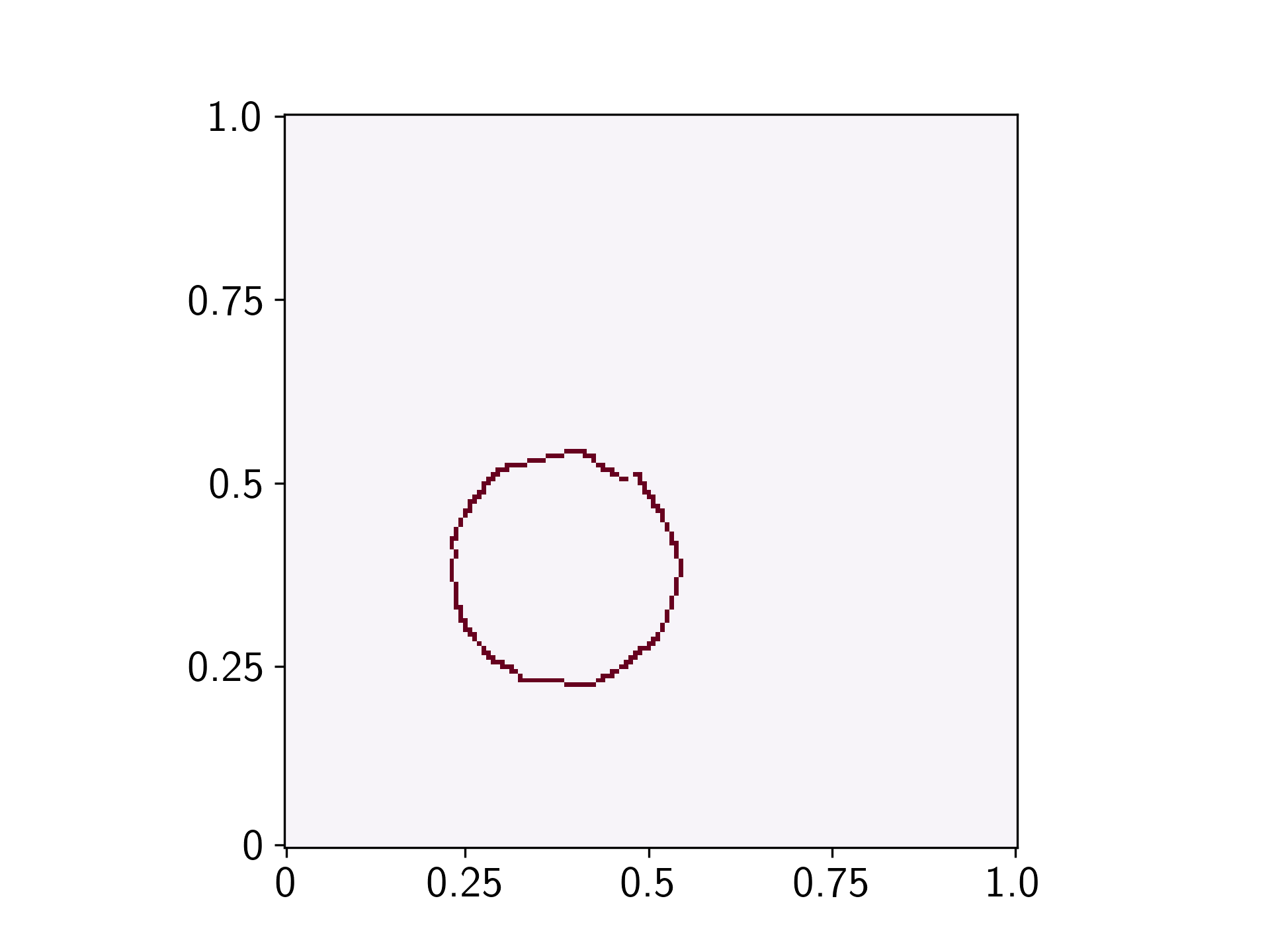}& 
        \includegraphics[trim=0.2cm 0 0.2cm 1cm, clip,height=0.18\textwidth]{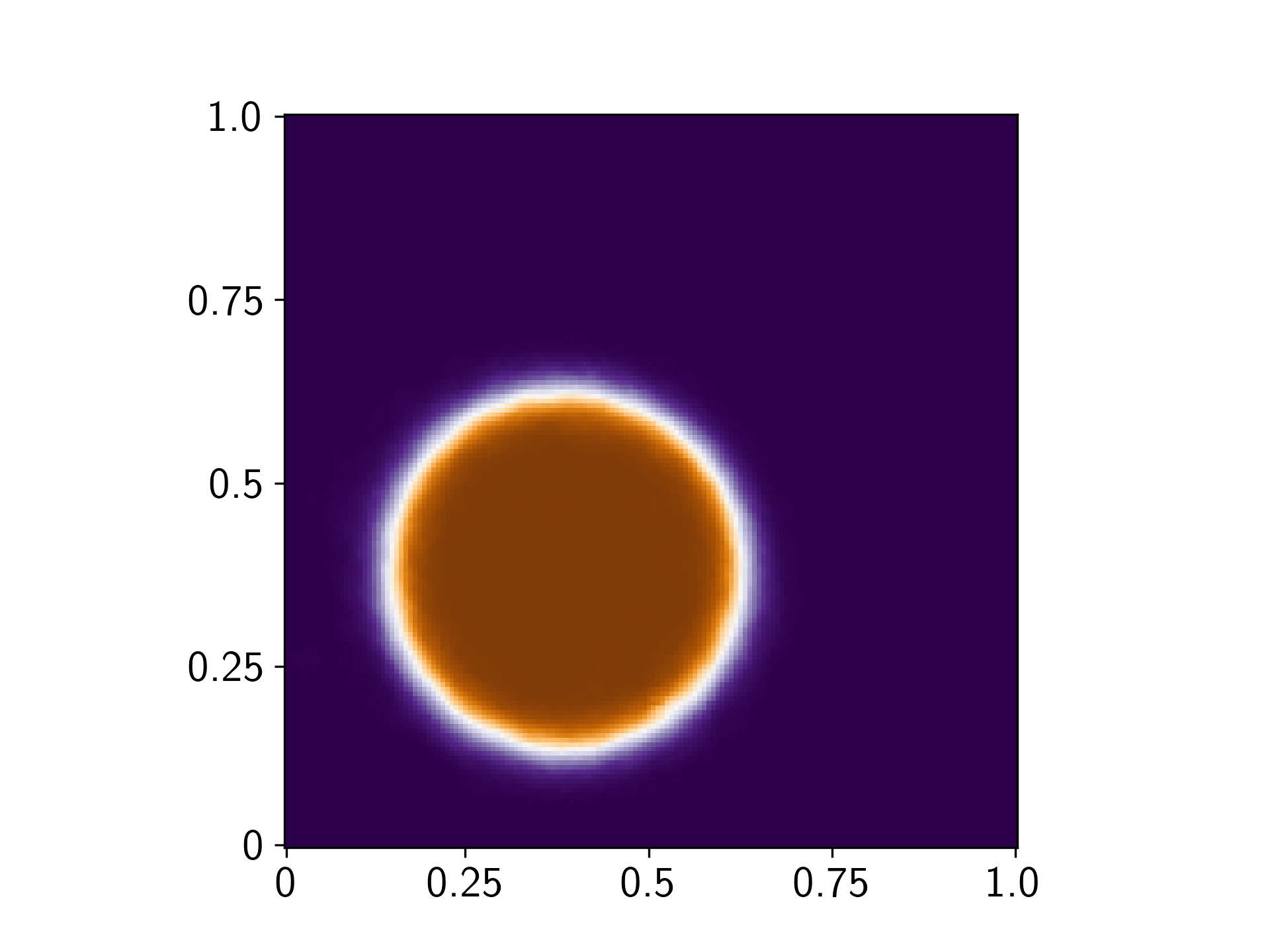}& \hspace{-1.25cm}
        \includegraphics[trim=0.2cm 0 0.2cm 1cm, clip,height=0.18\textwidth]{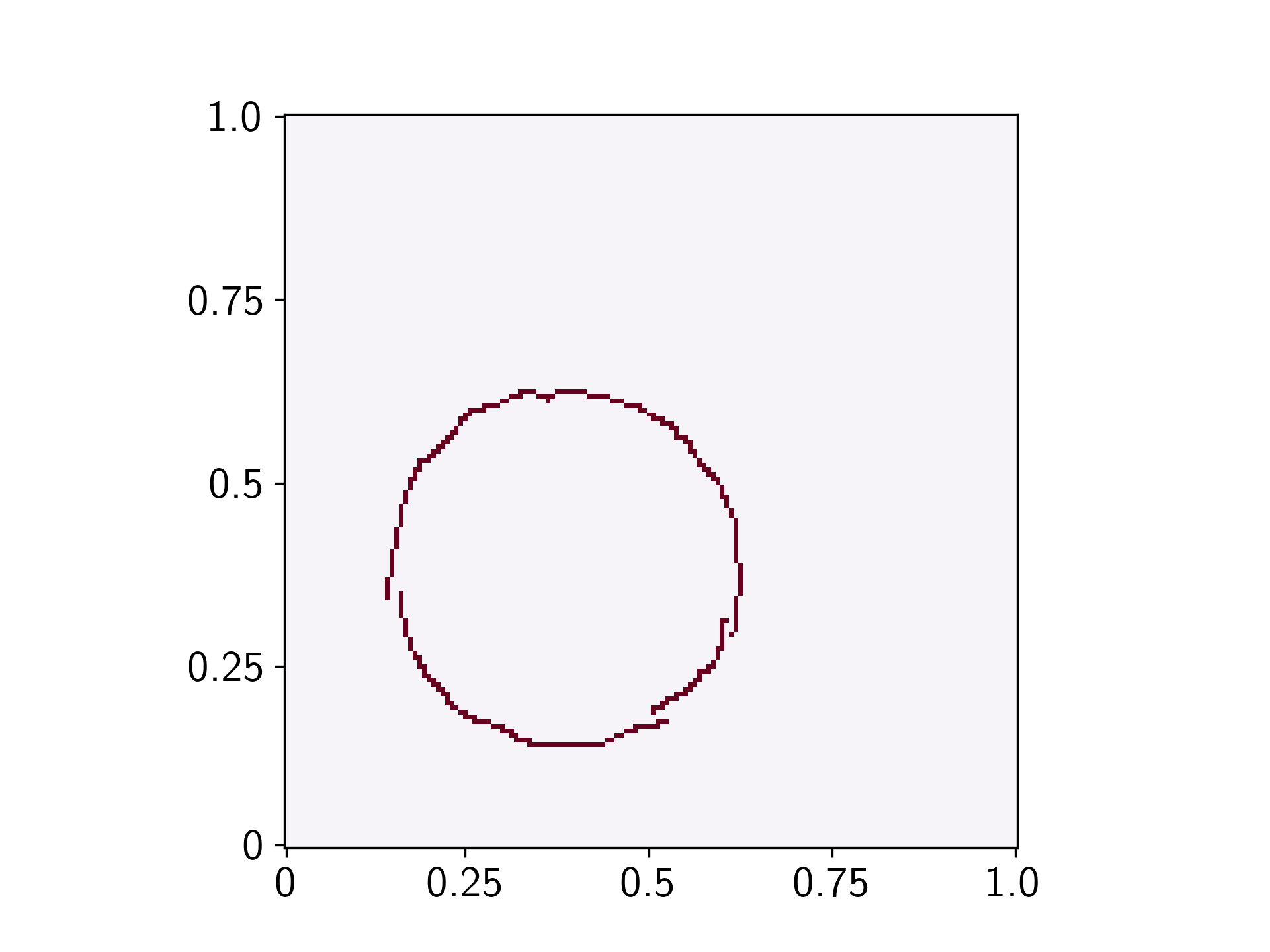}\\ 
        \raisebox{2.0em}{\rotatebox[origin=lb]{90}{\parbox{2cm}{\centering \footnotesize{POD-LaSDI}}}}& \hspace{-.5cm}
        \includegraphics[trim=0.2cm 0 0.2cm 1cm, clip,height=0.18\textwidth]{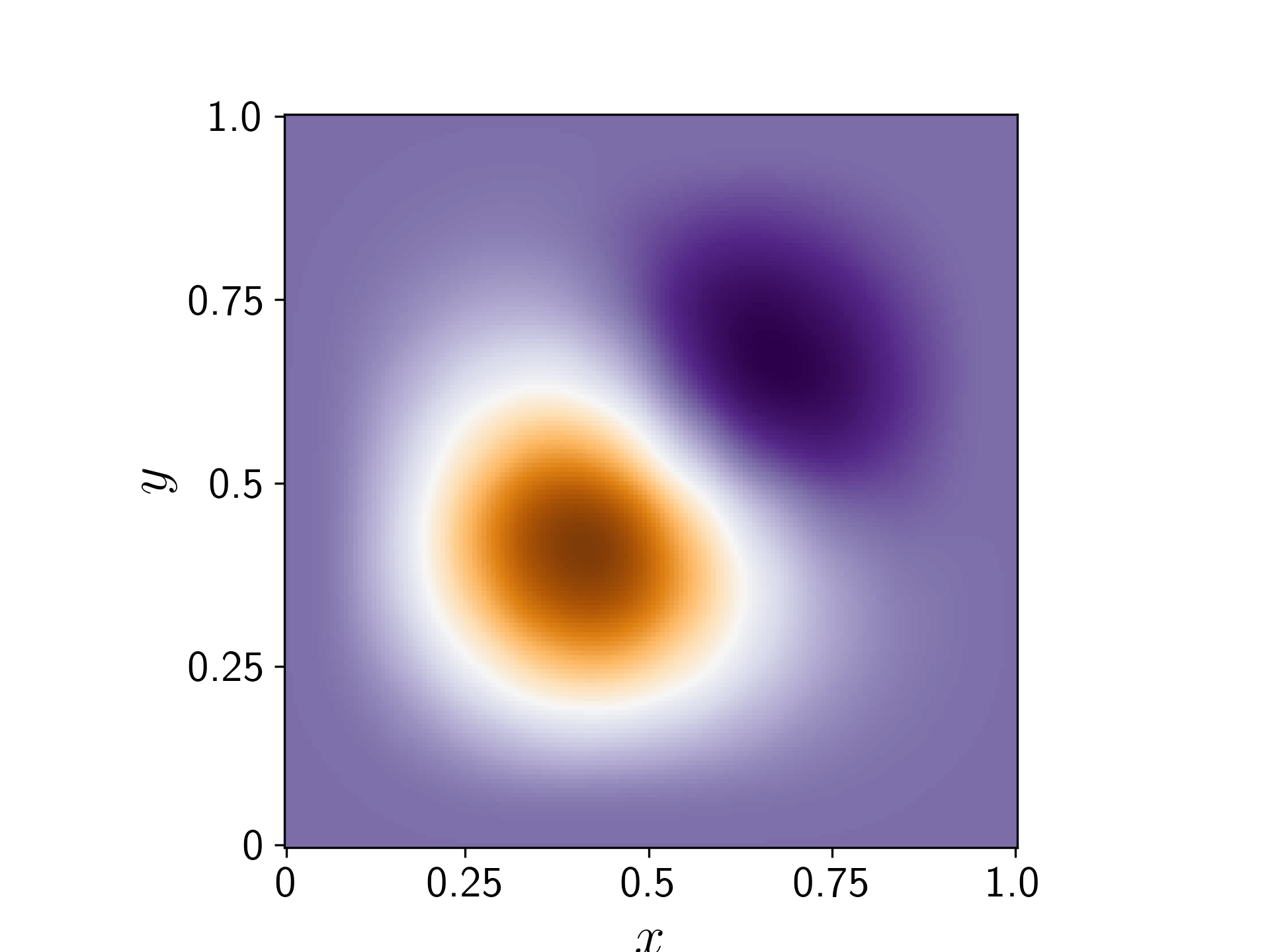}& \hspace{-1.25cm}
        \includegraphics[trim=0.2cm 0 0.2cm 1cm, clip,height=0.18\textwidth]{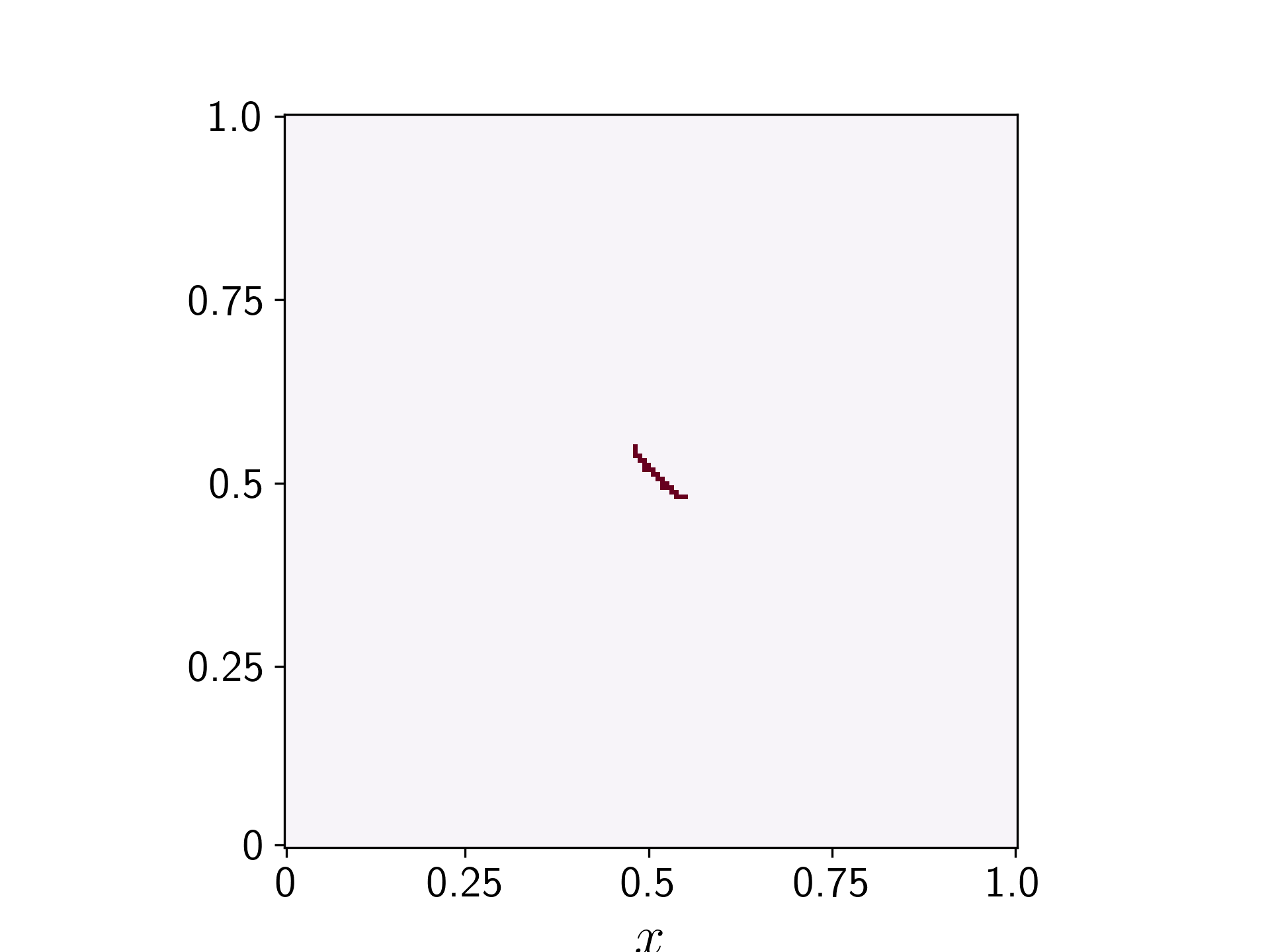}& 
        \includegraphics[trim=0.2cm 0 0.2cm 1cm, clip,height=0.18\textwidth]{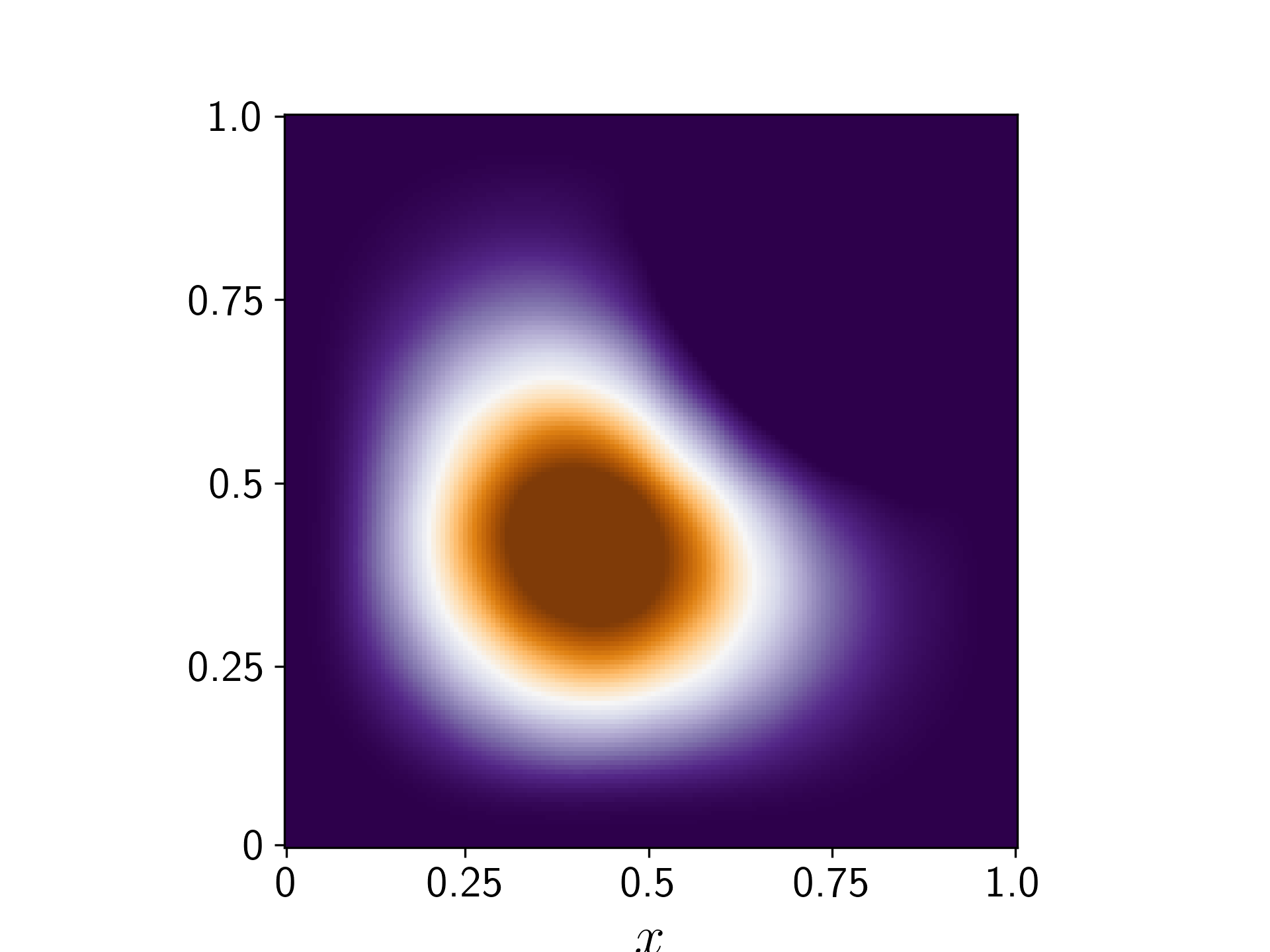}& \hspace{-1.25cm}
        \includegraphics[trim=0.2cm 0 0.2cm 1cm, clip,height=0.18\textwidth]{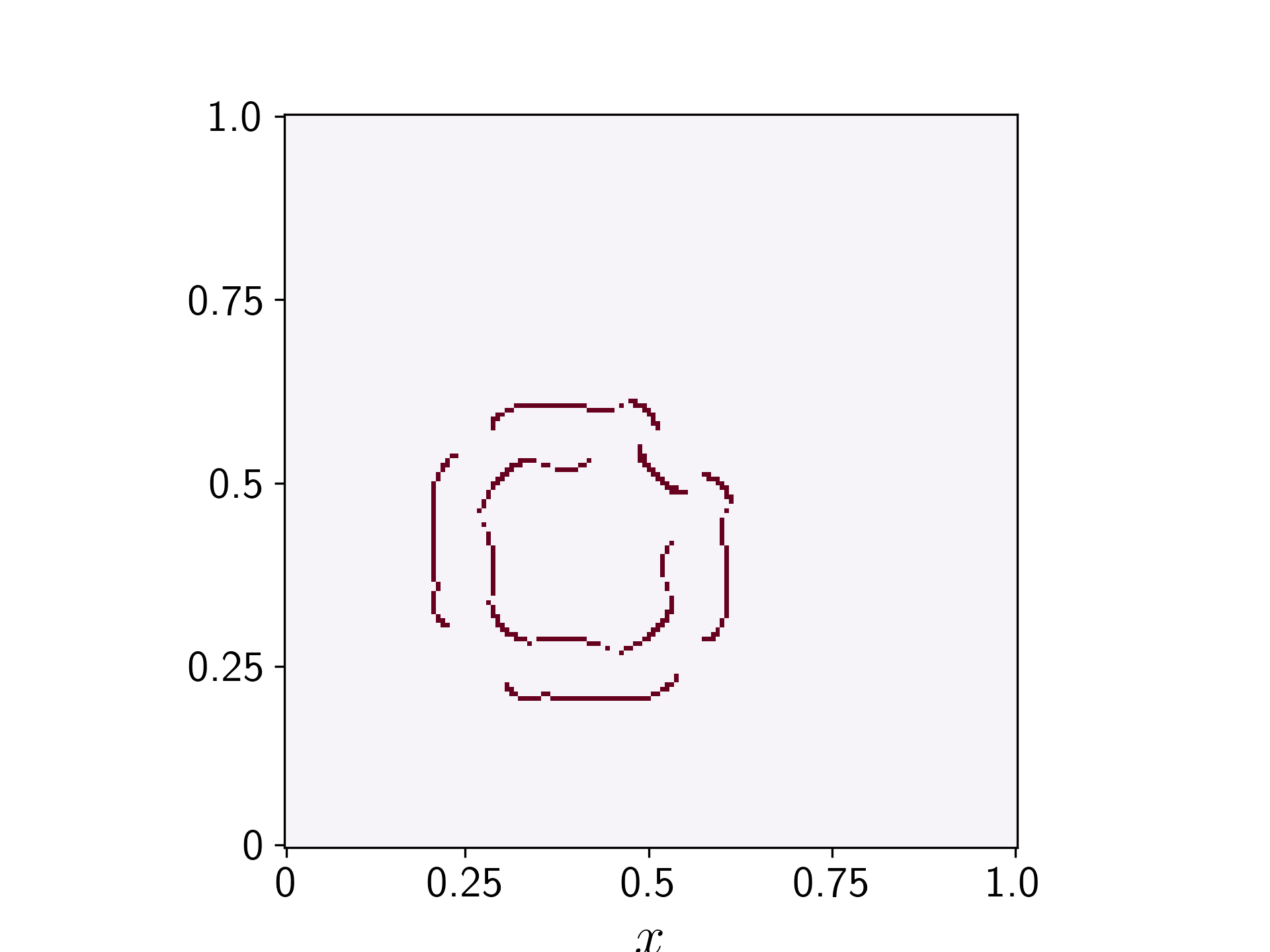}\\
        &  \multicolumn{4}{c}{\hspace{-0.25cm}\includegraphics[scale=.45]{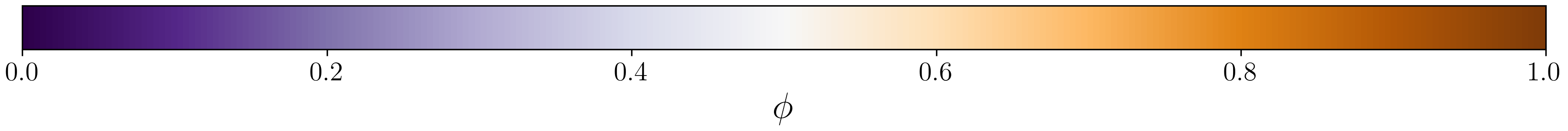}}
    \end{tabular}
    \captionsetup{justification=centering}
    \caption{Order parameter state solutions and corresponding point cloud representations of the solid-liquid interface obtained via Canny edge detection for the FOM (first row), GNN-LaSDI (second row), GD-LSPG (third row), and POD-LaSDI (fourth row) at times $t=0.005$ and $t=0.01$ for test parameter set $\boldsymbol{\mu} = (0.3875, 0.3875)$. All ROM results correspond to a latent state dimension of $n=3$. The point cloud representations indicate that GNN-LaSDI and GD-LSPG accurately capture the location and evolution of the moving solid-liquid interface, producing interfaces that closely resemble that of the FOM. In contrast, POD-LaSDI yields a diffuse solution and fails to preserve the interface geometry. The shared colorbar corresponds to the phase field order parameter, $\phi$, while dark red markers in the point cloud plots denote points identified as belonging to the interface. (Online version in color.)}
    \label{fig:ac_P_pc}
\end{figure}

Next, the point cloud error metric is used to assess the accuracy of each ROM in predicting the location of the solid-liquid interface. For this numerical experiment, the characteristic length in error \eqref{eq:pc_error} is chosen as the height or width of the spatial domain (i.e., $L=1$). The point cloud representation of the interface, $\mathcal{X}^m$, is obtained by applying Canny edge detection \cite{canny1986edge} to the phase-field order parameter state solution. Canny edge detection has been used in computer vision to detect edges in pixel images. The procedure consists of noise filtering via a Gaussian filter (chosen here to have a standard deviation $\sigma=1$), gradient computation, and selection of points corresponding to local maxima in the gradient magnitude. In this setting, these detected edge points provide a point cloud representation of the solid-liquid interface. Figure \ref{fig:ac_P_pc} presents the order parameter state solutions and their corresponding point cloud representations for the FOM, GNN-LaSDI, GD-LSPG, and POD-LaSDI. Results are shown at $t=0.005$ and $t=0.01$ for the test parameter instance $\boldsymbol{\mu} = (0.3875, 0.3875)$. ROM results are presented for latent state dimension $n=3$. The results demonstrate that GNN-LaSDI and GD-LSPG accurately capture the location of the moving solid-liquid interface, whereas POD-LaSDI fails to preserve a sharp interface and produces a noticeably more diffuse solution than the FOM. Additionally, Canny edge detection produces a meaningful point cloud representation of the interface. As expected, the FOM's point cloud representation of the interface is circular, with a radius that increases over time. GNN-LaSDI and GD-LSPG yield similar circular point cloud representations of the interface, with minor deviations from the FOM. In contrast, the POD-LaSDI point cloud is visibly distorted due to the diffuse nature of POD-LaSDI state solutions.

\begin{figure}[ht!]
    \centering
     \begin{tabular}{cccc}
        & {\footnotesize{$n=3$}} & {\footnotesize{$n=5$}} & {\footnotesize{$n=7$}} \\
        \raisebox{2.25em}{\rotatebox[origin=lb]{90}{\parbox{2.5cm}{\centering \footnotesize{$\boldsymbol{\mu}=(0.3875,0.3875)$}}}}&
        \includegraphics[trim=0.2cm 0 0.2cm 1cm, clip,height=0.22\textwidth]{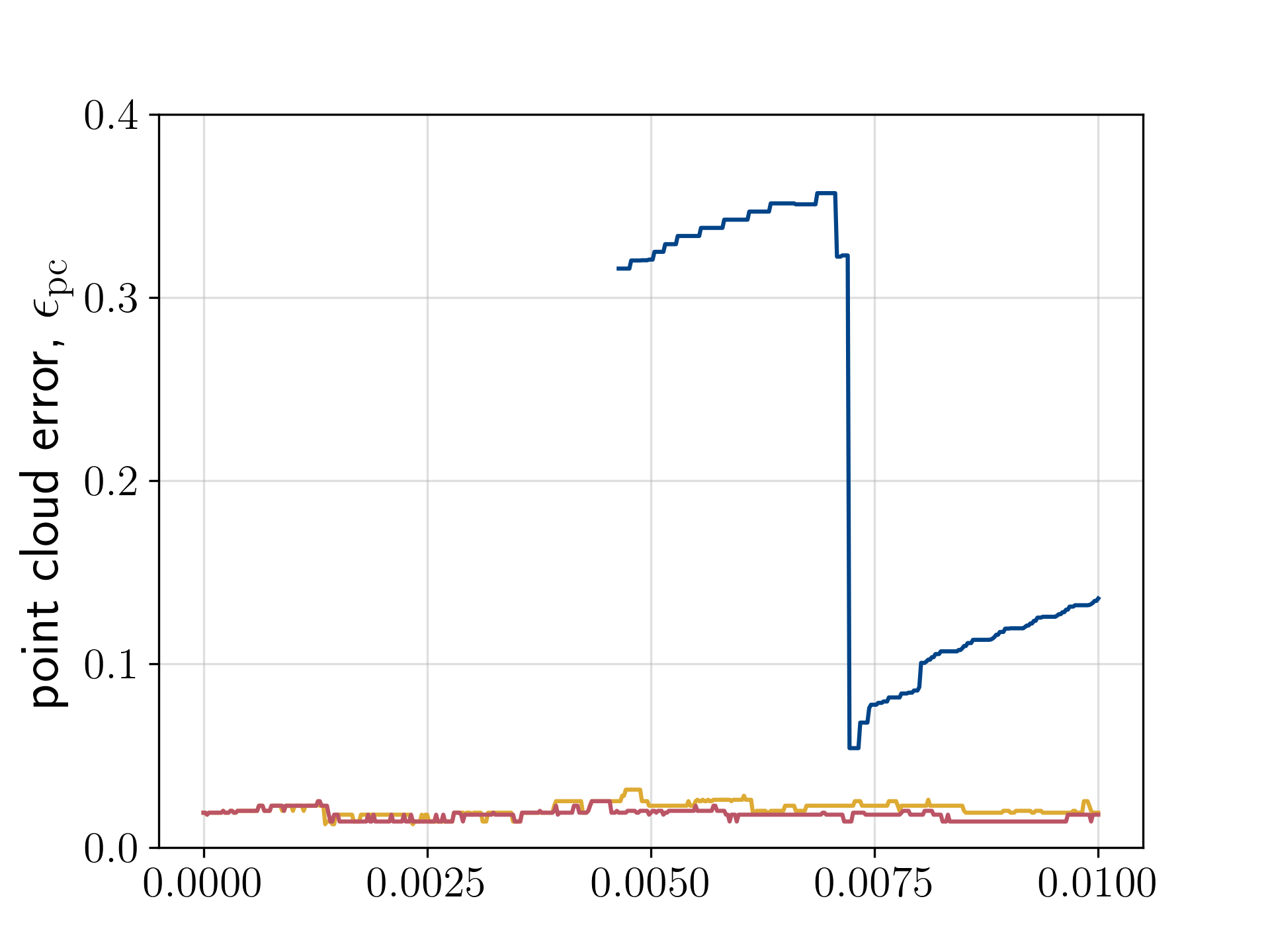}&\hspace{-.5cm}
        \includegraphics[trim=0.2cm 0 0.2cm 1cm, clip,height=0.22\textwidth]{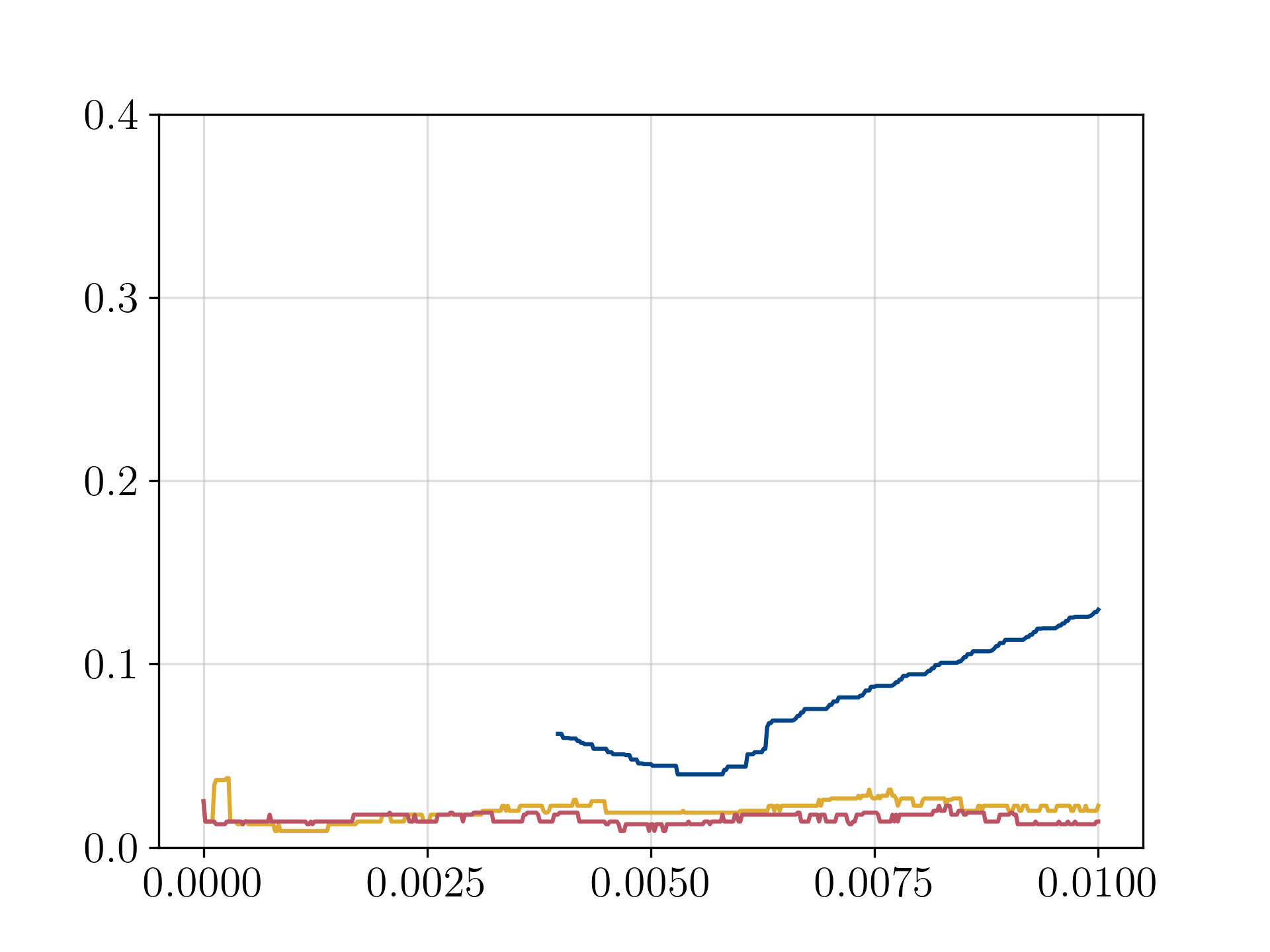}&\hspace{-.5cm}
        \includegraphics[trim=0.2cm 0 0.2cm 1cm, clip,height=0.22\textwidth]{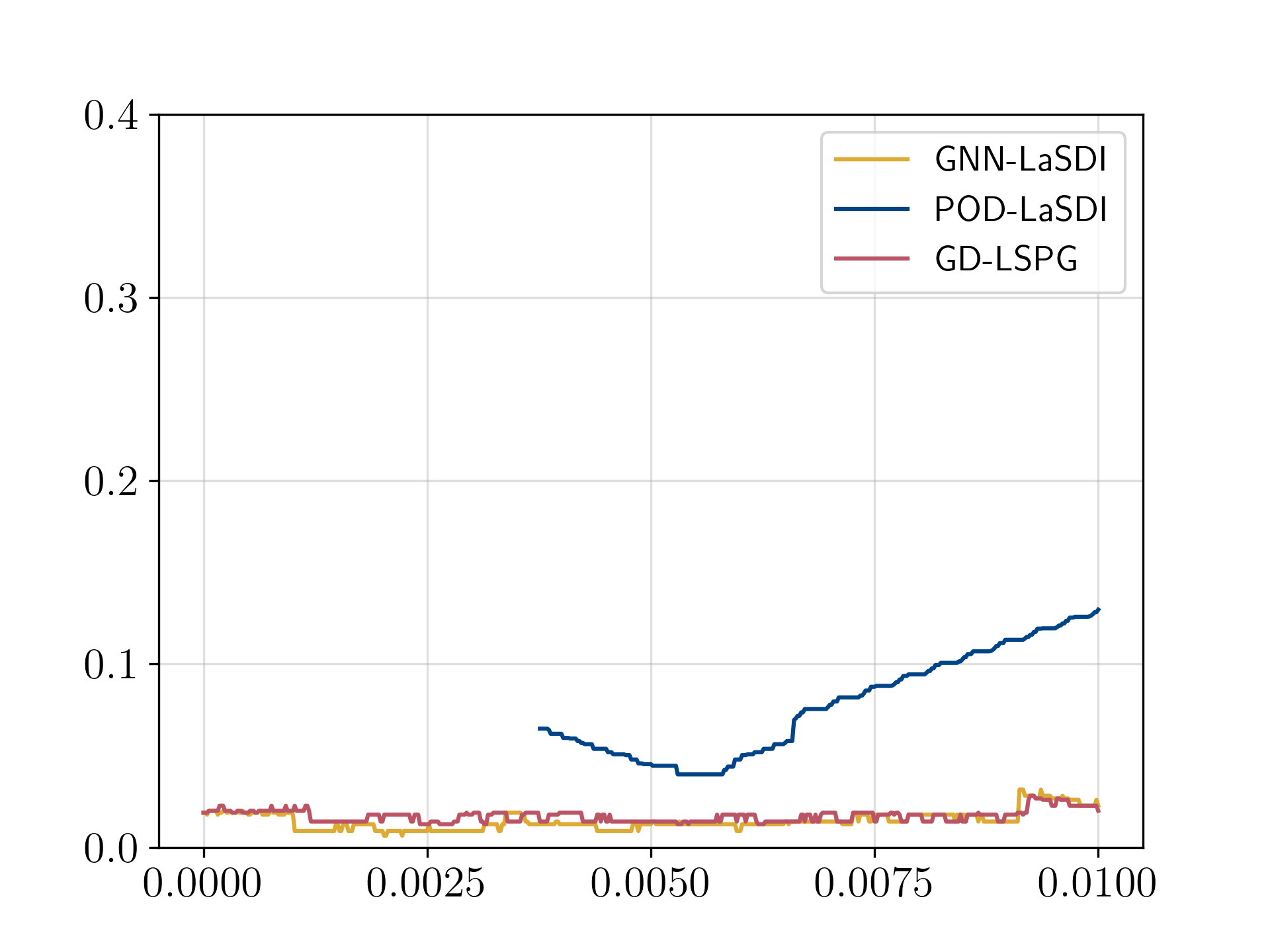}\\
        \raisebox{2.25em}{\rotatebox[origin=lb]{90}{\parbox{2.5cm}{\centering \footnotesize{$\boldsymbol{\mu}=(0.3875,0.6125)$}}}}&
        \includegraphics[trim=0.2cm 0 0.2cm 1cm, clip,height=0.22\textwidth]{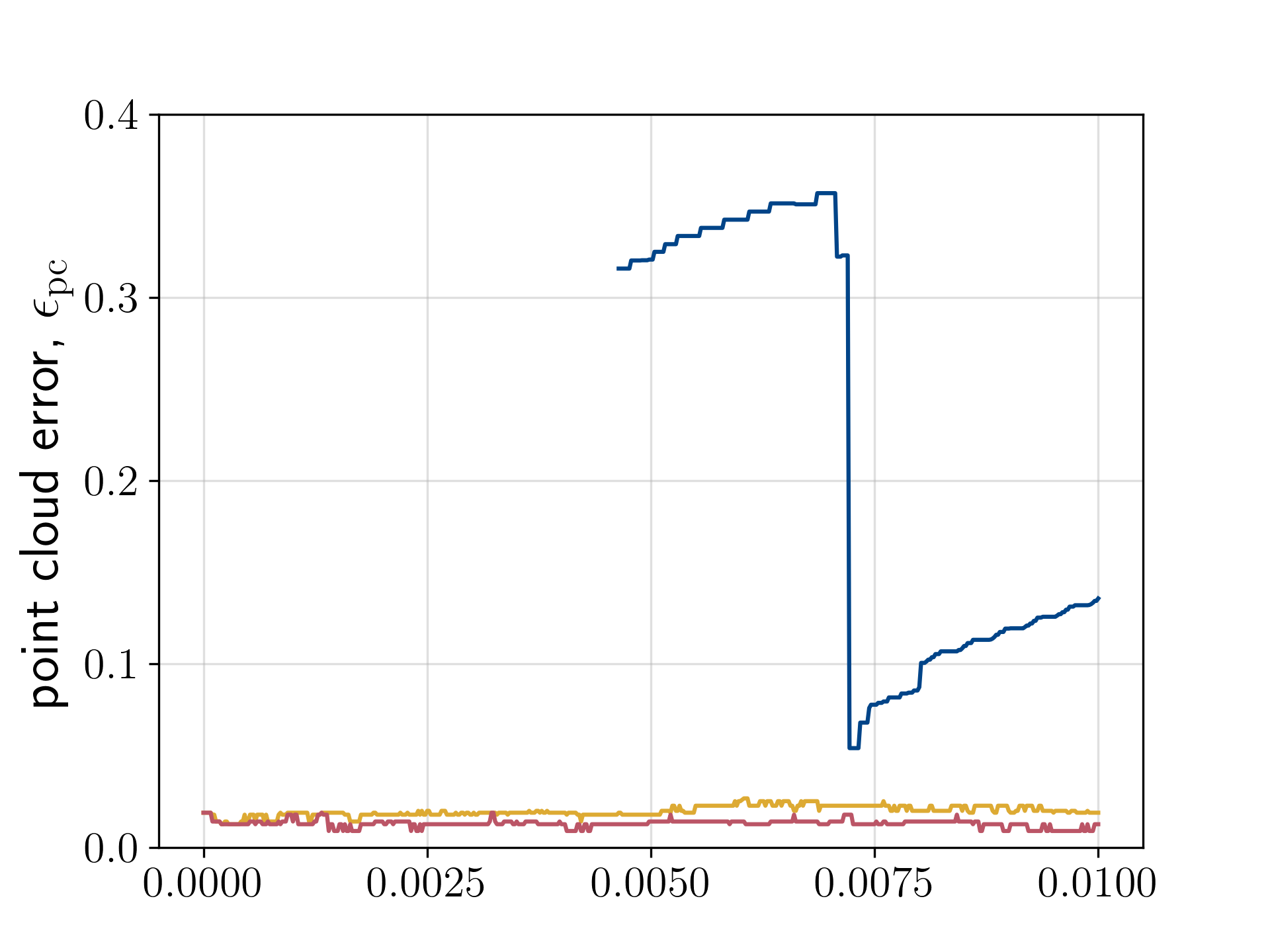}&\hspace{-.5cm}
        \includegraphics[trim=0.2cm 0 0.2cm 1cm, clip,height=0.22\textwidth]{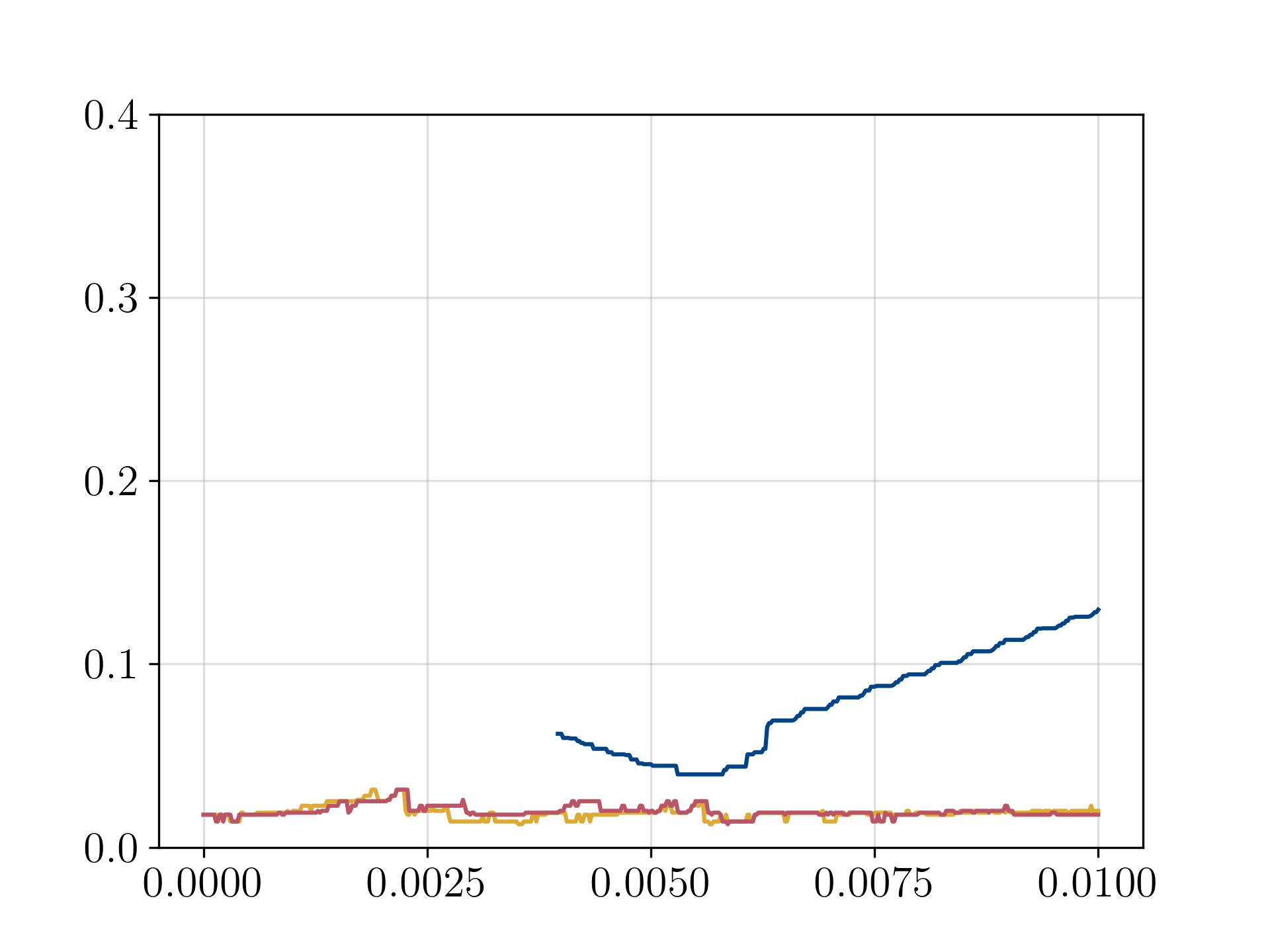}&\hspace{-.5cm}
        \includegraphics[trim=0.2cm 0 0.2cm 1cm, clip,height=0.22\textwidth]{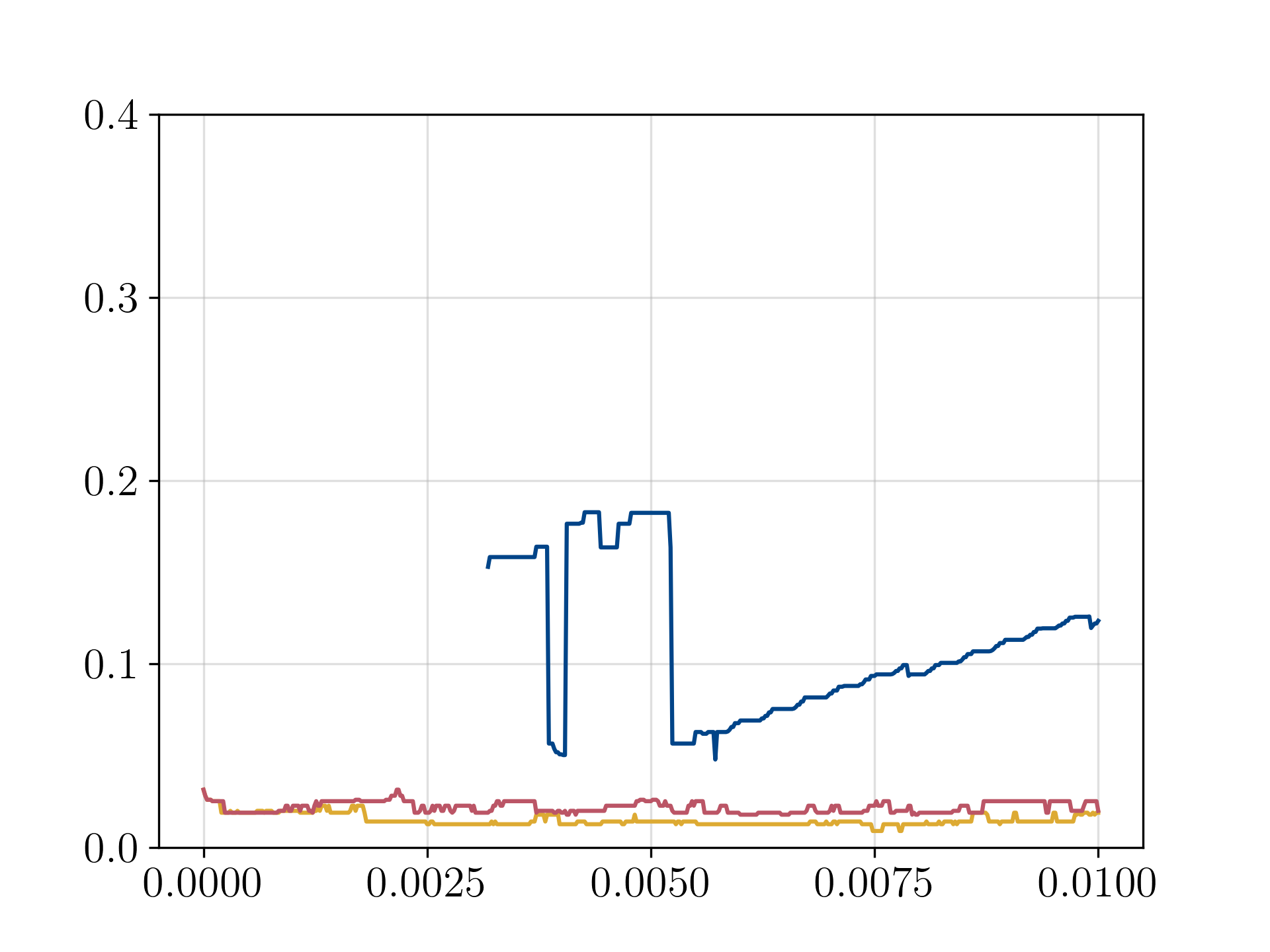}\\
        \raisebox{2.25em}{\rotatebox[origin=lb]{90}{\parbox{2.5cm}{\centering \footnotesize{$\boldsymbol{\mu}=(0.6125,0.3875)$}}}}&
        \includegraphics[trim=0.2cm 0 0.2cm 1cm, clip,height=0.22\textwidth]{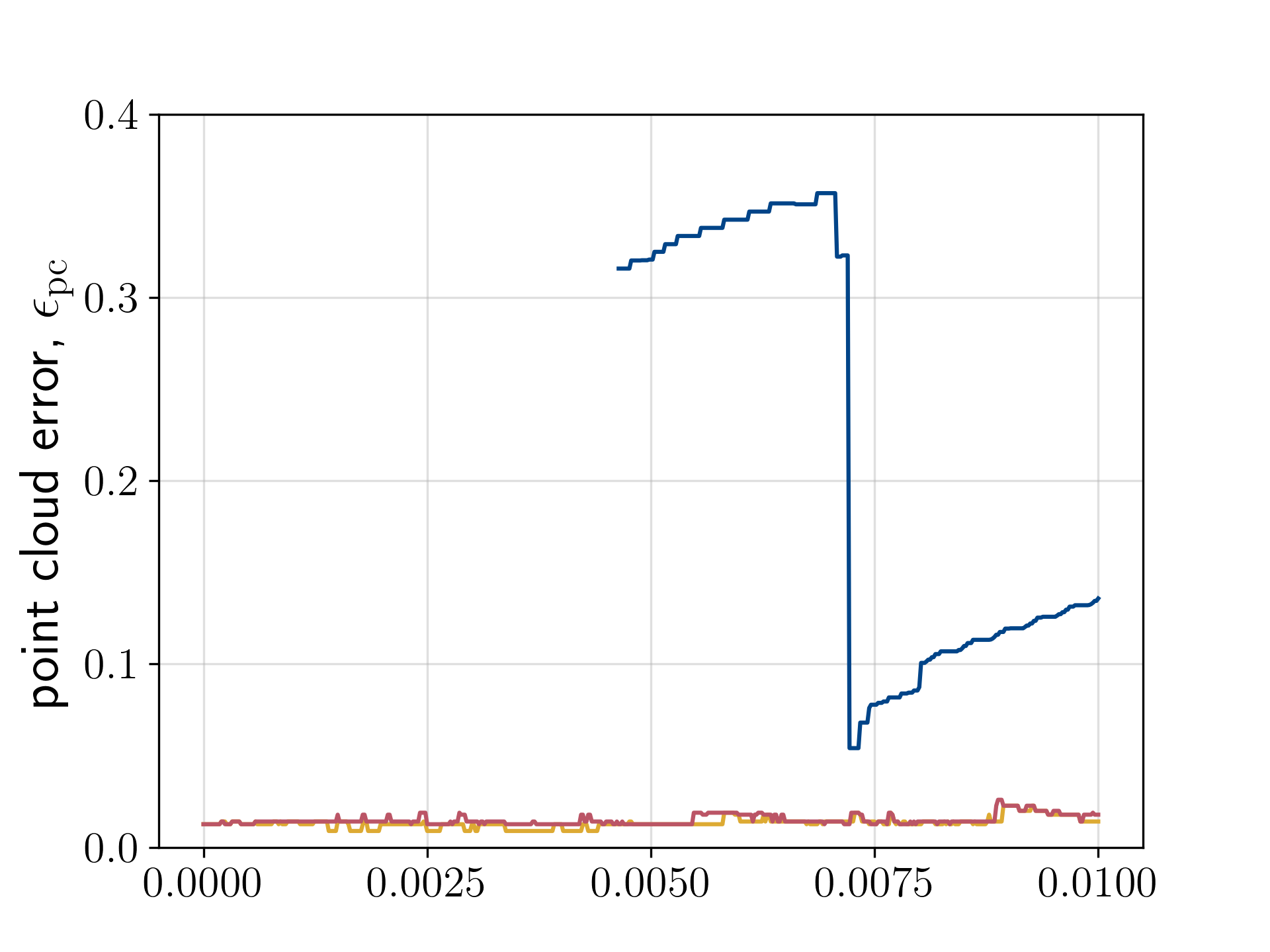}&\hspace{-.5cm}
        \includegraphics[trim=0.2cm 0 0.2cm 1cm, clip,height=0.22\textwidth]{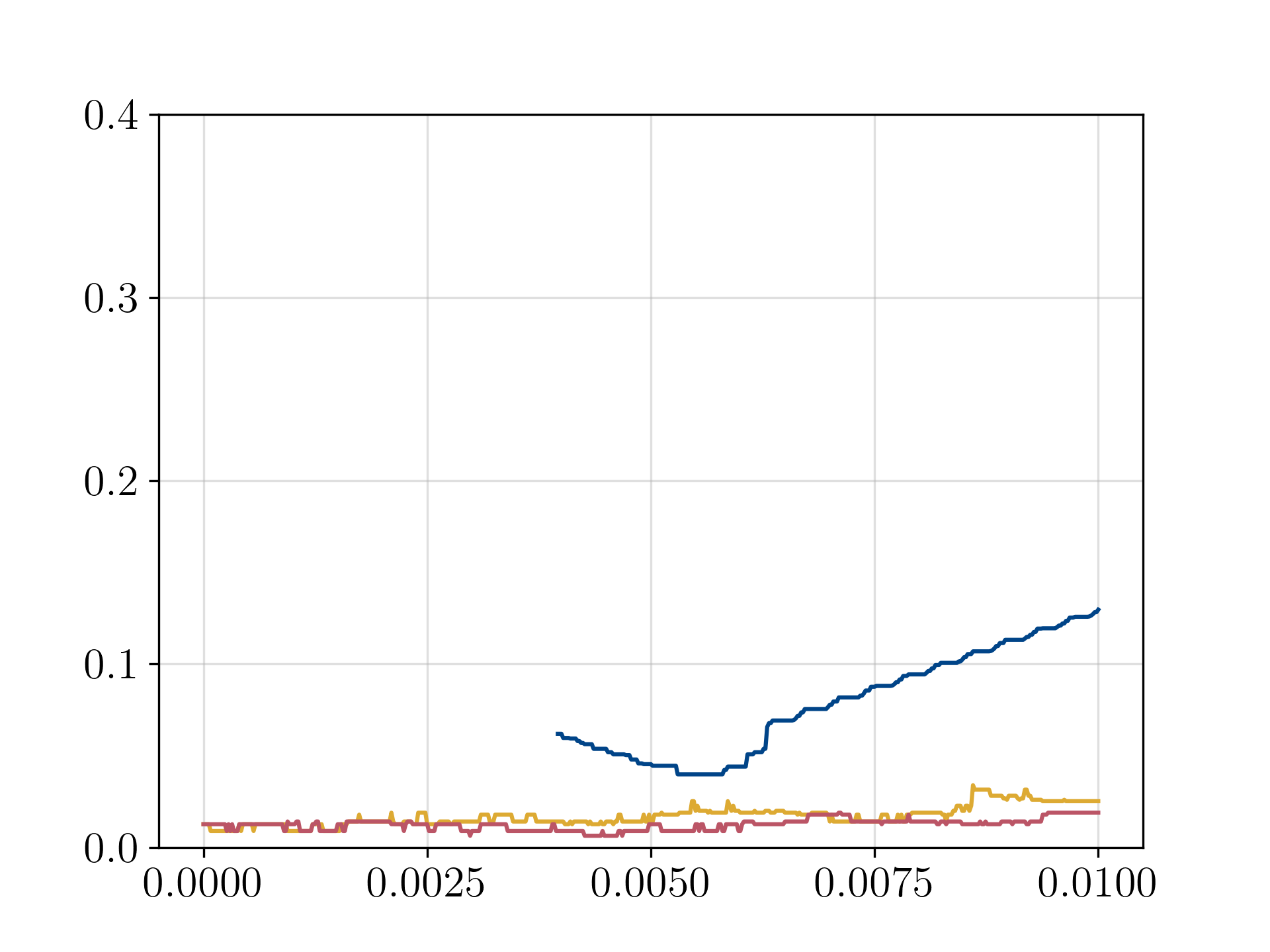}&\hspace{-.5cm}
        \includegraphics[trim=0.2cm 0 0.2cm 1cm, clip,height=0.22\textwidth]{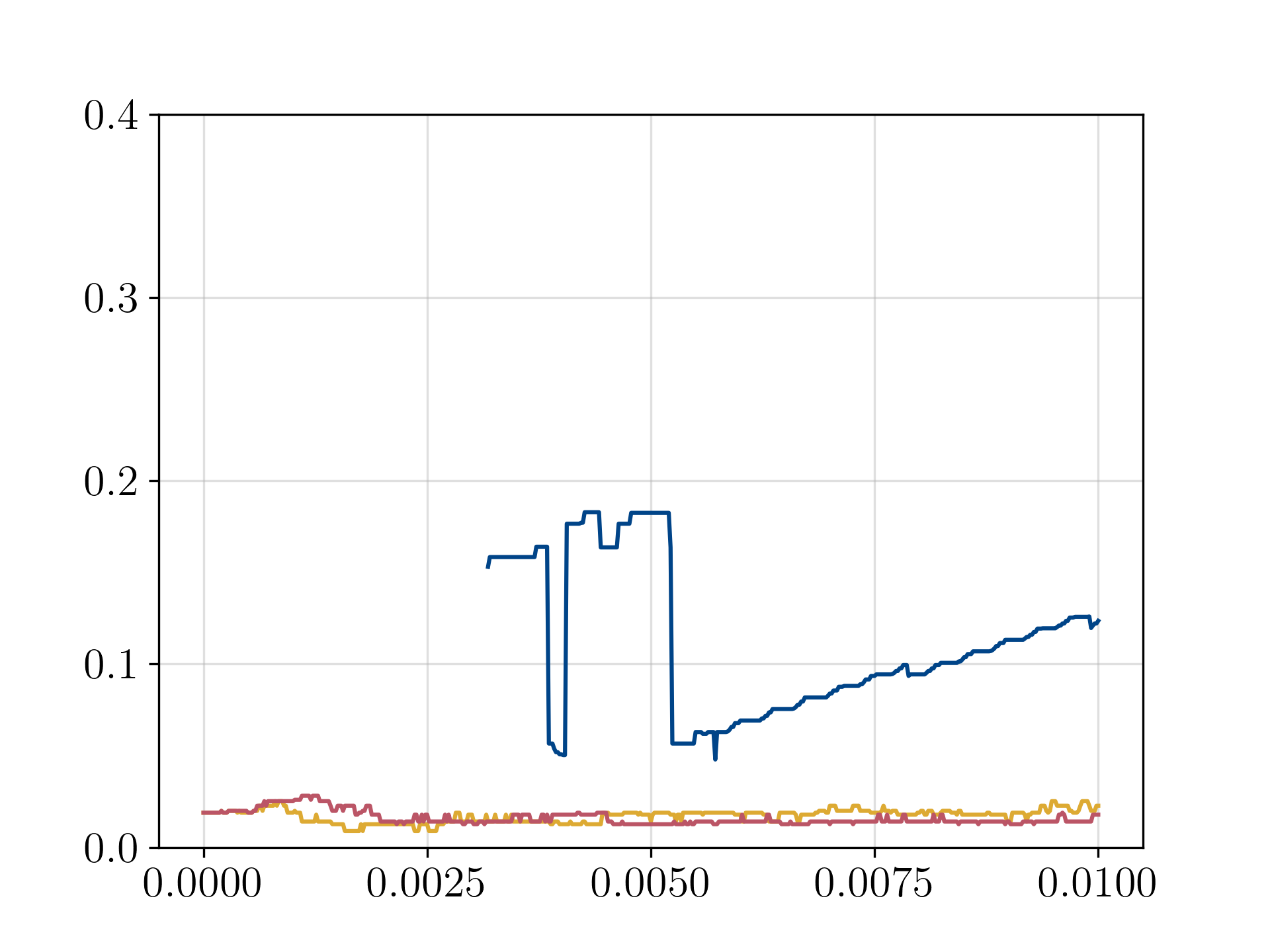}\\
        \raisebox{2.25em}{\rotatebox[origin=lb]{90}{\parbox{2.5cm}{\centering \footnotesize{$\boldsymbol{\mu}=(0.6125,0.6125)$}}}}&
        \includegraphics[trim=0.2cm 0 0.2cm 1cm, clip,height=0.22\textwidth]{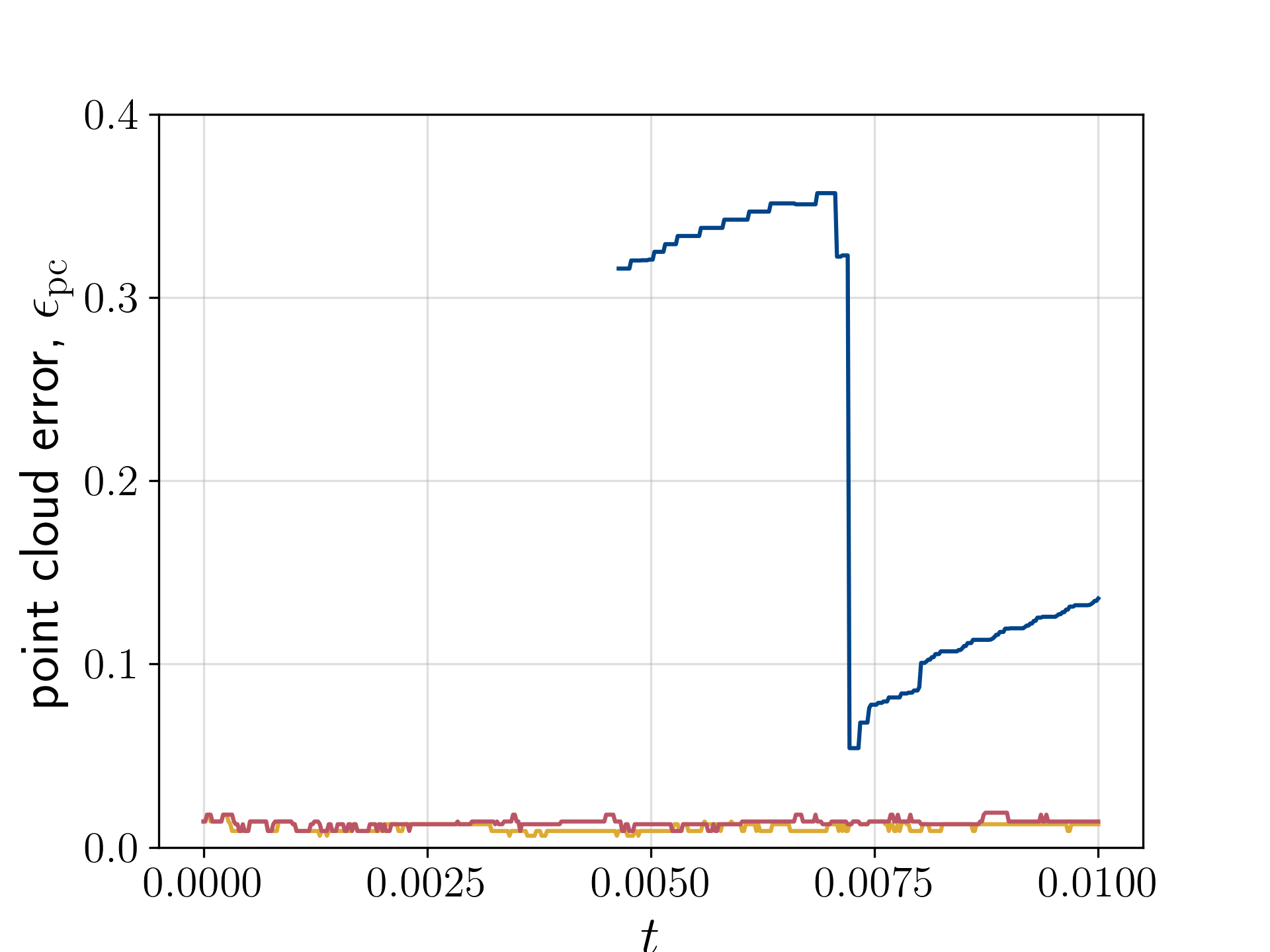}&\hspace{-.5cm}
        \includegraphics[trim=0.2cm 0 0.2cm 1cm, clip,height=0.22\textwidth]{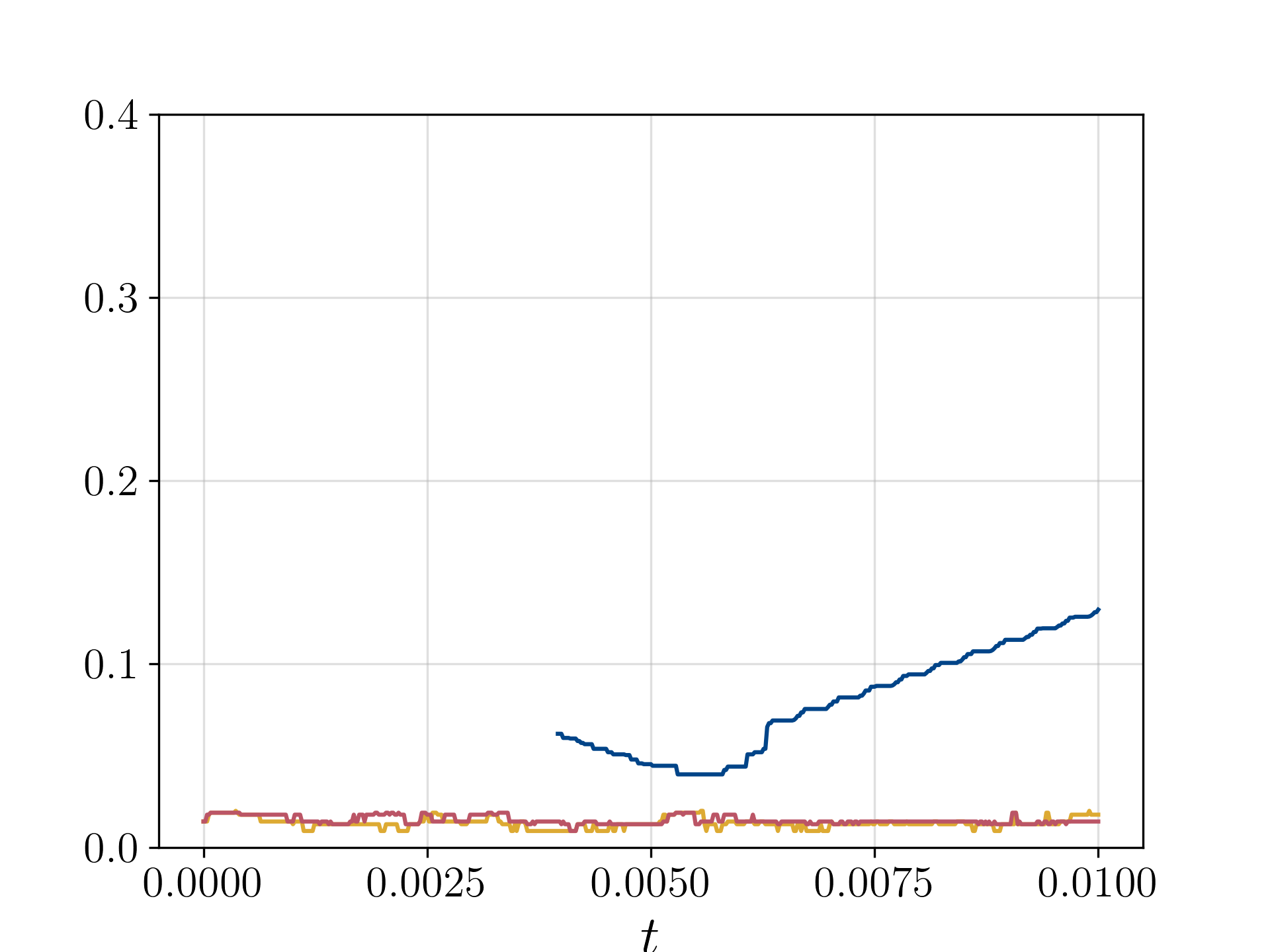}&\hspace{-.5cm}
        \includegraphics[trim=0.2cm 0 0.2cm 1cm, clip,height=0.22\textwidth]{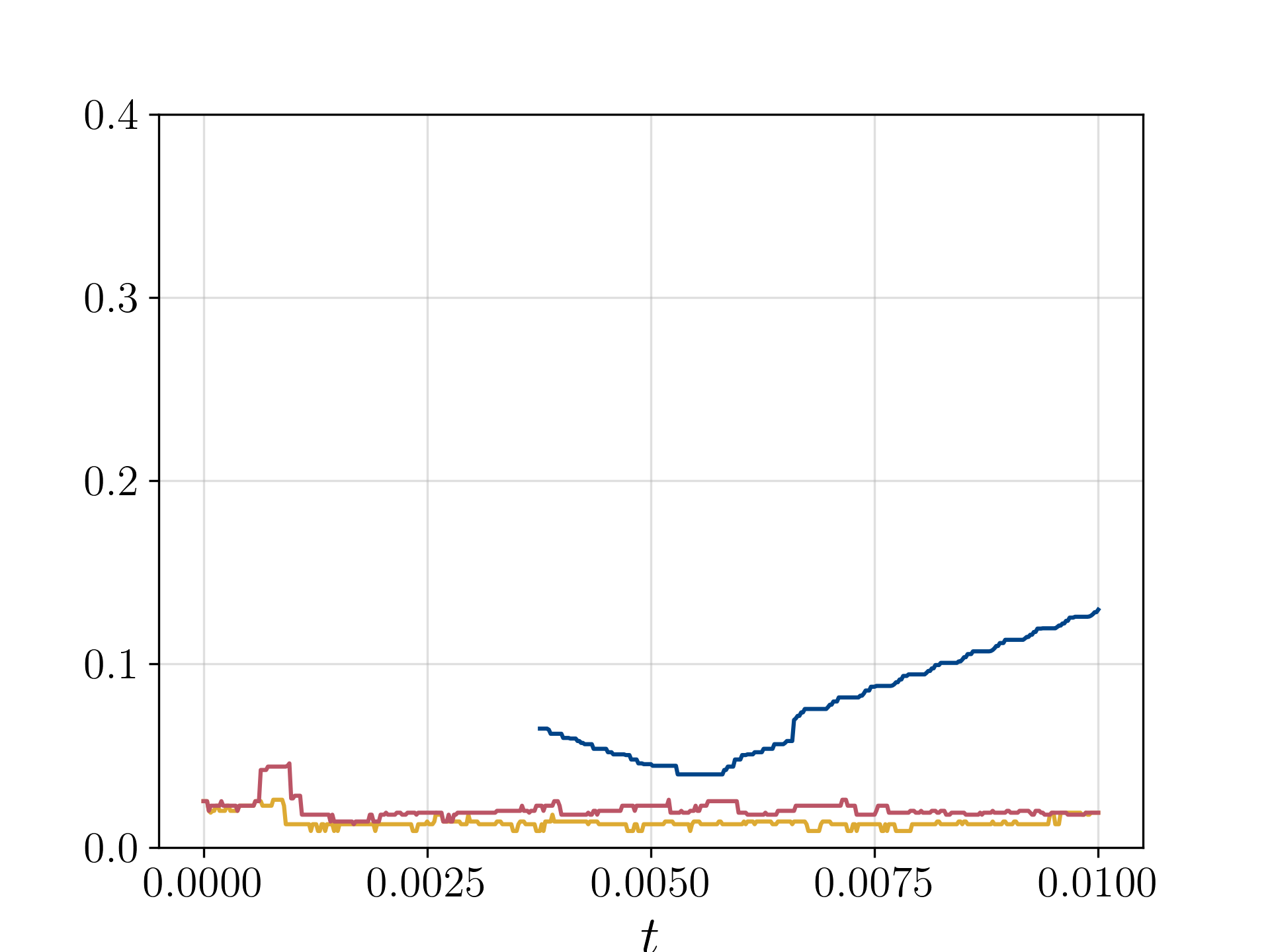}
    \end{tabular}
    \captionsetup{justification=centering}
    \caption{Point cloud errors \eqref{eq:pc_error} for GNN-LaSDI, GD-LSPG, and POD-LaSDI as a function of time. Columns correspond to latent state dimensions $n=3$, $5$, and $7$, while rows correspond to the test parameter instances 
    $\boldsymbol{\mu}= (0.3875, 0.3875)$, $(0.3875, 0.6125)$, $(0.6125, 0.3875)$, and $(0.6125, 0.6125)$. GNN-LaSDI and GD-LSPG consistently achieve low point cloud errors across all latent dimensions and parameter instances, whereas POD-LaSDI generally produces substantially larger errors. Missing portions of the POD-LaSDI curves indicate times at which no interface could be identified due to the diffuse nature of the predicted solution. All figures share the same legend. (Online version in color.)}
    \label{fig:ac_pc_error}
\end{figure}

Figure \ref{fig:ac_pc_error} presents the point cloud error plotted as a function of time for GNN-LaSDI, GD-LSPG, and POD-LaSDI at latent state dimensions $n=3$, $5$, and $7$. Across all considered latent dimensions, GNN-LaSDI and GD-LSPG consistently achieve low and nearly identical point cloud errors throughout the simulation, indicating that both methods accurately capture the evolution of the solid-liquid interface. In contrast, POD-LaSDI produces substantially larger point cloud errors, with performance exhibiting a stronger dependence on the latent state dimension. Furthermore, in early times, the POD-LaSDI solutions were often diffuse to the point that a distinct interface could not be identified. Consequently, a point cloud representation cannot be generated, leading to the missing data points observed in Figure \ref{fig:ac_pc_error}. These results indicate that GNN-LaSDI and GD-LSPG are significantly more effective at preserving the interface location and evolution than POD-LaSDI, which is consistent with the comparisons of the solutions and their corresponding point clouds presented in Figure \ref{fig:ac_P_pc}. Unlike traditional global error metrics, the point cloud error metric directly quantifies errors in the predicted interface location and therefore offers a more physically relevant assessment of model performance for applications with sharp gradients.

\begin{figure}[ht!]
    \centering
     \begin{tabular}{c}
        \includegraphics[height=0.3\textwidth]{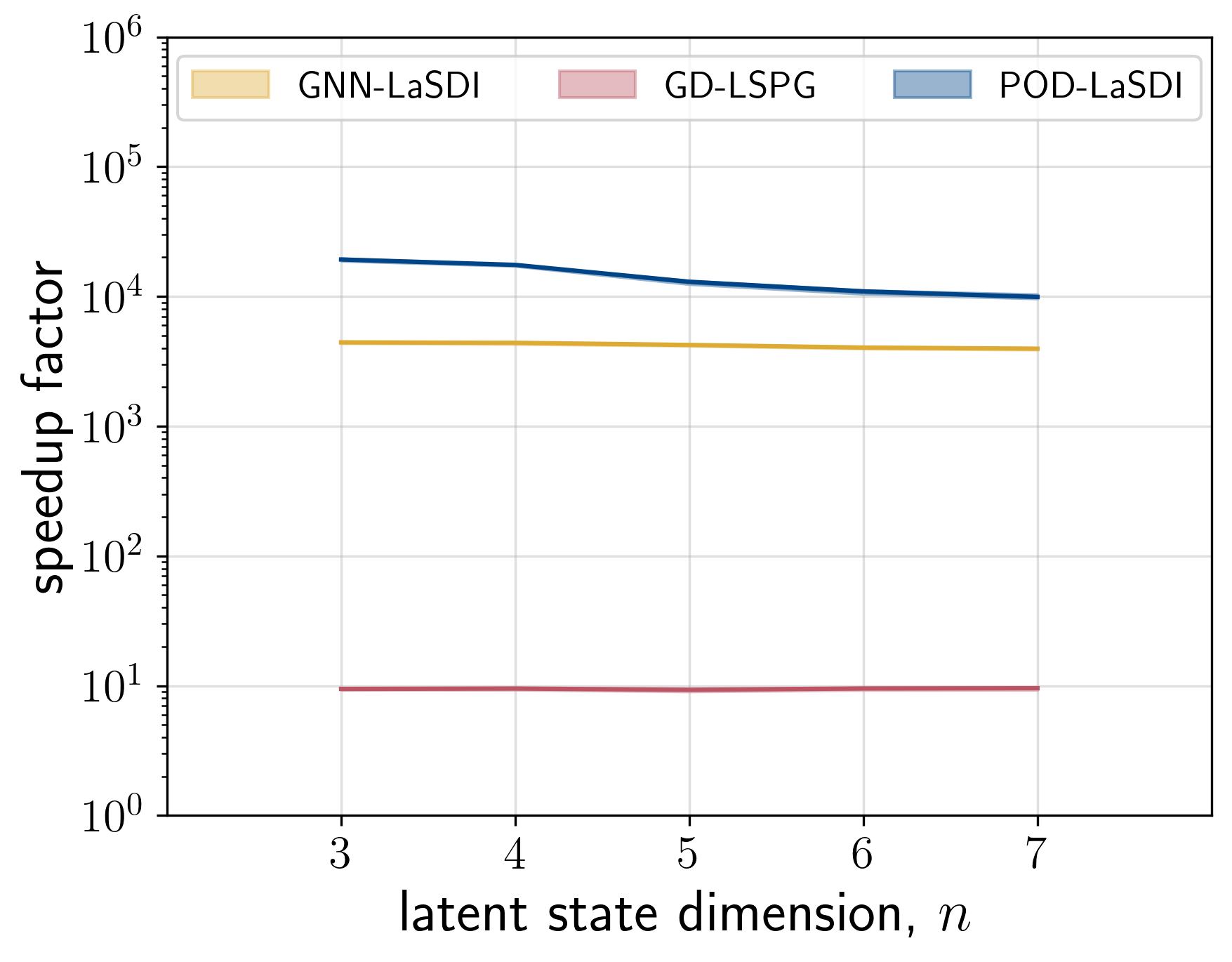}
    \end{tabular}
    \captionsetup{justification=centering}
    \caption{Range of speedup factors for GNN-LaSDI, GD-LSPG, and POD-LaSDI across all 16 test parameter sets as a function of latent state dimension. Solid lines denote the median speedup factor. The range of speedup factors across the test parameters is sufficiently narrow that it is visually indistinguishable from the median line. Note that speedup factors are calculated with respect to the FOM wall-clock time for each specific test parameter set. POD-LaSDI achieves the highest speedup factors for all latent state dimensions, whereas GD-LSPG yields the lowest. GNN-LaSDI offers an intermediate level of computational performance, with speedup factors more than two orders of magnitude higher than those of GD-LSPG and approximately half an order of magnitude lower than those of POD-LaSDI. The average FOM wall-clock time varied between $748.05$s and $803.95$s, depending on the test parameter set. (Online version in color.)}
    \label{fig:ac_speedupRange}
\end{figure}

Lastly, the computational efficiency of each ROM is evaluated. Figure \ref{fig:ac_speedupRange} presents the range of speedup factors across all test parameter instances as a function of the latent state dimension. POD-LaSDI consistently achieves the largest speedup factors for all latent state dimensions, owing to the computational efficiency of the LaSDI framework and the inexpensive reconstruction associated with the linear POD basis. In contrast, GD-LSPG yields the lowest speedup factors due to the high costs associated with the nonlinear LSPG time integration scheme and decoding process. GNN-LaSDI provides an intermediate level of computational performance. In bypassing the repeated evaluation of the Jacobian of the decoder, GNN-LaSDI achieves speedup factors approximately two orders of magnitude greater than those of GD-LSPG. However, its speedup factors remain roughly half an order of magnitude lower than those of POD-LaSDI because reconstruction through the graph autoencoder decoder is significantly more expensive than reconstruction using a linear POD basis.

While POD-LaSDI delivers the highest speedup factors, it also produces the largest state prediction and point cloud errors. On the other hand, GD-LSPG achieves the lowest state prediction and point cloud errors but at the highest computational cost. Finally, GNN-LaSDI offers a balance between the predictive accuracy of GD-LSPG and the computational efficiency of POD-LaSDI.

\subsection{2D bow shock generated by flow past a cylinder}

\begin{figure}[!htb]
    \centering
    \centerline{\includegraphics[scale=.3]{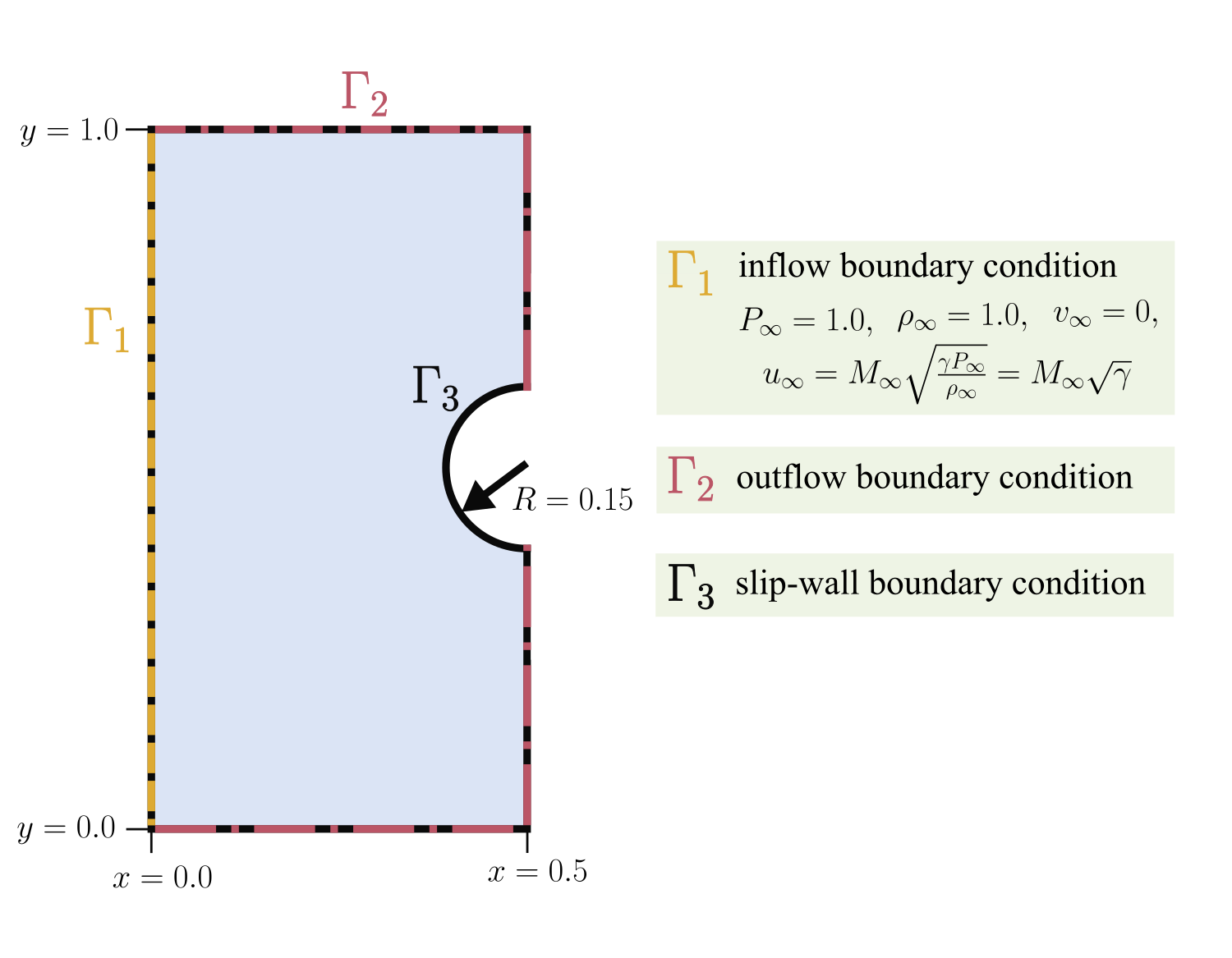}}
    \captionsetup{justification=centering}
    \caption{Computational domain and boundary conditions for the 2D bow shock problem governed by the Euler equations and solved using a Riemann solver. The computational domain consists of a rectangular channel with a half-cylinder along the right boundary and centered vertically within the domain. (Online version in color.)} 
    \label{fig:cyl_setup}
\end{figure}

In the second numerical experiment, GNN-LaSDI, GD-LSPG, and POD-LaSDI are applied to a Riemann solver for the 2D Euler equations that models the bow shock generated by flow past a cylinder. The governing equations are given by
\begin{gather}
    \frac{\partial \mathbf{U}}{\partial t} + \frac{\partial \mathbf{F}}{\partial x} + \frac{\partial \mathbf{G}}{\partial y} = \mathbf{0}, \label{eq:eulerEq1} \\
    \mathbf{U} = \begin{bmatrix}
        \rho \\
        \rho u \\
        \rho v \\
        \rho E
    \end{bmatrix}, \quad
    \mathbf{F} = \begin{bmatrix}
        \rho u \\
        \rho u^2 + P \\
        \rho u v \\
        \rho u H
    \end{bmatrix}, \quad
    \mathbf{H} = \begin{bmatrix}
        \rho v \\
        \rho u v \\
        \rho v^2 + P \\
        \rho v H
    \end{bmatrix}, \label{eq:eulerEq2}
\end{gather}
where $\mathbf{\rho}(x,y,t;\boldsymbol{\mu})\in\mathbb{R}_+$ denotes density, $u(x,y,t;\boldsymbol{\mu}) \in \mathbb{R}$ and $ v(x,y,t;\boldsymbol{\mu}) \in \mathbb{R}$ denote velocity in the $x$ and $y$ directions, respectively, $P(x,y,t;\boldsymbol{\mu}) \in \mathbb{R}_+$ denotes pressure. The specific total energy and enthalpy are defined as $E(x,y,t;\boldsymbol{\mu})=\frac{1}{\gamma-1}\frac{P}{\rho}+\frac{1}{2}(u^2+v^2) \in \mathbb{R}_+$ and $H(x,y,t;\boldsymbol{\mu})=\frac{\gamma}{\gamma-1}\frac{P}{\rho}+\frac{1}{2}(u^2+v^2) \in \mathbb{R}_+$, respectively, where $\gamma \in \mathbb{R}_+$ is the specific heat ratio taken here as $\gamma=1.4$. The physical domain, shown in Figure \ref{fig:cyl_setup}, consists of a rectangular channel with a cylinder positioned along the right boundary. The system is parameterized by the freestream Mach number, i.e., $\boldsymbol{\mu} = M_{\infty}$. Inflow conditions are prescribed along the left boundary
\begin{equation}
    \rho_{\infty} = 1.0, \quad P_{\infty}=1.0, \quad u_{\infty} = M_{\infty} \sqrt{\frac{\gamma P_{\infty}}{\rho_{\infty}}}= M_{\infty}\sqrt{\gamma}, \quad v_{\infty}=0,
\end{equation}
while outflow conditions are prescribed on the top, bottom, and right boundaries. A slip-wall boundary condition is imposed on the cylinder surface. As the freestream Mach number varies, the resulting bow shock exhibits different propagation characteristics. The initial conditions are taken to be 
\begin{equation}
    \rho^0(x,y;\boldsymbol{\mu}) = \rho_{\infty}, \quad u^0 (x,y;\boldsymbol{\mu}) = u_{\infty}, \quad v^0(x,y;\boldsymbol{\mu})=v_{\infty} \quad P^0(x,y;\boldsymbol{\mu})=P_{\infty}.
\end{equation}

In this study, the rotated Roe, Harten, Lax, and van Leer (R-RHLL) Riemann solver \cite{nishikawa2008rhrs} is employed to obtain approximate solutions of \eqref{eq:eulerEq1}-\eqref{eq:eulerEq2}. Spatial discretization using an unstructured mesh containing 4818 finite volume cells (i.e., $N_c=4818$) yields a semi-discrete system of the form \eqref{eq:ode1}, with the concatenated state vector $\mathbf{x}(t;\boldsymbol{\mu}) = \left[ (\boldsymbol{\rho}(t;\boldsymbol{\mu}))^\top, (\boldsymbol{\rho u}(t;\boldsymbol{\mu}))^\top, (\boldsymbol{\rho v}(t;\boldsymbol{\mu}))^\top, (\boldsymbol{\rho E}(t;\boldsymbol{\mu}))^\top \right]^\top$, where $\boldsymbol{\rho}$, $\boldsymbol{\rho u}$, $\boldsymbol{\rho v}$, and $\boldsymbol{\rho E}$ denote the semi-discrete representations of the conserved variables $\rho$, $\rho u$, $\rho v$, and $\rho E$, respectively. Consequently, the system contains $n_q=4$ state variables and has a dimension $N = n_q N_c = 19272$. Time integration is performed using a forward Euler scheme with a time step size $\Delta t = 0.001$ and $N_t=500$ (i.e., $T_f = 0.5$). 

To generate the training data used for graph autoencoder training, POD basis construction, and operator learning, FOM solutions are computed for 11 training parameters $\boldsymbol{\mu}=M_{\infty}=1+0.1i$, with $i=0,\ldots,10$ (i.e., $n_p=11$). In the online stage, the performance of GNN-LaSDI, GD-LSPG, and POD-LaSDI is evaluated using 10 test parameters given by $\boldsymbol{\mu}=M_{\infty}=1.05+0.1i$, with $i=0,\ldots,9$. The graph autoencoders and POD basis are trained using the conserved variables ($\rho$, $\rho u$, $\rho v$, and $\rho E$). Accordingly, all reconstruction errors and state prediction errors reported in this section are evaluated with respect to the conserved variables. Once the graph autoencoder is trained and the POD basis is obtained, the low-dimensional operators for the GNN-LaSDI and POD-LaSDI are identified using the procedure described in Section \ref{sec:lasdi}. For consistency with both the FOM and GD-LSPG, the online time integration for the GNN-LaSDI and POD-LaSDI ROMs is also performed using a forward Euler scheme. The encoded trajectories $\hat{\mathbf{X}}$ include the initial condition, yielding $N_t+1$ snapshots per training parameter. To construct $\dot{\hat{\mathbf{X}}}$, temporal derivatives are approximated using a fourth-order central difference scheme for the interior time instances. Consistent with the treatment adopted for Kobayashi's solidification model, the first two and last two time instances are handled separately using lower-order finite difference approximations. As a result, the training dataset contains a total of $11(N_t+1)=5511$ latent state snapshots and corresponding latent space velocity samples.

For this numerical example, ensuring the physical admissibility of the reconstructed state is critical. In particular, the graph autoencoder is tailored to preserve positive pressure throughout the domain and constrain the reconstructed state variables to their respective ranges in the training data. The following discussion describes how these constraints are embedded directly within the graph autoencoder architecture. In the na\"{\i}ve implementation of the GNN developed in \cite{magargal2025projection}, the postprocessing layer outputs an approximation of the concatenated state vector $\mathbf{x}$ and is given by
\begin{equation}\label{eq:postProcVanilla}
    \mathbf{g}_{n_\ell} : (\mathbf{Y}^{n_\ell-1}; \boldsymbol{\Omega}_{n_\ell})\mapsto \begin{bmatrix}
        \widetilde{\boldsymbol{\rho}} \\
        \widetilde{\boldsymbol{\rho u}} \\
        \widetilde{\boldsymbol{\rho v}} \\
        \widetilde{\boldsymbol{\rho E}} \\
    \end{bmatrix}, 
\end{equation}
where
\begin{equation}\label{eq:tildeBarVanilla}
    \begin{bmatrix}
        \widetilde{\boldsymbol{\rho}} \\
        \widetilde{\boldsymbol{\rho u}} \\
        \widetilde{\boldsymbol{\rho v}} \\
        \widetilde{\boldsymbol{\rho E}} \\
    \end{bmatrix} =\begin{bmatrix}
        (\rho^{\mathrm{max}} - \rho^{\mathrm{min}})\overline{\boldsymbol{\rho}} + \rho^{\mathrm{min}} \\
        ((\rho u)^{\mathrm{max}} - (\rho u)^{\mathrm{min}})\overline{\boldsymbol{\rho u}} + (\rho u)^{\mathrm{min}} \\
        ((\rho v)^{\mathrm{max}} - (\rho v)^{\mathrm{min}})\overline{\boldsymbol{\rho v}} + (\rho v)^{\mathrm{min}} \\
        ((\rho E)^{\mathrm{max}} - (\rho E)^{\mathrm{min}})\overline{\boldsymbol{\rho E}} + (\rho E)^{\mathrm{min}}
    \end{bmatrix}, \quad\text{with }\, \begin{bmatrix}
        \overline{\boldsymbol{\rho}} \\
        \overline{\boldsymbol{\rho u}} \\
        \overline{\boldsymbol{\rho v}} \\
        \overline{\boldsymbol{\rho E}} \\
    \end{bmatrix} = \begin{bmatrix}
        (\mathbf{Y}^{n_\ell-1})_1 \\
        (\mathbf{Y}^{n_\ell-1})_2 \\
        (\mathbf{Y}^{n_\ell-1})_3 \\
        (\mathbf{Y}^{n_\ell-1})_4 \\
    \end{bmatrix}. 
\end{equation}
Here, the superscripts $(.)^\mathrm{max}$ and $(.)^\mathrm{min}$ denote the maximum and minimum values for the respective variables observed in the training set, $\mathbf{Y}^{n_\ell-1} \in \mathbb{R}^{N_c \times 4}$ denotes input to the postprocessing layer, $\boldsymbol{\Omega}_{n_{\ell}} = \emptyset$, as there are no trainable weights or biases in the postprocessing layer, the subscript $(.)_i$ denotes the $i^{\mathrm{th}}$ column of a matrix, $\widetilde{(.)}$ denotes the approximate state variable, and $\overline{(.)}$ denotes its scaled representation. As in \cite{magargal2025projection}, the state variables are scaled to $[0,1]$ within the preprocessing layer of the encoder (i.e., $\mathbf{h}_0$) and are subsequently scaled back in the postprocessing layer of the decoder, $\mathbf{g}_{n_{\ell}}$ (refer to \ref{ssec:graphAEdetails} and \cite{magargal2025projection} for further information on the graph autoencoder architecture). Note that no activation function is applied in \eqref{eq:tildeBarVanilla}. Under this formulation, the reconstructed state variables are not restricted to the ranges observed in the training data, and the positivity of the pressure field is not guaranteed. It can be shown that the preservation of positive pressure may be enforced by imposing the constraint,
\begin{equation}\label{eq:pressureConstraint}
    \rho E > \frac{1}{2\rho}\left((\rho u)^2 + (\rho v)^2\right).
\end{equation}
To enforce \eqref{eq:pressureConstraint}, a modified postprocessing layer, $\mathbf{g}'_{n_\ell}(\hspace{1mm}\boldsymbol{\cdot}\hspace{1mm};\boldsymbol{\Omega}_{n_\ell})$, is employed in this numerical experiment. Rather than directly predicting the total energy, the graph autoencoder seeks to produce the closure term,
\begin{equation}\label{eq:closure}
    R = \rho E - \frac{1}{2 \rho} \left( (\rho u)^2 + (\rho v)^2 \right).
\end{equation}
As a result, the modified postprocessing layer  deployed in this numerical experiment is written as
\begin{equation}\label{eq:postProcModified}
    \mathbf{g}'_{n_\ell} : (\mathbf{Y}^{n_\ell-1}; \boldsymbol{\Omega}_{n_\ell})\mapsto \begin{bmatrix}
        \widetilde{\boldsymbol{\rho}} \\
        \widetilde{\boldsymbol{\rho u}} \\
        \widetilde{\boldsymbol{\rho v}} \\
        \frac{1}{2 \widetilde{\boldsymbol{\rho}}}\left( (\widetilde{\boldsymbol{\rho u}})^2 + (\widetilde{\boldsymbol{\rho v}})^2 \right) + \left( R^{\mathrm{max}} - R^{\mathrm{min}} \right) \overline{\boldsymbol{R}} + R^{\mathrm{min}} \\ 
    \end{bmatrix}, 
\end{equation}
with
\begin{equation}\label{eq:barTildeModified}
    \begin{bmatrix}
        \overline{\boldsymbol{\rho}} \\
        \overline{\boldsymbol{\rho u}} \\
        \overline{\boldsymbol{\rho v}} \\
        \overline{\boldsymbol{R}} \\
    \end{bmatrix} = S\left(\begin{bmatrix}
        (\mathbf{Y}^{n_\ell-1})_1 \\
        (\mathbf{Y}^{n_\ell-1})_2 \\
        (\mathbf{Y}^{n_\ell-1})_3 \\
        (\mathbf{Y}^{n_\ell-1})_4 \\
    \end{bmatrix}\right), 
\end{equation}
where $\widetilde{\boldsymbol{\rho}}$, $\widetilde{\boldsymbol{\rho u}}$, and $\widetilde{\boldsymbol{\rho v}}$ are defined analogously to \eqref{eq:tildeBarVanilla} using the values $\overline{\boldsymbol{\rho}}$, $\overline{\boldsymbol{\rho u}}$, and $\overline{\boldsymbol{\rho v}}$ obtained from \eqref{eq:barTildeModified}. The sigmoid activation function, $S: \mathbb{R} \rightarrow (0,1)$, constrains $\overline{\boldsymbol{\rho}}$, $\overline{\boldsymbol{\rho u}}$, $\overline{\boldsymbol{\rho v}}$, and $\overline{\boldsymbol{R}}$ to the interval $(0,1)$. Consequently, $\widetilde{\boldsymbol{\rho}}$, $\widetilde{\boldsymbol{\rho u}}$, and $\widetilde{\boldsymbol{\rho v}}$ are restricted to the ranges observed in the training data after rescaling. Note that the constraint imposed by the sigmoid activation function is stronger than enforcing positivity alone, as it bounds each reconstructed state variable between prescribed minimum and maximum values derived from the training dataset. The positivity of the closure term guarantees satisfaction of the pressure constraint \eqref{eq:pressureConstraint}. In practice, because the sigmoid output asymptotically approaches, but never attains, the values $0$ and $1$, the prescribed maximum and minimum values are taken to be roughly 5-10\% of the range observed in training data, $((.)^{\mathrm{max}}-(.)^{\mathrm{min}})$, above and below than the observed extrema, respectively. This adjustment ensures that the graph autoencoder remains sufficiently expressive to represent the training solutions while preserving pressure positivity. Unless otherwise noted, all graph autoencoder results reported in this numerical experiment are obtained using the constrained architecture described above. No analogous positivity constraint is imposed on the POD-LaSDI ROM. Additional implementation details regarding the graph autoencoder are provided in \ref{sec:aeTraining}.

\begin{figure}[ht!]
    \centering
     \begin{tabular}{cc}
        \includegraphics[height=0.3\textwidth]{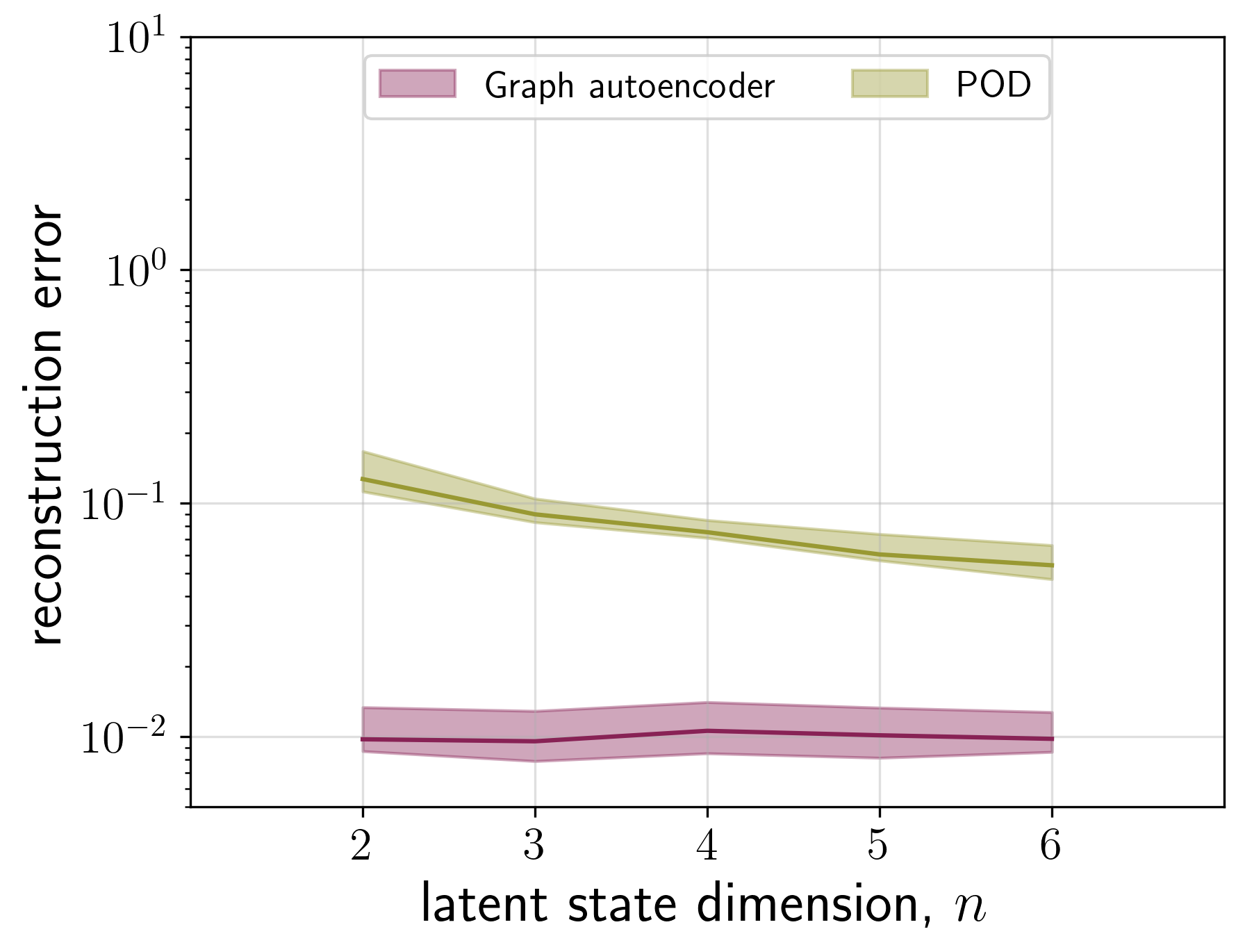}
        & \includegraphics[height=0.3\textwidth]{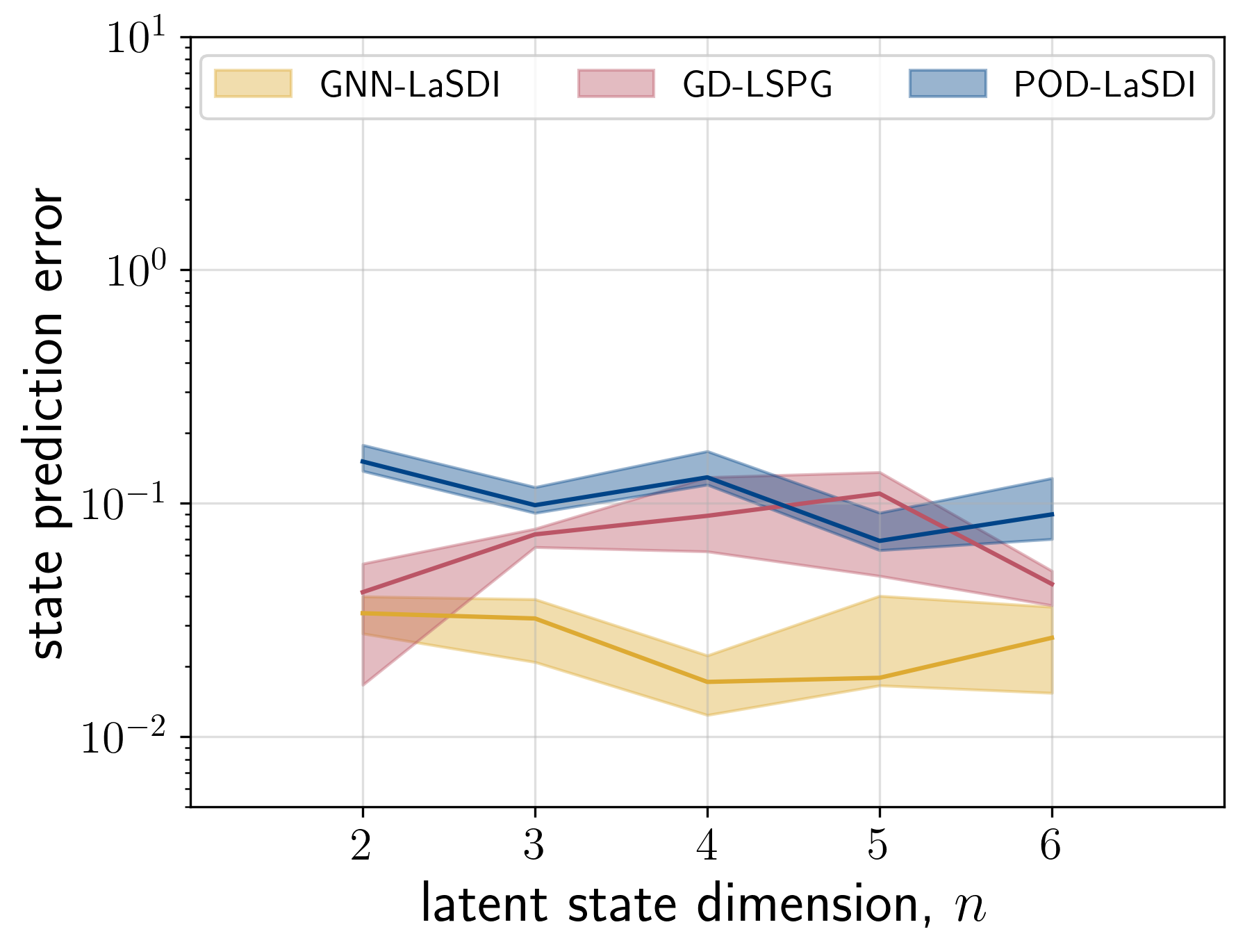}
    \end{tabular}
    \captionsetup{justification=centering}
    \caption{Range of graph autoencoder and POD reconstruction errors (left) and state prediction errors for GNN-LaSDI, GD-LSPG, and POD-LaSDI (right) across all 10 test parameters, plotted with respect to latent state dimension. Solid lines indicate median values, and shaded regions denote the range of errors across all test parameters. The graph autoencoder consistently achieves reconstruction errors approximately one order of magnitude lower than POD, while GNN-LaSDI produces state prediction errors roughly one order of magnitude lower than POD-LaSDI. GD-LSPG yields substantially larger state prediction errors than GNN-LaSDI. (Online version in color.)}
    \label{fig:cyl_errorRange}
\end{figure}

The graph autoencoder and POD reconstruction errors, together with the GNN-LaSDI, GD-LSPG, and POD-LaSDI state prediction errors for all 10 test parameters are reported in Figure \ref{fig:cyl_errorRange}. Across the range of latent state dimensions and test parameters, the graph autoencoder consistently achieves reconstruction errors approximately one order of magnitude lower than those obtained with POD. Similarly, GNN-LaSDI produces state prediction errors that are consistently about an order of magnitude lower than those of POD-LaSDI. Interestingly, GD-LSPG often yields state prediction errors that exceed both the graph autoencoder reconstruction errors and the corresponding GNN-LaSDI state prediction errors.

\begin{figure}[ht!]
    \centering
     \begin{tabular}{ccc}
        {\footnotesize{GNN-LaSDI}} & \hspace{-.5cm} {\footnotesize{GD-LSPG}} & \hspace{-.5cm}  {\footnotesize{POD-LaSDI}} \\
        \includegraphics[trim = 0 0 0 1.2cm, clip,height=0.25\textwidth]{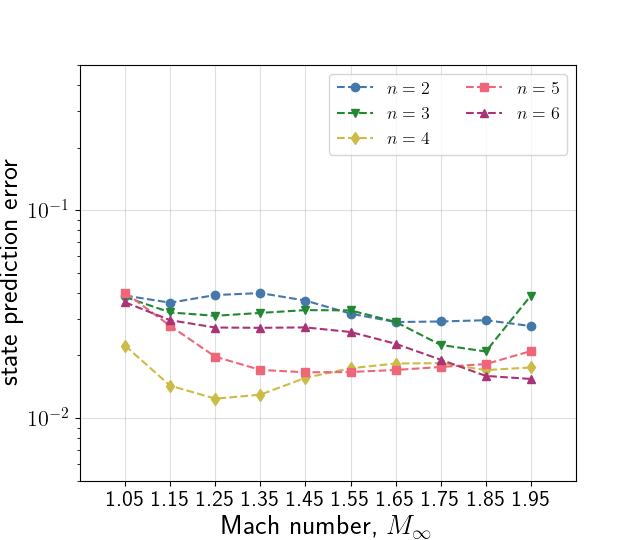}&\hspace{-.5cm}
        \includegraphics[trim = 0 0 0 1.2cm, clip,height=0.25\textwidth]{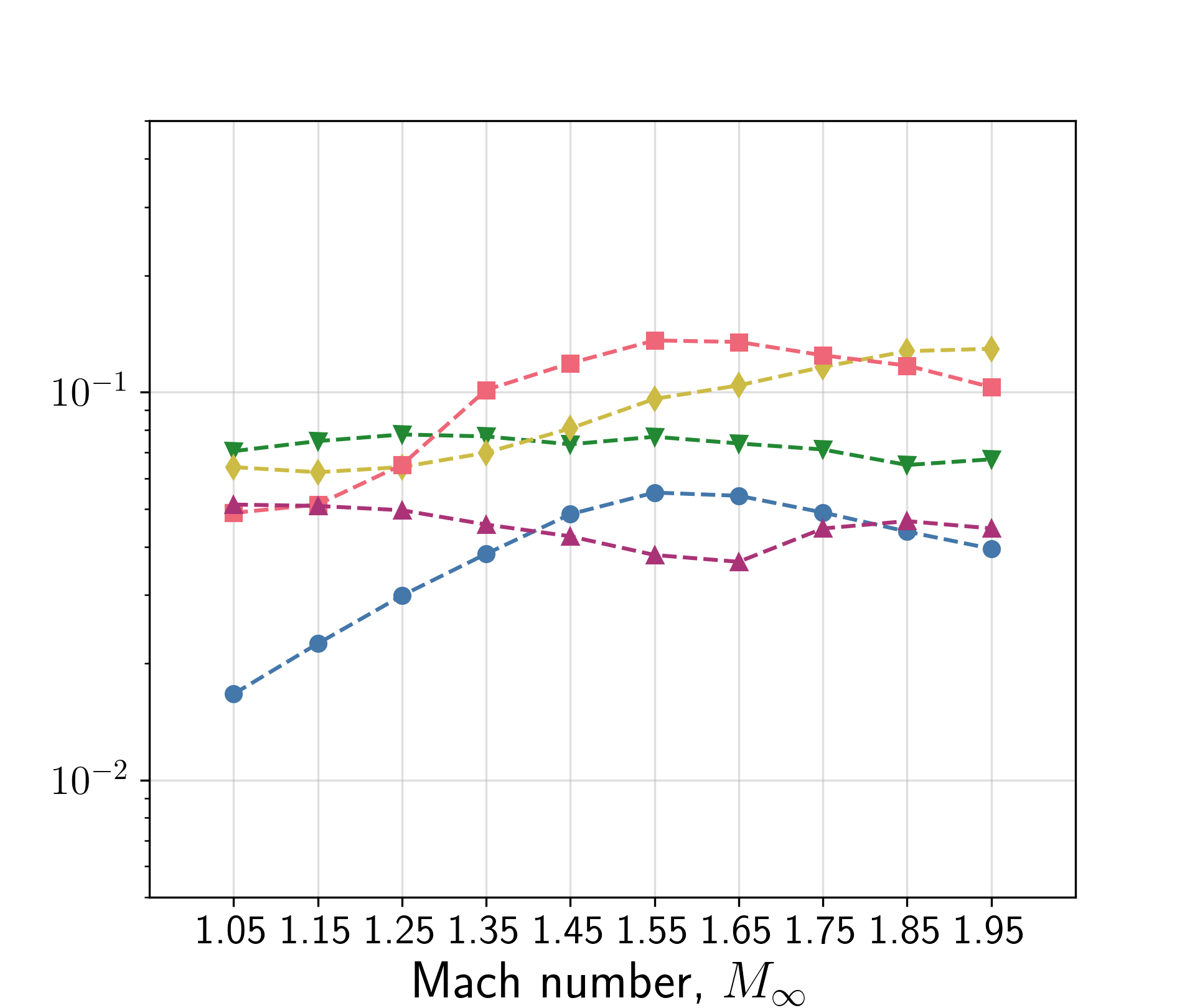}&\hspace{-.5cm}
        \includegraphics[trim = 0 0 0 1.2cm, clip,height=0.25\textwidth]{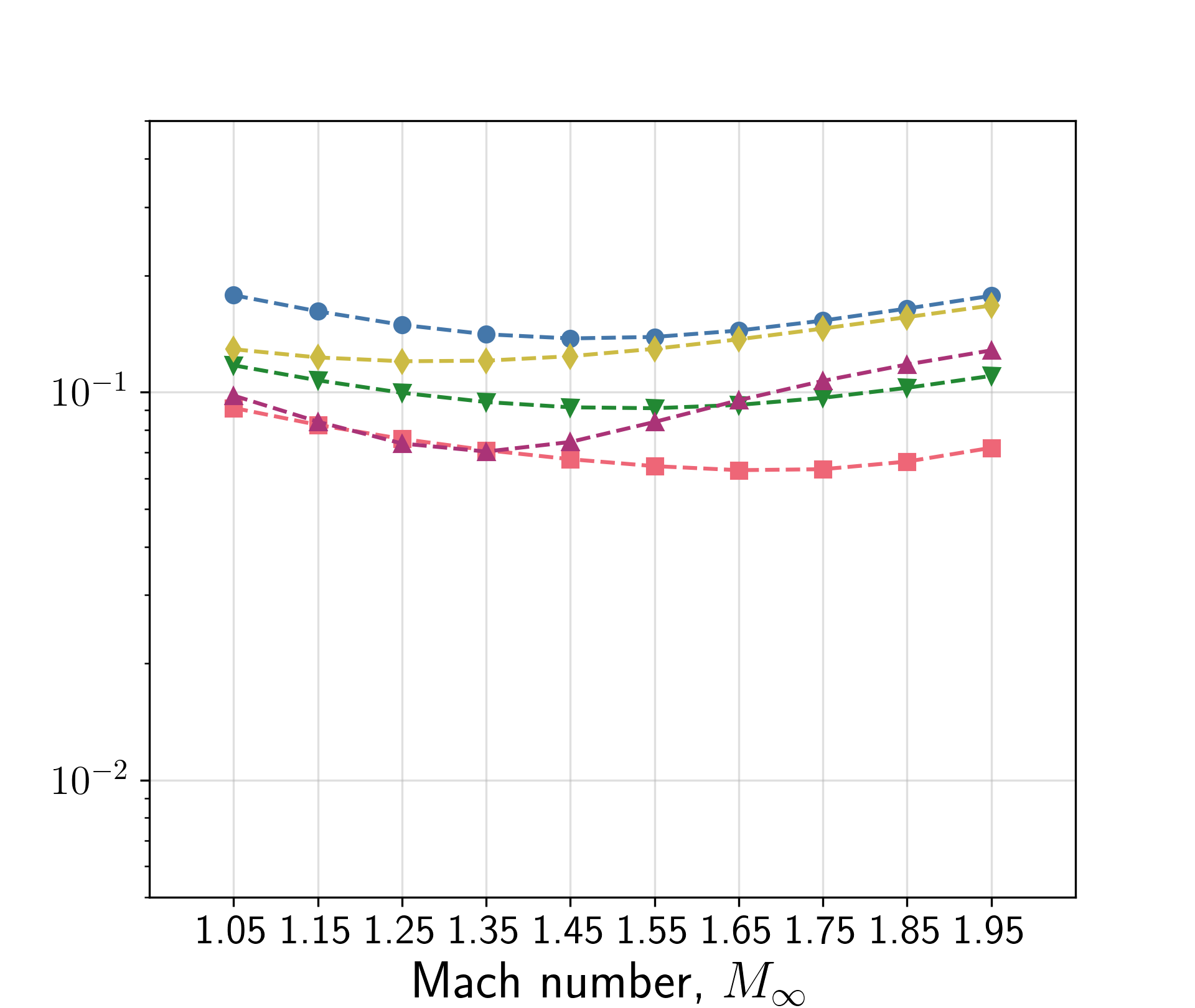}\\
    \end{tabular}
    \captionsetup{justification=centering}
    \caption{State prediction errors plotted with respect to Mach number, $M_{\infty}$, for GNN-LaSDI (left), GD-LSPG (middle), and POD-LaSDI (right) for latent state dimensions $n=2,\ldots,6$. GNN-LaSDI consistently achieves the lowest errors across the parameter domain, while all methods exhibit relatively uniform error levels with respect to $M_{\infty}$. All subplots share the same legend. (Online version in color.)}
    \label{fig:cyl_sp_mach}
\end{figure}

To investigate ROM accuracy across the parameter domain, Figure \ref{fig:cyl_sp_mach} reports the state prediction errors of GNN-LaSDI, GD-LSPG, and POD-LaSDI for latent state dimensions $n=2,\ldots,6$ as a function of Mach number, $M_{\infty}$. GNN-LaSDI achieves the lowest state prediction errors throughout the parameter space. Furthermore, the state prediction errors of all three methods exhibit slight variation with respect to $M_{\infty}$, indicating that their predictive performance remains relatively consistent across the range of test parameters studied.

Next, the point cloud error metric is used to evaluate the accuracy of each ROM in predicting the location of the bow shock. For this numerical example, the characteristic length in \eqref{eq:pc_error} is taken as the height of the computational domain ($L=1$). Following the approach of \cite{cucchiara2024convexDisplacement}, point cloud representations of the bow shock are generated using a Ducros sensor \cite{ducros1999sensor,modesti2017lowDissipative}. Specifically, the Ducros sensor at the $i^{\mathrm{th}}$ cell and $m^{\mathrm{th}}$ time step is defined as
\begin{equation}
    D_i^m = \frac{(-\nabla \cdot \mathbf{v}_i^m)^+}{\sqrt{(\nabla \cdot \mathbf{v}_i^m)^2 + \vert\vert \nabla \times \mathbf{v}_i^m\vert\vert_2^2 + (c_i^m)^2}} \frac{\vert\vert \nabla P_i^m \vert\vert_2}{P_i^m + \epsilon_{\mathrm{duc}}} \vert\vert \mathbf{v}_i^m \vert\vert_2,
\end{equation}
where $\mathbf{v}_i^m = (u_i^m, v_i^m) \in \mathbb{R}^{n_d}$, $P_i^m \in \mathbb{R}_+$, and $c_i^m=\sqrt{\frac{\gamma P_i^m}{\rho_i^m}}\in\mathbb{R}_+$ denote the velocity vector, pressure, and speed of sound at the $i^{\mathrm{th}}$ collocation point, respectively, $\epsilon_{\mathrm{duc}} \in \mathbb{R}_+$ is a user-prescribed tolerance (taken in this study to be $0.01$), and $(\cdot)^+ = \mathrm{max}(0,\cdot)$. Collocation points with large values of $D_i^m$ indicate the presence of a shock, due to a strong negative divergence and a significant pressure gradient. A point cloud representation of the shock is then obtained by selecting the collocation points whose Ducros sensor values exceed a threshold,
\begin{equation}
    \mathcal{X}^m =\bigcup_{i = 1,\ldots,N_c} \{ \boldsymbol{\zeta}_i ; \quad D_i^m > \epsilon_{\mathrm{threshold}}^m \},
\end{equation}
where $\boldsymbol{\zeta} \in \mathbb{R}^{N_c \times n_d}$ contains the coordinates of the cell centroids of the cells in the mesh, and $\boldsymbol{\zeta}_i \in \mathbb{R}^{n_d}$ denotes the vector containing the coordinates of the centroid of the $i^\mathrm{th}$ cell. The threshold is defined adaptively at each time step as
\begin{equation}
    \epsilon_{\mathrm{threshold}}^m = \mathrm{quantile}\left( \mathbf{D}^m, \epsilon^{\mathrm{quantile}} \right),
\end{equation}
where $\mathbf{D}^m$ contains the Ducros sensor values at time step $m$. In this study, the threshold is chosen as the  $95^{\mathrm{th}}$ percentile, corresponding to $\epsilon^{\mathrm{quantile}}=0.95$. 

\begin{figure}[ht!]
    \centering
     \begin{tabular}{ccc|cc}
        & \multicolumn{2}{c|}{\footnotesize{$t=0.25$}} & \multicolumn{2}{c}{\footnotesize{$t=0.5$}} \\
        \hline & & & &\\
        & \footnotesize{state solution} & \footnotesize{point cloud} & \footnotesize{state solution} & \footnotesize{point cloud} \\
        \raisebox{2.7em}{\rotatebox[origin=lb]{90}{\parbox{2cm}{\centering \footnotesize{FOM}}}}&
        \includegraphics[trim=0 0 0 1cm, clip,height=0.22\textwidth]{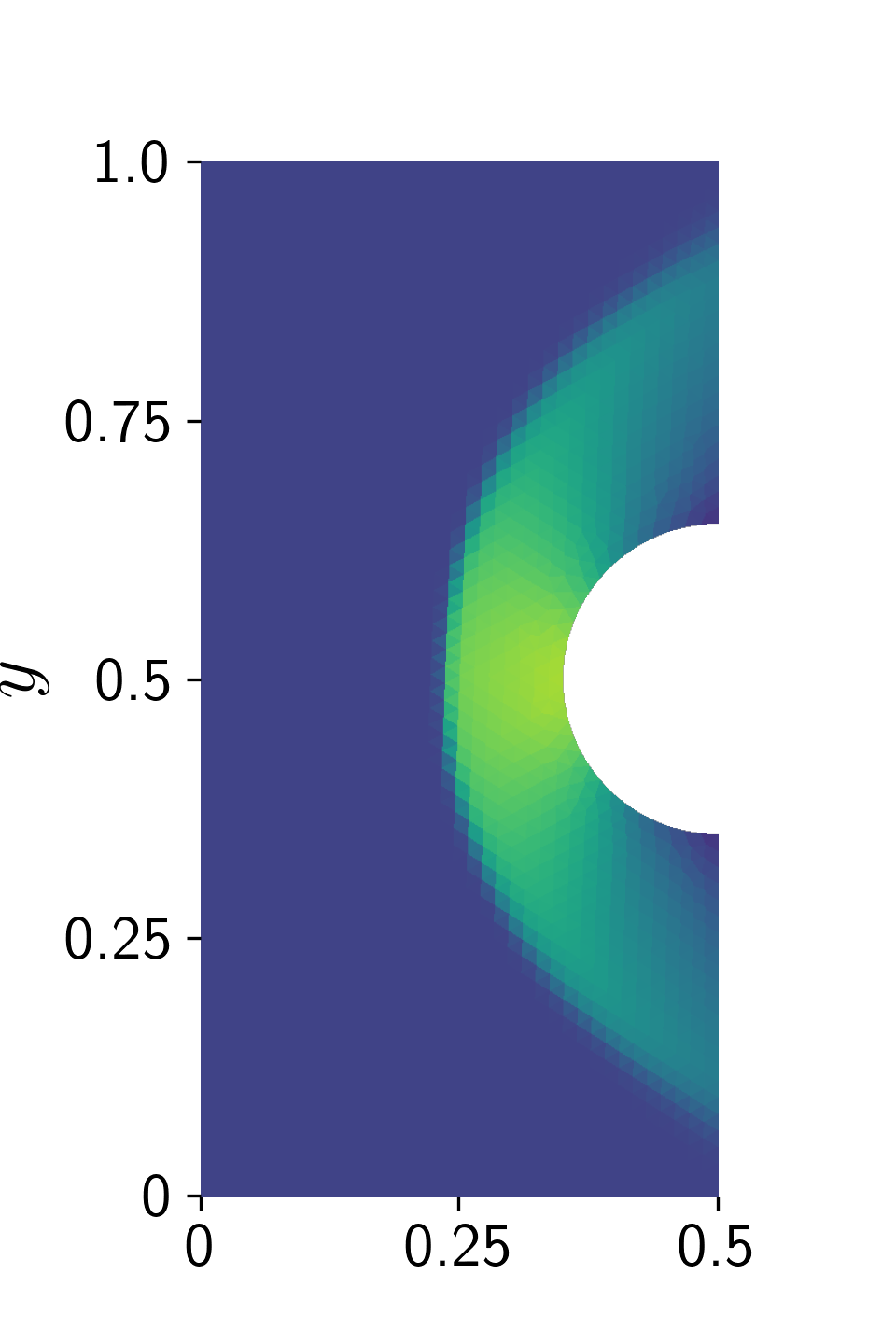}&
        \includegraphics[trim=0 0 0 1cm, clip,height=0.22\textwidth]{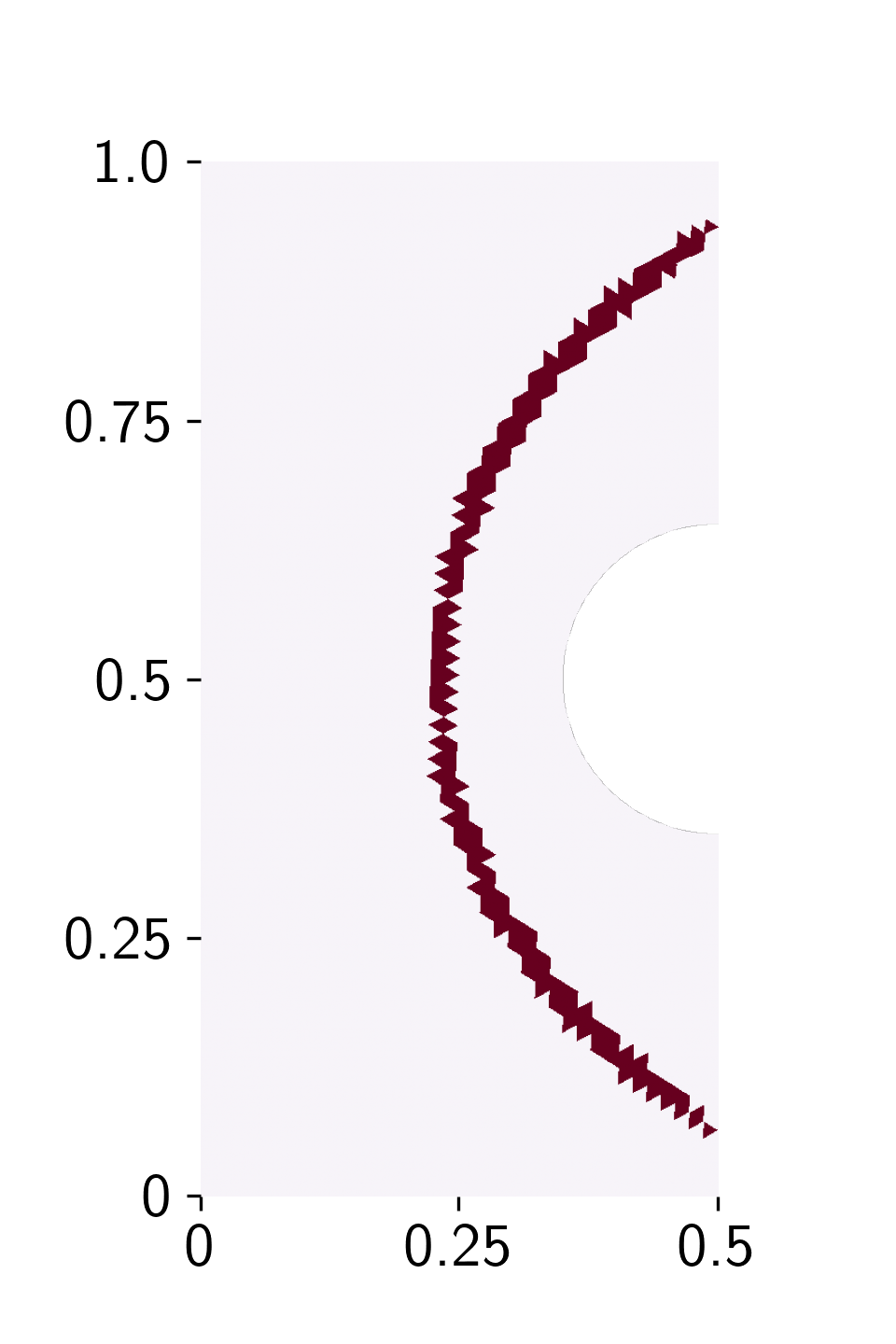}& 
        \includegraphics[trim=0 0 0 1cm, clip,height=0.22\textwidth]{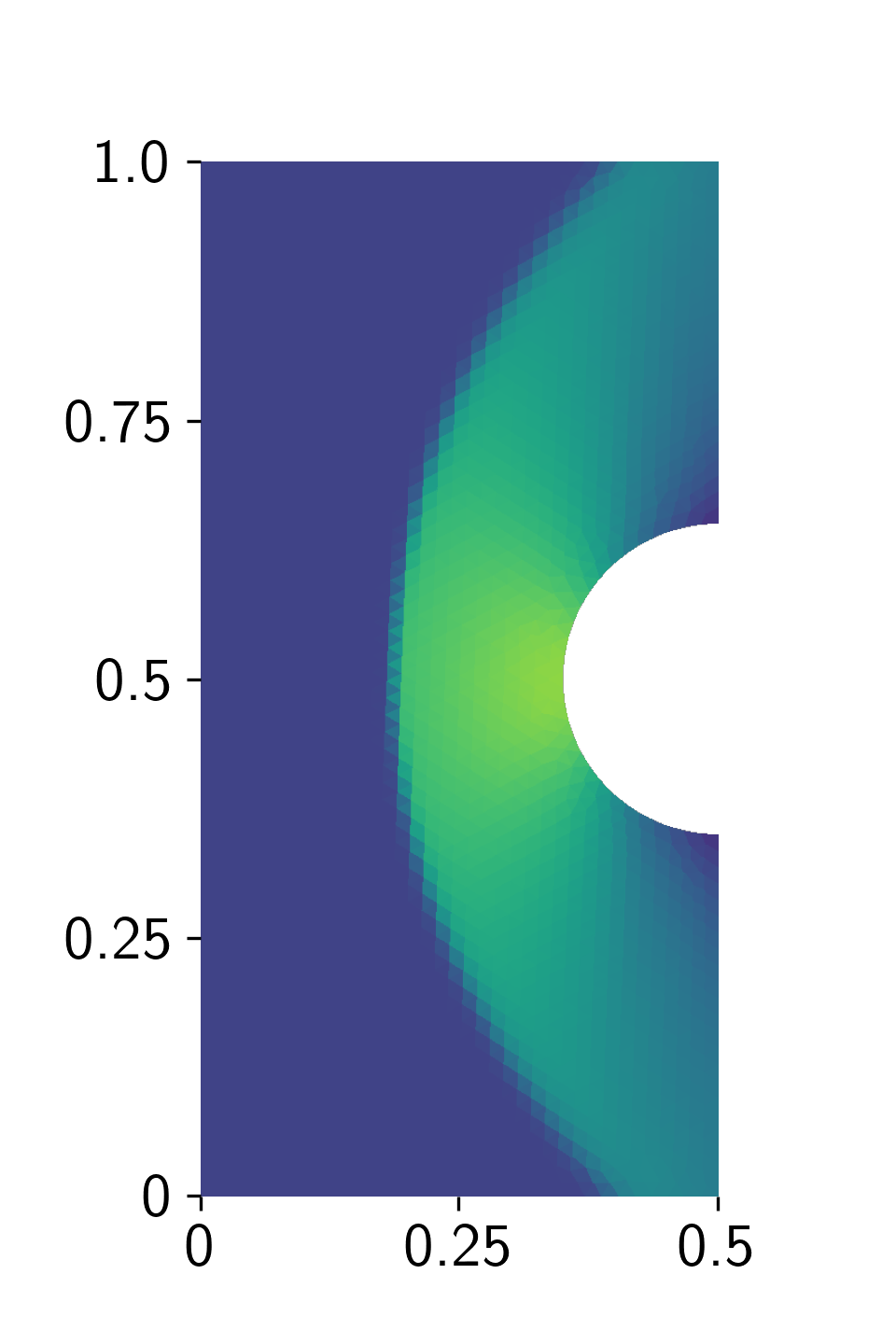}&
        \includegraphics[trim=0 0 0 1cm, clip,height=0.22\textwidth]{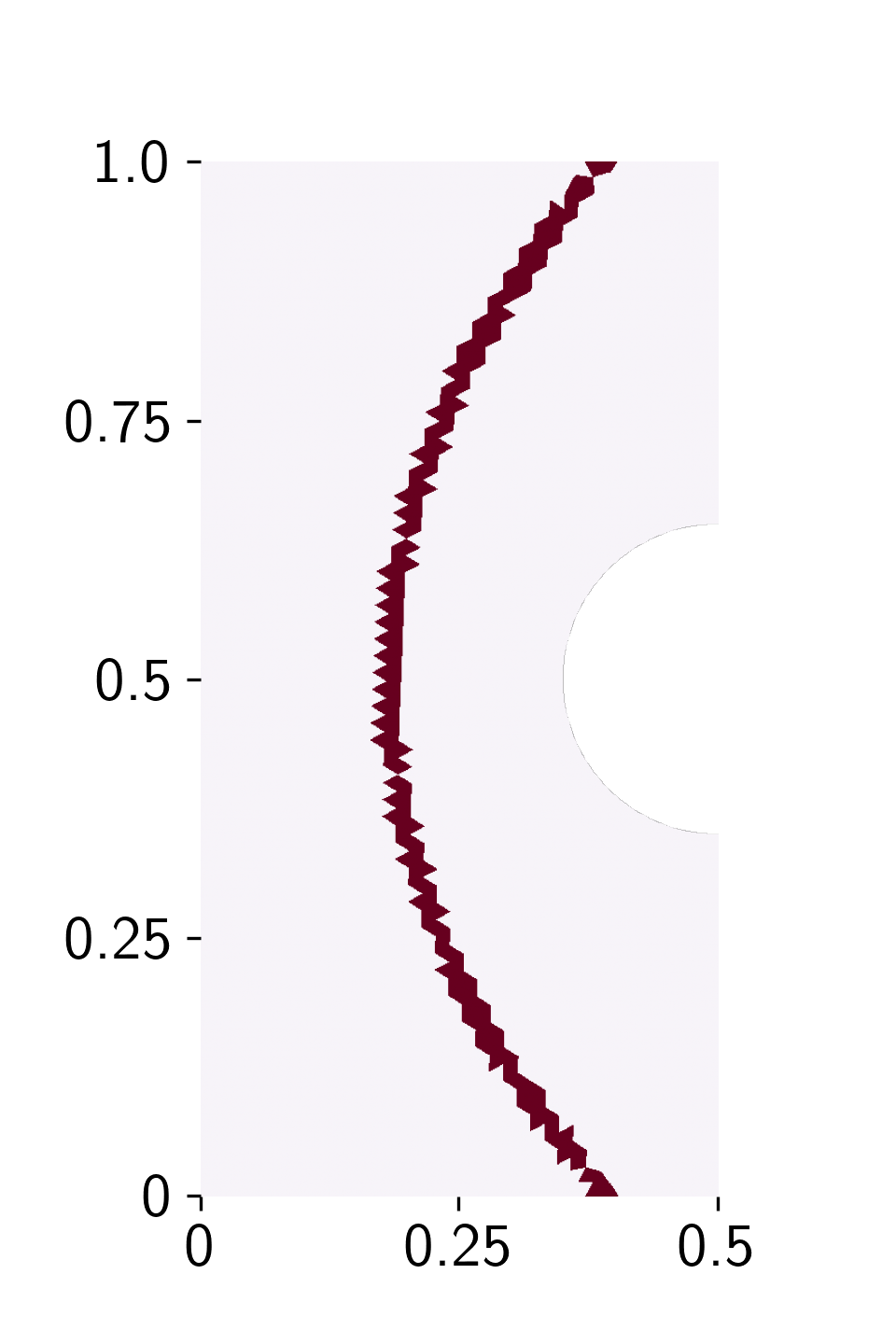}\\
        \raisebox{2.7em}{\rotatebox[origin=lb]{90}{\parbox{2cm}{\centering \footnotesize{GNN-LaSDI}}}}&  
        \includegraphics[trim=0 0 0 1cm, clip,height=0.22\textwidth]{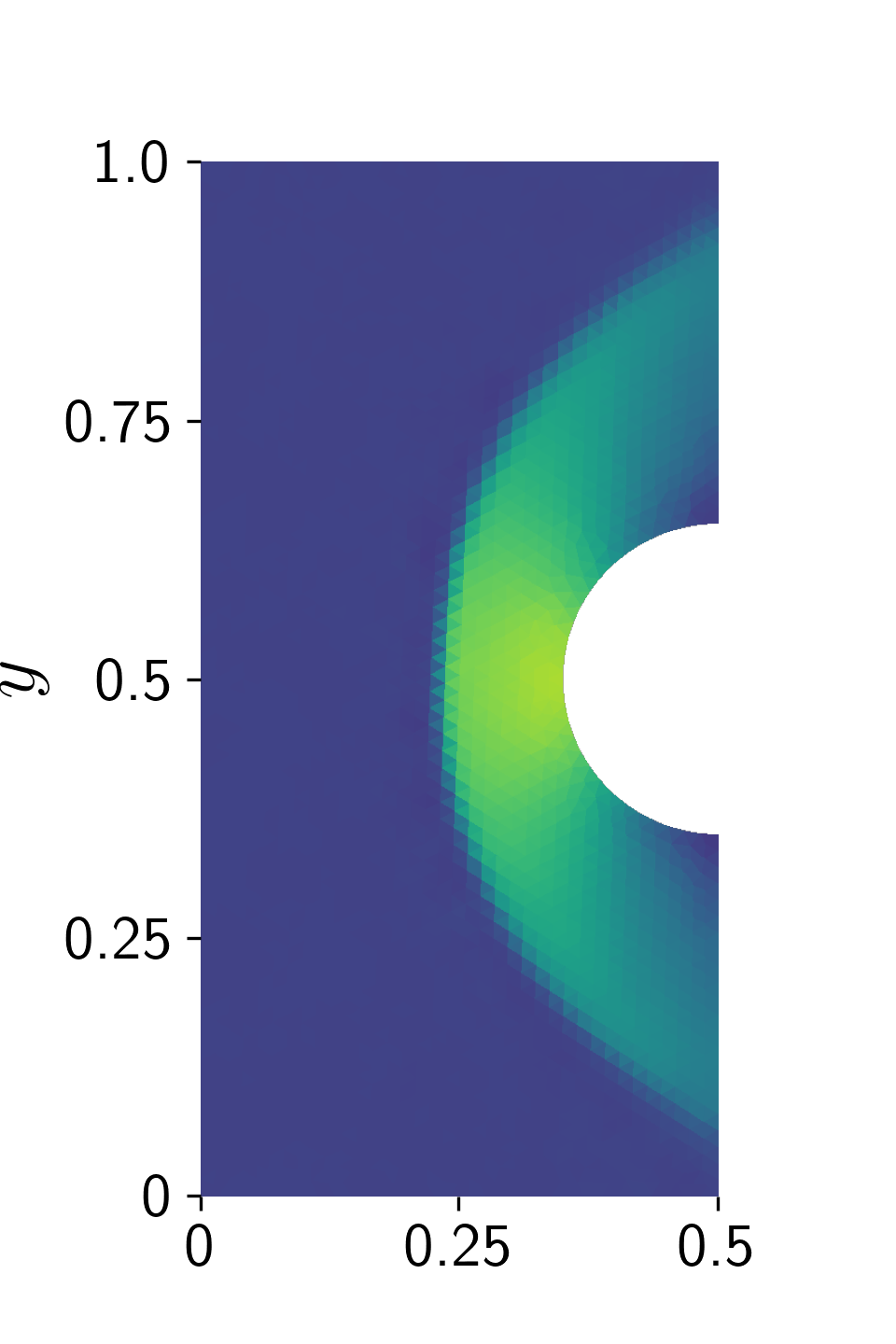}& 
        \includegraphics[trim=0 0 0 1cm, clip,height=0.22\textwidth]{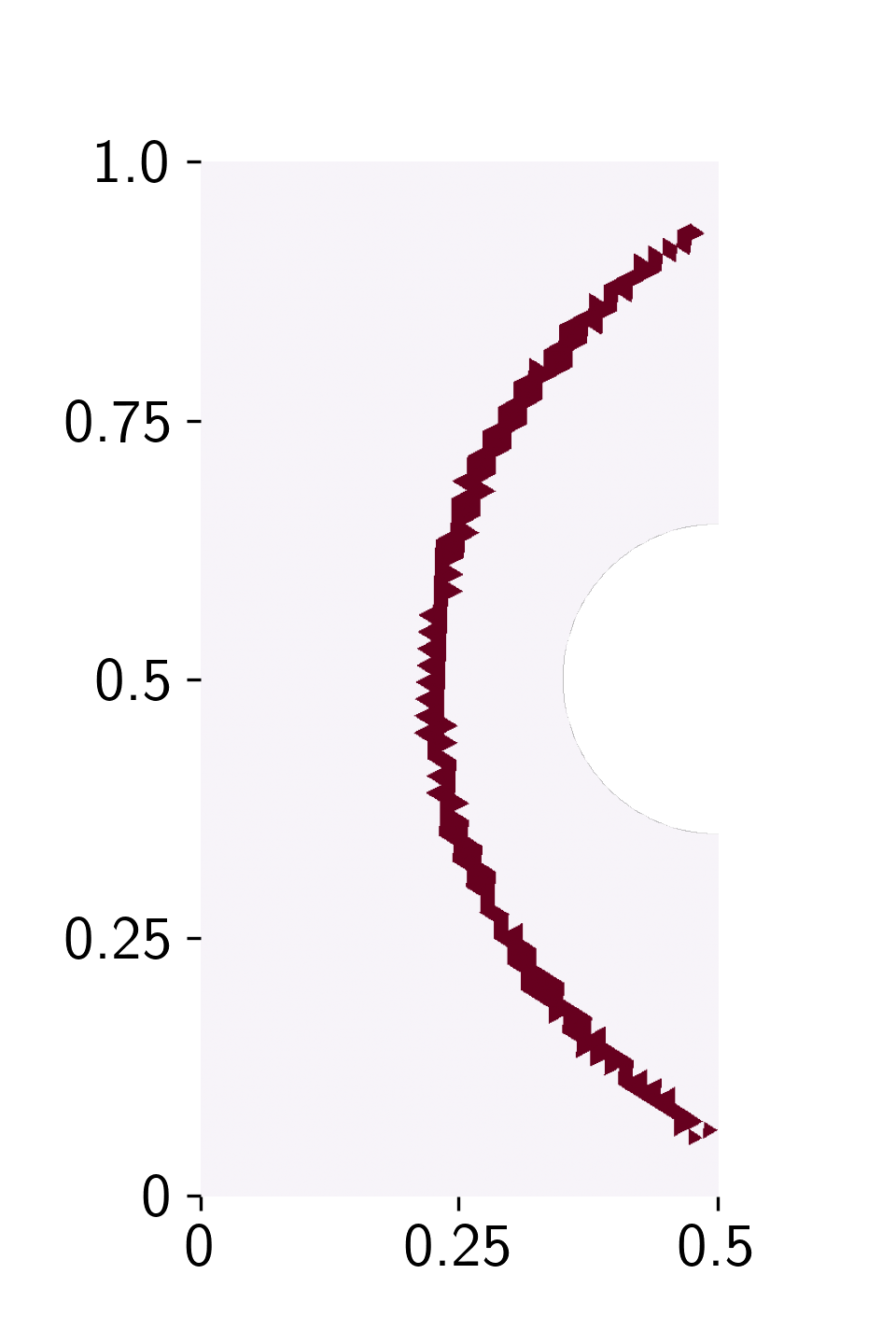}& 
        \includegraphics[trim=0 0 0 1cm, clip,height=0.22\textwidth]{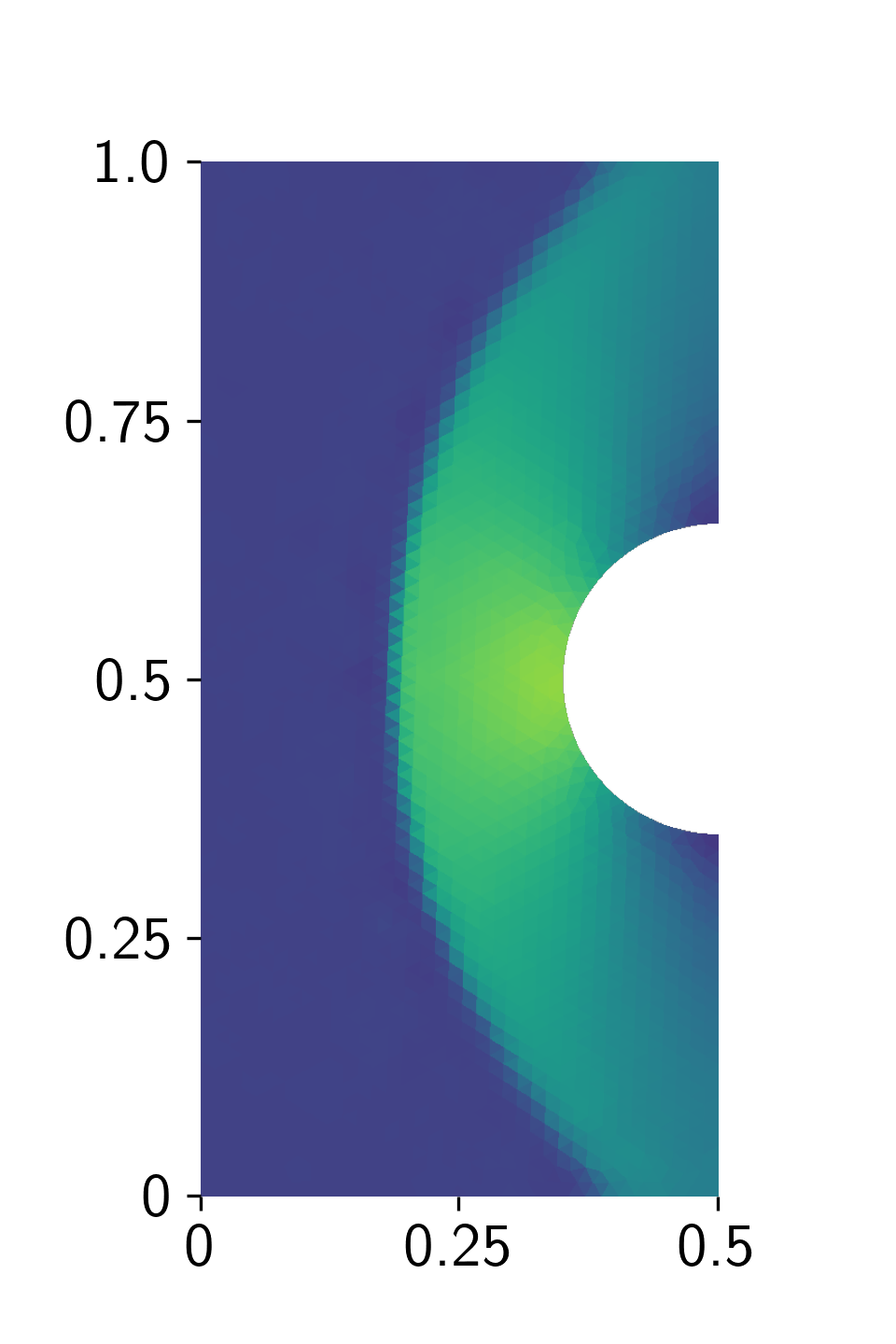}& 
        \includegraphics[trim=0 0 0 1cm, clip,height=0.22\textwidth]{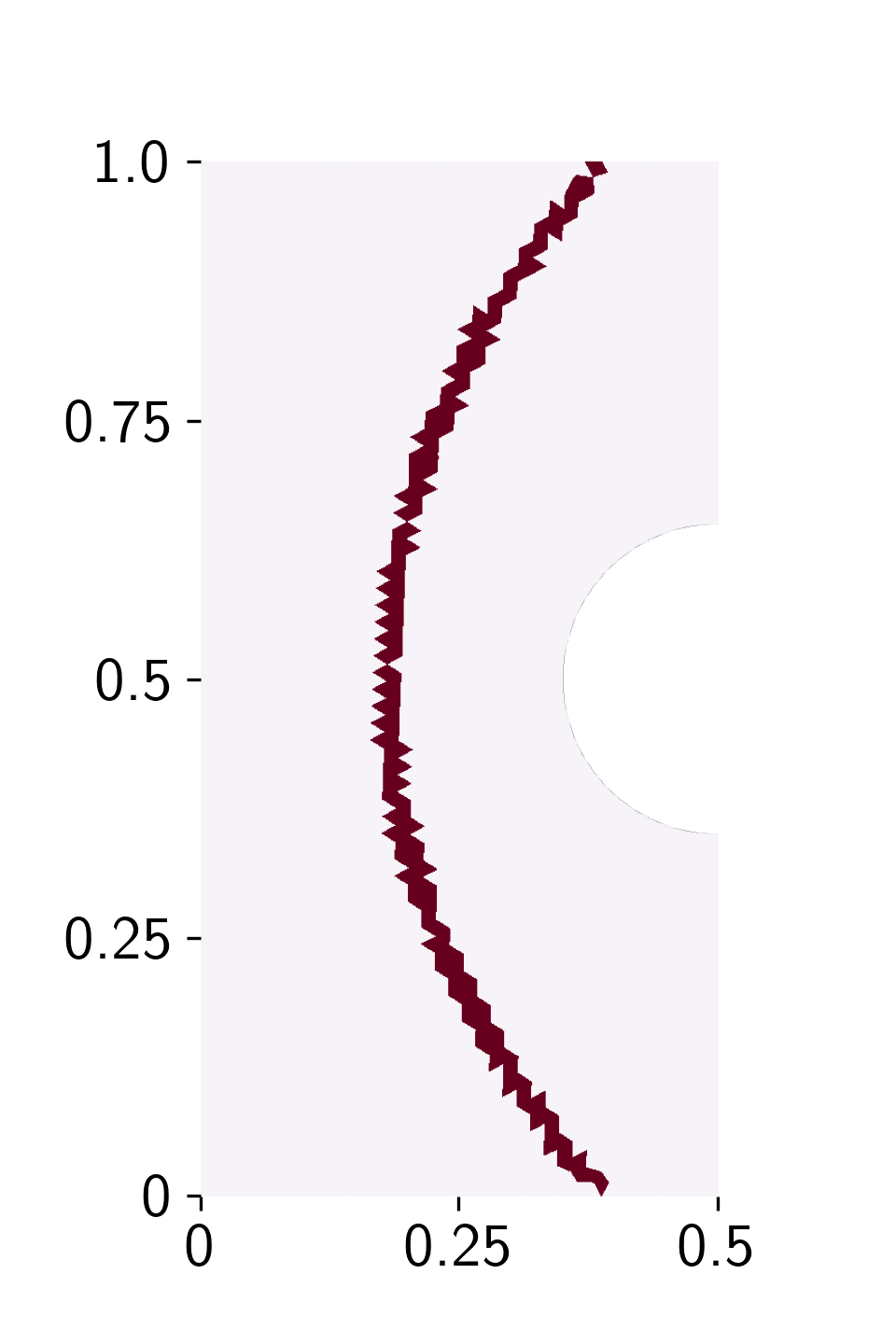}\\
        \raisebox{2.7em}{\rotatebox[origin=lb]{90}{\parbox{2cm}{\centering \footnotesize{GD-LSPG}}}}& 
        \includegraphics[trim=0 0 0 1cm, clip,height=0.22\textwidth]{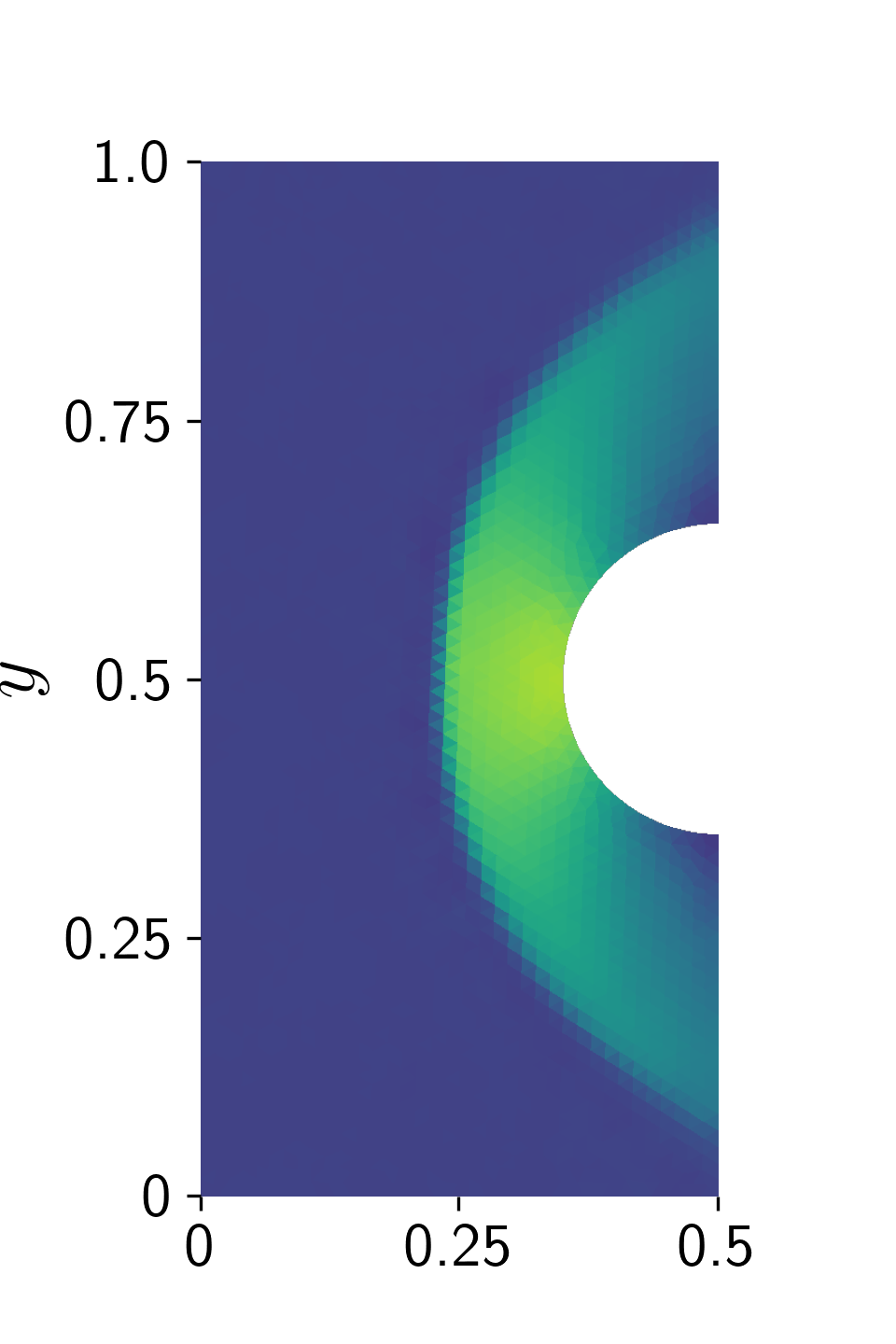}& 
        \includegraphics[trim=0 0 0 1cm, clip,height=0.22\textwidth]{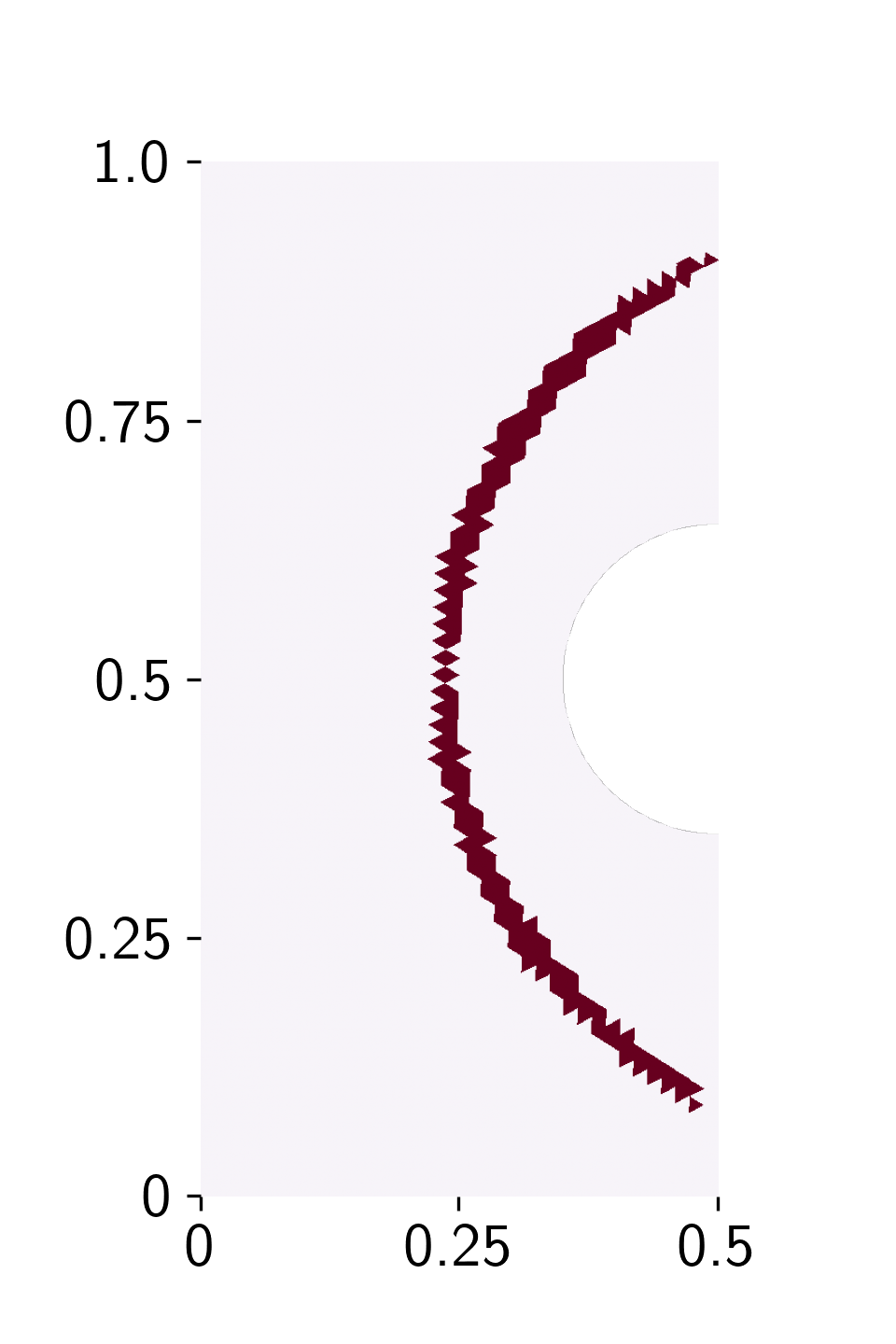}&
        \includegraphics[trim=0 0 0 1cm, clip,height=0.22\textwidth]{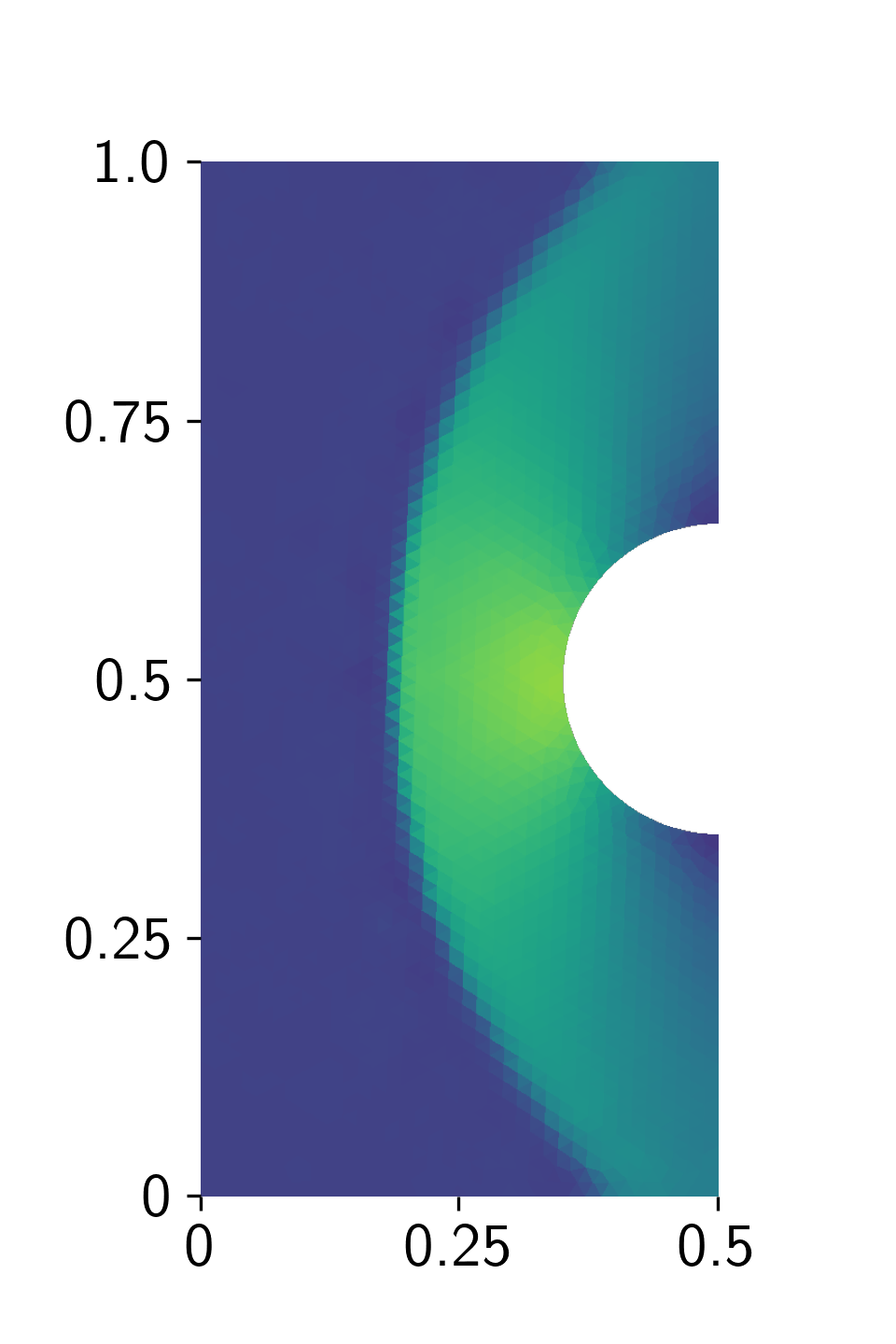}&
        \includegraphics[trim=0 0 0 1cm, clip,height=0.22\textwidth]{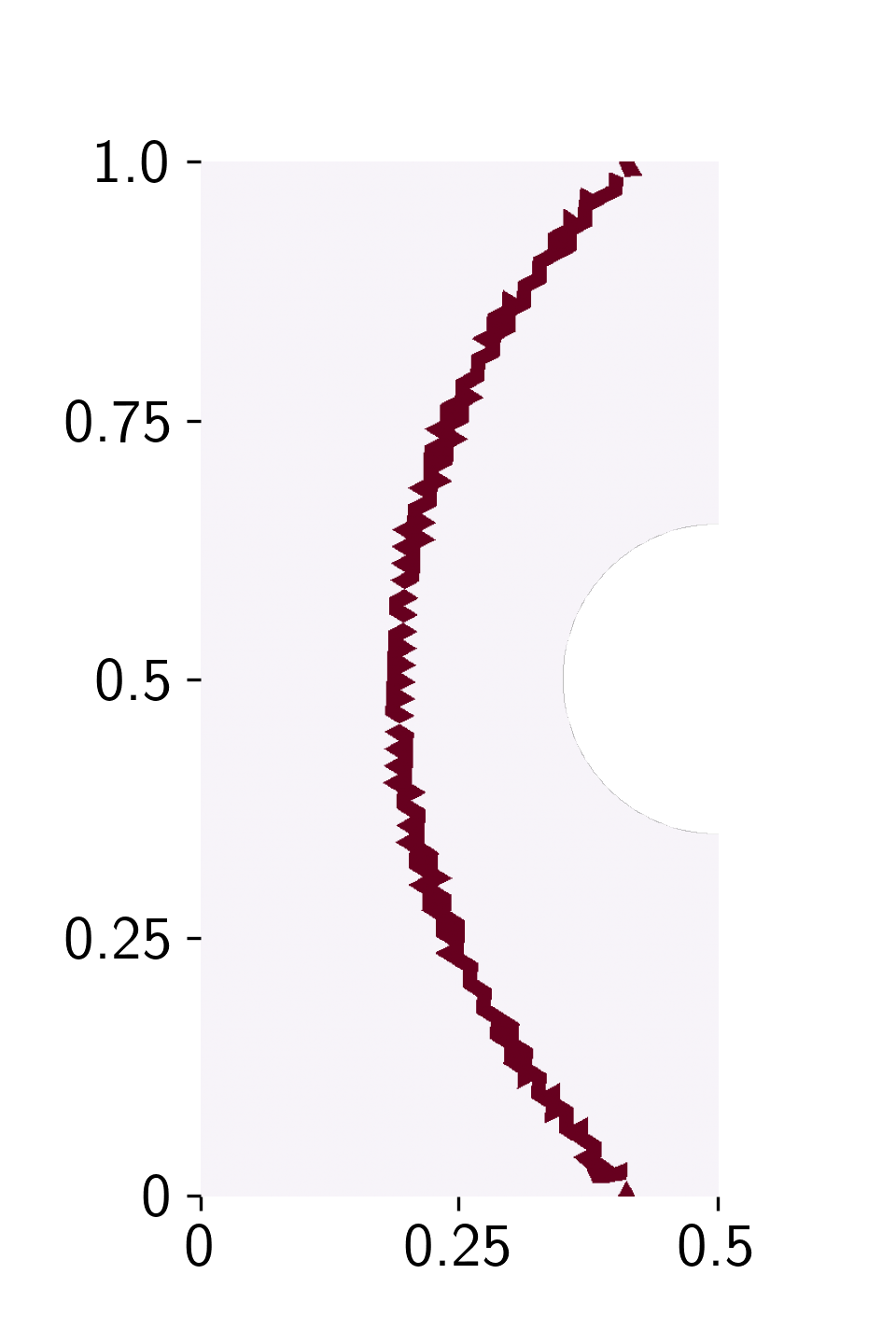}\\
        \raisebox{2.7em}{\rotatebox[origin=lb]{90}{\parbox{2cm}{\centering \footnotesize{POD-LaSDI}}}}& 
        \includegraphics[trim=0 0 0 1cm, clip,height=0.22\textwidth]{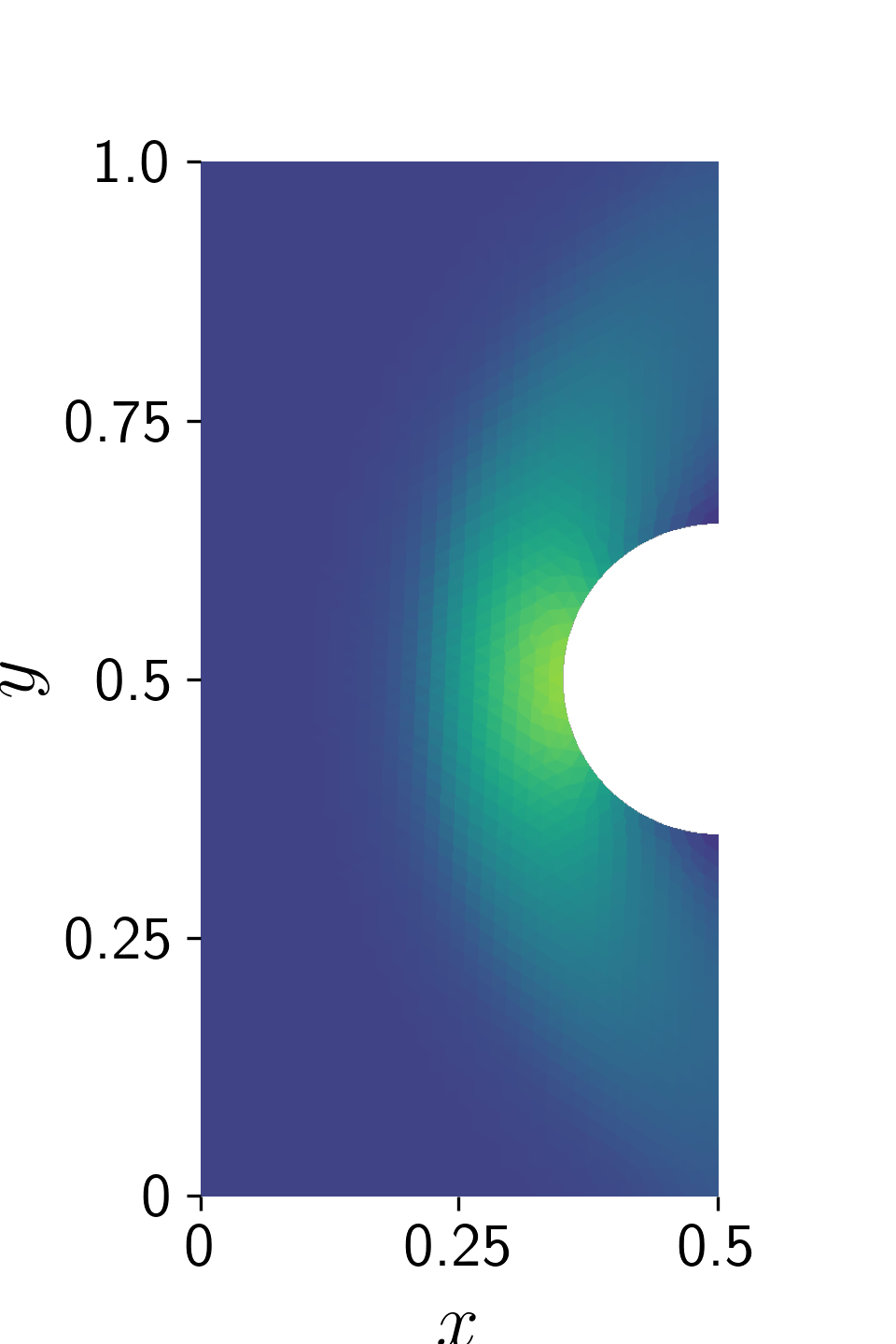}& 
        \includegraphics[trim=0 0 0 1cm, clip,height=0.22\textwidth]{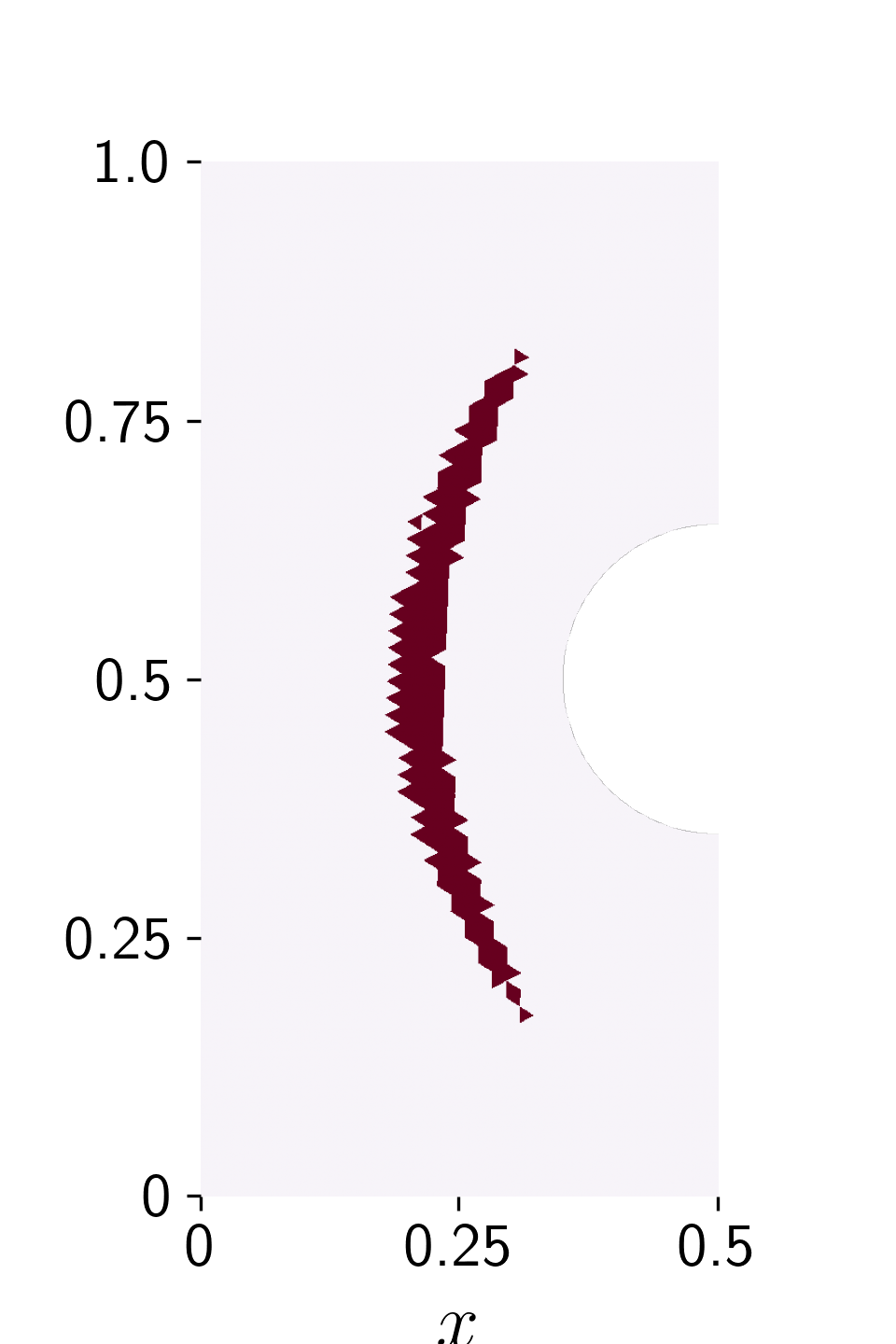}& 
        \includegraphics[trim=0 0 0 1cm, clip,height=0.22\textwidth]{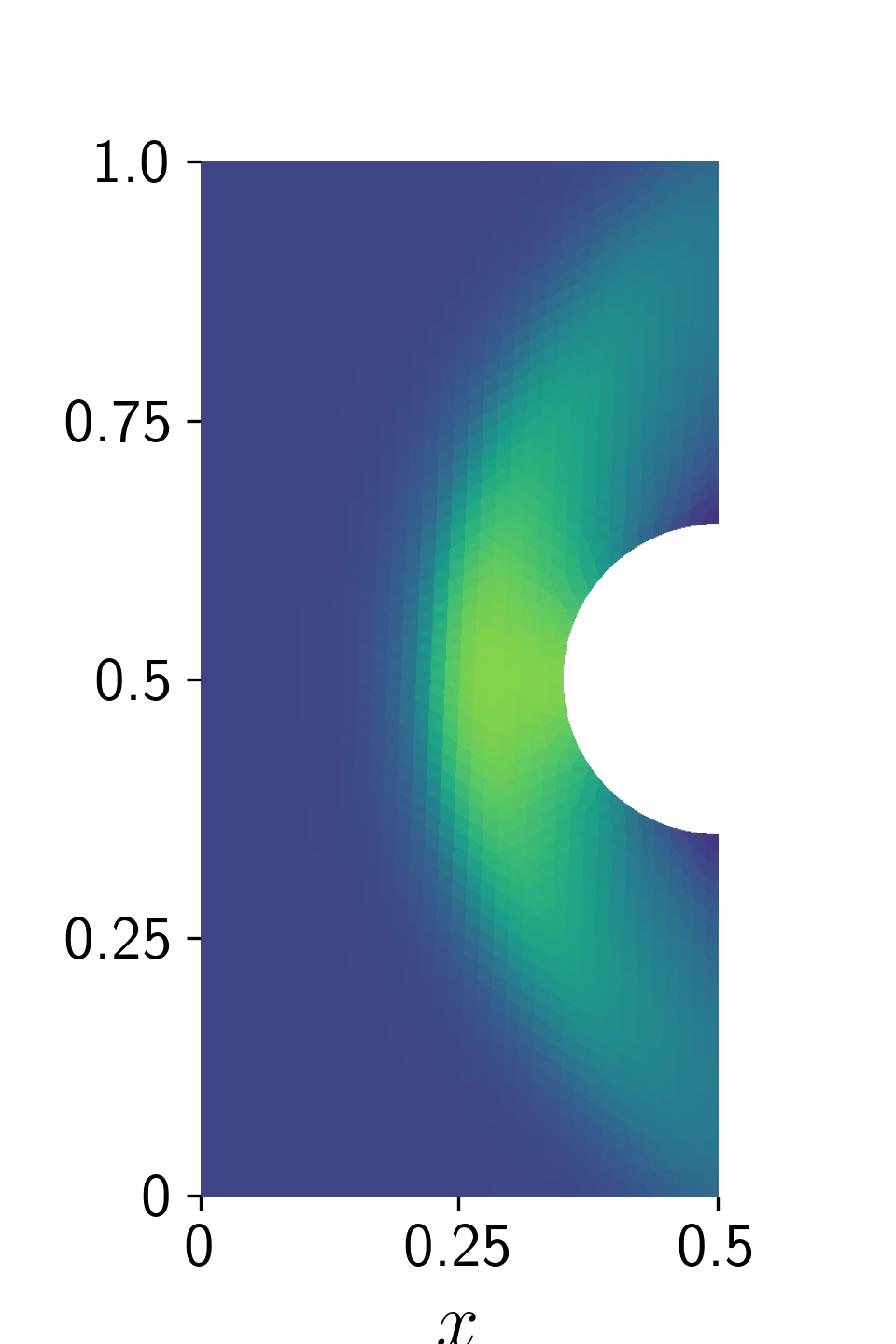}&
        \includegraphics[trim=0 0 0 1cm, clip,height=0.22\textwidth]{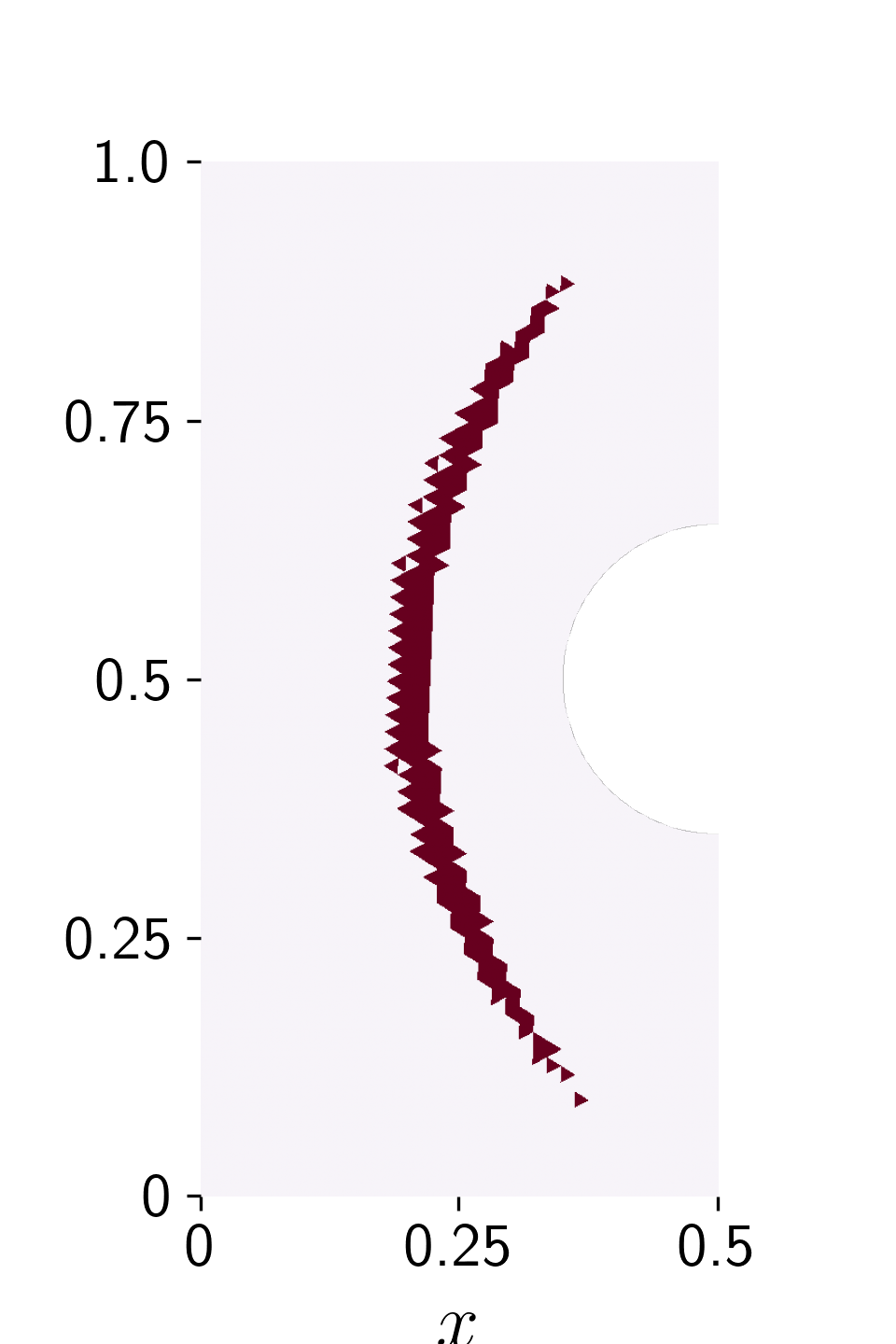} \\
        & \multicolumn{4}{c}{\includegraphics[scale=.3]{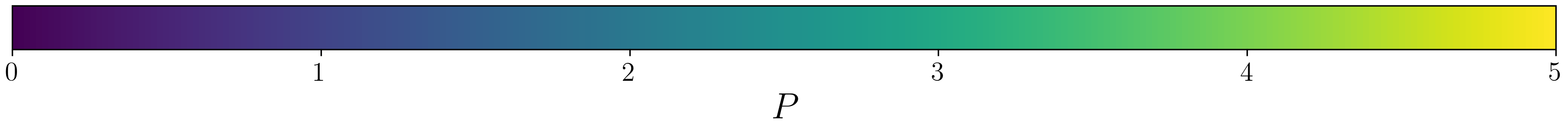}}
    \end{tabular}
    \captionsetup{justification=centering}
    \caption{Pressure field solution and corresponding point cloud representations of the bow shock obtained from the Ducros sensor for the FOM (top row), GNN-LaSDI (second row), GD-LSPG (third row), and POD-LaSDI (fourth row) at times $t=0.25$ and $t=0.5$ for test parameter $M_{\infty} = 1.65$. All ROM solutions are generated using a latent state dimension $n=2$. GNN-LaSDI and GD-LSPG accurately capture the moving shock, whereas POD-LaSDI produces a more diffuse solution that does not preserve a sharp shock boundary. The shared colorbar corresponds to the pressure field, while dark red markers in the point cloud plots denote collocation points identified as part of the shock. (Online version in color.)}
    \label{fig:cyl_sol_pc_n2}
\end{figure}

Figure \ref{fig:cyl_sol_pc_n2} presents the state solutions and their corresponding point cloud representations of the bow shock for the FOM, GNN-LaSDI, GD-LSPG, and POD-LaSDI at $M_{\infty}=1.65$, with all ROM solutions constructed using a latent state dimension of $n=2$. Qualitatively, both GNN-LaSDI and GD-LSPG accurately capture the temporal evolution of the shock at the leading edge of the cylinder. In contrast, the POD-LaSDI solution is more diffuse and fails to accurately preserve the moving shock boundary. The point clouds generated using the Ducros sensor clearly identify the shock location in the FOM, GD-LSPG, and GNN-LaSDI solutions. Conversely, the point cloud extracted from the POD-LaSDI solution differs substantially from that of the FOM, reflecting the inability of the linear manifold to accurately capture the moving shock.

\begin{figure}[ht!]
    \centering
     \begin{tabular}{cccc}
        & {\footnotesize{$n=2$}} & {\footnotesize{$n=4$}} & {\footnotesize{$n=6$}} \\
        \raisebox{2.1em}{\rotatebox[origin=lb]{90}{\parbox{2.5cm}{\centering \footnotesize{$M_{\infty}=1.05$}}}}&
        \includegraphics[trim=0.2cm 0 0.4cm 1cm, clip,height=0.23\textwidth]{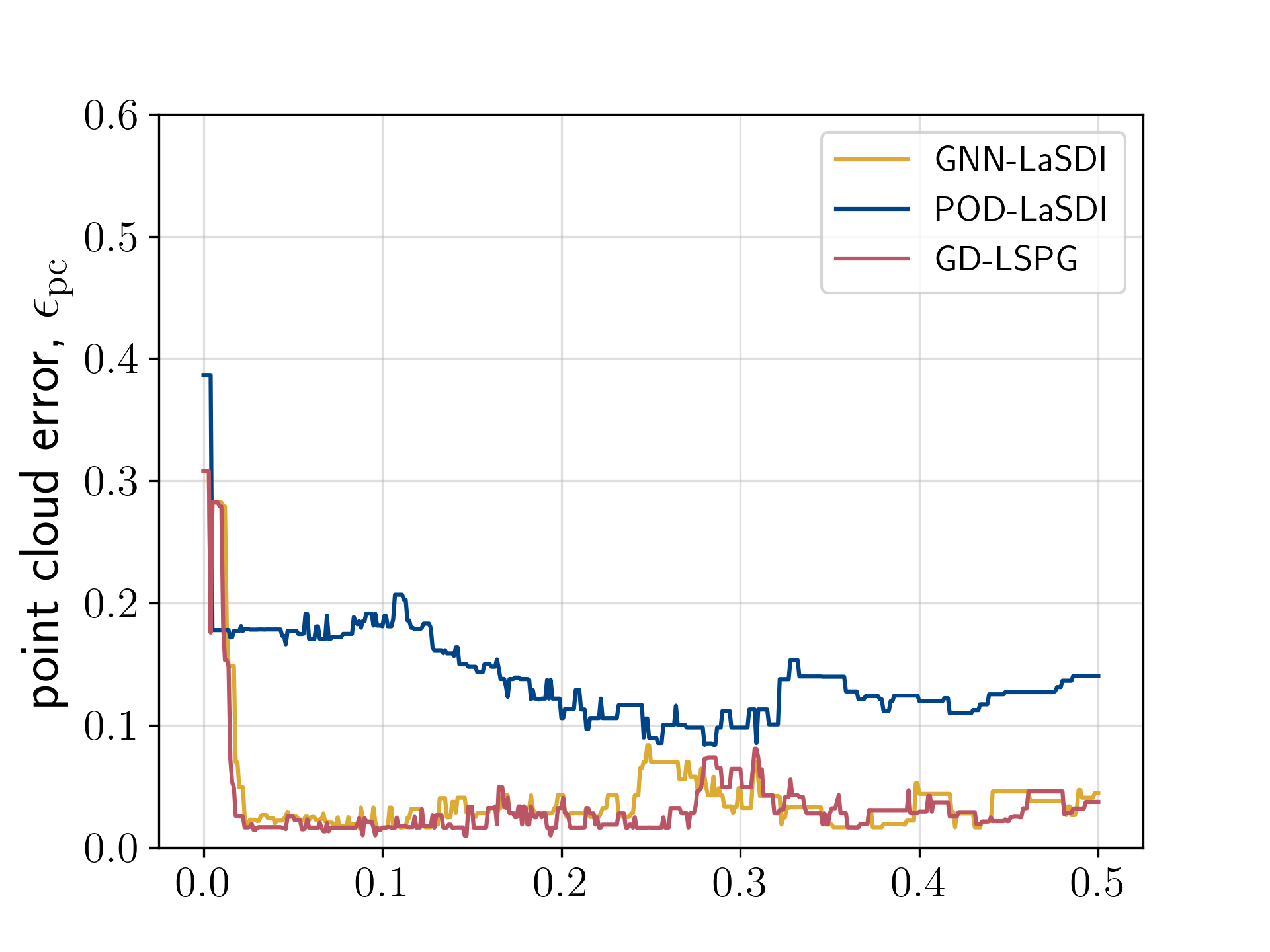}&\hspace{-.5cm}
        \includegraphics[trim=0.4cm 0 0.4cm 1cm, clip,height=0.23\textwidth]{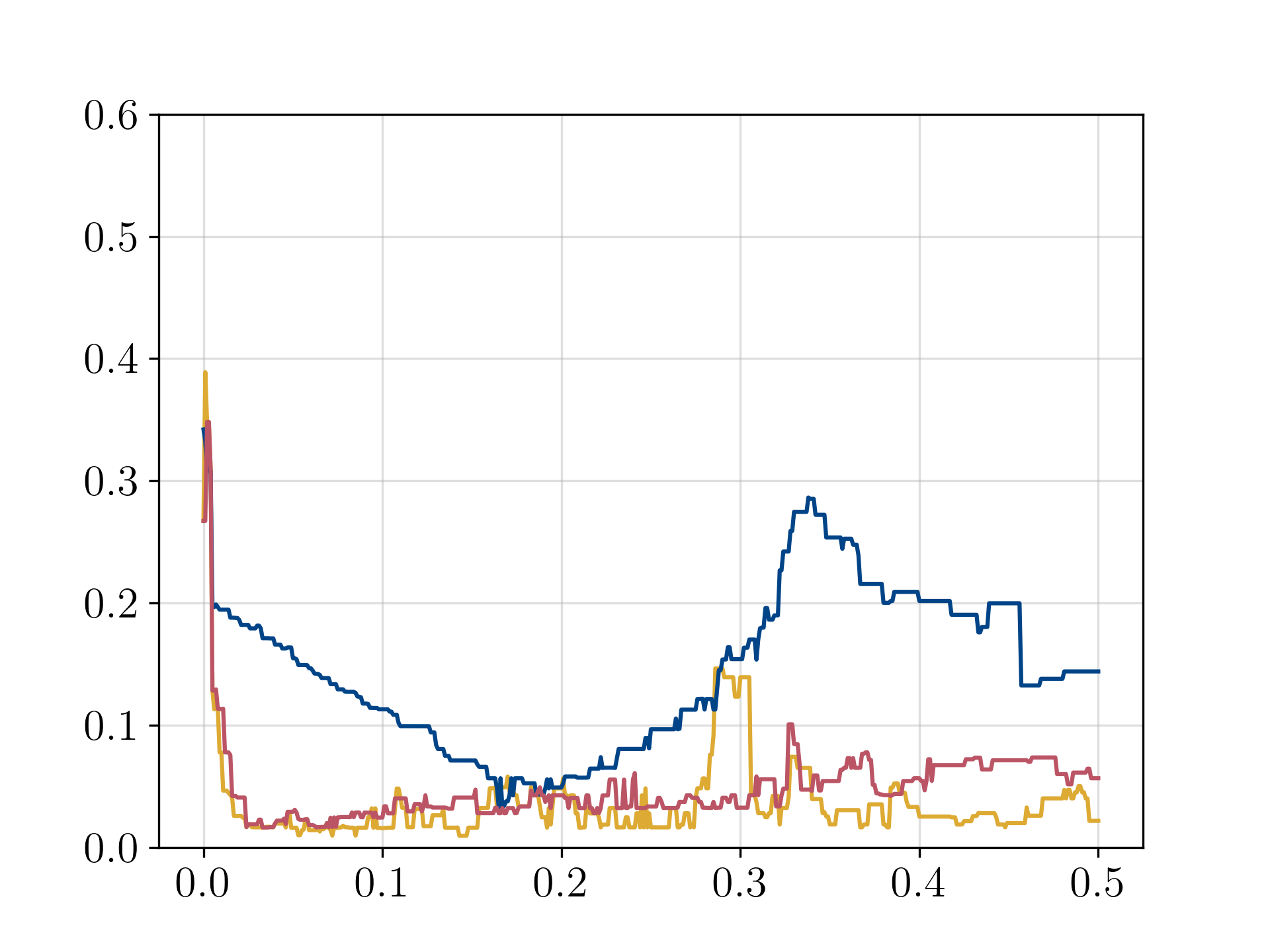}&\hspace{-.5cm}
        \includegraphics[trim=0.4cm 0 0.4cm 1cm, clip,height=0.23\textwidth]{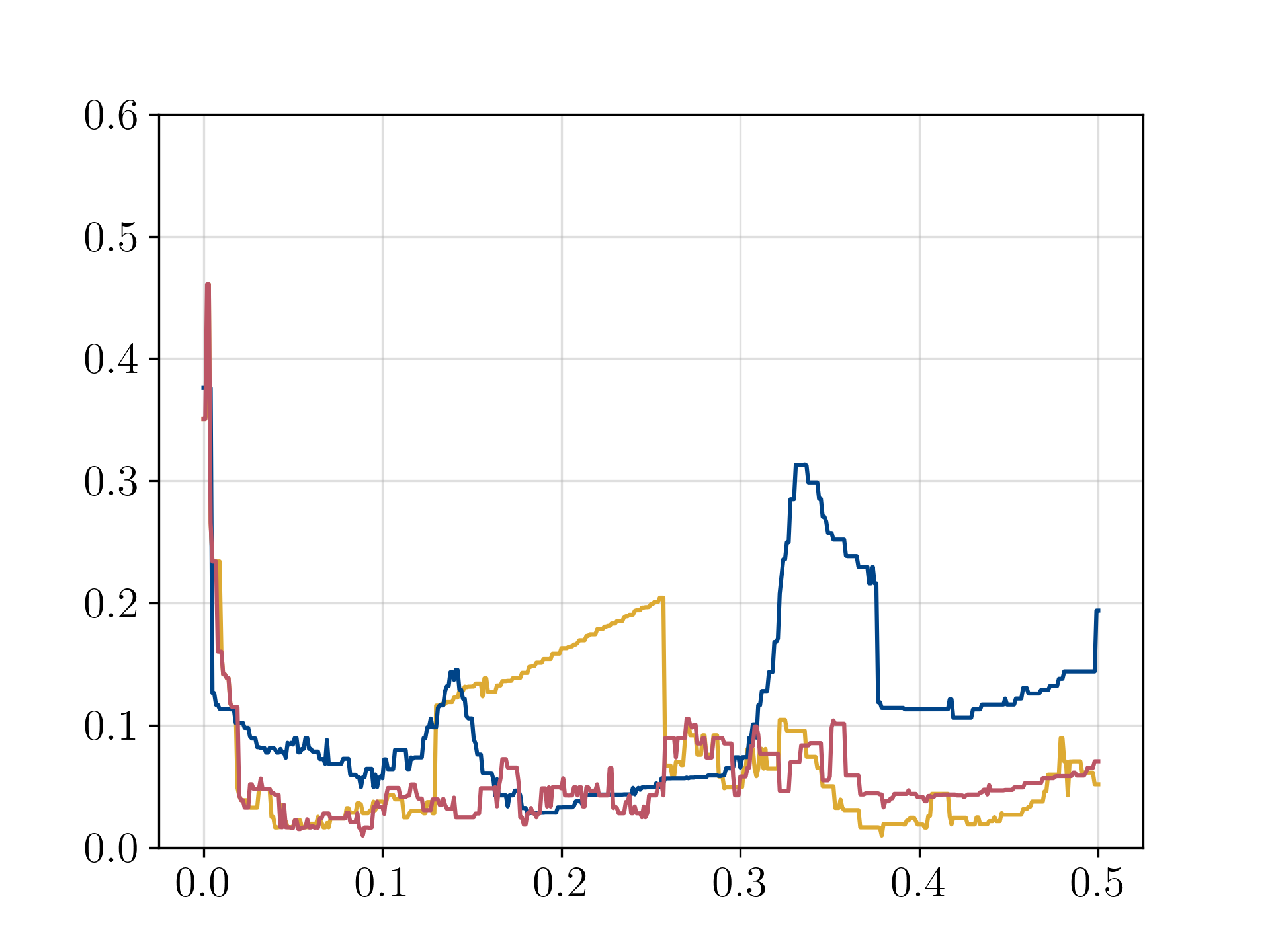}\\
        \raisebox{2.1em}{\rotatebox[origin=lb]{90}{\parbox{2.5cm}{\centering \footnotesize{$M_{\infty}=1.35$}}}}&
        \includegraphics[trim=0.2cm 0 0.4cm 1cm, clip,height=0.23\textwidth]{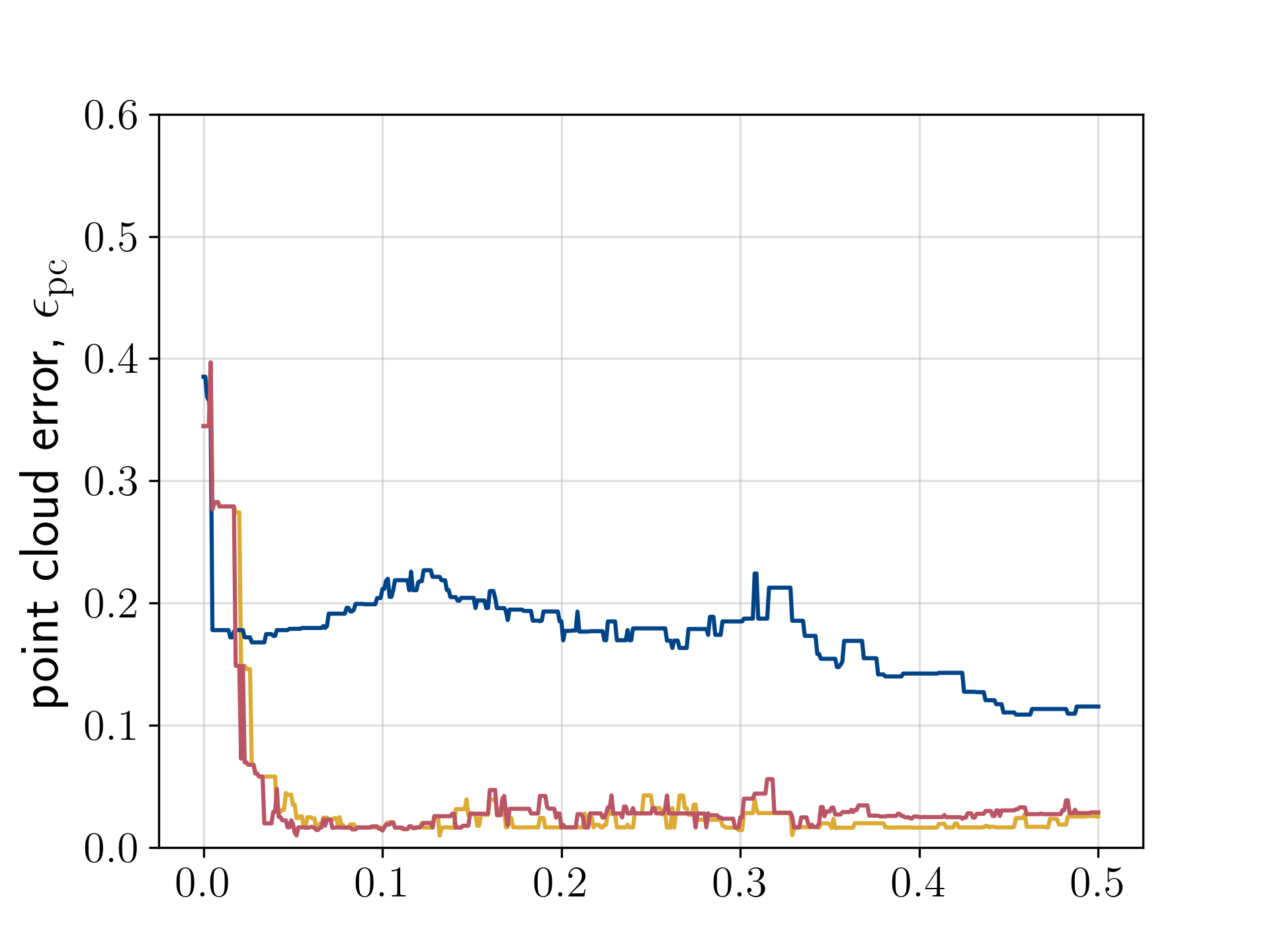}&\hspace{-.5cm}
        \includegraphics[trim=0.4cm 0 0.4cm 1cm, clip,height=0.23\textwidth]{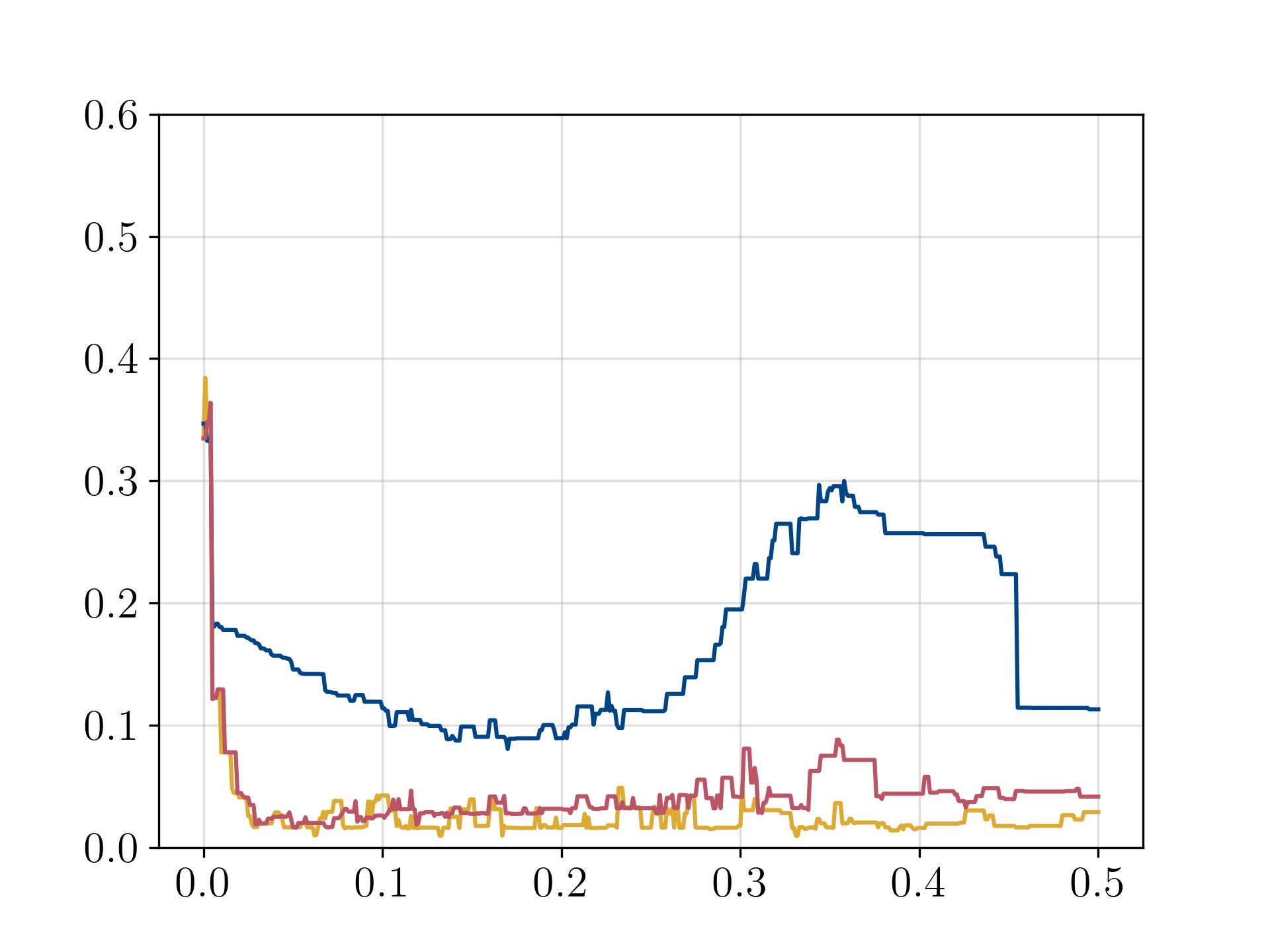}&\hspace{-.5cm}
        \includegraphics[trim=0.4cm 0 0.4cm 1cm, clip,height=0.23\textwidth]{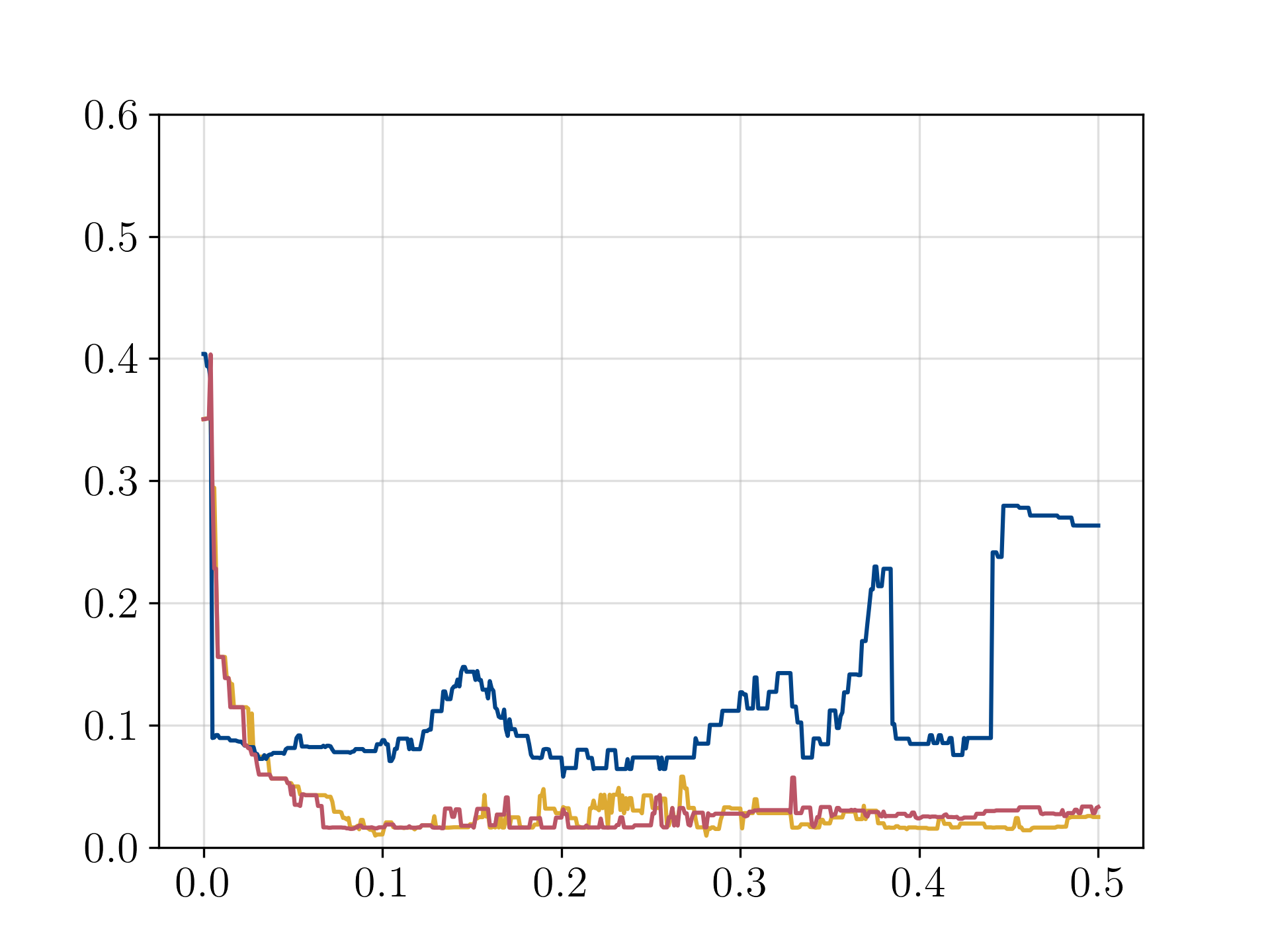}\\
        \raisebox{2.1em}{\rotatebox[origin=lb]{90}{\parbox{2.5cm}{\centering \footnotesize{$M_{\infty}=1.65$}}}}&
        \includegraphics[trim=0.2cm 0 0.4cm 1cm, clip,height=0.23\textwidth]{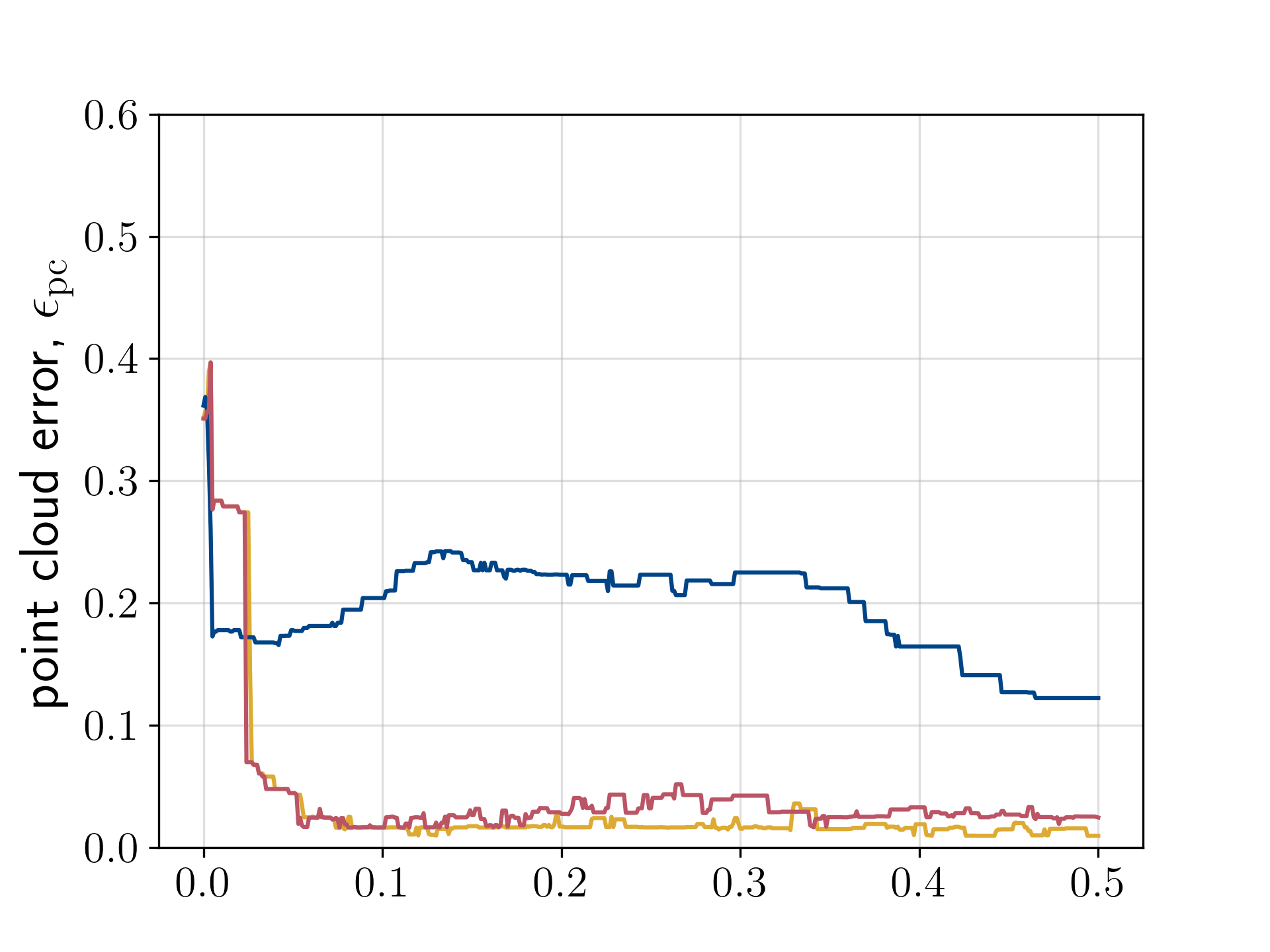}&\hspace{-.5cm}
        \includegraphics[trim=0.4cm 0 0.4cm 1cm, clip,height=0.23\textwidth]{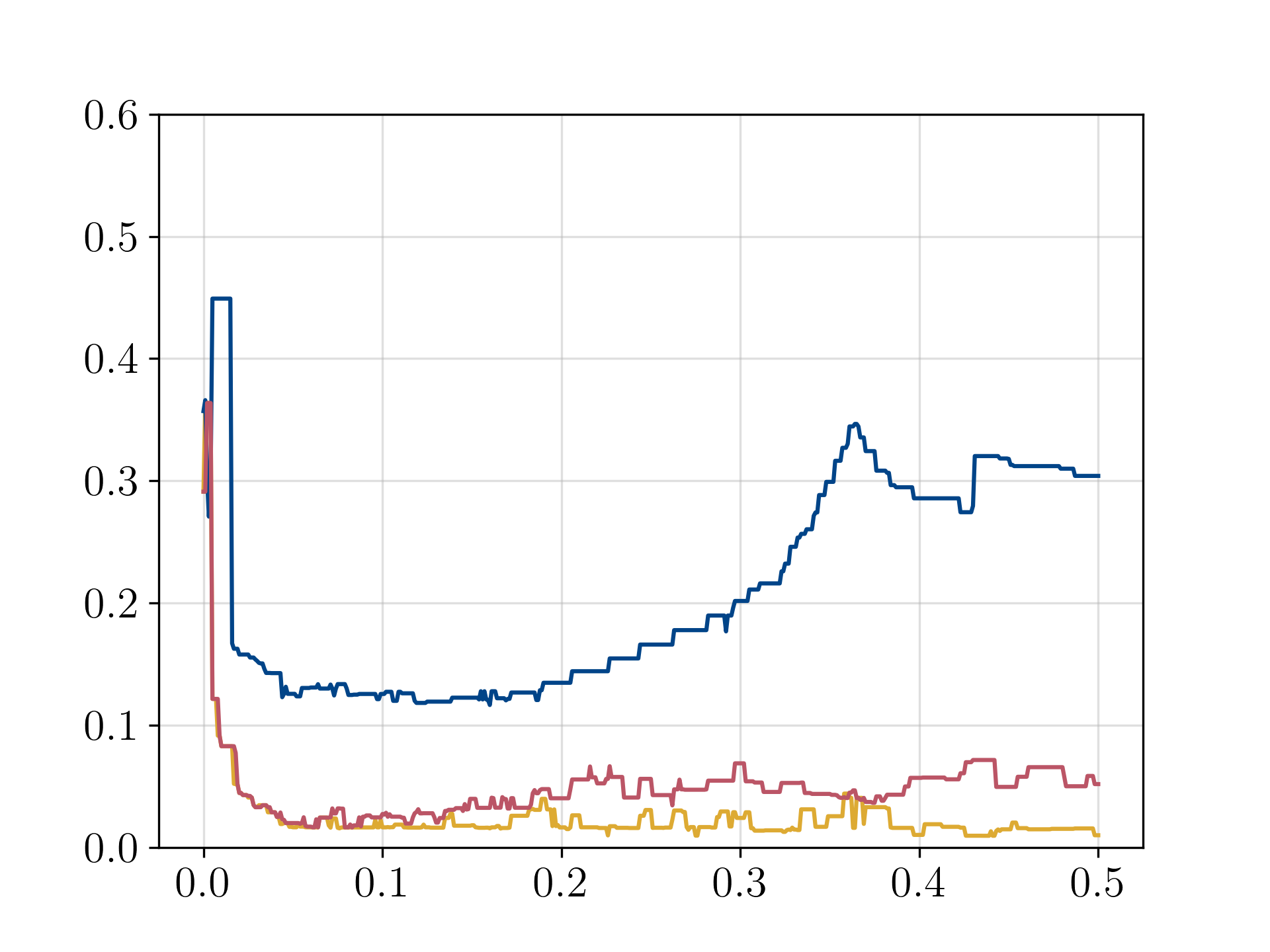}&\hspace{-.5cm}
        \includegraphics[trim=0.4cm 0 0.4cm 1cm, clip,height=0.23\textwidth]{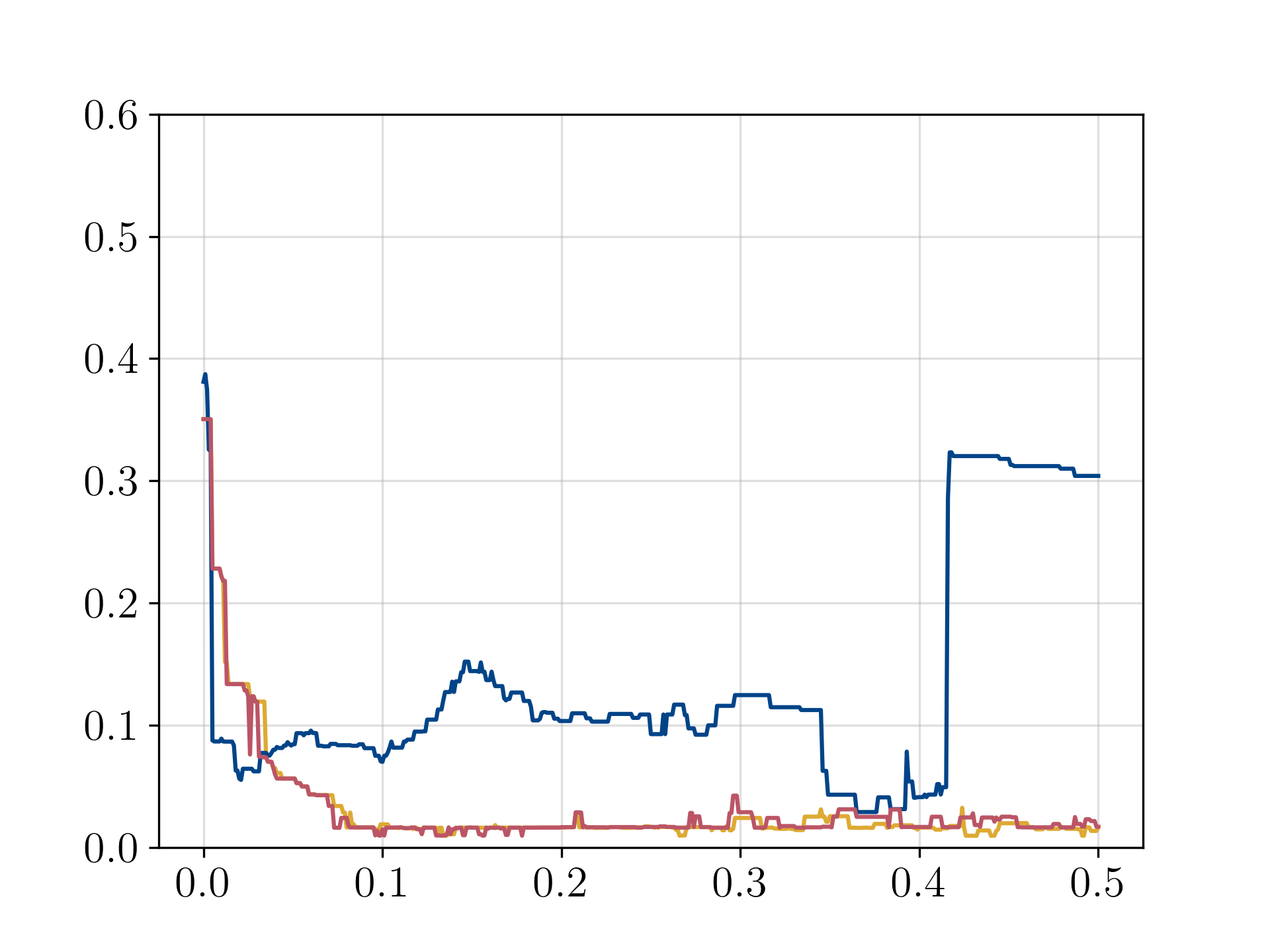}\\
        \raisebox{2.1em}{\rotatebox[origin=lb]{90}{\parbox{2.5cm}{\centering \footnotesize{$M_{\infty}=1.95$}}}}&
        \includegraphics[trim=0.2cm 0 0.4cm 1cm, clip,height=0.23\textwidth]{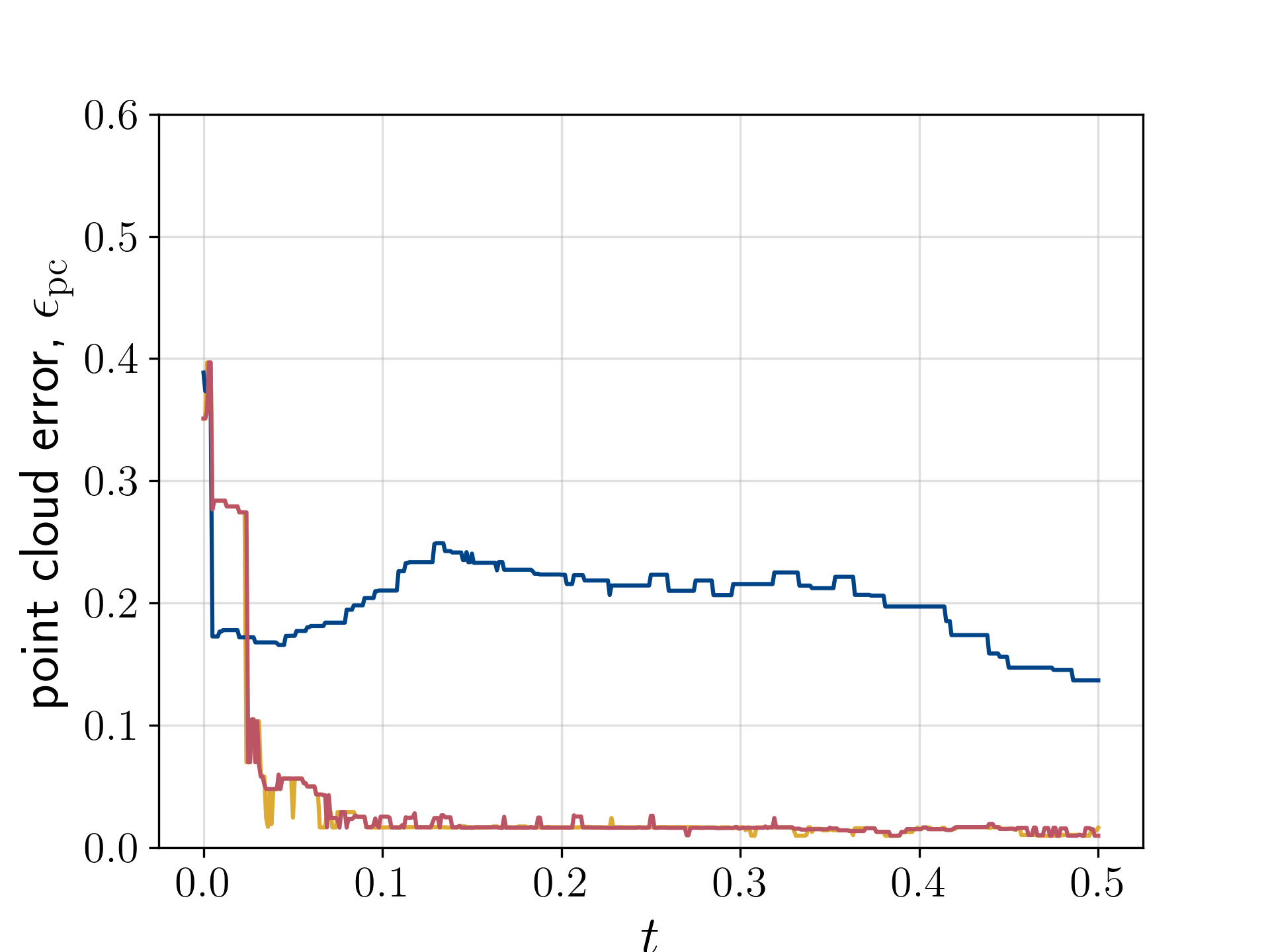}&\hspace{-.5cm}
        \includegraphics[trim=0.4cm 0 0.4cm 1cm, clip,height=0.23\textwidth]{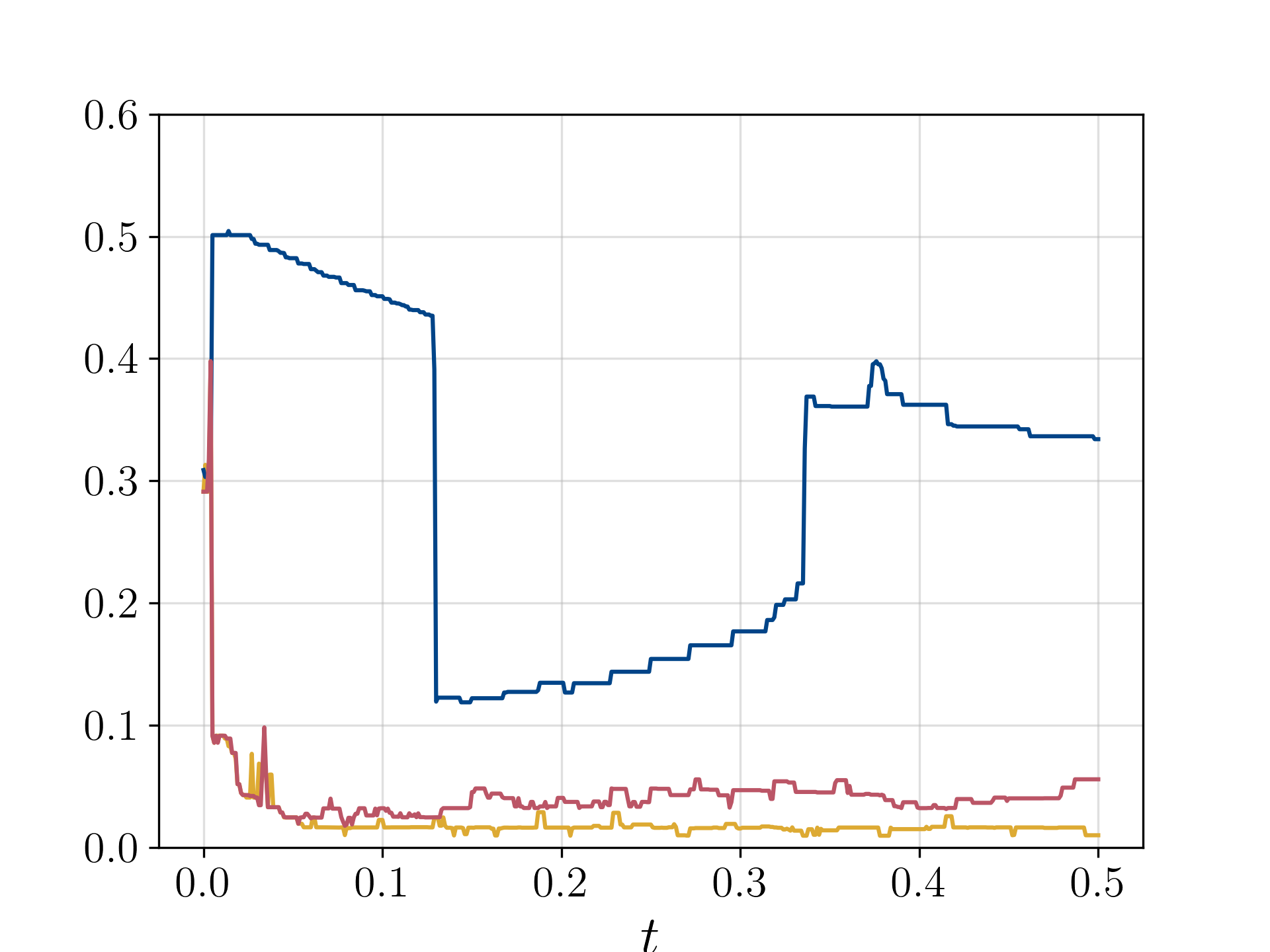}&\hspace{-.5cm}
        \includegraphics[trim=0.4cm 0 0.4cm 1cm, clip,height=0.23\textwidth]{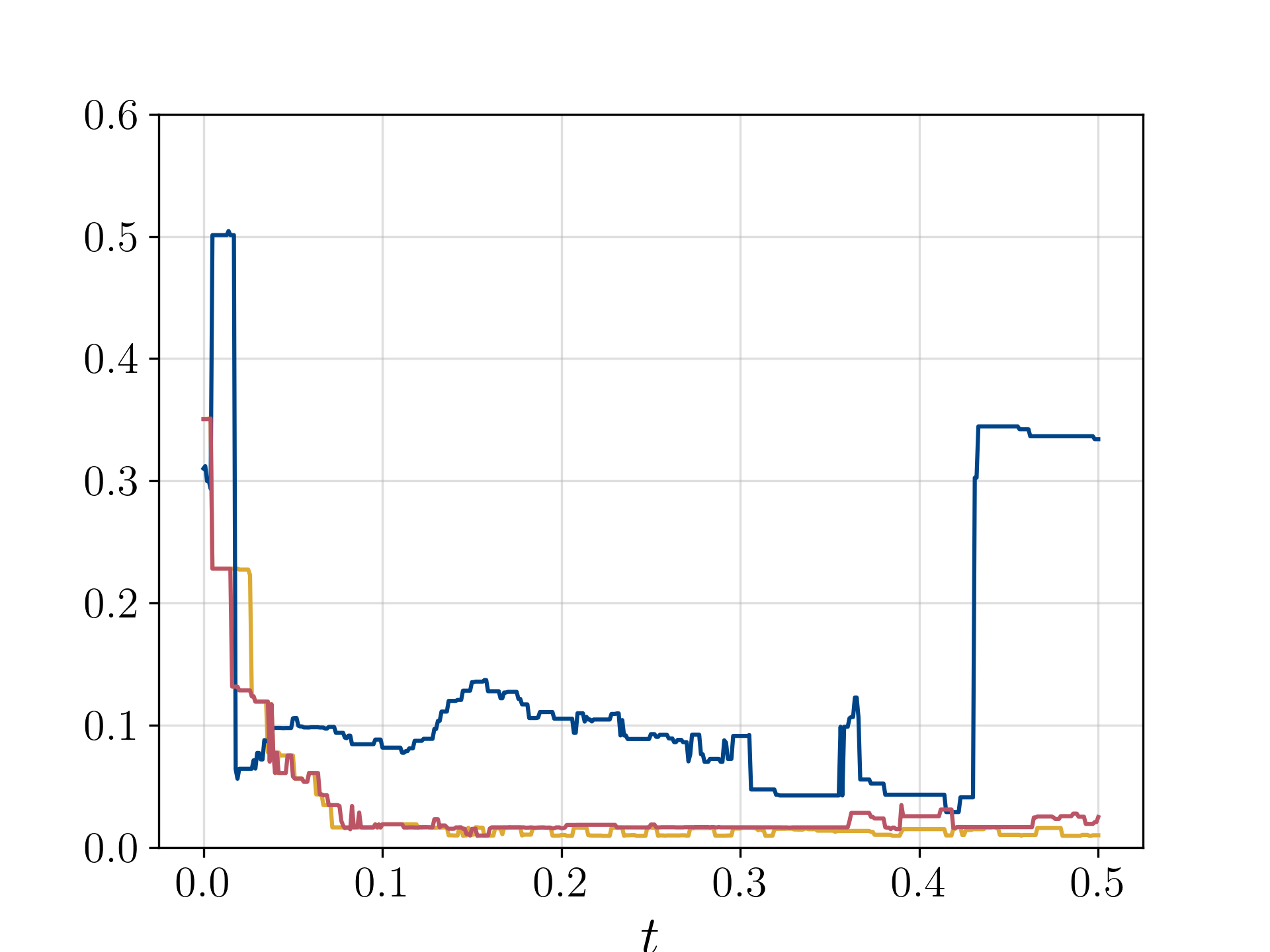}
    \end{tabular}
    \captionsetup{justification=centering}
    \caption{Point cloud errors \eqref{eq:pc_error} for GNN-LaSDI, GD-LSPG, and POD-LaSDI plotted with respect to time. The left, middle, and right columns correspond to latent state dimensions $n=2$, $4$, and $6$, respectively. The rows correspond to test parameters $M_{\infty}=1.05$, $1.35$, $1.65$, and $1.95$ from top to bottom. For all studied test parameters and latent state dimensions, GNN-LaSDI and GD-LSPG provide considerably lower point cloud errors than POD-LaSDI. Note that the bow shock is not yet fully developed in early time steps, resulting in elevated errors for all methods. All subplots share the same legend. (Online version in color.)}
    \label{fig:cyl_pc_error}
\end{figure}

To compare the ability of the studied ROMs to predict the shock location, Figure \ref{fig:cyl_pc_error} presents the point cloud errors for GNN-LaSDI, GD-LSPG, and POD-LaSDI for latent state dimensions $n=2$, $4$, and $6$, and test parameters $M_{\infty}=1.05$, $1.35$, $1.65$, and $1.95$. For all studied latent state dimensions and test parameters, GNN-LaSDI and GD-LSPG consistently achieve lower point cloud errors than POD-LaSDI. This result is consistent with the qualitative results presented in Figure \ref{fig:cyl_sol_pc_n2}, where the shock representations produced by GNN-LaSDI and GD-LSPG closely resemble those of the FOM. Importantly, although GD-LSPG exhibits considerably larger state prediction errors than GNN-LaSDI (Figure \ref{fig:cyl_errorRange}), the two methods produce nearly identical point cloud errors across the parameter domain. This finding further highlights the value of the point cloud error metric for systems exhibiting sharp gradients, as it more directly measures the accuracy of the predicted shock location than conventional state-based error metrics.

\begin{figure}[ht!]
    \centering
     \begin{tabular}{ccc|cc}
        & \multicolumn{2}{c|}{\footnotesize{$M_\infty=1.05$}} & \multicolumn{2}{c}{\footnotesize{$M_\infty=1.55$}} \\
        \hline & & & &\\
        & \footnotesize{state solution} & \footnotesize{point cloud} & \footnotesize{state solution} & \footnotesize{point cloud} \\
        \raisebox{2.3em}{\rotatebox[origin=lb]{90}{\parbox{2cm}{\centering \footnotesize{FOM}}}}&
        \includegraphics[trim=0 0 0 1cm, clip,height=0.22\textwidth]{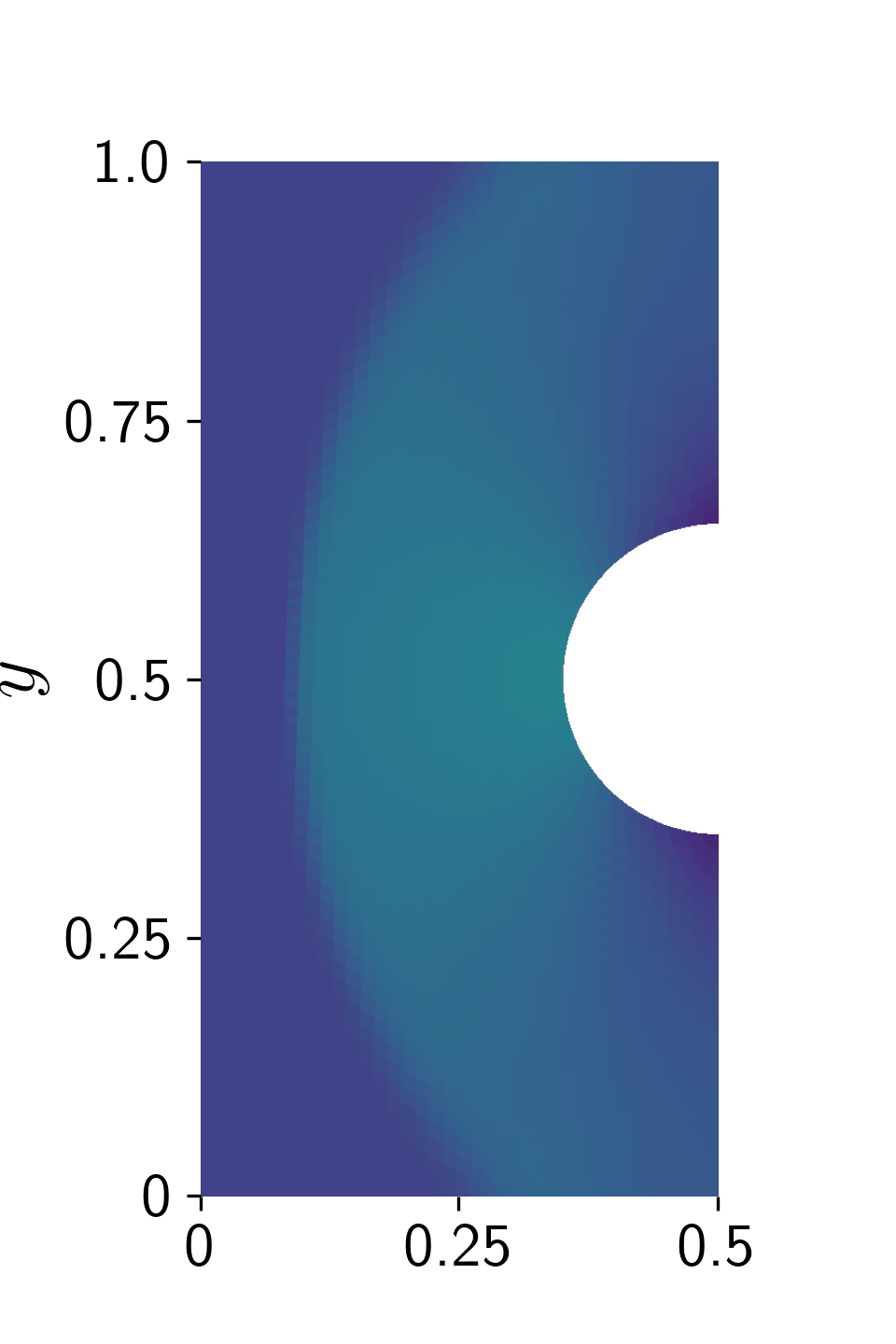}&
        \includegraphics[trim=0 0 0 1cm, clip,height=0.22\textwidth]{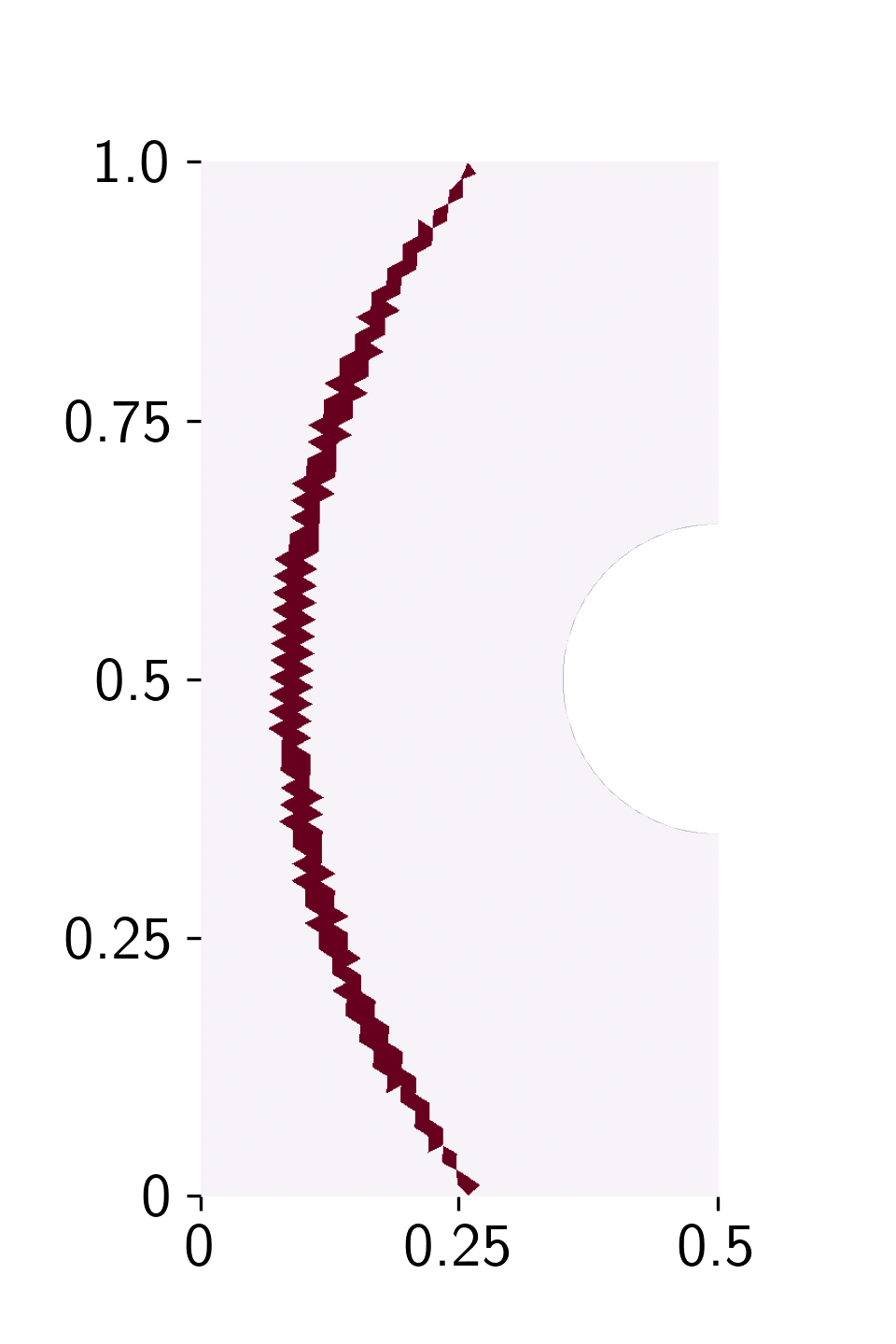}&
        \includegraphics[trim=0 0 0 1cm, clip,height=0.22\textwidth]{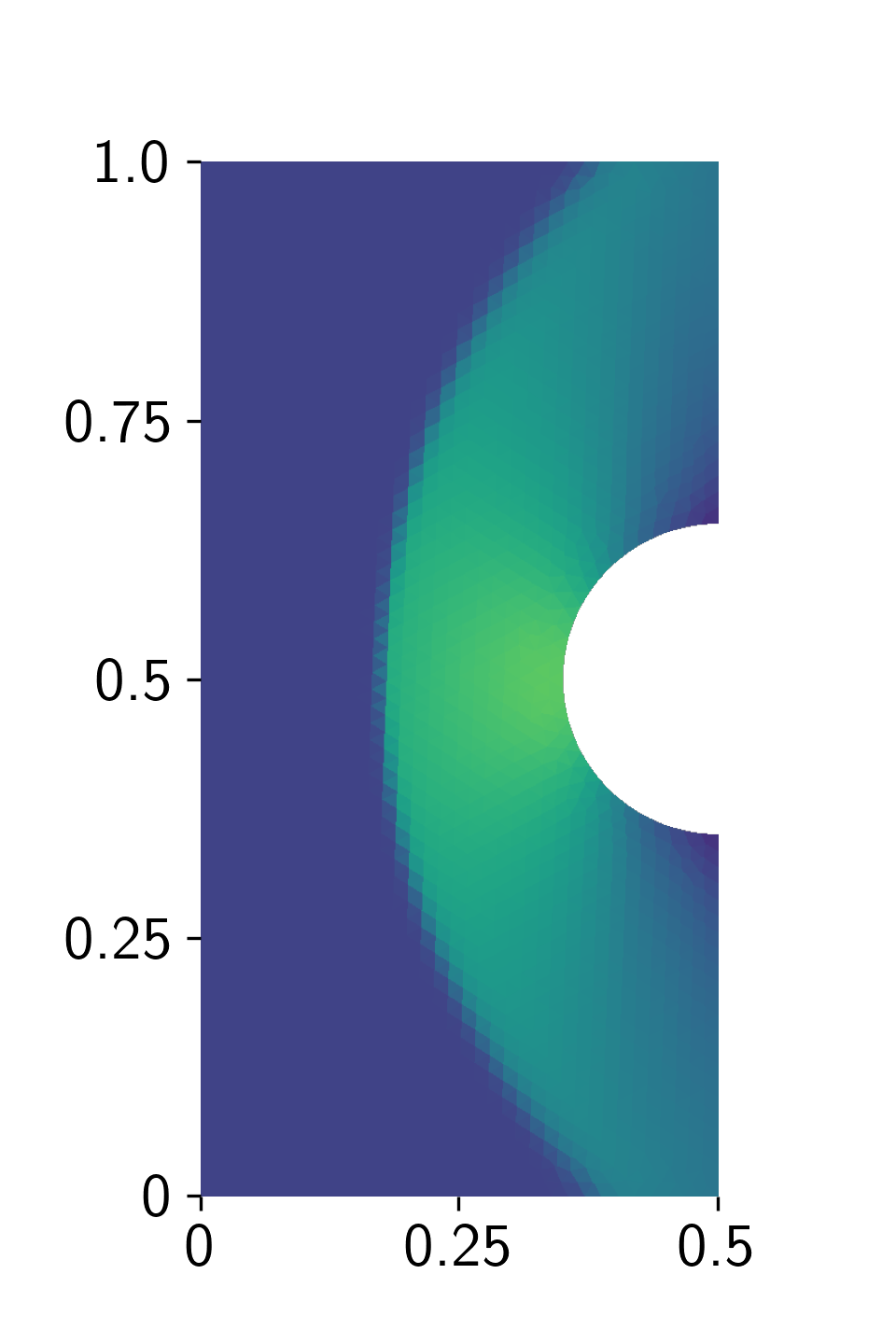}&
        \includegraphics[trim=0 0 0 1cm, clip,height=0.22\textwidth]{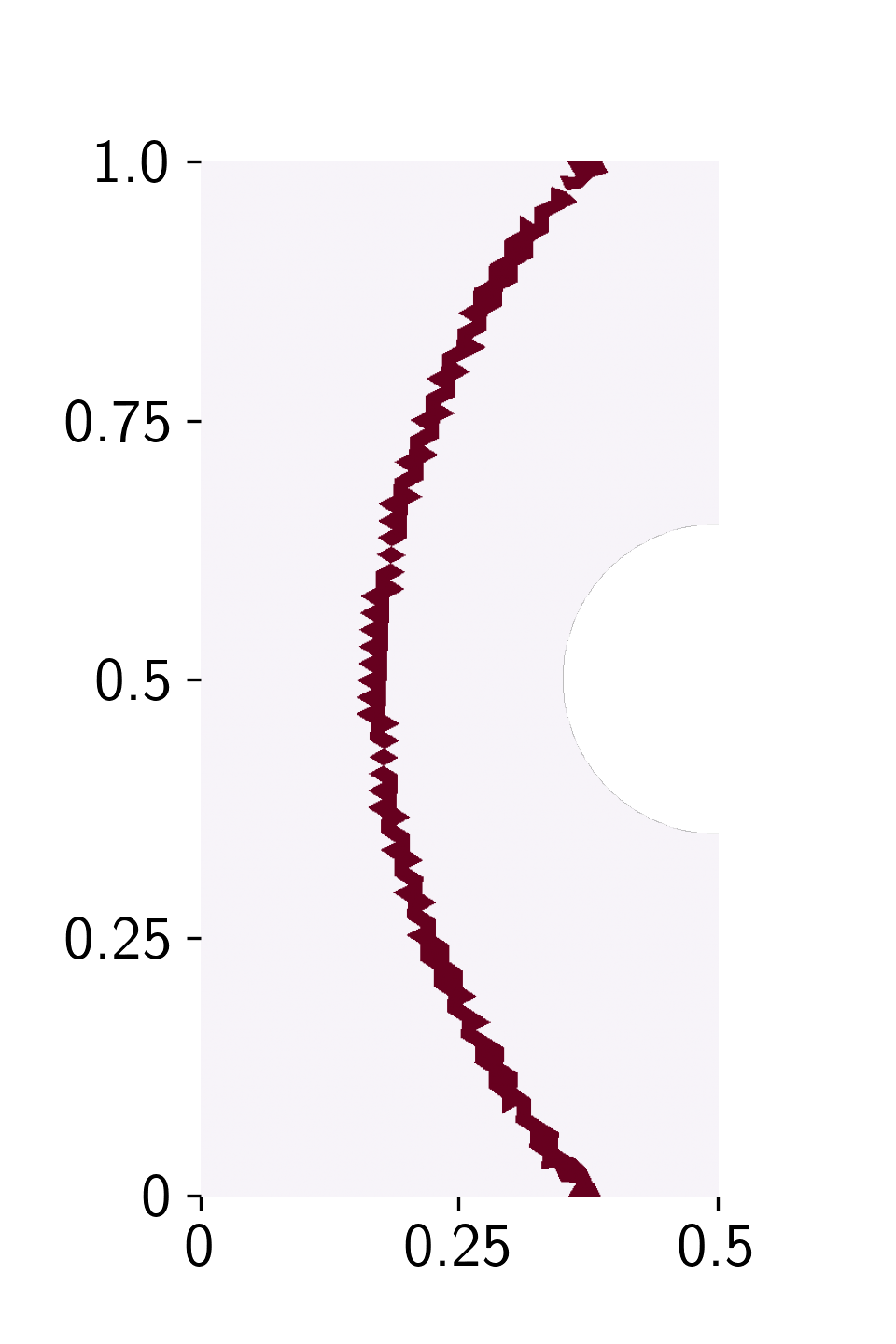}\\
        \raisebox{2.3em}{\rotatebox[origin=lb]{90}{\parbox{2cm}{\centering \footnotesize{GNN-LaSDI}}}}&
        \includegraphics[trim=0 0 0 1cm, clip,height=0.22\textwidth]{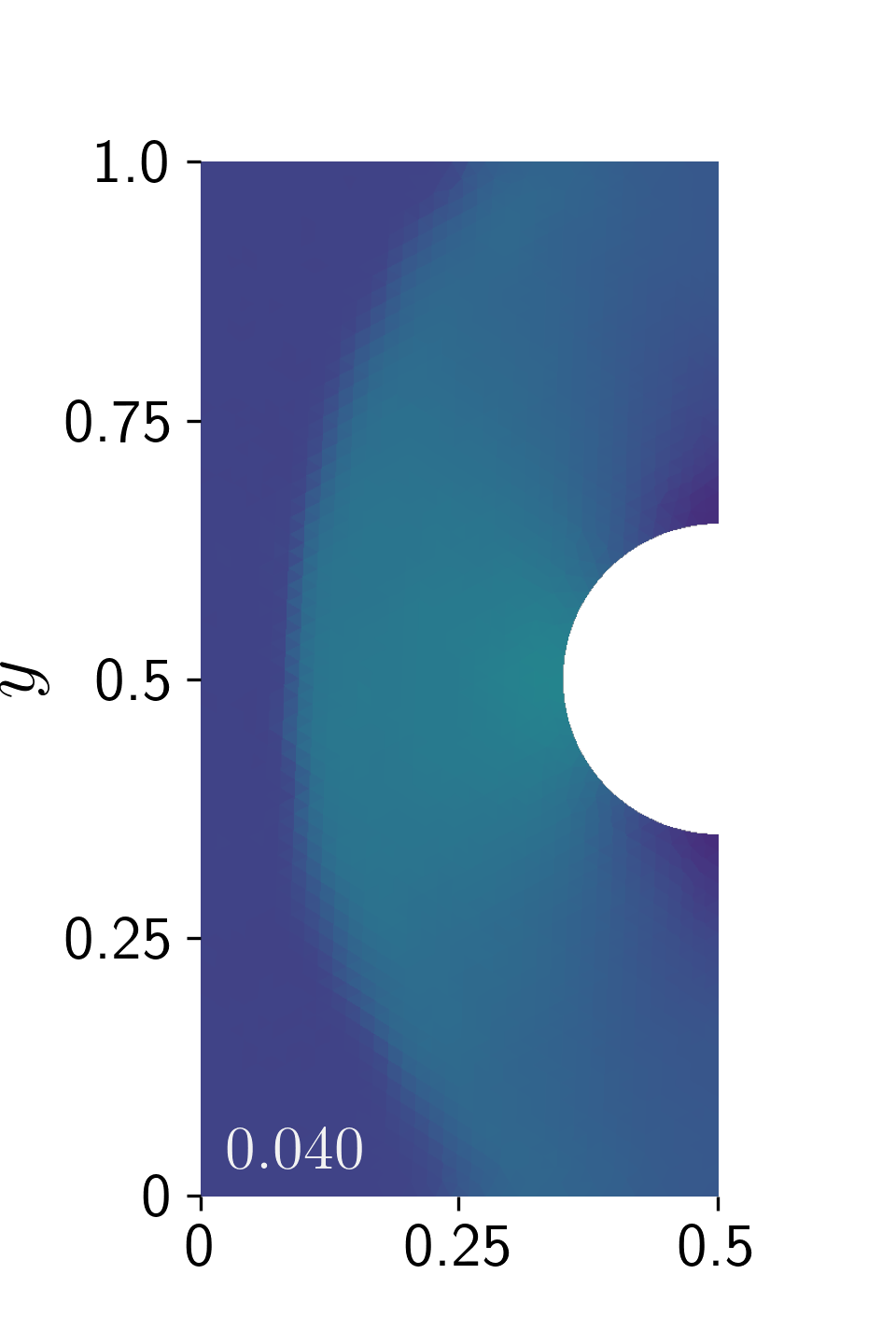}&
        \includegraphics[trim=0 0 0 1cm, clip,height=0.22\textwidth]{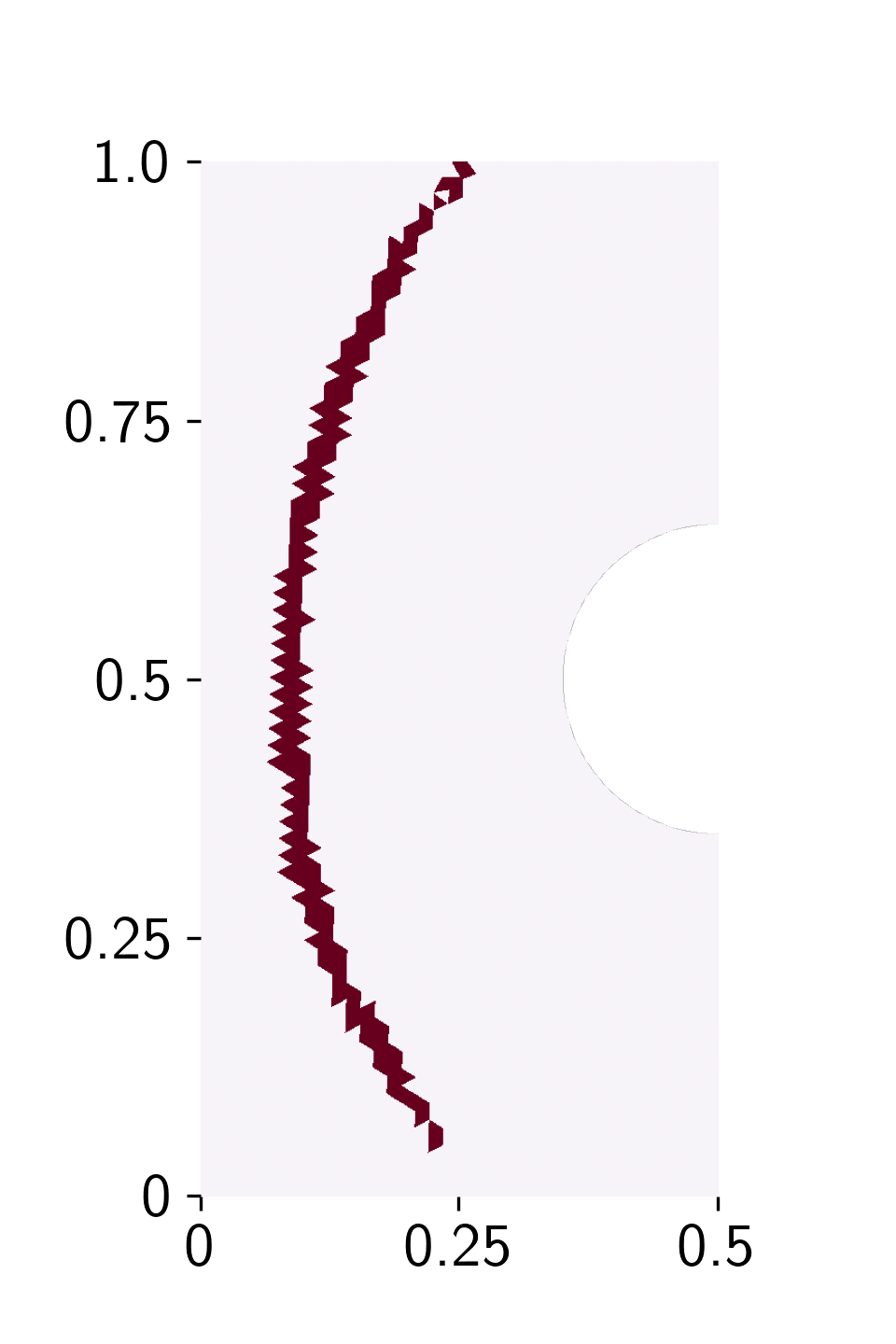}&
        \includegraphics[trim=0 0 0 1cm, clip,height=0.22\textwidth]{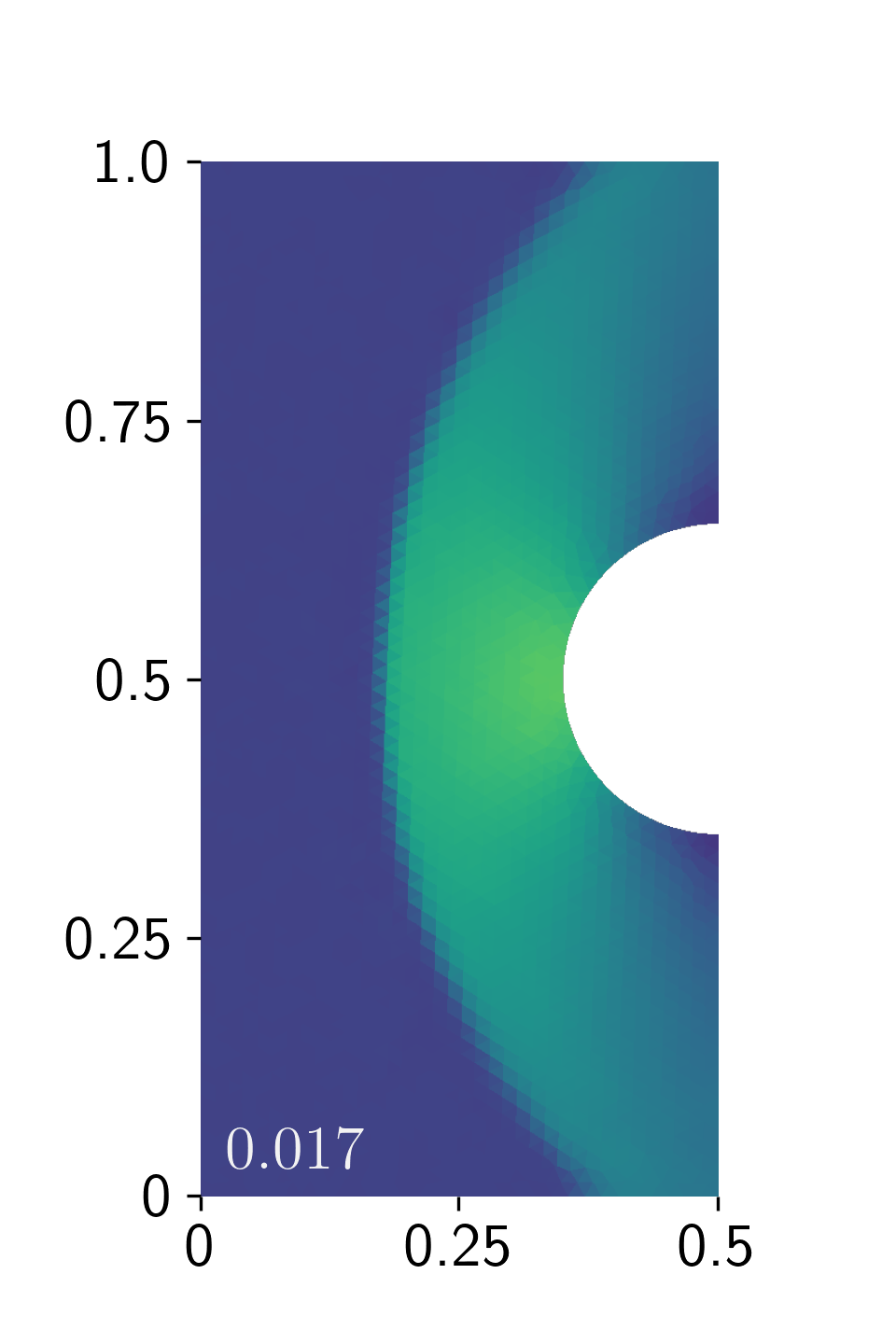}&
        \includegraphics[trim=0 0 0 1cm, clip,height=0.22\textwidth]{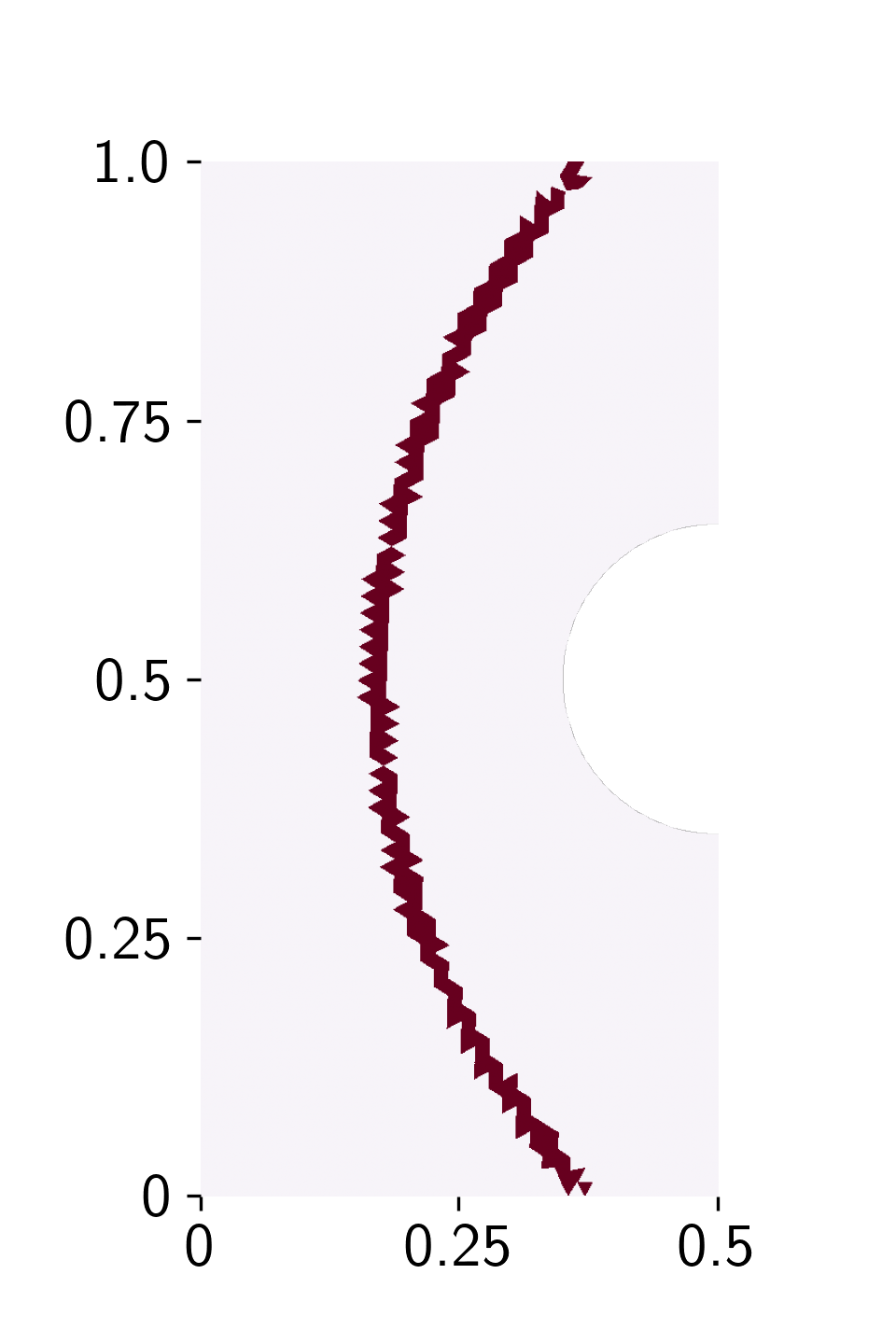}\\
        \raisebox{2.3em}{\rotatebox[origin=lb]{90}{\parbox{2cm}{\centering \footnotesize{GD-LSPG}}}}&
        \includegraphics[trim=0 0 0 1cm, clip,height=0.22\textwidth]{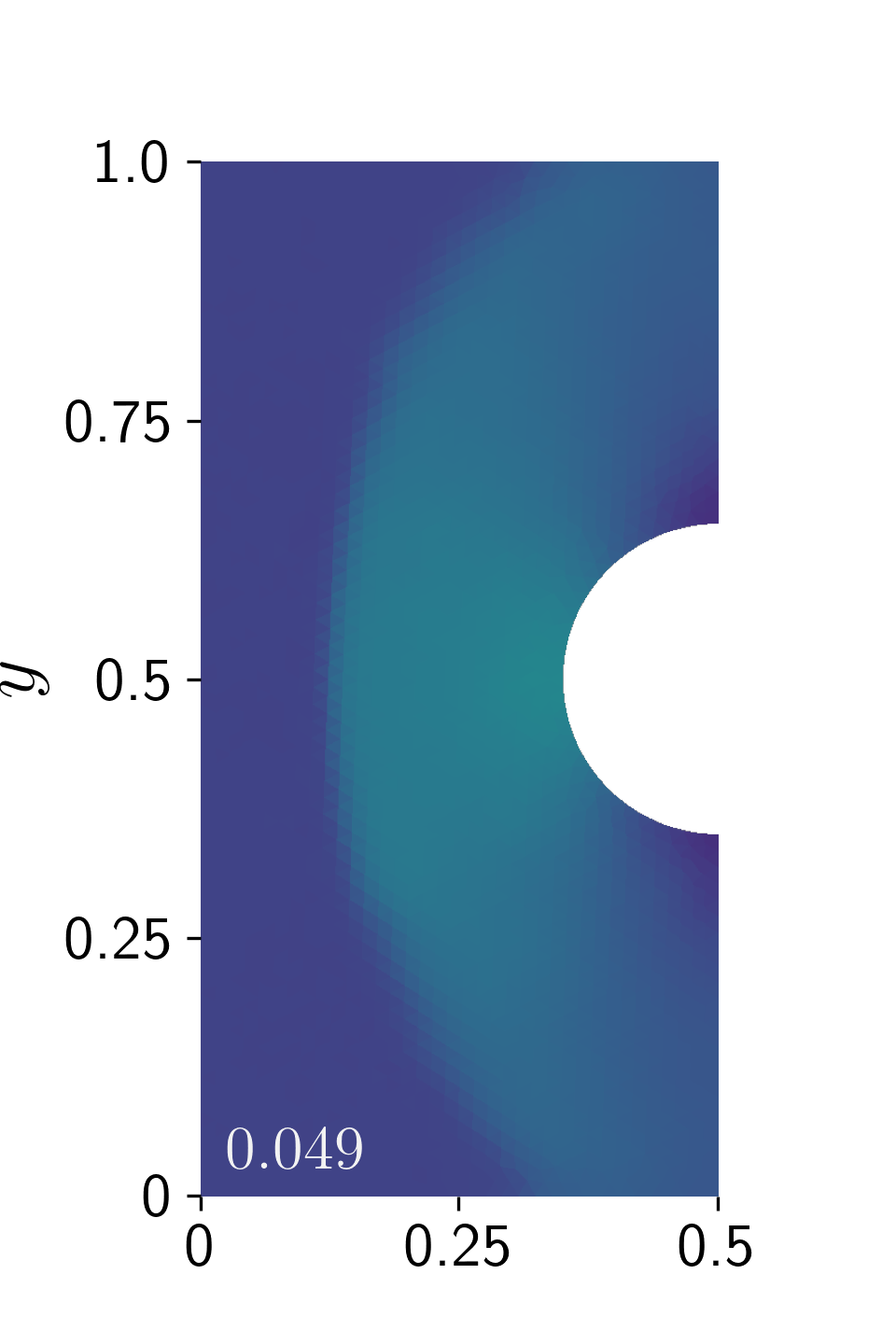}&
        \includegraphics[trim=0 0 0 1cm, clip,height=0.22\textwidth]{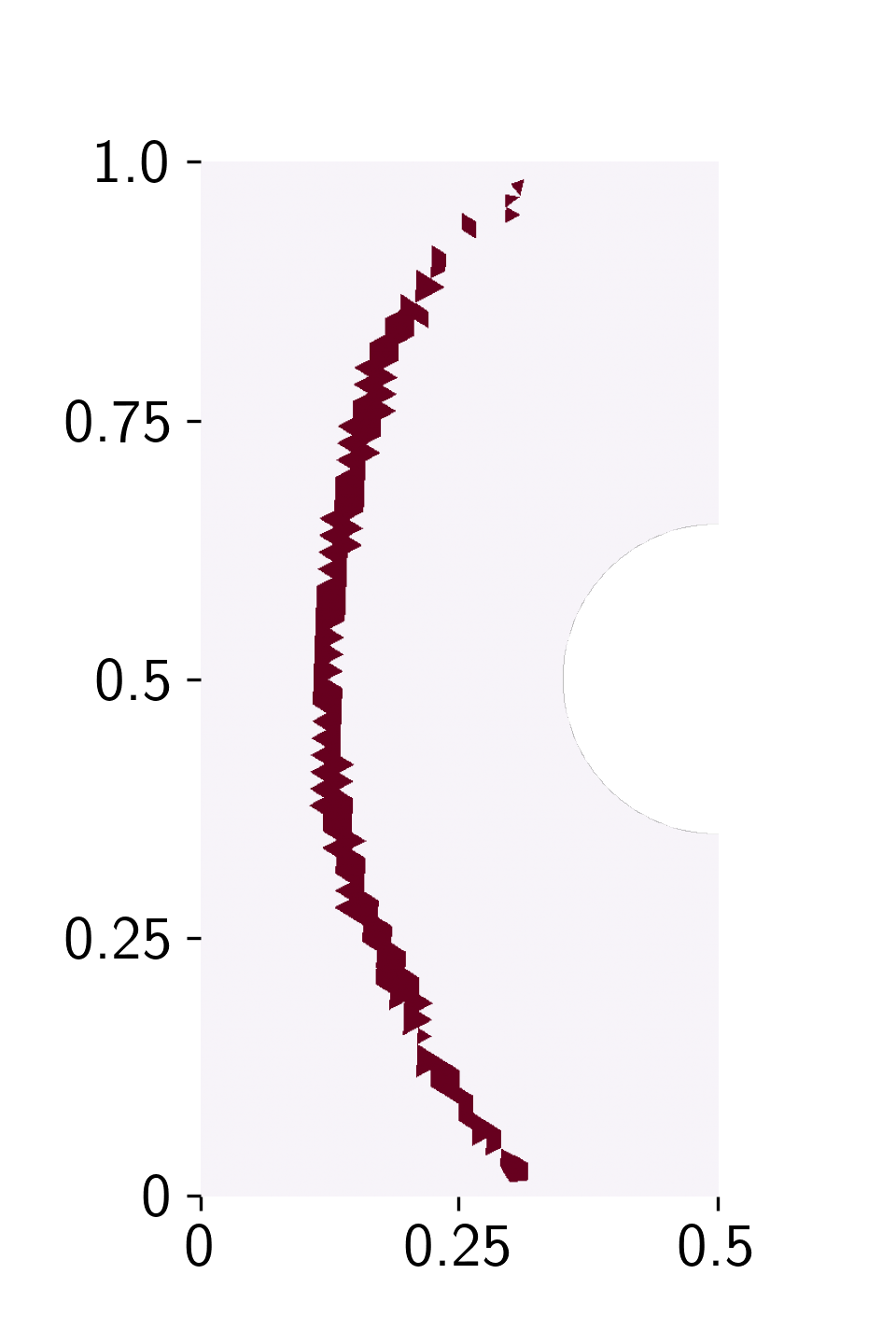}&
        \includegraphics[trim=0 0 0 1cm, clip,height=0.22\textwidth]{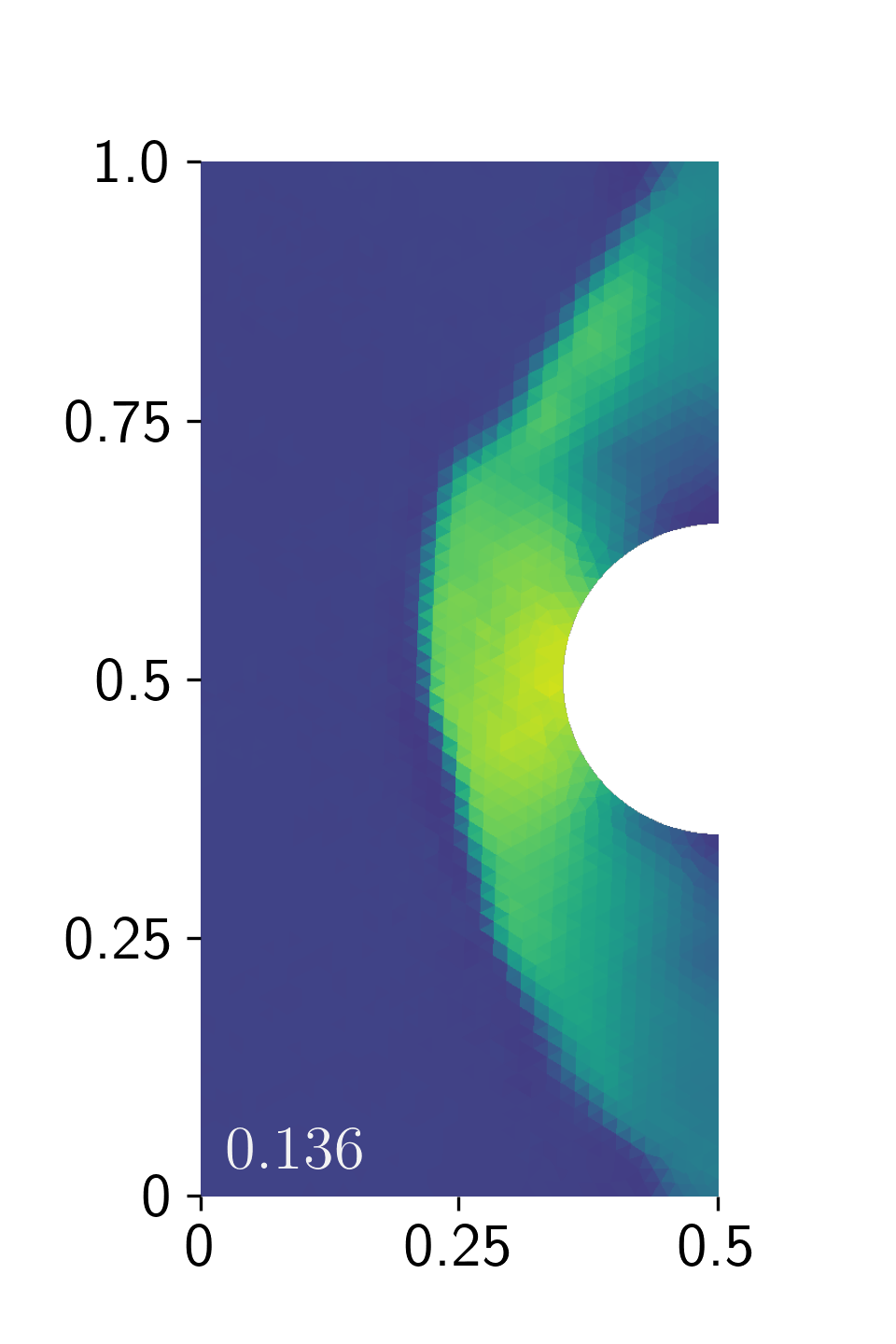}&
        \includegraphics[trim=0 0 0 1cm, clip,height=0.22\textwidth]{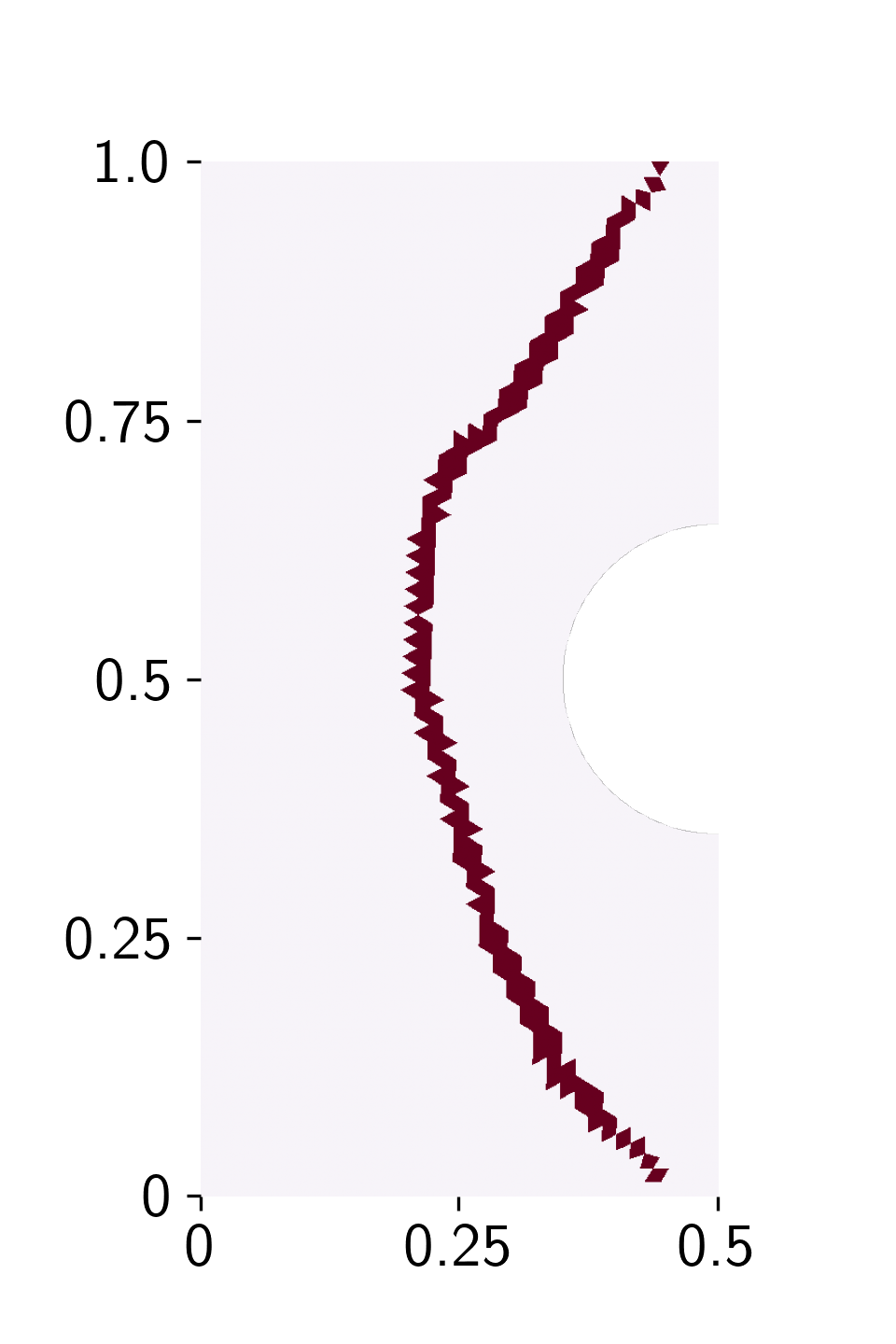}\\
        \raisebox{2.3em}{\rotatebox[origin=lb]{90}{\parbox{2cm}{\centering \footnotesize{POD-LaSDI}}}}&
        \includegraphics[trim=0 0 0 1cm, clip,height=0.22\textwidth]{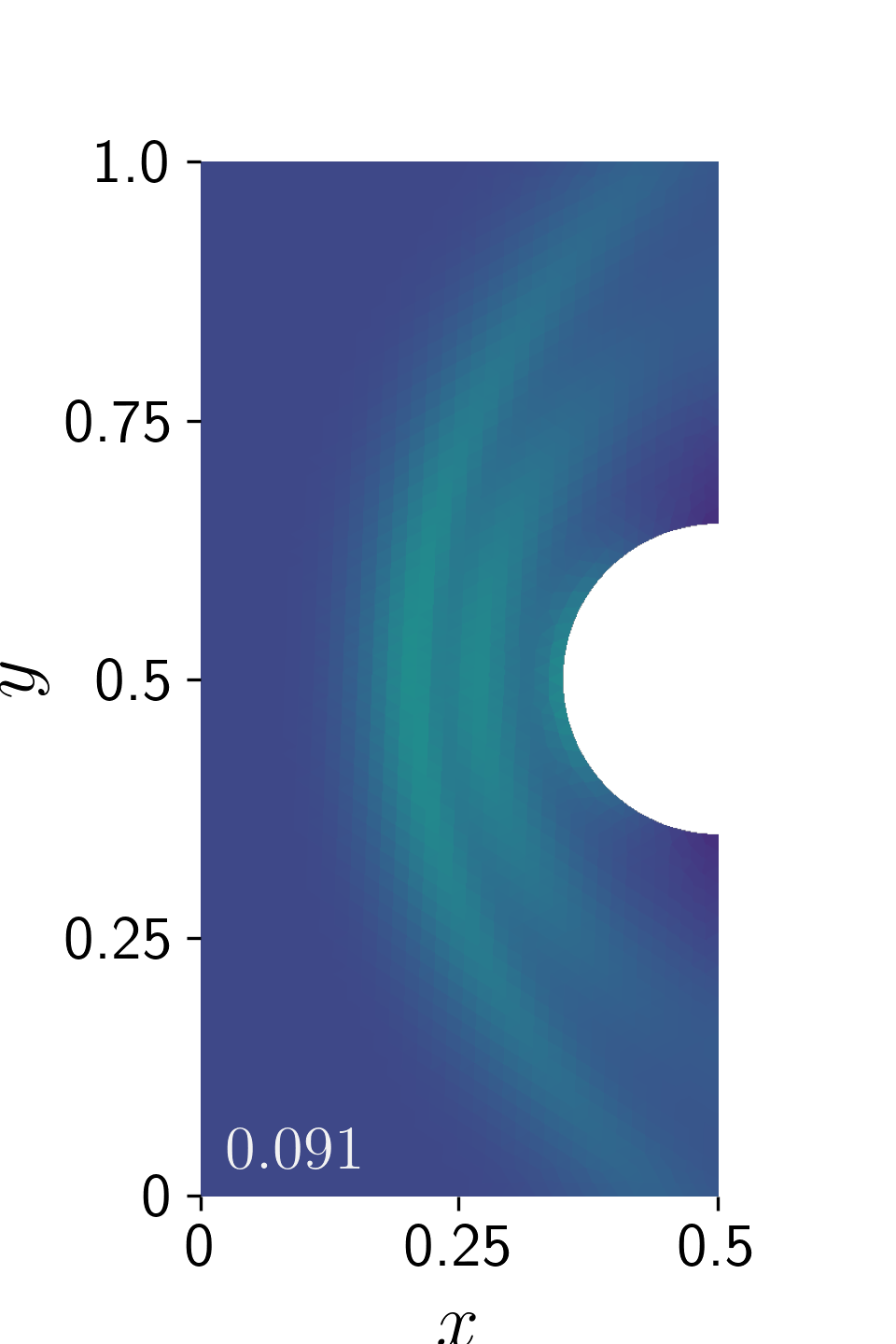}&
        \includegraphics[trim=0 0 0 1cm, clip,height=0.22\textwidth]{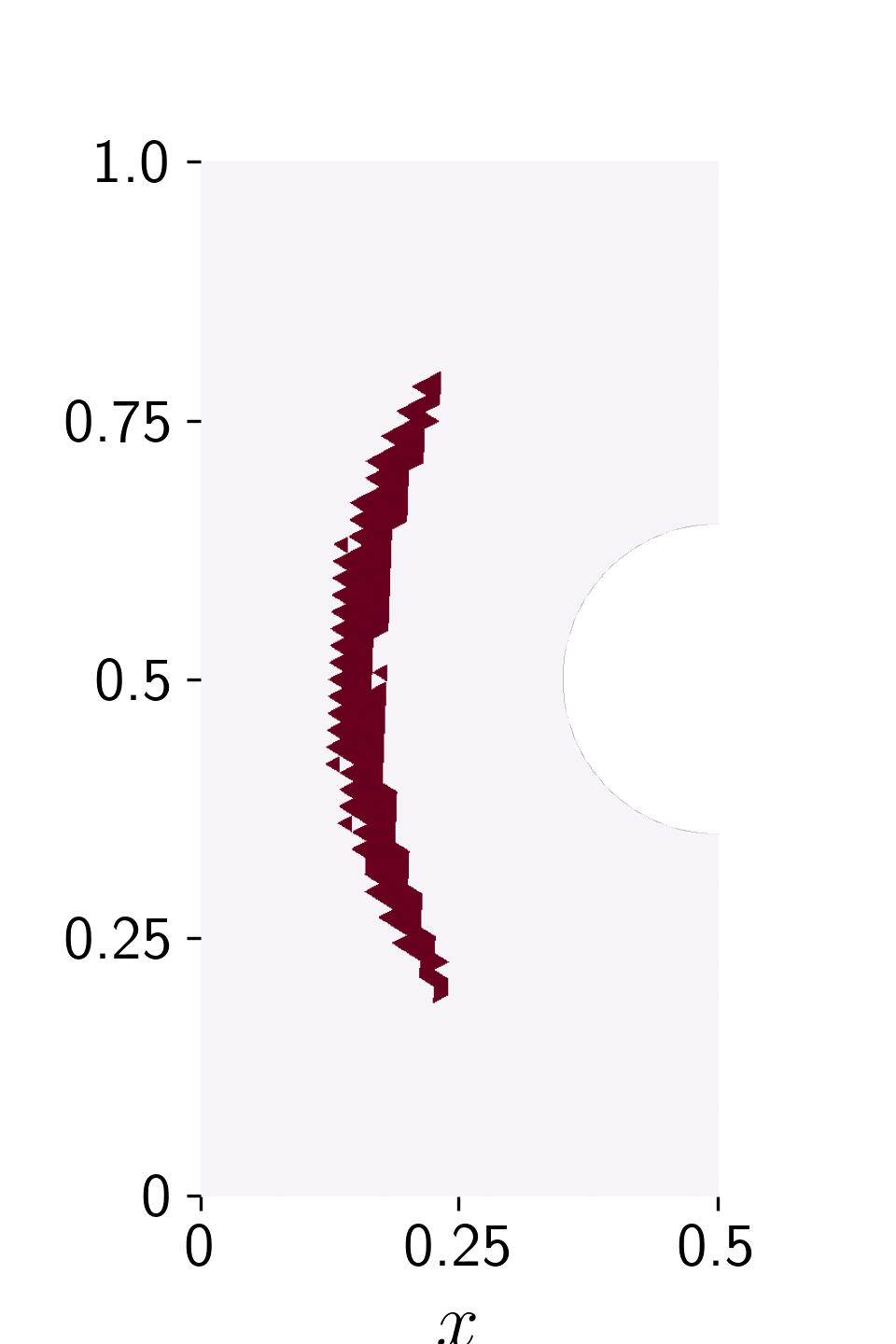}&
        \includegraphics[trim=0 0 0 1cm, clip,height=0.22\textwidth]{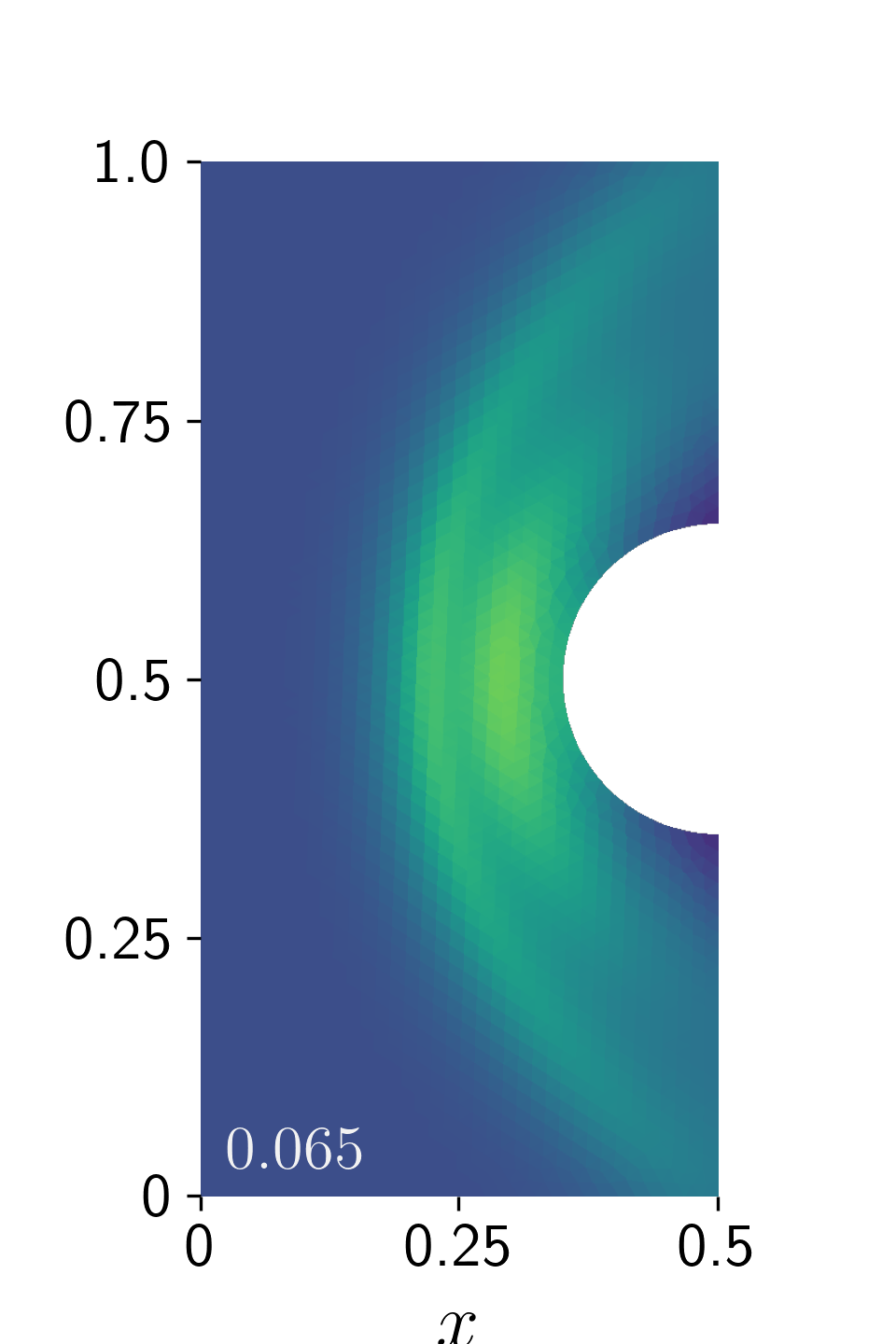}&
        \includegraphics[trim=0 0 0 1cm, clip,height=0.22\textwidth]{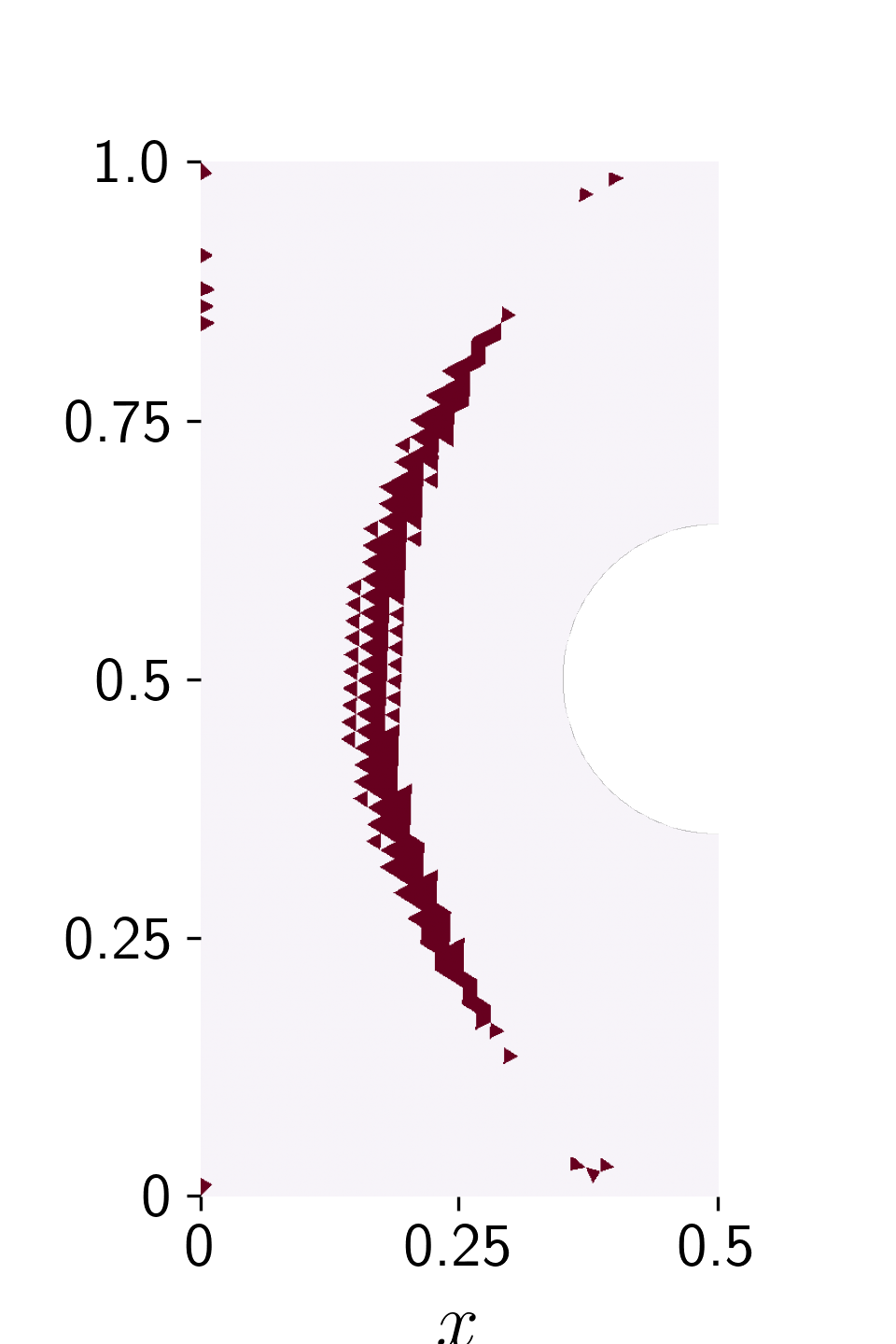}\\
        & \multicolumn{4}{c}{\includegraphics[scale=.3]{cyl_pressureColorbar.png}}
    \end{tabular}
    \captionsetup{justification=centering}
    \caption{Pressure field solutions and corresponding point cloud representations of the bow shock for the test parameters yielding the largest state prediction errors at latent state dimension $n=5$, shown at $t=0.5$. State prediction errors are indicated in the bottom left corner of the GNN-LaSDI, GD-LSPG, and POD-LaSDI pressure field solutions. The left two columns correspond to the worst-case test parameter for GNN-LaSDI and POD-LaSDI (both at $M_\infty=1.05$), while the right two columns correspond to the worst-case test parameter for GD-LSPG ($M_\infty=1.55$). Rows correspond to the FOM, GNN-LaSDI, GD-LSPG, and POD-LaSDI, respectively. Although the worst-case GD-LSPG solution exhibits a larger state prediction error than the worst-case POD-LaSDI solution, it preserves a sharp, well-defined bow shock, whereas the POD-LaSDI solution exhibits substantial shock diffusion. (Online version in color.)}
    \label{fig:cyl_sol_pc_worstCase}
\end{figure}

To further investigate the source of errors for each ROM, Figure \ref{fig:cyl_sol_pc_worstCase} presents the pressure field solutions and corresponding bow shock point cloud representations for the test parameters associated with the largest state prediction errors at latent state dimension $n=5$. Among the test parameters considered, the maximum state prediction error for GNN-LaSDI and POD-LaSDI occur at $M_\infty=1.05$, whereas the maximum state prediction error for GD-LSPG occurs at $M_\infty=1.55$. The corresponding state prediction errors are reported in the lower left corner of each pressure field solution in Figure \ref{fig:cyl_sol_pc_worstCase}. Even in its worst-case scenario, GNN-LaSDI accurately captures the pressure field and bow shock evolution, further presenting its robustness as a MOR framework. In contrast, the largest state prediction error observed for GD-LSPG is primarily associated with the distortion of the bow shock. Finally, the POD-LaSDI error is mainly attributable to excessive diffusion of the shock structure.

\begin{figure}[ht!]
    \centering
     \begin{tabular}{cc}
        \footnotesize{$M_\infty = 1.05$} & \footnotesize{$M_\infty = 1.55$} \\
        \includegraphics[trim=0 0 0 1cm, clip,height=0.26\textwidth]{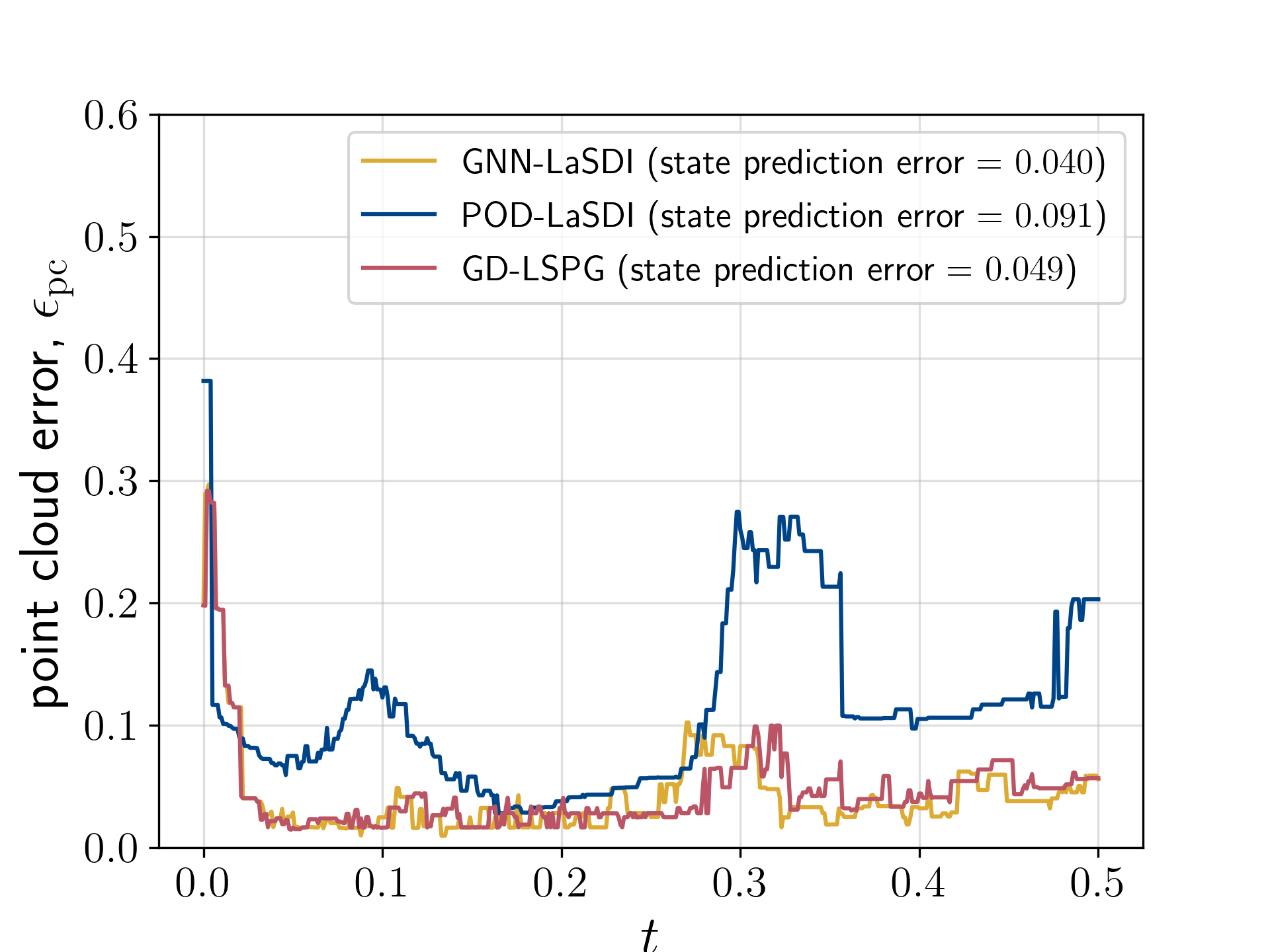} & \includegraphics[trim=0 0 0 1cm, clip,height=0.26\textwidth]{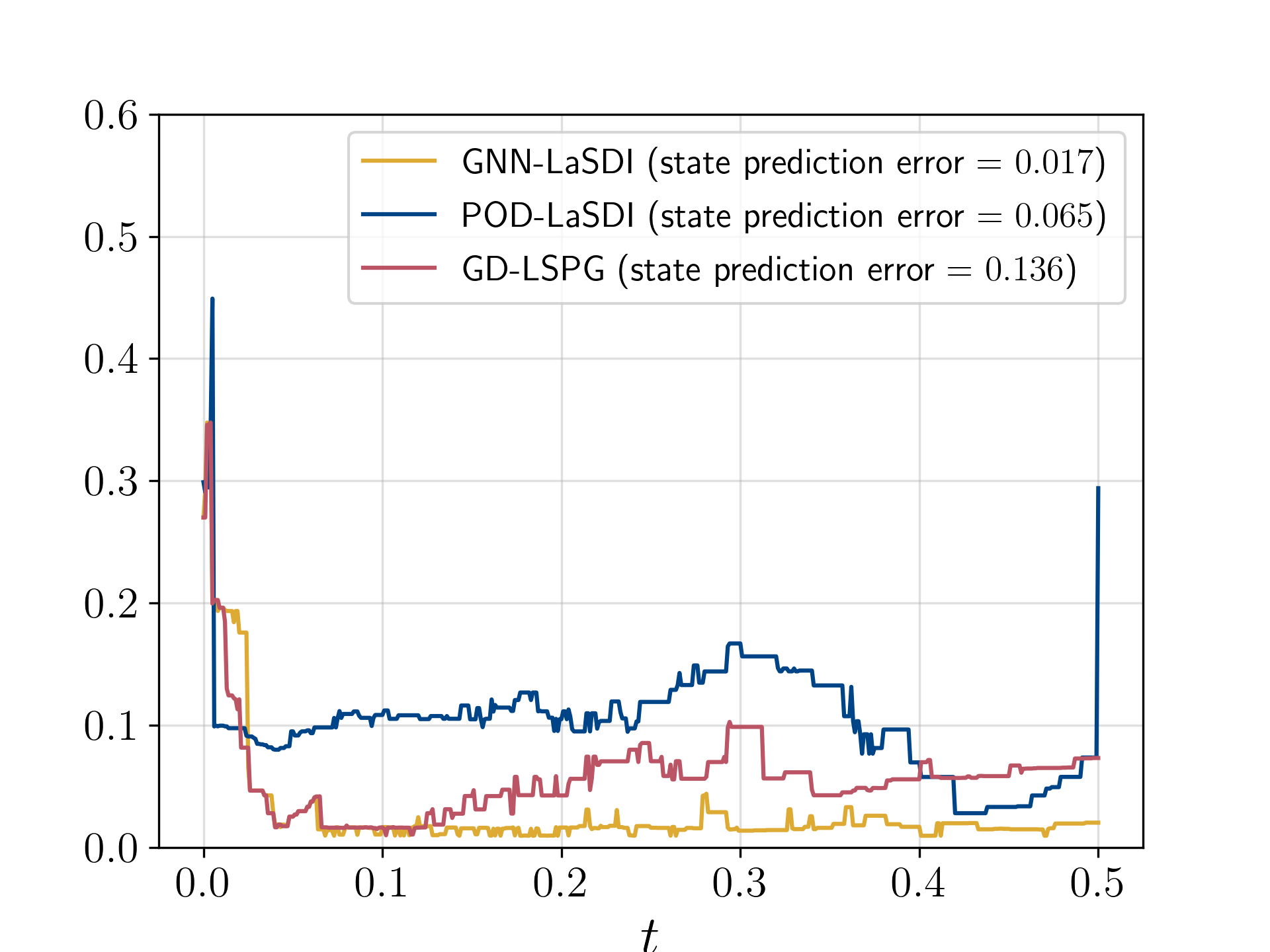}
    \end{tabular}
    \captionsetup{justification=centering}
    \caption{Point cloud errors \eqref{eq:pc_error} plotted with respect to time for the GNN-LaSDI, GD-LSPG, and POD-LaSDI solutions at latent state dimension $n=5$. Legends denote each ROM's corresponding state prediction error. The left figure corresponds to the worst-case test parameter for GNN-LaSDI and POD-LaSDI ($M_\infty=1.05$), while the right figure corresponds to the worst-case test parameter for GD-LSPG ($M_\infty=1.55$). Although GD-LSPG yields a larger state prediction error than POD-LaSDI for $M_\infty=1.55$, it generally achieves lower point cloud errors, indicating a more accurate prediction of the evolution of the bow shock. (Online version in color.)}
    \label{fig:cyl_pc_error_worstCase}
\end{figure}

To more directly assess the accuracy of the predicted bow shock location, Figure \ref{fig:cyl_pc_error_worstCase} presents the corresponding point cloud errors for the worst-case ROM solutions shown in Figure \ref{fig:cyl_sol_pc_worstCase}. Notably, despite the larger state prediction error reported for the worst-case GD-LSPG solution in Figure \ref{fig:cyl_sol_pc_worstCase}, POD-LaSDI generally produces larger point cloud errors. This indicates that GD-LSPG more accurately captures the evolution of the bow shock compared to POD-LaSDI, despite its larger global state-based error.

\begin{figure}[ht!]
    \centering
     \begin{tabular}{c}
        \includegraphics[height=0.3\textwidth]{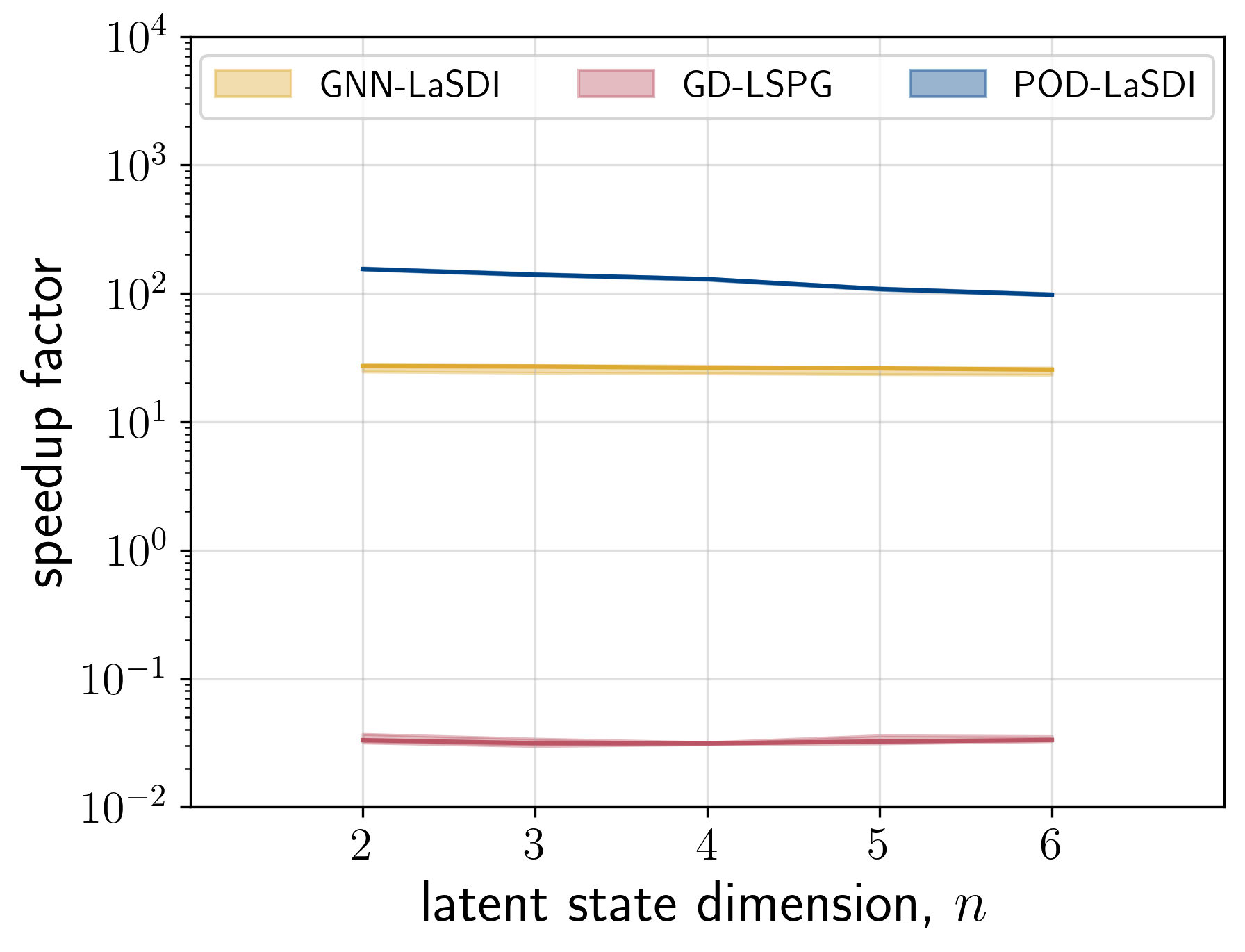}
    \end{tabular}
    \captionsetup{justification=centering}
    \caption{Range of speedup factors for GNN-LaSDI, GD-LSPG, and POD-LaSDI across all 10 test parameters, plotted with respect to the latent state dimension. Solid lines indicate the median speedup factor. Note that speedup factors are calculated with respect to the FOM wall-clock time for each specific test parameter. POD-LaSDI consistently achieves the largest speedup factors, whereas GD-LSPG yields the smallest and remains more computationally expensive than the FOM across all latent state dimensions considered. GNN-LaSDI achieves speedup factors more than two orders of magnitude higher than GD-LSPG, while remaining approximately one order of magnitude lower than POD-LaSDI. The average FOM wall-clock time varied between 1.22s and 1.26s, depending on the test parameter. (Online version in color.)}
    \label{fig:cyl_speedupRange}
\end{figure}

To assess the computational efficiency of the studied methods, Figure \ref{fig:cyl_speedupRange} presents the range of speedup factors achieved by GNN-LaSDI, GD-LSPG, and POD-LaSDI. Across all latent-state dimensions considered, POD-LaSDI consistently achieves the largest speedup factors, while GD-LSPG yields the smallest. Unlike the 2D isotropic nucleus growth experiment, GD-LSPG fails to realize any computational savings in this case, as evidenced by speedup factors below unity. Consistent with the previous numerical experiment, GNN-LaSDI achieves speedup factors that are more than two orders of magnitude greater than those of GD-LSPG, while remaining approximately one order of magnitude lower than those of POD-LaSDI (due to the relatively higher computational cost of decoding a graph autoencoder compared with the linear POD reconstruction used in POD-LaSDI).

\begin{figure}[ht!]
    \centering
     \begin{tabular}{ccc}
        & {\footnotesize{unconstrained graph autoencoder}} & \hspace{-.5cm} {\footnotesize{constrained graph autoencoder}} \\ \vspace{-.5cm}
        \raisebox{2.5em}{\rotatebox[origin=lb]{90}{\parbox{2cm}{\centering \footnotesize{GNN-LaSDI}}}}&
        \includegraphics[trim=0 0 0 1cm, clip,height=0.22\textwidth]{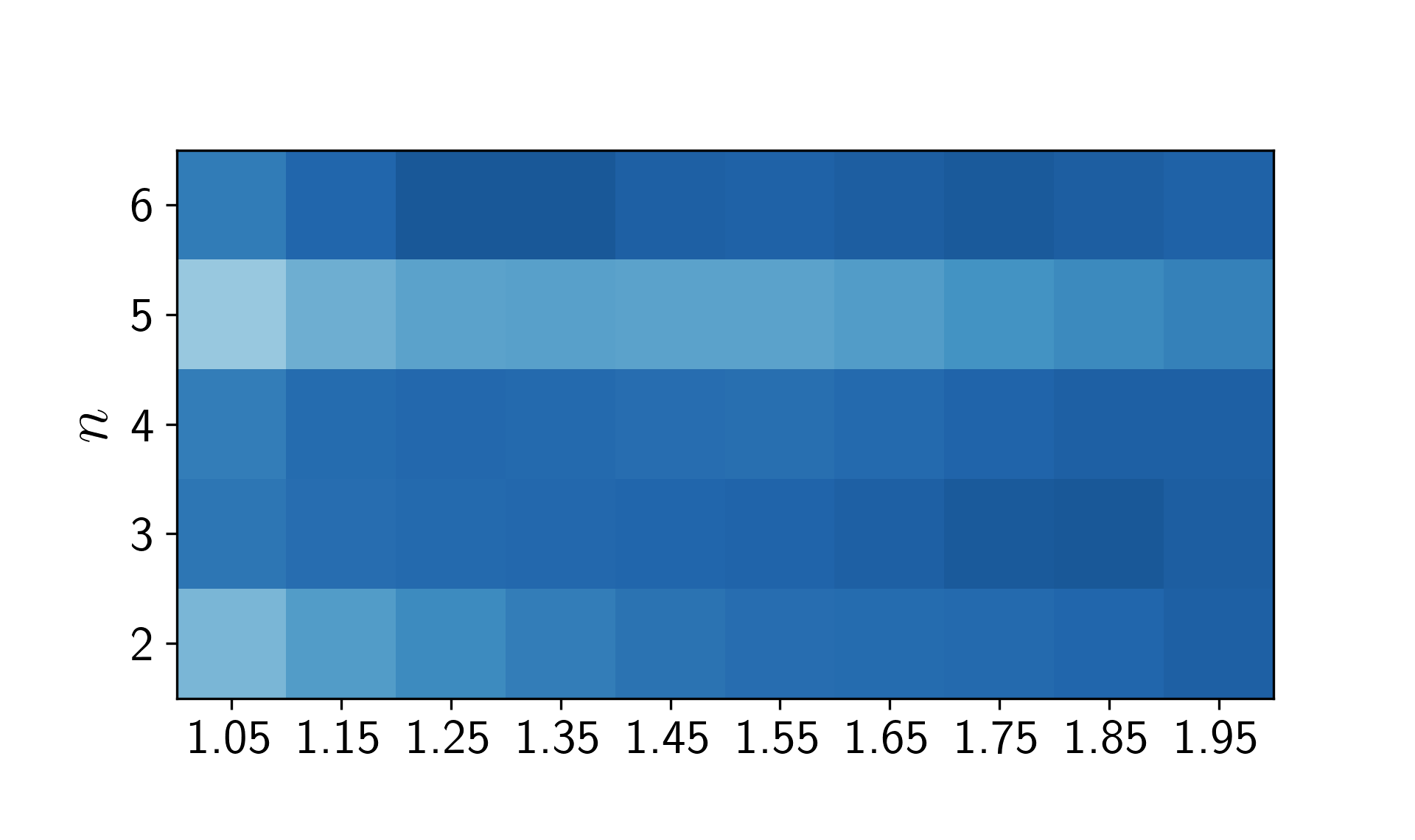}&\hspace{-.5cm}
        \includegraphics[trim=0 0 0 1cm, clip,height=0.22\textwidth]{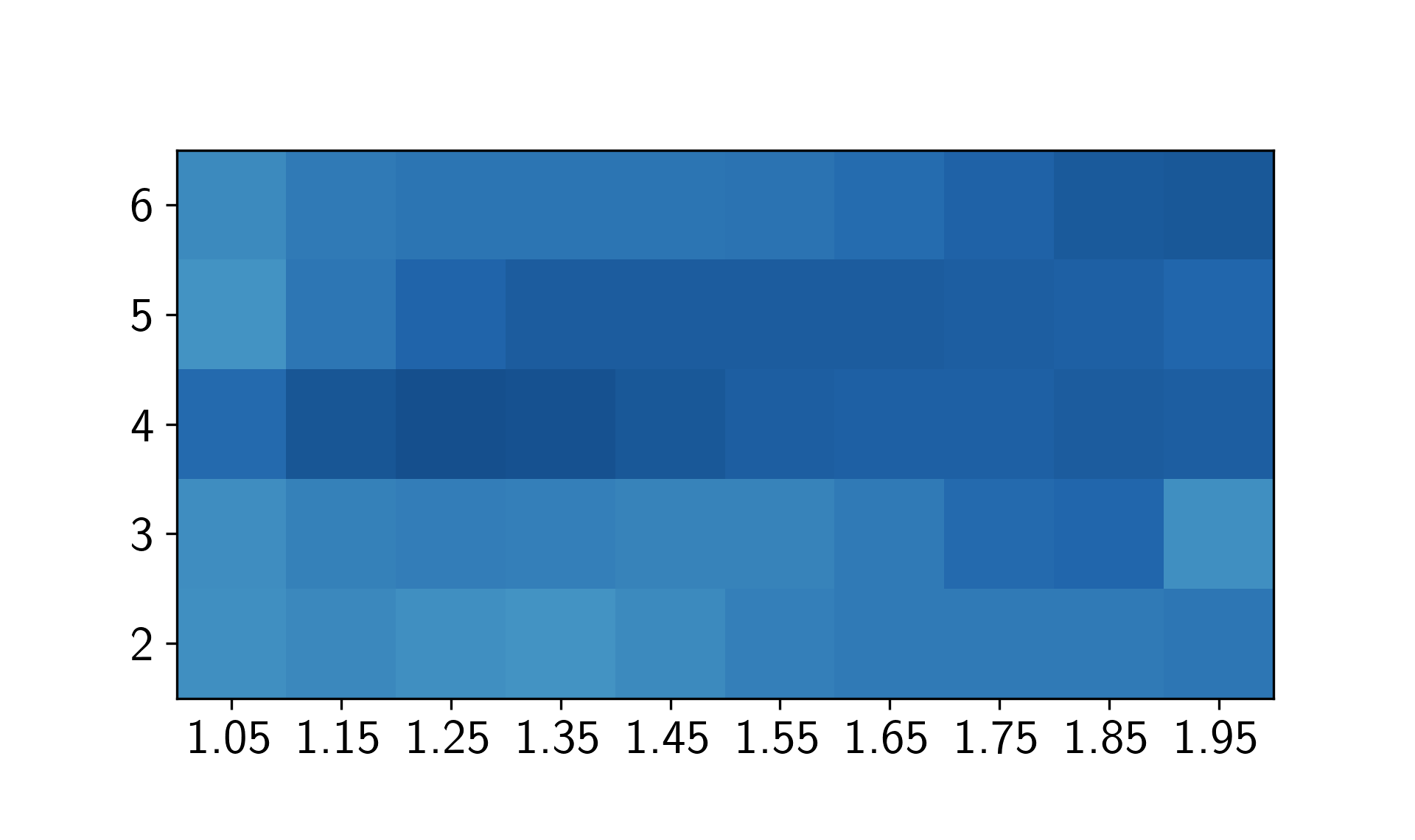}\\ 
        \raisebox{2.5em}{\rotatebox[origin=lb]{90}{\parbox{2cm}{\centering \footnotesize{GD-LSPG}}}}&
        \includegraphics[trim=0 0 0 1cm, clip,height=0.22\textwidth]{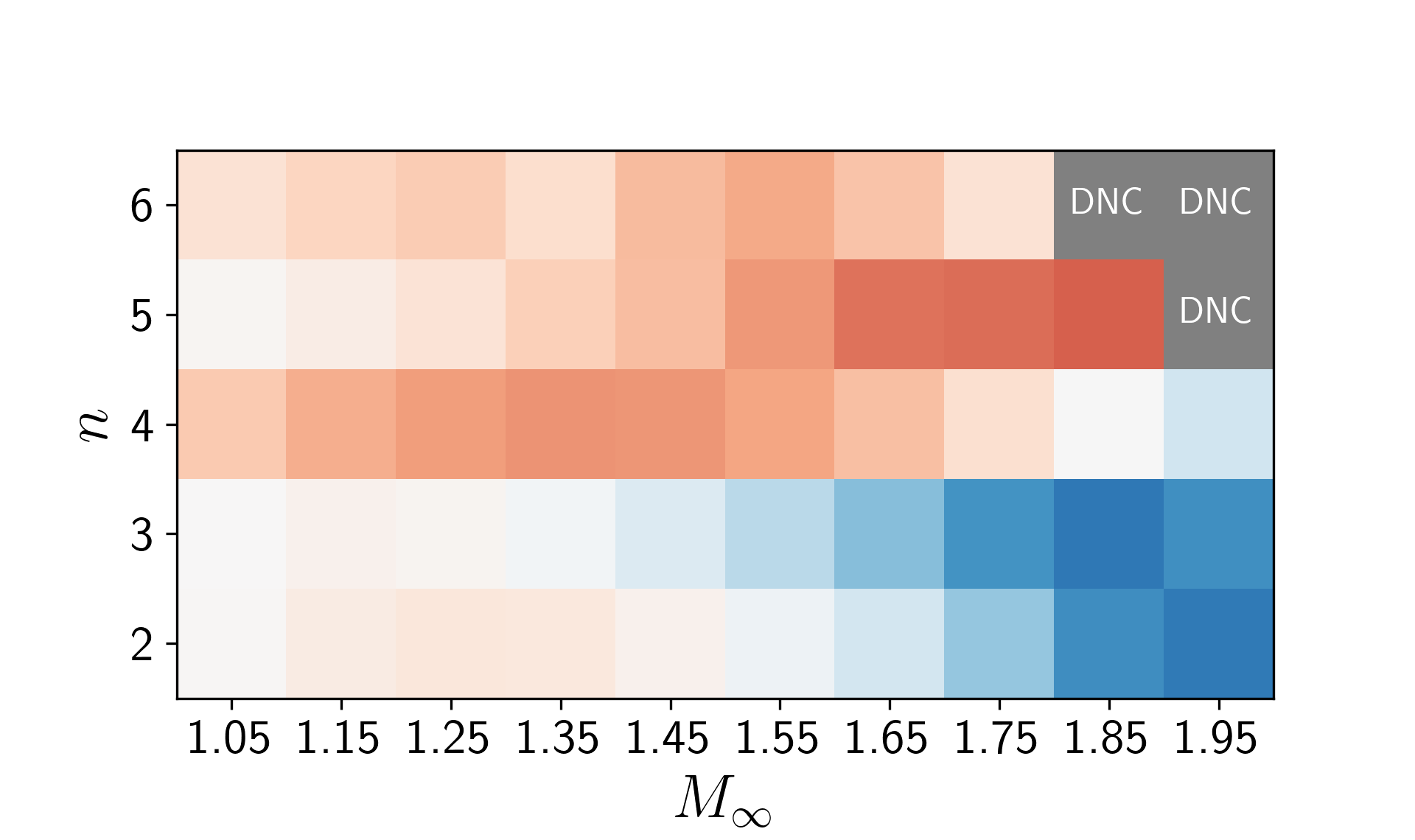}&\hspace{-.5cm}
        \includegraphics[trim=0 0 0 1cm, clip,height=0.22\textwidth]{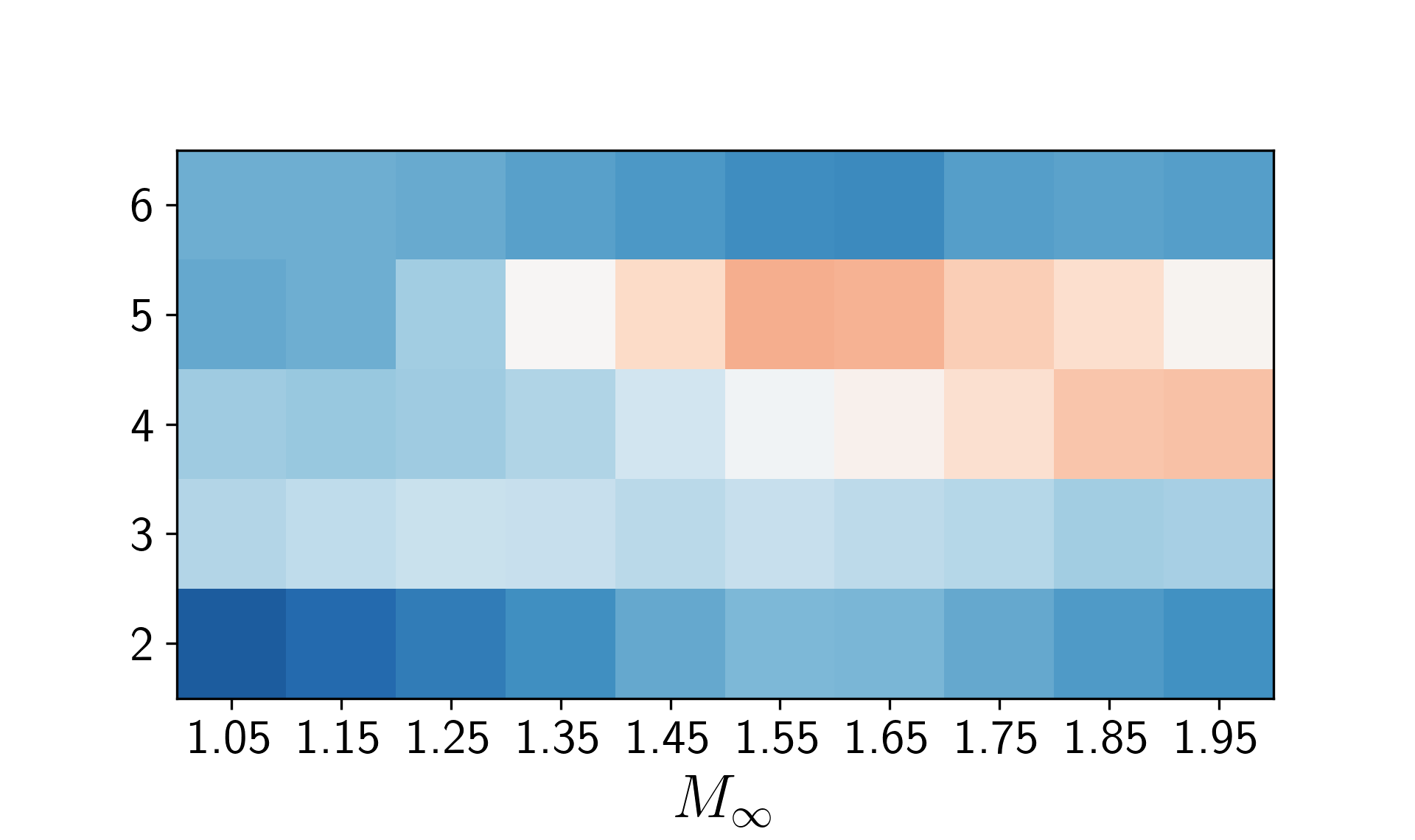}\\
        & \multicolumn{2}{c}{\includegraphics[scale=.35]{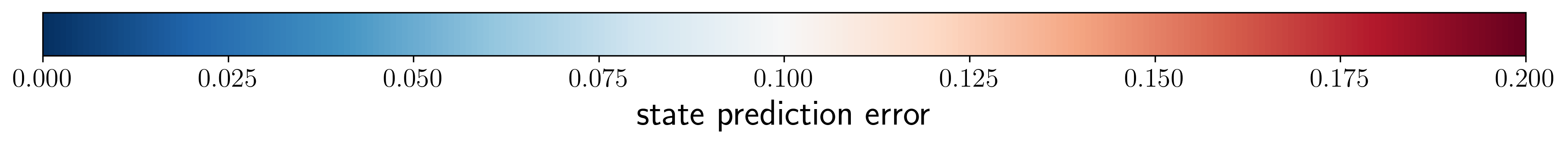}}
    \end{tabular}
    \captionsetup{justification=centering}
    \caption{State prediction errors as functions of latent state dimension $n$ and freestream Mach number $M_{\infty}$. The left and right columns correspond to ROMs constructed using the unconstrained and constrained graph autoencoders, respectively, where the former employs the postprocessing layer defined by \eqref{eq:postProcVanilla}-\eqref{eq:tildeBarVanilla} and the latter uses the modified postprocessing layer proposed in \eqref{eq:postProcModified}-\eqref{eq:barTildeModified}. The top row presents GNN-LaSDI results, while the bottom row presents GD-LSPG results. Cells labeled with ``DNC'' indicate cases for which the solver failed to converge. GNN-LaSDI converges for all combinations of $n$ and $M_{\infty}$ and consistently achieves low state prediction errors, whereas GD-LSPG fails to converge for several parameter combinations for the unconstrained graph autoencoder. (Online version in color.)}
    \label{fig:constrained}
\end{figure}

Lastly, to evaluate the effect of the proposed constraint-enforcing postprocessing layer introduced in this experiment, Figure \ref{fig:constrained} compares the state prediction errors for GNN-LaSDI and GD-LSPG obtained using both the unconstrained and constrained graph autoencoders. The unconstrained architecture employs the postprocessing layer defined by \eqref{eq:postProcVanilla}-\eqref{eq:tildeBarVanilla}, whereas the constrained architecture utilizes the modified postprocessing layer defined by \eqref{eq:postProcModified}-\eqref{eq:barTildeModified}. The results show that GNN-LaSDI maintains low state prediction errors regardless of whether the physical admissibility constraint is enforced. In contrast, GD-LSPG benefits substantially from the constrained architecture, achieving significantly lower state prediction errors than the unconstrained graph autoencoder. Moreover, the unconstrained GD-LSPG ROM fails to converge for several combinations of the latent state dimension $n$ and the freestream Mach number $M_{\infty}$. These results indicate that embedding the physical admissibility constraints within the graph autoencoder improves the reliability of the GD-LSPG framework for this numerical example.

Once again, POD-LaSDI achieves the largest speedup factors among the methods considered, but does it at the expense of larger state prediction and point cloud errors. In contrast, GNN-LaSDI produces state prediction and point cloud errors that are comparable to, and often lower than, those of GD-LSPG, while delivering speedup factors that are several orders of magnitude greater than GD-LSPG. These results demonstrate that GNN-LaSDI maintains a balance between predictive fidelity and online performance.

\section{Conclusions and discussion}\label{sec:conclusions}

This study introduced GNN-LaSDI, a MOR framework that uses operator learning to overcome the computational bottleneck associated with evolving the latent representation of graph autoencoders via GD-LSPG \cite{magargal2025projection}. By directly evaluating the temporal evolution of the latent state through operator learning, GNN-LaSDI eliminates the need for the repeated evaluation of the Jacobian of the decoder required by GD-LSPG, enabling efficient temporal evolution of the low-dimensional representation while retaining the nonlinear state compression capabilities of the graph autoencoder. In addition, this study introduced a point cloud error metric that directly quantifies errors in the predicted locations of sharp gradient features. Unlike conventional state-based error metrics, which can be dominated by small spatial shifts in interfaces and shocks, the proposed metric evaluates the spatial accuracy of the extracted feature location through a point cloud representation. Consequently, it provides a complementary measure of ROM fidelity that is well suited for transport-dominated systems.

The proposed framework was assessed on two nonlinear systems characterized by evolving sharp gradients: Kobayashi's phase field model for 2D solidification and the 2D Euler equations modeling a bow shock generated by flow past a cylinder. Across both numerical experiments, GNN-LaSDI and GD-LSPG consistently outperformed POD-LaSDI in preserving the evolution of interfaces and shocks. In the solidification problem, GNN-LaSDI achieved state prediction and point cloud errors comparable to those obtained with GD-LSPG while providing speedup factors approximately two orders of magnitude larger. Similar trends were observed for the bow shock problem, where GNN-LaSDI accurately captured shock evolution over the parameter domain and maintained point cloud errors comparable to GD-LSPG. The proposed constraint incorporated into the graph autoencoder architecture improved the GD-LSPG framework by ensuring physically admissible reconstructions for the second numerical example. A consistent observation across both numerical examples is that POD-LaSDI achieved the largest speedup factors, but at the expense of significantly larger state prediction and point cloud errors. Conversely, GD-LSPG generally yielded the highest computational cost. GNN-LaSDI, on the other hand, demonstrated a balance between predictive accuracy and computational performance. 

Although the proposed point cloud error metric proved effective for evaluating the accuracy of interface and shock locations, it does not assess the accuracy of the state values on either side of the identified shock feature. Consequently, it should be viewed as a complementary metric rather than a replacement for conventional state-based error metrics. Future work will investigate developing feature-aware error metrics that simultaneously account for both feature accuracy and state reconstruction fidelity. Another avenue for future work is to investigate a richer library of functions utilized by LaSDI. Many dynamical systems admit parsimonious representations in appropriately chosen coordinates. Inspired by SINDy \cite{brunton2016discovering}, future work investigates combining Tikhonov and Lasso regularizations \cite{tibshirani1996lasso} within the LaSDI framework to facilitate the selection of representative terms from a richer library of candidate functions.

\paragraph{Acknowledgments}
L. K. Magargal is supported by the Department of Defense (DoD) through the National Defense Science \& Engineering Graduate (NDSEG) Fellowship Program. This material is based upon work supported by the Air Force Office of Scientific Research under award number FA9550-23-F-0014 in the amount \$$136{,}175.09$. P. Khodabakhshi acknowledges the support of the National Science Foundation under award 2450804. Portions of this research were conducted on Lehigh University's Research Computing infrastructure, partially supported by NSF Award 2019035.

\paragraph{Availability of data and materials} The dataset supporting the conclusions of this article is available in the GNN-LaSDI repository, \url{https://github.com/liam-magargal/GNN-LaSDI}.

\bibliographystyle{elsarticle-num}
\bibliography{bibliography}

\appendix

\section{Fundamentals of Geometric deep least-squares Petrov-Galerkin} \label{sec:gdlspg}

The deep LSPG framework \cite{lee2020deeplspg} employed a nonlinear manifold LSPG projection using a CNN-based autoencoder. While CNN-based autoencoders demonstrated exceptional dimensionality reduction capabilities for systems exhibiting sharp gradients \cite{lee2020deeplspg,fresca2021comprehensive,maulik2021advectionrom}, they were not readily extendable to unstructured meshes. To overcome this limitation, the authors' previous work \cite{magargal2025projection} introduced a graph autoencoder (also used in this study), enabling nonlinear manifold MOR  on unstructured meshes through GNN architectures. A brief overview of GD-LSPG is presented here. Specifically, \ref{ssec:graphAEdetails} summarizes the encoder and decoder architectures presented in Section \ref{ssec:ae}, while \ref{ssec:nmlspg} provides a summary of the nonlinear manifold LSPG projection used to compute the temporal evolution of the latent variables.

\subsection{Graph autoencoder architecture details}\label{ssec:graphAEdetails}

The general architecture of a graph autoencoder is presented in Figure \ref{fig:ae_arch}. In the following, the details of each layer are described. The interested reader is referred to \cite{magargal2025projection} for further information. 

The zeroth layer in the encoder, $\mathbf{h}_0 : \mathbb{R}^{N_c n_q} \rightarrow \mathbb{R}^{N_c \times n_q}$, is a preprocessing layer that reshapes the input state vector into a node feature matrix, and scales each node feature to the interval $[0, 1]$ in an element-wise manner. Since the zeroth layer of the encoder contains no trainable parameters, $\mathbf{\Theta}_0 = \emptyset$. Layers $i=1,\ldots,n_{\ell}-1$ of the encoder are message passing and pooling (MPP) layers designed to emulate the filtering and compressive capabilities of CNNs:
\begin{equation}
    \mathbf{h}_{i}: \left( \mathbf{X}^{i-1}; \mathbf{\Theta}_i \right) \mapsto \mathrm{\mathbf{Pool}}^{i} \left( \cdot \right) \circ \mathrm{\mathbf{MP}}^{i}_{\mathrm{enc}} \left( \mathbf{X}^{i-1} ; \boldsymbol{\Theta}_i \right),
\end{equation}
where $\mathrm{\mathbf{MP}}^{i}_\mathrm{enc}: \mathbb{R}^{\vert \mathcal{V}^{i-1} \vert \times N_F^{i-1}} \rightarrow \mathbb{R}^{\vert \mathcal{V}^{i-1} \vert \times N_F^i}$ denotes a trainable MP operation in which nodes exchange information between neighboring nodes in the graph using weights $\boldsymbol{\Theta}_i$, and $\mathrm{\mathbf{Pool}}^{i}: \mathbb{R}^{\vert \mathcal{V}^{i-1} \vert \times N_F^i} \rightarrow \mathbb{R}^{\vert \mathcal{V}^{i} \vert \times N_F^i}$ denotes a pooling operation that coarsens the graph by aggregating groups of nodes in the previous layer into a single node in the next layer. Here, $N_F^{i-1}$, and $N_F^i \in \mathbb{N}$ denote the number of input and output node features in $\mathbf{h}_i$, respectively. The final layer in the encoder, $\mathbf{h}_{n_\ell}: \mathbb{R}^{\vert \mathcal{V}^{n_\ell-1}\vert \times N_F^{n_\ell-1}} \rightarrow \mathbb{R}^n$, maps the graph representation to the low-dimensional representation, $\hat{\mathbf{x}}$, by flattening the node feature matrix $\mathbf{X}^{n_\ell-1}$ to a vector and applying an FC layer using weights $\boldsymbol{\Theta}_{n_\ell}$. 

The decoder mirrors the encoder. Its first layer, $\mathbf{g}_0: \mathbb{R}^{n} \rightarrow \mathbb{R}^{\vert \mathcal{V}^{n_\ell-1} \vert \times N_F^{n_\ell-1}}$, maps the latent state $\hat{\mathbf{x}}$ to a node feature matrix $\mathbf{Y}^{n_\ell-1} \in \mathbb{R}^{\vert \mathcal{V}^{n_\ell-1} \vert \times N_F^{n_\ell-1}}$ using an FC layer (using weights $\boldsymbol{\Omega}_{n_\ell}$) and reshaping the its output as a matrix. Layers $i=1,\ldots,n_\ell-1$ consist of unpooling and message passing (UMP) layers that increase the number of nodes in the graph at each level. Specifically, layers $i=1,\ldots,n_{\ell}-1$ are defined as,
\begin{equation}
    \mathbf{g}_{i}: \left( \mathbf{Y}^{i-1}; \mathbf{\Omega}_i \right) \mapsto \mathrm{\mathbf{MP}}^{i}_{\mathrm{dec}}\left( \cdot ; \boldsymbol{\Omega}_i \right)  \circ \mathrm{\mathbf{Unpool}}^{i} \left( \mathbf{Y}^{i-1} \right),
\end{equation}
where $\mathrm{\mathbf{Unpool}}^{i}: \mathbb{R}^{\vert \mathcal{V}^{n_\ell-i} \vert \times N_F^{n_\ell-i}} \rightarrow \mathbb{R}^{\vert \mathcal{V}^{n_\ell-i-1} \vert \times N_F^{n_\ell-i}}$ denotes an unpooling operation that increases the number of nodes in the graph and obtains their respective features, and $\mathrm{\mathbf{MP}}^i_{\mathrm{dec}}: \mathbb{R}^{\vert \mathcal{V}^{n_\ell-i-1} \vert \times N_F^{n_\ell-i}} \rightarrow \mathbb{R}^{\vert \mathcal{V}^{n_\ell-i-1} \vert \times N_F^{n_\ell-i-1}}$ denotes a trainable MP operation where neighboring nodes in the graph exchange information using weights $\boldsymbol{\Omega}_i$. The final layer in the decoder, $\mathbf{g}_{n_\ell}: \mathbb{R}^{N_c \times n_q} \rightarrow \mathbb{R}^{N_c n_q}$, serves as a postprocessing layer that rescales each node feature to its original physical range and reshapes the resulting node feature matrix into the approximate state solution vector, $\tilde{\mathbf{x}}$. Because these operations contain no trainable parameters, $\boldsymbol{\Omega}_{n_\ell} = \emptyset$. 

In this study, all MP operations are taken as the mean-aggregation MP layer from \cite{hamilton2018graphsage}. Pooling operation is performed using mean-pooling, while unpooling operation is implemented using $K-$nearest neighbors interpolation with $K=3$.

\subsection{Nonlinear manifold least-squares Petrov-Galerkin projection}\label{ssec:nmlspg}

A brief overview of the use of the nonlinear manifold LSPG in the GD-LSPG framework is presented in this section. To illustrate, the solution to \eqref{eq:ode1} is sought at each time step by minimizing the residual
\begin{equation} \label{eq:residual}
    \mathbf{r}: \left( \boldsymbol{\xi}; \boldsymbol{\mu} \right) \mapsto \alpha_0 \boldsymbol{\xi} + \sum_{i=1}^{{\tau}} \alpha_i \mathbf{x}^{m-i} + \beta_0 \mathbf{f}\left(\boldsymbol{\xi}, t; \boldsymbol{\mu}, \Delta t \right) + \sum_{i=1}^{\tau} \beta_i \mathbf{f}\left(\mathbf{x}^{m-i}, t; \boldsymbol{\mu}, \Delta t \right),
\end{equation}
where $\alpha_i, \beta_i \in \mathbb{R}$, $i=0,1,\ldots,\tau$, are constants defined by the chosen time integration scheme, and $\boldsymbol{\xi} \in \mathbb{R}^N$ denotes the unknown state vector. The solution at time step $m$ is sought by minimizing \eqref{eq:residual}, i.e.,
\begin{equation} \label{eq:ode2argmin}
    \mathbf{x}^{m} = \underset{\boldsymbol{\xi} \in \mathbb{R}^N}{\argmin} \Big\vert \Big\vert \mathbf{r} \left( \boldsymbol{\xi}; \boldsymbol{\mu} \right) \Big\vert \Big\vert _2, \quad m=1,\cdots,N_t,
\end{equation}
where $\boldsymbol{\xi}$ denotes the sought-after state solution that minimizes the residual at time step $m$, and the superscripts in \eqref{eq:residual} and \eqref{eq:ode2argmin} denote the time step.

To approximate the temporal evolution of latent variables, the nonlinear manifold LSPG framework first approximates the full-order state solution using the decoder for the graph autoencoder introduced in Section \ref{sec:problemFormulation} 
\begin{equation} \label{eq:approx}
    \mathbf{\tilde{x}}\left(t; \boldsymbol{\mu} \right) \approx \mathbf{Dec}\left(\mathbf{\hat{x}}\left(t ; \boldsymbol{\mu} \right)\right).
\end{equation}
Substituting \eqref{eq:approx} into \eqref{eq:residual} and projecting the residual onto a test basis, $\boldsymbol{\Psi}: \mathbb{R}^n \times \mathcal{D} \rightarrow \mathbb{R}^{N \times n}$, yields the low-dimensional residual
\begin{equation} \label{eq:lowDimRes}
    \mathbf{\hat{r}}: (\boldsymbol{\hat{\xi}}; \boldsymbol{\mu}) \mapsto \left(\mathbf{\Psi} \left( \boldsymbol{\hat{\xi}}; \boldsymbol{\mu}\right)\right)^\top \mathbf{r} \left(\mathbf{Dec}(\boldsymbol{\hat{\xi}}); \boldsymbol{\mu} \right),
\end{equation}
where $\hat{\boldsymbol{\xi}} \in \mathbb{R}^n$ denotes the sought-after low-dimensional state solution at time step $m$. Analogous to the FOM formulation, the low-dimensional state solution at each time step is determined by minimizing the low-dimensional residual
\begin{equation} \label{eq:res_min_full}
    \mathbf{\hat{x}}^{m} = \underset{\boldsymbol{\hat{\xi}} \in \mathbb{R}^n}{\argmin} \Big\vert \Big\vert (\mathbf{\Psi}(\boldsymbol{\hat{\xi}}; \boldsymbol{\mu}))^\top \mathbf{r} \left( \mathbf{Dec}\left(\boldsymbol{\hat{\xi}}\left(t ; \boldsymbol{\mu} \right)\right) \right) \Big\vert \Big\vert ^2_2, \quad m=1,\cdots,N_t.
\end{equation}

The choice of test basis, $\boldsymbol{\Psi}$, defines the projection scheme. In the GD-LSPG framework, a nonlinear manifold LSPG projection \cite{lee2020deeplspg} is employed, where the test basis is chosen as
\begin{equation} \label{eq:dlspg_psi}
    \mathbf{\Psi}: (\boldsymbol{\hat{\xi}}; \boldsymbol{\mu}) \mapsto \left(\frac{\partial \mathbf{r}}{\partial \mathbf{x}} \Big \vert_{\mathbf{Dec}\left(\boldsymbol{\hat{\xi}}\left(t ; \boldsymbol{\mu} \right)\right)} \right) \left( \frac{\mathrm{d}\mathbf{Dec}}{\mathrm{d}\boldsymbol{\hat{\xi}}} \Big \vert_{ \boldsymbol{\hat{\xi}}\left(t ; \boldsymbol{\mu}\right)} \right),
\end{equation}
where $\dfrac{\mathrm{d}\mathbf{Dec}}{\mathrm{d}\boldsymbol{\hat{\xi}}} \Big \vert_{ \boldsymbol{\hat{\xi}}\left(t ; \boldsymbol{\mu}\right)}$ is the Jacobian of the decoder. To solve \eqref{eq:res_min_full}, a Newton solver of the form
\begin{equation} \label{eq:gn_update}
    \left(\mathbf{\Psi}( \mathbf{\hat{x}}^{m(j)}; \boldsymbol{\mu})\right)^\top \mathbf{\Psi}( \mathbf{\hat{x}}^{m(j)}; \boldsymbol{\mu}) \left(\mathbf{\hat{x}}^{m(j+1)} - \mathbf{\hat{x}}^{m(j)}\right) = - \eta^{(j)} \left(\mathbf{\Psi} \left( \mathbf{\hat{x}}^{m(j)}; \boldsymbol{\mu}\right)\right)^\top \mathbf{r} \left(\mathbf{Dec}(\mathbf{\hat{x}}^{m(j)}); \boldsymbol{\mu} \right),
\end{equation}
is employed, where the superscript $m(j)$ denotes the $j^{\mathrm{th}}$ Newton iteration at time step $m$ and $\eta^{(j)} \in \mathbb{R}_+$ is the step length. The iterative solver is repeated until convergence is achieved according to
\begin{equation} \label{eq:convergence}
    \frac{\Big \vert \Big \vert \mathbf{\hat{r}}(\mathbf{\hat{x}}^{m(j)}; \boldsymbol{\mu}) \Big \vert \Big \vert_2}{\Big \vert \Big \vert \mathbf{\hat{r}}(\mathbf{\hat{x}}^{1(0)}; \boldsymbol{\mu}) \Big \vert \Big \vert_2} \leq \kappa,
\end{equation}
where $\kappa\in (0,1)$ is a user-defined tolerance parameter, the initial guess at each time step is taken as the converged solution from the previous time step, i.e., $\mathbf{\hat{x}}^{m(0)} = \mathbf{\hat{x}}^{m-1}$, and $\mathbf{\hat{x}}^{1(0)}$ denotes the initial guess at time step $m=1$ (i.e., the encoded initial conditions).

It is important to note that while \eqref{eq:gn_update} is a linear system of $n$ equations, computing its left- and right-hand sides have an operation count complexity that is dependent on the dimension of the FOM, $N$, specifically due to computing $\left(\mathbf{\Psi}( \mathbf{\hat{x}}^{m(j)}; \boldsymbol{\mu})\right)^\top \mathbf{\Psi}( \mathbf{\hat{x}}^{m(j)}; \boldsymbol{\mu})$ and $\left(\mathbf{\Psi} \left( \mathbf{\hat{x}}^{m(j)}; \boldsymbol{\mu}\right)\right)^\top \mathbf{r} \left(\mathbf{Dec}(\mathbf{\hat{x}}^{m(j)}); \boldsymbol{\mu} \right)$.

\section{Implementation details and graph autoencoder training} \label{sec:aeTraining}

This appendix provides the details for the FOM implementation, as well as the graph autoencoder architecture used for both numerical experiments, training procedures, and deployment within the GNN-LaSDI and GD-LSPG frameworks. The graph autoencoders are constructed using $n_\ell=5$ layers following the framework presented in \cite{magargal2025projection}. Consequently, the architectures consist of MPP layers, FC layers, and UMP layers. To generate the hierarchy of graphs for the autoencoders of both examples, an $\epsilon$-neighborhood graph is used to generate the edge set, $\mathcal{E}^i$, at each layer in the hierarchy, consistent with the authors' previous work \cite{magargal2025projection}. Specifically, the edge set at level $i$ is constructed using a radius $r^i= \sqrt{\frac{9}{\pi \vert \mathcal{V}^i\vert}}$. Table \ref{table:GNN} summarizes the graph autoencoder architectures and the number of tunable parameters for both the Kobayashi's solidification model, and the Euler equations to model bow shock generated by flow past a cylinder. All graph autoencoder models are trained using PyTorch \cite{paszke2019pytorch} and PyTorch-Geometric \cite{fey2019pyg} on an NVIDIA L40S GPU. The weights and biases of all models are initialized with Xavier initialization \cite{glorot2010initialization}. For both experiments, the loss function defined in \eqref{eq:loss} is minimized using the Adam optimizer \cite{kingma2014adam} with a learning rate of $10^{-4}$. The MP operation is implemented using the SAGEConv layer \cite{hamilton2018graphsage}. Exponential linear unit (ELU) \cite{clevert2016elu} activation functions are applied after the FC layer in the decoder and after each MP operation in both the encoder and the decoder, with the exception of the final MP operation in the decoder. In the experiment using Kobayashi's solidification model, no activation function is applied after the final MP operation in the decoder. Alternatively, the modified postprocessing layer deployed in the 2D Euler equations experiment applies a sigmoid activation function after the final MP operation in the decoder. 

The FOM implementation for Kobayashi's solidification model for 2D homogeneous phase evolution is executed on four Intel\textsuperscript{\textregistered} Xeon\textsuperscript{\textregistered} Platinum 8358 CPU @ 2.60GHz ICE LAKE CPU cores using NumPy \cite{harris2020numpy}. Since backward Euler time integration is employed, a Newton solver is used to minimize the high-dimensional residual. Sparse linear algebra routines from SciPy \cite{virtanen2020scipy} are used to efficiently solve the resulting linear system repeatedly. The FOM implementation for the Riemann solver for the 2D Euler equations is likewise executed on four Intel\textsuperscript{\textregistered} Xeon\textsuperscript{\textregistered} Platinum 8358 CPU @ 2.60GHz ICE LAKE CPU cores using NumPy \cite{harris2020numpy} together with Numba's just-in-time compiler \cite{lam2015numba}.

For the Kobayashi's solidification problem, the graph autoencoder is trained using a training/validation split consisting of $12500$ snapshots and $25$ validation snapshots obtained form the FOM. For the bow shock problem governed by the 2D Euler equations, the graph autoencoder is trained using $5000$ training snapshots, and $511$ validation snapshots. The snapshots are represented using the conserved variables ($\rho,\rho u,\rho v,$ and $\rho E$). In both experiments, the network is trained for $2500$ epochs using a batch size of $20$. To construct the POD basis for the POD-LaSDI ROM, a thin singular value decomposition (SVD) is performed on the snapshot matrix containing all FOM snapshots (i.e., $12525$ for the first experiment, and $5511$ for the second experiment).

The hyperparameters used for the GD-LSPG simulations are summarized next. For Kobayashi's solidification model, the Newton update in \eqref{eq:gn_update} uses a constant step length $\eta^j=1.0$ for all updates. For the bow shock problem, the Newton update \eqref{eq:gn_update} is computed using a step length of $\eta^j=1.0$, however, if convergence is not achieved, the step length is reduced by 90\% for every 10 iterations. Both experiments use the covergence tolerance $\kappa = 10^{-3}$ in \eqref{eq:convergence}.
 
Each ROM is executed using hardware chosen to optimize online computational performance. For GNN-LaSDI and POD-LaSDI, the temporal evolution of the latent space representation is computed using NumPy \cite{harris2020numpy} on four Intel\textsuperscript{\textregistered} Xeon\textsuperscript{\textregistered} Platinum 8358 CPU @ 2.60GHz ICE LAKE CPU cores. Alternatively, GD-LSPG computes the temporal evolution of the low-dimensional representation using the Newton solver defined by the nonlinear manifold LSPG projection. In this setting, evaluation of both the approximate high-dimensional state solution and the Jacobian of the decoder is performed on an NVIDIA L40S GPU using PyTorch \cite{paszke2019pytorch} and PyTorch-Geometric \cite{fey2019pyg}. The Newton updates in \eqref{eq:gn_update} are likewise computed on the GPU. The approximate high-dimensional state solution trajectories are obtained by decoding the latent state trajectories. For GNN-LaSDI and GD-LSPG, decoding is performed on the NVIDIA L40S GPU using PyTorch \cite{paszke2019pytorch} and PyTorch-Geometric \cite{fey2019pyg}. In contrast, decoding of POD-LaSDI latent state solutions is performed on four Intel\textsuperscript{\textregistered} Xeon\textsuperscript{\textregistered} Platinum 8358 CPU @ 2.60GHz ICE LAKE CPU cores.

\begin{table}[h]
\begin{center}
\caption{Graph autoencoder architectures used for the Kobayashi's solidification example and the bow shock experiment. Here, $i$ denotes the layer number, and $\vert \mathcal{V}^i \vert$ represents the number of nodes in the $i^{\mathrm{th}}$ graph in the hierarchy. $N_F^i$ denotes the number of features for each node in graph $i$, and $n$ is the latent space dimension. Note that parentheses in the ``\# of MP operations'' column indicate the number of features in the output of the intermediate MP operations (refer to \cite{magargal2025projection} for more detail), and the column labeled ``vector length'' denotes the vector length of the output vector of layers, wherein N/A indicates layers that do not produce a single vector output.}
\begingroup
\footnotesize
\begin{tabular}{*{9}c}
    \toprule
    & & $i$ & layer type & $\vert \mathcal{V}^i \vert$ & $N_F^i$ & \# of MP operations & vector length & \# of tunable parameters\\
    \midrule
    \multirow{12}{0em}{{{\rotatebox[origin=lb]{90}{Kobayashi's solidification model}}}} & \multirow{6}{0em}{{\rotatebox[origin=lb]{90}{encoder}}} & 0 & preprocessing & 25600 & 2 & N/A & N/A & N/A\\
    & & 1 & MPP & 512 & 16 & 2 (16) & N/A & 576 \\
    & & 2 & MPP & 64 & 64 & 2 (64) & N/A & 10240 \\
    & & 3 & MPP & 8 & 128 & 2 (128) & N/A & 49152 \\
    & & 4 & MPP & 2 & 256 & 2 (256) & N/A & 196608 \\
    & & 5 & $\mathrm{FC}$ & N/A & N/A & N/A &  $n$ & 512$n$ \\
    \cmidrule(l){2-9}
    & \multirow{6}{0em}{\rotatebox[origin=lb]{90}{decoder}} &  0 & $\mathrm{FC}$ & 2 & 256 & N/A & N/A & 512$n$\\
    & & 1 & UMP & 8 & 128 & 2 (256) & N/A & 196608 \\
    & & 2 & UMP & 64 & 64 & 2 (128) & N/A & 49152 \\
    & & 3 & UMP & 512 & 16 & 2 (64) & N/A & 10240 \\
    & & 4 & UMP & 25600 & 2 & 2 (16) & N/A & 576 \\
    & & 5 & postprocessing & N/A & N/A & N/A & 25600 $\times$ 2 & N/A \\  
    \bottomrule
    \multirow{12}{0em}{\rotatebox[origin=lb]{90}{bow shock problem}} & \multirow{6}{0em}{{\rotatebox[origin=lb]{90}{encoder}}} & 0 & preprocessing & 4148 & 4 & N/A & N/A & N/A\\
    & & 1 & MPP & 512 & 16 & 2 (16) & N/A & 640 \\
    & & 2 & MPP & 64 & 64 & 2 (64) & N/A & 10240 \\
    & & 3 & MPP & 8 & 128 & 2 (128) & N/A & 49152 \\
    & & 4 & MPP & 2 & 256 & 2 (256) & N/A & 196608 \\
    & & 5 & $\mathrm{FC}$ & N/A & N/A & N/A &  $n$ & 512$n$ \\
    \cmidrule(l){2-9}
    & \multirow{6}{0em}{\rotatebox[origin=lb]{90}{decoder}} & 0 & $\mathrm{FC}$ & 2 & 256 & N/A & N/A & 512$n$\\
    & & 1 & UMP & 8 & 128 & 2 (256) & N/A & 196608 \\
    & & 2 & UMP & 64 & 64 & 2 (128) & N/A & 49152 \\
    & & 3 & UMP & 512 & 16 & 2 (64) & N/A & 10240 \\
    & & 4 & UMP & 4148 & 4 & 2 (16) & N/A & 640 \\
    & & 5 & postprocessing & N/A & N/A & N/A & 4148 $\times$ 4 & N/A \\
    \bottomrule
\end{tabular}
\endgroup
\label{table:GNN}
\end{center}
\end{table}

\end{document}